\documentclass[10pt,twocolumn,preprintnumbers,amsmath,amssymb,nofootinbib
,superscriptaddress,aps,prd,usenatbib]{revtex4} 
\usepackage{graphicx,amsmath,amssymb,amstext}
\usepackage{amsbsy,amsfonts,amsthm,color}
\usepackage{xcolor}
\usepackage[colorlinks,linkcolor=red,citecolor=blue,urlcolor=blue ]{hyperref}
\usepackage{xspace}
\usepackage{mathptmx} 

\newcommand{\hmcode}{\textsc{hmcode}\xspace}
\newcommand{\eg}{e.g.,\xspace}

\def\TX{$T_\mathrm{X}$}

\newcommand{\Mpch}{\,h^{-1}\mathrm{Mpc}}
\newcommand{\iMpch}{\,h\,\mathrm{Mpc}^{-1}}

\def \prd {{\it Phys. Rev. D}}

\def \apj {{\it ApJ}}
\def \apjl {{\it ApJ Let.}}
\def \apjs {{\it ApJ Sup.}}

\def \aap {{\it A\&A}}

\def \mnras {{\it MNRAS}}
\def \nat {{\it Nature}}
\def \procspie {{\it Proc.~SPIE}}

\def\jcap{{\it JCAP}}
\def\physrep{{\it Phys.~Rep.}}
\def\ssr{{\it Space Science Rev.}}

\begin{document}

\title{Modelling baryonic feedback for survey cosmology}

\author{Nora Elisa Chisari}
\affiliation{Sub-department of Astrophysics, Denys Wilkinson Building, Keble Road, Oxford OX1 3RH, UK}
\email{elisa.chisari@physics.ox.ac.uk}
\author{Alexander J. Mead}
\affiliation{Department of Physics and Astronomy, University of British Columbia, 6224 Agricultural Road, Vancouver, BC V6T 1Z1, Canada}
\affiliation{Institut de Ci\`encies del Cosmos, Universitat de Barcelona, Mart\'i Franqu\`es 1, E08028 Barcelona, Spain}
\author{Shahab Joudaki}
\affiliation{Sub-department of Astrophysics, Denys Wilkinson Building, Keble Road, Oxford OX1 3RH, UK}
\author{Pedro Ferreira}
\affiliation{Sub-department of Astrophysics, Denys Wilkinson Building, Keble Road, Oxford OX1 3RH, UK}
\author{Aurel Schneider}
\affiliation{Institute for Particle Physics and Astrophysics, ETH Zurich, Wolfgang Pauli Strasse 27, 8093 Zurich, Switzerland}
\author{Joseph Mohr}
\affiliation{Faculty of Physics, Ludwig-Maximilians-Universitaet, Scheinerstrasse 1, 81679 Munich, Germany}
\affiliation{Max Planck Institute for Extraterrestrial Physics, Giessenbachstrasse, 85748, Garching, Germany}
\affiliation{Excellence Cluster Origins, Boltzmannstrasse 2, 85748, Garching, Germany}
\author{Tilman Tr\"oster}
\affiliation{Institute for Astronomy, University of Edinburgh, Royal Observatory, Blackford Hill, Edinburgh, EH9 3HJ, UK}
\author{David Alonso}
\affiliation{Sub-department of Astrophysics, Denys Wilkinson Building, Keble Road, Oxford OX1 3RH, UK}
\author{Ian G. McCarthy}
\affiliation{Astrophysics Research Institute, Liverpool John Moores University, 146 Brownlow Hill, Liverpool L3 5RF, UK}
\author{Sergio Martin-Alvarez}
\affiliation{Sub-department of Astrophysics, Denys Wilkinson Building, Keble Road, Oxford OX1 3RH, UK}
\author{Julien Devriendt}
\affiliation{Sub-department of Astrophysics, Denys Wilkinson Building, Keble Road, Oxford OX1 3RH, UK}
\author{Adrianne Slyz}
\affiliation{Sub-department of Astrophysics, Denys Wilkinson Building, Keble Road, Oxford OX1 3RH, UK}
\author{Marcel P. van Daalen}
\affiliation{Institute for Astronomy, University of Edinburgh, Royal Observatory, Blackford Hill, Edinburgh, EH9 3HJ, UK}
\affiliation{Leiden Observatory, Leiden University, P.O. Box 9513, 2300 RA Leiden, The Netherlands}

\begin{abstract}
Observational cosmology in the next decade will rely on probes of the distribution of matter in the redshift range between $0<z<3$ to elucidate the nature of dark matter and dark energy. In this redshift range, galaxy formation is known to have a significant impact on observables such as two-point correlations of galaxy shapes and positions, altering their amplitude and scale dependence beyond the expected statistical uncertainty of upcoming experiments at separations under $10$ Mpc. Successful extraction of information in such a regime thus requires, at the very least, unbiased models for the impact of galaxy formation on the matter distribution, and can benefit from complementary observational priors. This work reviews the current state of the art in the modelling of baryons for cosmology, from numerical methods to approximate analytical prescriptions, and makes recommendations for studies in the next decade, including a discussion of potential probe combinations that can help constrain the role of baryons in cosmological studies. We focus, in particular, on the modelling of the matter power spectrum, $P(k,z)$, as a function of scale and redshift, and of the observables derived from this quantity. This work is the result of a workshop held at the University of Oxford in November of 2018.
\end{abstract}

\maketitle


\section{Introduction} 

Over the past two decades, observational cosmology has become an effective tool for learning about the past, present and fate of the Universe. From the analysis of the cosmic microwave background (CMB), and measurements of the large-scale structure (LSS) of gas and galaxies, it has been possible to constrain the density and expansion rate of the Universe as well key aspects of its initial conditions \cite{Aghanim:2018eyx}.

In the next decade, ambitious campaigns will aim to pin down the source of the accelerated expansion of the Universe -- whether it is a cosmological constant, an exotic source of dark energy or modifications to general relativity -- as well as the total mass and structure of the neutrino sector -- whether there is a normal- or inverted-mass hierarchy -- and the detailed statistics of the initial seeds of structure formation -- whether they are Gaussian or not, or if these properties change unexpectedly with scale. To be able to tackle these challenges, the scientific community will undertake far more precise and accurate measurements of the large-scale distribution of matter in the Universe \cite{Weinberg13} by means of the next generation of optical and infrared surveys of the sky, such as the Large Synoptic Survey Telescope \cite{Ivezic08}, {\it Euclid} \cite{Laureijs11} or {\it WFIRST} \cite{Spergel13}. These will use, among others, the combination of the clustering and gravitational lensing of galaxies as cosmological probes. 

To increase statistical power, it is necessary to both consider larger volumes of distribution of matter and also to be able to model their details with finer precision, i.e., to smaller wavelengths, $\lambda_{\rm max}$. The enhancement in statistical power from being able to measure smaller physical scales (and thus increasing the number of independent modes which can be used) is counteracted by the fact that one is now probing the non-linear regime of gravitational clustering, where astrophysical systematics inevitably affect theoretical predictions. A number of methods have been developed for calculating gravitational clustering in the non-linear regime, from perturbation theory \cite{Bernardeau:2001qr} and phenomenological models \cite{Cooray:2002dia,Baumann:2010tm} to high-resolution numerical simulations \cite{Agertz:2006qb}. 

A crucial and unavoidable improvement required in the modelling of the large-scale structure is the description of the impact of galaxy formation on the distribution of matter \cite{vanDaalen11}. For forthcoming surveys, it is no longer sufficient to rely on models of the Universe as composed of solely gravitationally interacting particles. Processes that heat and cool gas, re-distribute it or transform it into stars, have to be included in the models. Most importantly, this generation of cosmological studies has flagged the need to better model the impact of feedback from Active Galactic Nuclei (AGN) on the distribution of matter at the scales and for the accuracy level needed for future studies.

How well will we need to model the non-linear regime? The goal is to measure the equation of state of dark energy with a precision of $1\%$ \cite{Slosar19}, the sum of neutrino masses with a precision of order $15$ meV \cite{Allison15}, or the degree of non-Gaussianity, $f_{\rm NL}$, to order $O(1)$, to rule out alternatives to inflation \cite{Alvarez14,Meerburg19}. 
The current focus in the analysis of cosmological observables is on two-point statistics. Specifically, if we define  $\delta_{\rm M}(t,{\bf x})$ to be the inhomogeneous density contrast in matter at a given time, its three-dimensional power spectrum, $P(k)$, is defined via
\begin{eqnarray}
\langle\delta^*_{\rm M}({\bf k}')\delta_{\rm M}({\bf k})\rangle = (2\pi)^3P(k)\delta_D^{(3)}({\bf k}'-{\bf k})
\end{eqnarray}
where $\langle \cdots \rangle$ indicates an ensemble average and $\delta^{(3)}$ is the three-dimensional Dirac delta function. Previous works have indicated that, at the statistical level expected from future experiments, $P(k)$ needs to be accurately modelled to a precision such that the uncertainty in the power spectrum, $\Delta P$, is constrained to being $\sim~1\%$ out to scales of $k_{\rm max}=2\pi/\lambda_{\rm max}\simeq 10\iMpch$ or even larger \cite{Huterer05,Hearin12}. 

\begin{figure}
  	\centering
	\includegraphics[width=\columnwidth]{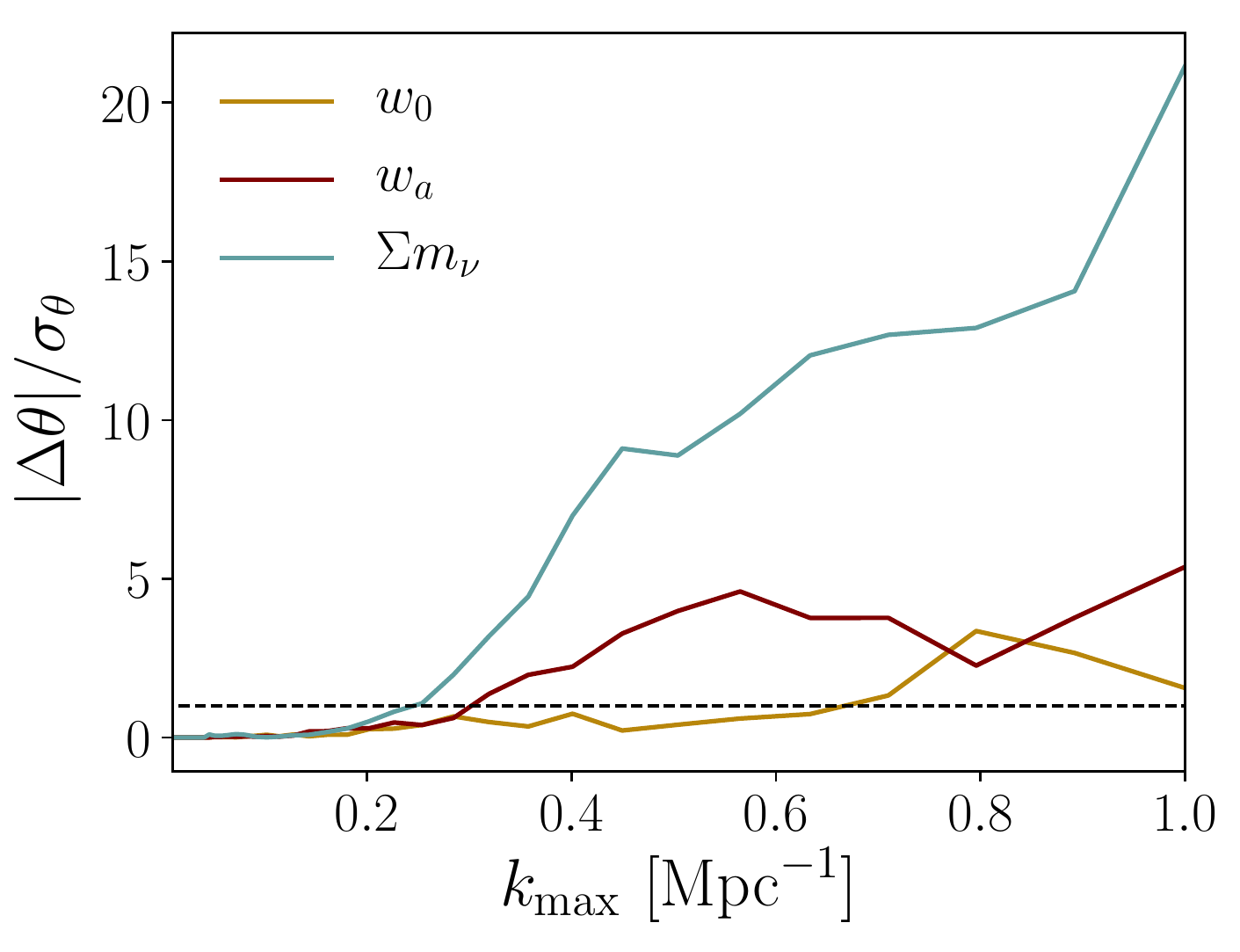}
	\caption{Expected bias in the parameters of the equation of state of dark energy, $w=w_0+w_a(1-a)$ (red and yellow), where $a$ is the scale factor, and the sum of neutrino masses, $\Sigma m_\nu$ (blue), relative to the statistical uncertainty in each parameter that would result from ignoring the impact of baryons on the distribution of matter. The figure assumes the information is directly extracted from the matter power spectrum up to a given maximum wavenumber $k_{\rm max}$ in the redshift range $0<z<3$, and that the impact of baryons is given by the upper limit of the $1\sigma$ scatter of predictions available from different cosmological simulations \cite{Chisari18}. The dashed line indicates where the bias is equal to the $1\sigma$ statistical uncertainty.}
	\label{fig:bias}
\end{figure}

The uncertainty in the impact of baryonic processes on $P(k)$ greatly exceeds the $1\%$ threshold mentioned above \cite{vanDaalen11,Chisari18} and can lead to significant biases in cosmological parameters if unaccounted for \cite{Semboloni11,Eifler15,Huang19}. Figure \ref{fig:bias} illustrates this statement by showing the expected bias in some cosmological parameters of interest to forthcoming experiments (those parameterising the equation of state of dark energy \cite{Chevallier01,Linder03}, and the sum of neutrino masses) relative to the statistical uncertainty as a function of $k_{\rm max}$. This figure assumes the information is directly extracted from the matter power spectrum, evaluated in the expected median redshift of future optical surveys ($0<z<3$). The results from a simple Fisher forecast indicate that even at $k_{\rm max}=0.3\,{\rm Mpc}^{-1}$, the bias on the derived cosmological parameters is significant. For this exercise, the impact of baryons is assumed to be at most consistent with the $1\sigma$ scatter among predictions from existing cosmological hydrodynamical simulations \cite{Chisari18}.

Multiple strategies have been proposed to account for this astrophysical phenomenon. One possibility is to adopt an effective analytical parametrisation of the effect on the matter power spectrum \cite{Mead15,Harnois15,Schneider15} that can be marginalized over \cite[e.g., as in][]{Hildebrandt19}. Another option is to build a model of the Universe where matter resides in haloes, and these have three components: gas, stars and dark matter, with their own abundances and density profile - a ``baryonic halo model" \cite{Semboloni11,Semboloni13,Fedeli2014a}. Such a model is typically calibrated to cosmological hydrodynamical simulations. In the case of \cite{Semboloni13}, this was the first time that several observables (two- and three-point functions) were combined to break the degeneracy between baryons and cosmology, and to show that it is possible to obtain constraints on galaxy formation (i.e., gas fractions) and cosmology simultaneously given a physical model. A third approach proposes to model the Universe by modifying the profiles of dark matter haloes in cosmological $N$-body simulations with a prescription to displace particles based on observational constraints of the distributions and abundances of gas and stars in haloes \cite{Schneider15,Schneider18}. This approach can also be tested self-consistently in hydrodynamical simulations. Finally, the authors of \cite{Eifler15} have proposed the use of a Principal Component Analysis (PCA) method to identify the modes of the data vector most sensitive to baryonic effects, and to marginalize over them.

Our aim with this work is to summarize the current status and make suggestions for future progress in the modelling of baryonic effects on the distribution of matter, with the specific goal of supporting the effort of obtaining cosmological constraints from weak gravitational lensing of the large-scale structure in the next decade. We review the current status of the field (Section \ref{sec:current}) and make recommendations for the community moving forward to the next decade of survey cosmology (Section \ref{sec:next}). We conclude in Section \ref{sec:conclude}.

\section{Current status} 
\label{sec:current}

\begin{figure}
  	\centering
	\includegraphics[width=\columnwidth]{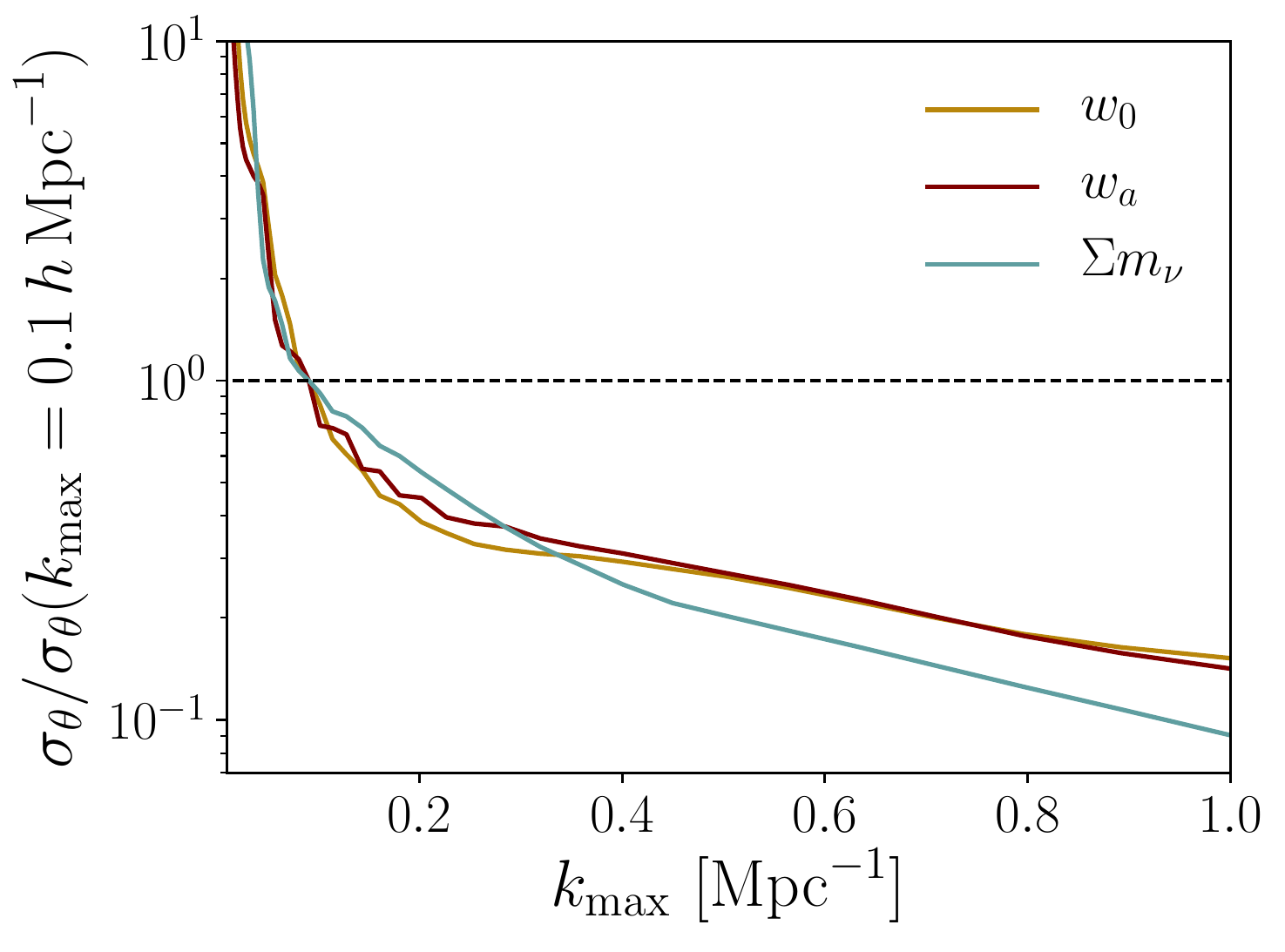}
	\caption{Expected fractional gain in constraining power of the parameters of the equation of state of dark energy, $w=w_0+w_a(1-a)$ (red and yellow), and the sum of neutrino masses, $\Sigma m_\nu$ (blue), as a function of maximum wavenumber $k_{\rm max}$ included in the analysis. The results were produced by a Fisher forecast in the same set-up as for Figure \ref{fig:bias}, where the information is extracted directly from the matter power spectrum. The uncertainty in each parameter is normalized relative to the $k_{\rm max}=0.1\iMpch$ case. The results indicate there is a significant amount of cosmological information to be gained from going to non-linear scales.}
	\label{fig:info}
\end{figure}

In analyses of weak-gravitational-lensing data, two distinct approaches to account for baryonic feedback have been considered to date. The first approach considers stringent cuts to the scales included in the lensing analysis, such that the impact of baryonic feedback on the remaining scales are negligible (e.g.~\cite{Heymans13,Planck15,Planck15mg,Troxel18}). While this approach avoids biases to the parameter constraints, it inevitably discards a large amount of cosmological information that exists on smaller scales. As shown in Figure \ref{fig:info}, this is particularly important in future attempts to constrain the sum of neutrino masses with weak lensing. The second approach includes non-linear scales in the analysis and instead attempts to model the impact of baryonic feedback on those scales (e.g.~\cite{MacCrann15,Harnois15,Joudaki17a}). While this approach in principle allows for cosmological information to be extracted on smaller scales, in practice this depends on the size of the baryonic feedback prior used in the analysis.

Notably, the effects of baryonic feedback on the non-linear matter power spectrum have been captured by \hmcode~(\cite{Mead15}; see Sec.~\ref{sec:halo_model} for a detailed description), through calibration of the halo model to the OverWhelmingly Large hydrodynamical Simulations (OWLS, \cite{Schaye10,vanDaalen11}). In the prescription of \cite{Mead15}, uncertainties in the modelling of the baryonic feedback are quantified with either one or two free parameters, which are related to the effect of baryons on halo profiles. These parameters are allowed to vary freely in cosmological analyses and have been propagated into analyses of weak gravitational lensing data from the Canada-France-Hawaii Telescope Lensing Survey (CFHTLenS, \cite{Joudaki17a}), the Kilo-Degree Survey  (KiDS, \cite{hildebrandt17, Hildebrandt19, J17b, Joudaki18, vanUitert18}), the Deep Lens Survey (DLS, \cite{Yoon19}), and the Subaru Hyper Suprime-Cam survey (HSC, \cite{hikage19}).  As an alternative to the \hmcode approach, in some analyses (e.g.~\cite{MacCrann15,Harnois15,Kohlinger16,Kohlinger17}) the nonlinear matter power spectrum is multiplicatively modified by $A\times P_\mathrm{hydro}(k)/P_\mathrm{DM}(k)$, where $P_\mathrm{hydro}(k)$ is the power spectrum measured in some hydrodynamical simulation, $P_\mathrm{DM}(k)$ is the power measured in a gravity-only simulation starting from the same initial conditions as the hydrodynamical simulation (thus the ratio smoothly tends to unity at small $k$) and $A$ is a free amplitude parameter.

As the feedback parameters can have similar effects on the matter power spectrum to changes in standard cosmological parameters, allowing them to be free in an analysis has a noticeable, degrading impact on the parameter constraints (compared to what would be achieved if feedback is tightly constrained in advance). In particular, they affect constraints on $S_8 = \sigma_8 \sqrt{\Omega_{\rm m}/0.3}$, both by shifting the posterior mean and increasing the uncertainty. As current lensing surveys primarily measure a correlation function amplitude, increasing the magnitude of feedback in a model suppresses this amplitude and allows data to be fit with a higher value of $S_8$. This picture becomes more complicated for future surveys as the shape of the correlation function becomes more constraining and the characteristic scale dependence of feedback can be disentangled from that induced by the primary cosmological parameters. It might be possible to also use the redshift evolution of the impact of baryons in favor of breaking the degeneracy with the growth of structure. However, there is little consensus at the moment on the expected redshift trend of the effect \citep{Chisari18}.

As a concrete example, if the amplitude of baryonic feedback is allowed to vary between the best-fit values to simulations with no feedback and strong AGN feedback, the uncertainty on the $S_8$ constraint from the combination of the KiDS and VIKING data for weak lensing degrades by $12\%$ and the posterior mean shifts to larger values by $0.5\sigma$ \cite{Hildebrandt19}. Naturally, the constraint on $S_8$ degrades further when allowing for a wider prior range on the feedback amplitude \cite{J17b,Joudaki18,Hildebrandt19}. Allowing for an uncertainty in the `bloating' of haloes in addition to the feedback amplitude (see Sec.~\ref{sec:halo_model}), with prior ranges on the two parameters that capture the full range of values favored by multiple simulation suites \cite{Huang19}, the $S_8$ uncertainty increases by another $15\%$ with a $<0.1\sigma$ shift in the posterior mean \cite{Hildebrandt19}. In other words, while allowing for more than one feedback parameter can be necessary for future surveys, it does not seem to significantly impact current survey results.

Current cosmic shear data have not been sufficiently powerful to provide a firm detection of baryonic feedback. Notably, the combined ${\rm KiDS \times (2dFLenS + BOSS)}$ analysis of cosmic shear, galaxy-galaxy lensing, and redshift-space galaxy clustering yielded a preference away from `no feedback' at 95\% confidence level \cite{Joudaki18}. Although this preference is not statistically significant, the posterior was found to peak at values corresponding to strong feedback. Reference \cite{Yoon19} similarly suggested a preference for strong feedback, beyond that given by the OWLS AGN case, from the combination of galaxy clustering and galaxy-galaxy lensing by the Deep Lens Survey (DLS) together with the cosmic microwave background from {\it Planck}. 

As the constraining power of weak lensing surveys increases, so does their sensitivity to physics on non-linear scales, in particular from baryonic feedback, modified gravity, and neutrino mass. As a result, the importance of accurate modelling of baryonic feedback increases given the otherwise adverse impact on the cosmological constraining power of wide and deep lensing surveys. However, this also implies that lensing surveys will have a greater ability to detect baryonic feedback and distinguish between different astrophysical scenarios \cite{Foreman16}. In \cite{Mead16}, the non-linear impact of baryonic feedback, neutrino mass, and modified gravity on the matter power spectrum were considered within the same halo model framework. Although it was found that baryonic feedback can be treated independently of massive neutrinos (see also \cite{Mummery}) and modified gravity (however, the latter two cannot be treated independently relative to one another, with biases at the level of 5\% for $k>0.5\iMpch$), their effects are correlated and the impact of baryonic feedback needs to be determined at sub-per cent precision to avoid biases in future constraints on the concordance model. 

\subsection{Simulations} 
\label{sec:sims}

The growth of structure is a highly non-linear problem that is most accurately tackled with direct numerical $N$-body(+hydrodynamics) simulations.  To be useful for survey cosmology, large volumes (typically with $L_{\rm box} \gtrsim 100$ comoving Mpc) must be simulated.  However, because feedback processes associated with galaxy and black hole formation can significantly influence the large-scale distribution of matter, relatively high resolution is also required to capture these effects which originate on small scales.  Thus, there is a severe dynamic range problem.  It is only within the past few years that simulations have reached sufficient box size while also having resolutions reasonable for modelling galaxy formation, albeit through ``sub-grid" modelling prescription for astrophysical processes below the resolution scale. Typical sub-grid processes that are included in such simulations include: star formation, stellar feedback and winds, black hole formation and merger, gas cooling, black hole accretion and active galactic nuclei (AGN) feedback (kinetic and thermal), for example. For the purpose of this review, we are mainly interested in the interplay between star formation and AGN feedback, which are the main processes that contribute to affecting the large-scale distribution of matter and its power spectrum \cite{vanDaalen11}.

Below we provide a brief description of the current state-of-the-art in large-scale-structure simulations, including how the simulations are initialized and how the feedback processes included are calibrated. We focus on the simulations listed in Table \ref{tab:sims}, which through different studies over the past few years have been used to explore the impact of baryons on the distribution of matter. Details on the calibration of each simulation are provided in the following sub-section.

\begin{itemize}
\item The Horizon suite is a set of three adaptive-mesh-refinement (AMR) simulations of a cubic volume $100\Mpch$ on a side featuring the same initial conditions and run using the {\sc Ramses} code \cite{Teyssier02}. It consists of a full hydrodynamical simulation with AGN feedback, an identical run without AGN feedback and a third $N$-body  counterpart which only models the dark matter component. The simulation suite is described in \cite{Dubois14} and \cite{Peirani17} and the quantification of the impact of baryons on the power spectrum from these runs was performed in \cite{Chisari18}. A lightcone constructed from this suite has been presented in \cite{Gouin19}, where the impact of baryons on weak lensing observables at small scales was quantified. Other connected works have explored the role of AGN in galaxy quenching \cite{Beckmann17} and on the density profile of haloes \cite{Peirani17}. Galaxy luminosity functions \cite{Kaviraj17} and morphologies \cite{Dubois16} predicted by Horizon-AGN are in reasonable agreement with observations. 
\item The MassiveBlack-II simulation \cite{Khandai15} is a cosmological simulation of the same volume as the Horizon-AGN run using a modified version of the GADGET\footnote{\url{http://www.mpa-garching.mpg.de/gadget/}} smoothed-particle-hydrodynamics (SPH) code \cite{Springel05}. It adopts the same sub-grid model as the original MassiveBlack simulation \cite{DiMatteo12}, including star formation, black hole growth and their associated feedback mechanisms. A control $N$-body run without baryons is available \cite{Tenneti15} at the same resolution and with the same initial conditions and cosmological parameters. MassiveBlack-II is successful in predicting a range of different observables (e.g., galaxy clustering and halo occupation distribution statistics), but shows evidence of insufficient quenching of star formation in massive galaxies.
\item The OverWhelmingly Large Simulations (OWLS) \cite{Schaye10} is a large suite, comprised of fifty simulations, used to explore the sensitivity of the cosmic star formation history to the calibration of sub-grid parameters that govern the evolution of this observable. The suite of boxes of different size and resolution was run using the  Tree-PM SPH code GADGET \cite{Springel05}.
\item cosmo-OWLS \cite{LeBrun14,McCarthy14} is a suite of simulations crafted to extend the OWLS project for cosmological applications. With boxes of $400\Mpch$ on each side, cosmo-OWLS provides a larger cosmological volume to explore the role of baryons in affecting cluster astrophysics and non-linear structure formation. These simulations were also updated to incorporate {\it WMAP}7 and {\it Planck}~2013 cosmologies, as opposed to {\it WMAP}3 used in OWLS.
\item The Virgo Consortium's EAGLE project \cite[][Evolution and Assembly of GaLaxies and their Environments]{Schaye15} is a suite of cosmological SPH simulations run using GADGET 3 and spanning box lengths from $25$ to $100\Mpch$. A range of resolutions (to zoom into individual galaxies or groups), multiple numerical techniques and sub-grid models are available. The sub-grid physics model was based on OWLS and cosmo-OWLS, but there are significant changes in the implementations of the star formation law, the energy feedback from star formation and the accretion of gas onto black holes. High resolution runs informed the fiducial model, which was chosen as the one that best fit the $z\sim 0$ GSMF from among all models which produced reasonable physical sizes for disc galaxies.
\item BAHAMAS \cite[BAryons and haloes of MAssive Systems][]{McCarthy17} follows on the OWLS and cosmo-OWLS sub-grid model to build a larger suite of ($400\Mpch$)$^3$ boxes in which the AGN feedback parameter is varied to study its impact on massive dark matter haloes. The most important aspect of this suite is the calibration of such parameter to existing observations, a procedure that we detail in the following sub-section.
\item The Illustris simulation\footnote{\url{http://www.illustris-project.org}} \cite{Vogelsberger14,Vogelsberger14b,Genel14} was the first cosmological simulation run using the moving-mesh code AREPO \cite{Springel10}. It consists of a set of cosmological boxes of $75\Mpch$ on a side run to $z=0$. Three of the simulations share the same and most complete sub-grid physics model \cite{Vogelsberger13} at different resolutions, and additional runs in a dark-matter-only and non-radiative (adiabatic) scenarios are provided for comparison.
\item The Next Generation Illustris simulation (IllustrisTNG)\footnote{\url{http://www.tng-project.org}}, similarly run with AREPO, comprises three tiers of simulations boxes (of $300$, $100$ and $50\Mpch$ on each side) at different resolutions. Compared to Illustris, it has seen developments in the sub-grid model \cite{Weinberger17,Pillepich18}, specifically in the treatment of kinetic AGN feedback, galactic winds and magnetic fields, as well as improvements in numerical implementation towards flexibility and better hydrodynamical convergence. 
\end{itemize}

\begin{table*}
  \centering
  \begin{tabular}{ l|c c c c}
    \hline
    Simulation & hydrodynamic scheme & Box length & DM mass resolution & Cosmology \\
    \hline
    Horizon-AGN & AMR ({\tt RAMSES}) & 100 $h^{-1}$ Mpc &  $8\times 10^7$ M$_\odot$ & {\it WMAP}7 \\
    MassiveBlack-II & SPH ({\tt GADGET}) & 100 $h^{-1}$ Mpc & $1.1\times 10^7$ $h^{-1}$ M$_\odot$ & {\it WMAP}7\\
    OWLS(*) & SPH ({\tt GADGET}) & 25-100 $h^{-1}$ Mpc & $5.1\times 10^7$-$2.6\times 10^{10}$ $h^{-1}$ M$_\odot$ & {\it WMAP}3\\
    cosmo-OWLS & SPH ({\tt GADGET}) & 400 $h^{-1}$ Mpc & $\simeq 4\times 10^{9}$ $h^{-1}$ M$_\odot$ & {\it WMAP}7 \& {\it Planck} 2013\\
    EAGLE & SPH ({\tt GADGET}) & $25-100$ Mpc &  $1.21\times 10^6$-$9.7\times 10^{6}$ M$_\odot$ & {\it Planck} 2013\\
    BAHAMAS & SPH ({\tt GADGET}) &  400 $h^{-1}$ Mpc  & $\simeq 4\times 10^{9}$ $h^{-1}$ M$_\odot$ & {\it WMAP}9 \&  {\it Planck} 2013\\
    Illustris & Moving Mesh ({\tt AREPO}) & $75$ $h^{-1}$ Mpc & $6.26\times 10^6$ $h^{-1}$ M$_\odot$ &  {\it WMAP}9\\
    IllustrisTNG & Moving Mesh ({\tt AREPO}) & $35-205$ $h^{-1}$ Mpc & $4.5\times 10^5$-$5.9\times 10^{7}$ M$_\odot$ & {\it Planck} 2016\\
    \hline
  \end{tabular}
  \caption{Table of cosmological hydrodynamical simulations for which constraints on the impact of baryons on $P(k)$ are available. (*) Only simulations run to $z=0$ are listed.}
  \label{tab:sims}
\end{table*}

\begin{figure}
  	\centering
	\includegraphics[width=\columnwidth]{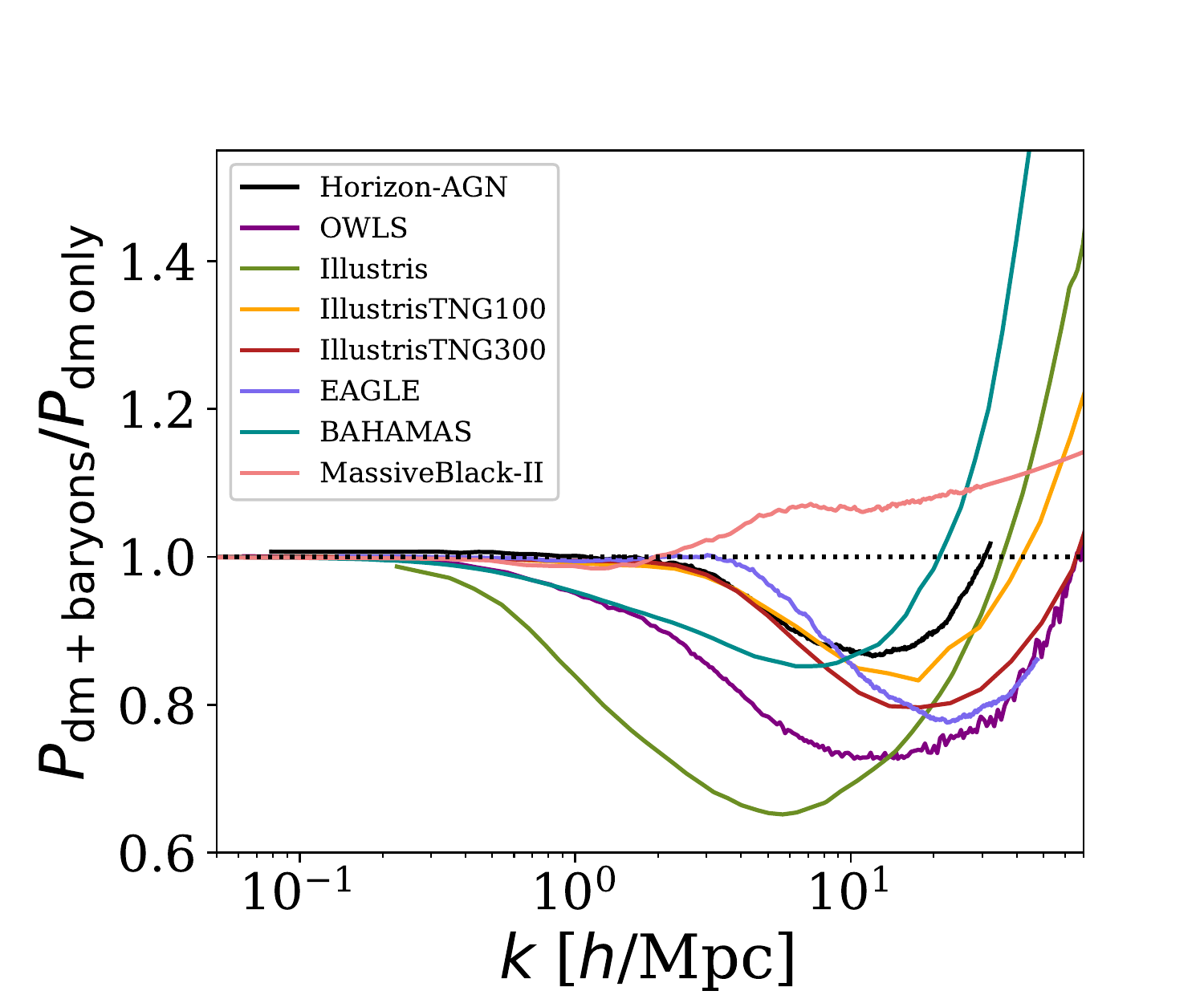}
	\caption{Fractional impact of baryons on the matter power spectrum at $z=0$ for all the simulations described in Section \ref{sec:sims} from which this quantity is available. The curves are collected from \cite{Chisari18}, \cite{Huang19} and Marcel van Daalen (private communication). The small scale upturn is representative of star formation and gas cooling, while the suppression at scale of a few $h$ Mpc$^{-1}$ is due to feedback redistributing gas and dark matter in the simulation.}
	\label{fig:allpow}
\end{figure}

A comparison of the impact of baryons on the matter power spectrum from the different simulations was performed by \cite{Chisari18} and \cite{Huang19}. At $z=0$, the simulation predictions differ on the amplitude and scale-dependence of the impact of baryons on the power spectrum, oscillating between a $10-30\%$ suppression of power at wavenumbers between a few and $\sim 20\,h$ Mpc$^{-1}$, as shown in Figure \ref{fig:allpow}. Such differences can be attributed to multiple factors. The choice of sub-grid model, resolution and the calibration strategy play a decisive role. The results concerning the impact of baryons on the matter power spectra for the simulations suites shown in Figure \ref{fig:allpow} were first presented in \cite[][OWLS]{vanDaalen11}, \cite[][Illustris]{Vogelsberger14}, \cite[][EAGLE]{Hellwing16}, \cite[][IllustrisTNG]{Springel18}, \cite[][Horizon-AGN]{Chisari18} and \cite[][MassiveBlack-II]{Huang19}. 

On the other hand, the hydrodynamical scheme (particle vs. grid) can also have an impact in the results. Reference \cite{Chisari18} found evidence of a difference in the distribution of matter at high redshift between OWLS and Horizon. At $z\simeq 5$, the impact of AGN and supernova feedback is negligible, and this difference was thus attributed to the numerical scheme. Most simulations neglect the impact of neutrinos in the large-scale distribution of matter. The exception is BAHAMAS, where a subset of runs now includes massive neutrinos \cite{McCarthy18}.  Using these runs, reference \cite{Mummery} find that the impact of AGN and massive neutrinos in the matter power spectrum is separable to $\approx$1-2 per cent accuracy in various statistics, including $P(k)$. They caution, however, that they have only verified this conclusion for a relatively small range of cosmologies and feedback models. Furthermore, this separability may not hold for other extensions of the standard model, such as modified gravity and dynamical dark energy.

We note that while all simulations shown in Figure \ref{fig:allpow} include feedback from SNe, this process by itself did not have a strong impact on the matter power spectrum. It is possible to adapt the implementation of SN feedback such that a suppression similar to that of AGN feedback can be achieved, e.g.\ by increasing the strength of SN feedback in high-density environments \citep[see][]{vanDaalen11}. However, only simulations that include AGN feedback have thus far been able to reproduce the observed properties of groups and clusters.

Other simulations that have achieved similar volumes and resolutions, which have not to this date been used to explore the impact of baryons on the matter power spectrum are MUFASA \cite{MUFASA}, Simba \cite{Simba}, Romulus \cite{Tremmel17} and Magneticum Pathfinder\footnote{\url{www.magneticum.org}}. 
\begin{itemize}
\item MUFASA is a simulation suite run using the {\sc Gizmo} meshless finite mass hydrodynamics code \cite{Hopkins15}. Its fiducial run, evolved until $z=0$, spans a cosmological volume of (50 $h^{-1}$ Mpc)$^3$. The simulation shows good agreement with the galaxy stellar mass function (GSMF) in the range $0<z<4$ and with the cosmic star formation history. Reference \cite{Simba} caution that the GSMF is sensitive to the hydrodynamics methodology within a factor of 2, but that this is sub-dominant compared to the impact of different choices in parameterising stellar feedback. MUFASA does not include AGN feedback in its sub-grid model but implemented a heuristic quenching prescription to mimic its effect \cite{Gabor15}. 
\item The Simba simulation suite follows on the MUFASA prescription but adopts an updated sub-grid model that explicitly includes AGN feedback \cite{Hopkins11,Angles15}. Only the 100 $h^{-1}$ Mpc box of the suite was run until $z=0$. Comparisons to a broad range of observations (stellar and gas content of group-sized haloes, star formation rates, black hole properties, etc.) demonstrated that the simulation successfully reproduces the observable Universe. Some points of tension remain. Notably, Simba has difficulties in reproducing the knee of the GSMF at $z=0$. There are key differences between Simba and other simulation suites in the AGN feedback prescription and the hydrodynamical scheme which suggest a comparison prediction of the distribution of matter in this simulation could be useful in extending the existing set \cite{Chisari18,Huang19}. To this end, a pure dark matter, control $N$-body simulation at the same resolution and with similar initial conditions would be needed. 
\item The Romulus simulation \cite{Tremmel17} was designed to study the co-evolution of galaxies and black holes and run using the SPH code ChaNGa \cite{Menon15}. The largest volume in the Romulus suite is (50 Mpc)$^3$, which so far limits is application for precision cosmology purposes. Predictions of the matter power spectrum are known to be subject to cosmic variance on such scale \cite{Chisari18}.
\item Magneticum Pathfinder is a suite of SPH simulations run using the GADGET code with sizes from 18 $h^{-1}$ Mpc to 2.6 $h^{-1}$ Gpc. Pure dark matter $N$-body control runs are available for several of the boxes. The cosmology adopted was {\it WMAP}7. Several works have used Magneticum Pathfinder for prediction or comparison to cosmological observables \cite[e.g.][]{Dolag16,Bocquet16,Castro18}, though none of them directly provide a quantification of the impact of baryons on the matter power spectrum so far. 
\end{itemize}

\subsubsection{Initial conditions and calibration of feedback physics} 
\label{ss:calibIC}

Most modern cosmological simulations are initialized at early times (typically $z\gtrsim100$), with the initial distribution of dark matter and baryons being set to reproduce the matter power spectrum calculated by Boltzmann codes (such as CAMB, \cite{camb1,camb2}, or CLASS, \cite{CLASS}) using cosmological parameter values determined from the cosmic microwave background \cite[e.g.][]{Planck18}.  In detail, small differences are present between different studies when computing the initial conditions. For example, in whether the baryons and dark matter are assumed to follow the same (total) power spectrum or if they (more accurately) use individual power spectra computed by Boltzmann codes \cite[e.g.][]{CLASS}. This process can affect the power spectrum above per cent accuracy on large scales \cite{Angulo13,Valkenburg17}. Other choices, such as whether one adopts the Zel'dovich approximation or one uses more accurate second-order Lagrangian perturbation theory to initialize the particle positions and velocities, can also affect present-day, large-scale structure diagnostics at the few per cent level. While not small in the context of future large-scale structure observations (and thus requiring some attention), at present these uncertainties are typically sub-dominant to those associated with uncertain baryon physics. Similarly, the choice of numerical scheme, box size and resolution can result in discrepancies in predictions for the matter power spectrum at the per cent level, even in dark-matter-only simulations \cite{Schneider16}.

In the case of cosmological hydrodynamical simulations, there is further freedom in the choice of sub-grid model parameters. For the purpose of this work, these pertain to the sub-grid mechanisms governing the efficiency of star and black hole formation, and of stellar and AGN feedback. In a more general context, if other physical processes such as radiation feedback or magnetic fields are incorporated in the simulations, these would also require their own calibration schemes. Amongst these additional physics, magnetic fields of primordial origin have the potential to intervene in processes of structure formation \cite{Kim96,Tsagas00}. However, only a subset of magnetogenesis models provide strong enough primordial magnetic fields \cite{Kandus11} for these effects to take place. The required primordial magnetization is remarkably close to the current observational upper limit \cite{Planck16} of the vast bracket of allowed values. As a result, when considering most primordial magnetic field scenarios, these are not expected to have a large impact on cosmological structure formation \cite{Ryu12,Vazza14,Marinacci16}. Cosmic rays are also interesting candidates to influence the baryonic distribution of matter. They have been reported to boost galactic outflows for low mass galaxies (i.e. below $\lesssim 10^{12} M_\odot$ \cite{Salem14,Jacob18}), and they are expected to be an important ingredient of AGN winds. However, their numerical modelling on cosmological scales remains computationally too expensive, and additional precursory work is required to determine how important they are for simulations attempting to model the matter power spectrum. Due to the uncertainty surrounding the impact of these additional physics on the distribution of matter at the redshifts of interest to optical and infrared surveys, in this work we mostly focus on AGN feedback.

The calibration of sub-grid processes in general is also sensitive to the size of the simulated volume, resolution and perhaps the choice of simulation technique (e.g., SPH vs. AMR). As evidenced from Table \ref{tab:sims}, typical cosmological hydrodynamical simulations probe volumes of approximately ($100$ Mpc)$^3$, and consist of at least a handful of runs that share the same initial conditions but vary in the implementation of sub-grid recipes, often providing a dark-matter-only (DMO) and different resolution runs for comparison. Research groups have also adopted different calibration strategies for the sub-grid recipes, which we summarize below.  

In the Horizon-AGN suite \cite{Dubois14}, the AGN feedback implementation follows the prescription set by \cite{Dubois12}, with a thermal isotropic and a kinetic bipolar mode, each triggered by the Bondi-Hoyle-Lyttleton accretion rate into black holes \cite{Booth09} when, respectively, above or below 1 per cent of the Eddington rate. The vast majority of AGN are in thermal (``quasar'') mode at $z>2$ and transition to kinetic (``jet'') mode at lower redshifts \cite{Beckmann17}.  The efficiency of the quasar mode energy deposition rate is calibrated to reproduce the scaling relations between black hole mass and galaxy properties (stellar mass, velocity dispersion) in our local Universe. There is evidence that the fraction of gas in groups is slightly over-predicted compared to existing observational constraints \cite{Chisari18}.

In the MassiveBlack-II simulation \cite{Khandai15}, the authors took a different approach at specifying their sub-grid physics model. To ensure continuity with previous work \cite{DiMatteo08,Croft09,Degraf10,DiMatteo12}, they adopted the same sub-grid modelling schemes and parameters. In the case of AGN feedback, the thermal coupling efficiency had been calibrated from galaxy merger simulations to reproduce the scaling relation between black hole mass and galaxy velocity dispersion. Their aim was to assess the performance of this already established model, rather than actively re-calibrating it for a cosmological box. Notably, some of the previous work was intended for high redshift investigation of the distribution of quasar properties, and thus it is not surprising that reference \cite{Khandai15} finds that the best agreement with AGN bolometric luminosity functions is obtained at $z>2$. Moreover, the lack of kinetic feedback in MassiveBlack-II eliminates the ``maintenance'' mode of AGN at low redshift. Consequently, the observed GSMF was poorly matched at $z<2$, with an overproduction of stars at large mass and an underproduction at low mass. Reference \cite{Khandai15} indeed found evidence of insufficient quenching of star formation by AGN feedback at low redshift and suggested this was a consequence of the sub-grid parameters being motivated by the calibration of simulated boxes of smaller volume, which missed the high-mass end of the halo mass function. This could be the origin of the enhancement of power evidenced for MassiveBlack-II in Figure \ref{fig:allpow}, although conclusive evidence cannot be obtained without an exploration of the parameter space of that sub-grid model. On the contrary, halo occupation distribution (HOD) statistics and the clustering of galaxies displayed a good match to observations.

In OWLS \cite{Schaye10}, a reference ({\it REF}) sub-grid model is used as a baseline from which sub-grid parameters are varied. This model does not necessarily provide the best fit to observations, as it does not consider the impact of AGN feedback. Variations of the {\it REF} model are described in Table 2 of \cite{Schaye10}, including, for example, changes in the initial stellar mass function (IMF), cooling, heating, reionization redshift, wind mass loading, stellar feedback, supernova feedback, etc. For the purpose of this work, we are mostly interested in the impact of AGN \cite{McCarthy10}, as the corresponding OWLS runs have been extensively used in cosmological studies of the impact of baryons on the distribution of matter. For such runs, the AGN feedback sub-grid recipe \cite{Booth09} implemented consists effectively of a single isotropic mode of feedback and its efficiency was tuned to reproduce observed black hole scaling relations. 

The cosmo-OWLS follow-up suite \cite{LeBrun14} extended the volume and parameter space of the OWLS simulation suite. With boxes of $400\Mpch$, the authors explored (at lower resolution) the role of a specific sub-grid parameter that is crucial in determining the impact of AGN feedback in the galaxy cluster regime. In the prescription proposed in \cite{Booth09}, $\Delta T_{\rm heat}$ is the temperature increase undergone by surrounding particles when an AGN releases the energy stored. \cite{McCarthy11} showed that the bulk of the gas ejection is done at high redshifts ($2 \lesssim z \lesssim 4$) in the progenitors of groups and clusters and that the present-day baryonic content of galaxy groups is particularly sensitive to the adopted AGN heating temperature.  

In the EAGLE project \cite{Schaye15}, the choice was to simplify the implementation of different modes of stellar feedback (stellar winds, radiation pressure, supernovae) into a single sub-grid prescription. Similarly, AGN feedback from jet and quasar modes were unified into a single mechanism that injects energy in thermal form without switching off radiative cooling or hydrodynamical forces. The efficiency of stellar feedback and black hole accretion were calibrated to broadly match the observed $z=0.1$ GSMF. Additional constraints were placed on the distribution of galaxy sizes, which were crucial to produce realistic galaxies and to obtain acceptable agreement with galaxy scaling relations. The calibration procedure and simulation runs beyond the reference EAGLE model are described in \cite{Crain15}. 

In the case of the Illustris project, the AGN feedback implementation is done via three modes: thermal quasar-mode, thermal-mechanical radio mode and radiative feedback. Their efficiency is calibrated to reproduce the cosmic star formation history. Reference \cite{Vogelsberger14b} gives details of the calibration process and the comparison to different galaxy observables. They argue against tuning sub-grid recipes to match galaxy observables such as the GSMF because of such tuning typically does not account for systematic errors in these observables. Reference \cite{Genel14} explains that the approach taken for Illustris was to establish the values of the $\sim 15$ free parameters of the sub-grid model \cite{Vogelsberger13} based on their physical meaning in as much as possible, but clarify that a subset of these had to be tuned based on smaller volume ($35.5$ Mpc on each side) simulations, whose predictions were compared against the history of cosmic star-formation rate density and the $z = 0$ stellar mass function. Despite this calibration, the fraction of baryons in haloes in the Illustris simulation is too low compared to observations \cite{Genel14}, an effect that was attributed to an exceedingly efficient AGN feedback prescription.

The more recent IllustrisTNG simulations \cite{Springel18} feature an updated thermal and kinetic AGN feedback prescription \cite{Dubois12,Weinberger17} that allows energy injection at low accretion rates, in addition to the incorporation of an ab-initio weak and uniform magnetic field, modelled through a divergence-cleaning scheme \cite{Pakmor11}, and a modified prescription for galactic winds \cite{Pillepich18}. Reference \cite{Weinberger17} demonstrates that the newly implemented AGN feedback prescription yields excellent agreement between simulated and observed stellar mass fractions without overly heating the gas. The inclusion of this mode of feedback is crucial, taking precedence over the choice of values of actual sub-grid parameters. For IllustrisTNG, the original choice of parameters of Illustris was varied to alleviate known tensions with observations \cite{Pillepich18}. Sub-grid parameters were kept fixed for boxes of different resolutions, with the exception of the gravitational softening lengths and the number of neighbouring cells used as input to the black hole feedback model \cite{Weinberger17}.

The simulations described above have mostly been designed to probe the statistical properties of galaxies in a representative volume of a $\Lambda$CDM universe at low redshift. While they were not tailored to probe the impact of baryonic processes on large-scale correlations, multiple works have used these suites to address this problem (see \cite{Chisari18} and references there in). The BAHAMAS suite \cite{McCarthy17}, on the contrary, has been designed to tackle this question specifically.  Building on the previous OWLS and cosmo-OWLS programs, BAHAMAS represents a first attempt to calibrate the feedback processes not just on the galaxy properties but also on the integrated gas fractions of galaxy groups and clusters.  X-ray observations have consistently demonstrated for well over a decade that galaxy groups and low-mass clusters are significantly deficient in their integrated baryon content with respect to the Universal mean.  As the hot gas dominates the baryon budget, McCarthy et al.~argue that calibration on this phase in particular is crucial. Reference \cite{McCarthy17} calibrated their stellar and AGN feedback to reproduce both the present-day GSMF and the amplitude of the gas fraction--halo mass relation of local groups and clusters, as determined from high resolution X-ray data.  They then a posteriori checked the calibration against a very wide range of independent data sets. In \cite{McCarthy18}, the authors explored the impact of the calibrated feedback on cosmological observables (cosmic shear, tSZ, and CMB lensing), showing that the effects can be significant.  They also quantified the degree of uncertainty in predictions by using variation models that bracket the observed baryon fractions. They also explored the possible degeneracies between cosmological and feedback parameters, demonstrating that these are not a significant source of error at present, though will likely become more important for the next generation of surveys. 

\subsection{Approximate Methods} 
\label{sec:model}
 
Approximate methods inspired by the halo model \cite{Seljak2000,Peacock2000,Cooray2002} consist of an alternative approach to quantify baryonic feedback effects on the large-scale structure of the universe. The general advantage of such methods compared to full hydrodynamical simulations is that they allow one to parametrise the baryonic effects on halo profiles and they make it possible to efficiently investigate the full parameter space. Due to their speed, they can also be directly used for cosmological parameter inference. The downside of approximate methods is their potentially limited accuracy, which means that they have to be tested or calibrated using hydrodynamical simulations.
 
\subsubsection{Halo model}
\label{sec:halo_model}

The idea that gravitational evolution aggregates matter into dense roughly spherical clumps of matter, known as haloes, can be used to make a model for the total matter power spectrum. In this model, the power is decomposed into a two-halo and a one-halo term, where the two-halo term, $P_{\rm 2H}$, accounts for power arising from clustering between haloes and the one-halo term, $P_{\rm 1H}$, for that arising from clustering within individual haloes. The two terms are:
\begin{equation}
P_{2\mathrm{H}}(k)=P_{\mathrm{lin}}(k)
\left[\int_0^\infty n(M)b(M)W(M,k)\;\mathrm{d}M\right]^2\ ,
\label{eq:two_halo}
\end{equation}
\begin{equation}
P_{1\mathrm{H}}(k)=
\int_0^\infty n(M)W^2(M,k)\;\mathrm{d}M\ ,
\label{eq:one_halo}
\end{equation}
where $M$ is the halo mass, $b(M)$ is the linear halo bias and $n(M)$ is the halo-mass function (the distribution function for the number-density of haloes as a function of mass, sometimes denoted $\mathrm{d}n/\mathrm{d}M$ in the literature). $W(M,k)$ is the spherical Fourier transform of the halo density profile:
\begin{equation}
W(M,k)=\frac{1}{\bar\rho}\int_0^\infty4\pi r^2\frac{\sin(kr)}{kr}\rho(M,r)\;\mathrm{d}r\ ,
\label{eq:halo_window_function}
\end{equation}
where $\rho(M,r)$ is the halo density profie, which is usually taken to be the \citeauthor*{Navarro1997} (NFW; \citeyear{Navarro1997}) density profile
\begin{equation}
\rho(r)=\frac{\rho_0}{r/r_\mathrm{s}(1+r/r_\mathrm{s})^2};\quad r\leq r_\mathrm{v}\ .
\label{eq:NFW_profile}
\end{equation}
Here, $\rho_0$ is a normalisation, which can be related to $M$, $r_\mathrm{s}$ is the halo scale radius and $r_\mathrm{v}$ the virial radius. These are related via the mass-dependent halo concentration parameter: $c=r_\mathrm{v}/r_\mathrm{s}$.

Given that baryonic feedback is known to redistribute matter within haloes it seems natural to explore the range of possible effects this may have on the matter power spectrum via equations~(\ref{eq:two_halo}) and (\ref{eq:one_halo}). Caution must be taken, however, because several approximations have gone into the derivations these equations. Specifically, it has been assumed that haloes are linearly biased with respect to the underlying linear matter field, that haloes are spherical, and that halo properties depend exclusively on the halo mass. It is also common to take a smooth halo profile for $\rho(r)$ (e.g., equation~\ref{eq:NFW_profile}), which is fitted to stacked data from simulations, and therefore may not capture the true granularity of the halo substructure.

Early work using the halo model to investigate the effect of baryons on the matter spectrum was mainly interested on how gas cooling might contract and deform the inner halo. Reference \cite{White2004} showed that gas cooling would increase the effective concentration of a halo and that this in turn would boost the small-scale signal in the lensing power spectrum. After hydrodynamical simulations demonstrated that feedback could redistribute significant amounts of gas from halo centres (e.g., \cite{Jing2006}), attention shifted to how this could affect the power spectrum: Rudd et al. \cite{Rudd2008} showed that the effect of baryonic feedback on the power spectrum could be accounted for by altering the concentration-mass relation that went into a standard halo-model calculation (equations \ref{eq:two_halo} and \ref{eq:one_halo}), with the general trend that more aggressive feedback required less concentrated haloes. In \cite{Semboloni11,Semboloni13}, \cite{Fedeli2014a} and \cite{Mohammed2014b} halo models were developed that explicitly account for the gas and stellar components of haloes. This requires modelling the gas and stellar density profiles of haloes in addition to their dark matter profile (equation \ref{eq:NFW_profile}). The fractional mass of each component with respect to the total halo mass $M$ also needs to be specified. 

In \cite{Semboloni11} bound gas was modelled using single-$\beta$ profile \cite[e.g.,][]{Osmond2004} and gas considered ejected from the halo was uniformly distributed in a spherical annulus between the halo virial radius and twice this radius. Despite this simple assumption, it was shown that the OWLS power spectra could be matched by tuning parameters associated with the halo baryonic component. The authors of \cite{Mohammed2014b} were able to match a range of simulation results using a model with a single free parameter that captures the transition in halo mass between feedback-dominated haloes, mostly devoid of gas, and gas rich haloes, in which AGN feedback effects weaken. The CDM-gas-stars halo model of \cite{Fedeli2014a} was compared to data from the OWLS simulations in \cite{Fedeli2014b} where it was shown that it could match the simulations reasonably well, but it was noted that the model failed in the transition region between the two- and one-halo term, and a `non-linear halo bias' was introduced to smooth the transition. In all of the cases discussed in this paragraph the response of the power spectrum from OWLS was matched by the halo models at the few per cent level.

In \cite{Mead15} the standard halo model calculation was modified with a mixture of physical and ad hoc parameters to improve the accuracy compared to gravity-only $N$-body simulations. The resulting model fitted simulated power spectra at the $5$ per cent level for a range of wCDM cosmological parameters. The impact of baryonic feedback was then added by considering the OWLS power spectra data and modifying the halo profiles. In the model of \cite{Mead15}, the halo-concentration relation from \cite{Bullock2001} was used:
\begin{equation}
    c(M,z)=\frac{B}{1+z_\mathrm{f}(M,z)}\ ,
    \label{eq:Bullock_concentration}
\end{equation}
where $z_\mathrm{f}(M,z)$ is a halo-formation redshift that is calculated using a prescription given by \cite{Bullock2001} and $B=4$ is a constant fitted by the authors to provide a good match to halo profiles measured in simulations. In \cite{Mead15} equation~(\ref{eq:Bullock_concentration}) was modified match to power spectrum data from simulations. First, to best match gravity-only simulation power spectra from the emulator of \cite{Lawrence10} it was found that $B=3.13$ was necessary. Then, to best match the OWLS power spectra it was found that $B=2.32$ provided the best match to the OWLS AGN simulation. In each case this change in $B$ is independent of halo mass and redshift. In addition a Fourier space halo-bloating parameter was added to match the detailed change in power induced by baryonic feedback. This was added at the level of the $W(M,k)$ functions in equation (\ref{eq:one_halo}) via
\begin{equation}
    W(M,k)\to W(\nu^\eta M,k)\ ,
    \label{eq:mead_bloating}
\end{equation}
where $\nu$ is a proxy for halo mass: $\nu = \delta_\mathrm{c}/\sigma(R)$ and $\sigma(R)$ is the square-root of the linear density field variance when smoothed on scale $R$ that encloses mass $M=4\pi\bar\rho R^3/3$. Usually $\delta_\mathrm{c}=1.686$ but this parameter is modified in the \cite{Mead15} approach to $\delta_\mathrm{c}=1.59+0.0314\sigma_8(z)$. $\eta$ that appears in equation~(\ref{eq:mead_bloating}) was fixed to the redshift-dependent $\eta=\eta_0-0.3\sigma_8(z)$ to best match gravity-only simulations with $\eta_0=0.603$ providing the best match. However, $\eta_0$ was modified further to best match the OWLS simulations, with $0.760$ providing the best match to the AGN simulation. In \cite{Mead15} a one-parameter baryon model was proposed that related the change in $\eta$ to the change in $B$
\begin{equation}
    \eta_0=1.03-0.11B\ ,
    \label{eq:Mead_one_parameter_1}
\end{equation}
but this was updated in \cite{Joudaki18}
\begin{equation}
    \eta_0=0.98-0.12B\ ,
    \label{eq:Mead_one_parameter_2}
\end{equation}
to ensure that the model passed through the best-fitting point to the gravity-only simulations, which would correspond to no baryonic feedback.

The consequence of adopting equation~(\ref{eq:mead_bloating}) can be better understood in real space \cite{Copeland2019}: the standard NFW profile is converted to
\begin{equation}
\rho(r)=\frac{\rho_0}{r/r_\mathrm{s}(\nu^\eta+r/r_\mathrm{s})^2};\quad r<\nu^\eta r_\mathrm{v}\ .
\label{eq:NFW_conversion}
\end{equation}
Note that the way the `halo bloating' is implemented effectively extends the virial radii of high-mass haloes (for example, $\nu=2$ halo virial radii increase by 50 per cent) while compressing low mass haloes. The advantage of the \cite{Mead15} approach over competing methods is that the base gravity-only halo model already provides a good description of the gravity-only power spectrum (due to the free parameters added) so the model was able to predict the full OWLS power spectra reasonably well, and not just the ratio of baryonic power spectrum to gravity-only power spectrum.

The prescription for baryonic feedback presented in \cite{Mead15} has been propagated into analyses of weak gravitational lensing data, which has not only allowed for more accurate cosmological inferences, but also provided constraints on the strength of feedback itself (e.g.~\cite{Joudaki17a,J17b,hildebrandt17,Hildebrandt19,vanUitert18,Joudaki18,hikage19,Yoon19}). Notably, the combined analysis of ${\rm KiDS \times (2dFLenS + BOSS)}$ found $B < 3.3$ at 95\% confidence level \cite{Joudaki18}, with a peak in the posterior at $B=1.6$, which can be compared to the `no feedback' value of $B=3.13$ and the `OWLS AGN feedback' value of $B=2.0$\footnote{We note that the OWLS AGN feedback value of $B=2.0$ is different from $B=2.32$ quoted in \cite{Mead15}, and is a result of the updated \hmcode $\eta_0-B$ parameterization given in \cite{Joudaki18}.}. The analysis of \cite{Yoon19} found an even stronger preference for feedback, where $B=1.19^{+0.51}_{-0.45}$ from the galaxy-galaxy lensing and galaxy clustering measurements of the Deep Lens Survey, and $B=1.07^{+0.31}_{-0.39}$ when these measurements were combined with {\it Planck}.

The approach of \cite{Mead15} was extended by \cite{Copeland2018} who added in the effect of a baryonic core on the halo-density profiles and showed how constraints on dark energy for the {\it Euclid} mission would be degraded and biased by a lack of knowledge of the underlying baryon model. In a similar spirit, \cite{MacCrann2017} provided forecasts for how well the current Dark Energy Survey (DES) could measure the baryonic parameters in the \cite{Mead15} model and therefore, forecasts for what could be learned about galaxy formation physics from weak lensing.

\subsubsection{Baryonification}
\label{sec:baryonif}

The method of \emph{baryonification} consists of modifying particle outputs of gravity-only $N$-body simulations based on a parametrisation of halo profiles. The approach is related to the halo model in the sense that baryon effects are implemented at the level of haloes. However, unlike the halo model, the \emph{baryonification} method allows one to directly work on non-linear density maps, which provide an accurate description of the gravity-only structure formation process, making it possible to go beyond two-point statistics in terms of the analysis. A general description of the method can be found in \cite{Schneider15}, while the detailed model parametrisation is described in \cite{Schneider18}.

\begin{figure}
  	\centering
	\includegraphics[width=\columnwidth,trim=0.3cm 0.2cm 1.2cm 1.0cm,clip]{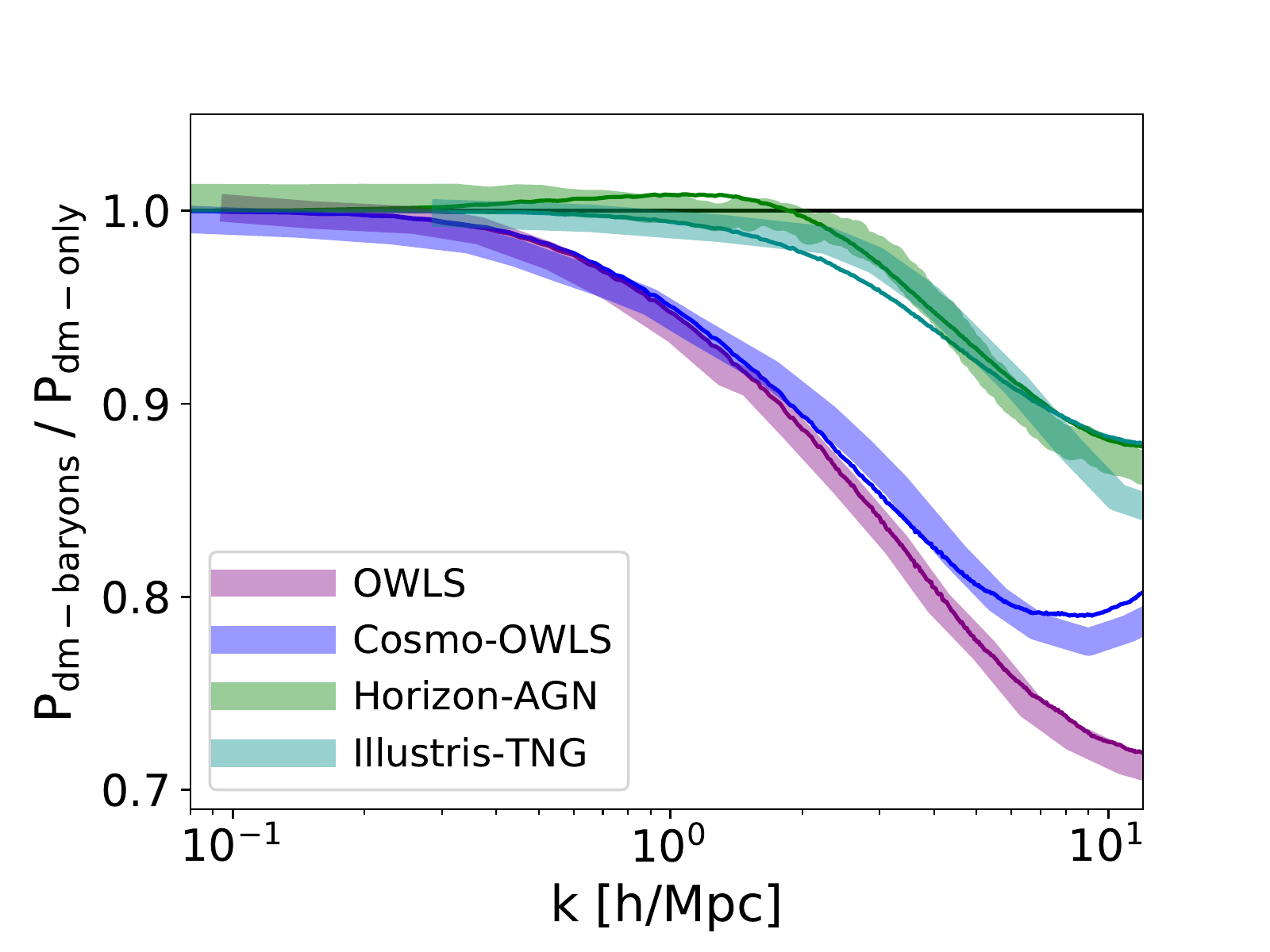}
	\caption{Ratio of the power spectra with and without baryon effects from the \emph{baryonification} model (coloured lines) and from several hydrodynamical simulations (coloured bands) at $z=0$. The \emph{baryonification} parameters have been tuned to match the gas and stellar fractions of the corresponding simulations. See Ref.~\cite{Schneider18} for more information about the comparison method.}
	\label{fig:ps_bfc}
\end{figure}

The baryonification method is based on displacement functions $d_j(r)$ for each halo $j$ in the simulated volume. Every simulation particle within the vicinity of a given halo is displaced according to $d_j(r)$. Since the displacement functions have non-zero values until far beyond the virial radius of their corresponding halo, a given simulation particles can be affected by multiple displacements. The displacement functions are defined as
\begin{equation}\label{displfct}
d_j(r_i)=r_{\rm dmb}(M_j)-r_i 
\end{equation}
where $M_j$ is the halo mass and $r_i$ the distance between halo centre and particle $i$. The function $r_{\rm dmb}$ is obtained by inverting the dark-matter-baryon (dmb) mass profile
\begin{equation}
M_{\rm dmb}(r)=4\pi\int_0^r ds s^2 \left[\rho_{\rm dm}(s)+\rho_{\rm gas}(s)+\rho_{\rm gas}(s)\right] +M_{\rm 2h}(r),
\end{equation}
which is composed of a dark matter ($\rho_{\rm dm}$), a gas ($\rho_{\rm gas}$), and a stellar ($\rho_{\rm star}$) density component plus a 2-halo mass term ($M_{\rm 2h}$). The individual components are parametrised using observationally motivated density profiles (see \cite{Schneider18} for more details). Overall, while the density profiles of the haloes are modified by the baryonification approach, it is assumed that their locations are unchanged with respect to the $N$-body box, an assumption supported by the findings presented in \cite{vanDaalen14}.

The goal of the \emph{baryonification} method is to perturb $N$-body simulations, approximating baryonic effects on the total matter distribution. This has the advantage of accurately capturing the large-scale structure of the universe, while allowing for a free model parametrisation for the baryon effects. In \cite{Schneider18} it has been shown that the method is able to recover the matter power spectrum of hydrodynamical simulations to 2 per cent or better, provided the model parameters are tuned to match the average gas and stellar fractions in haloes of the same simulation. A comparison of the \emph{baryonification} method with different hydrodynamical simulations is shown in Fig.~\ref{fig:ps_bfc}.

The \emph{baryonification} model can be calibrated against both direct observations or hydrodynamical simulations. Reference \cite{Schneider18} used observed X-ray gas fractions of individual galaxy groups and clusters to constrain the baryonic model parameters. They found that gas observations help to strongly reduce the current uncertainties in terms of weak lensing two-point statistics. Based on X-ray data (and including uncertainties related to the X-ray halo mass estimates), they found the angular power spectrum of the cosmic shear to be suppressed by up to 15-25 per cent beyond multipoles $l=100-500$.

Fitting functions to the fractional impact of baryons on the matter power spectrum can be derived from the \emph{baryonification} application to $N$-body simulations \cite{Schneider15}. This approach has been taken for example by reference \cite{Parimbelli19}, in which the authors explored the degeneracy between the massive neutrino imprint on weak lensing observables and the effect of baryonic feedback. The advantage of this approach is the possibility of making fast analytical predictions for the impact of baryons on cosmological constraints from future surveys via a Markov Chain Monte Carlo method. However, there are limitations that will need to be overcome for accurate predictions for upcoming surveys. Mainly, the redshift evolution of the fitting function presented in \cite{Schneider15} has not been validated and there are indeed indications that simulation results depart from this fit significantly \citep{Chisari18}.

\subsection{Data analysis}

The methods described in the previous section rely on having sufficiently accurate analytical expressions for the impact of baryons on the density profile of haloes. An alternative approach that is more agnostic on the choice of parametrisation is the PCA marginalization proposed by \cite{Eifler15}. This methodology towards mitigating systematics in the data does not only apply to the impact of baryons on the observables, but more widely to any nuisance effect.

Reference \cite{Eifler15} proposed to capture any systematic effects on the data vector through a principal component (PC) basis. In this case, the marginalization occurs not over a set of parameters of an analytical model, but over the amplitude of the PCs, each of which corresponds to a specific linear combination of the observables. In the case of \cite{Eifler15}, the observables were cosmic shear angular power spectra in multiple tomographic bins in the multipole range $30\leq \ell \leq 5000$, mimicking a Stage IV LSST or {\it Euclid} survey. Their findings suggest that removing 3-4 PCs is sufficient to avoid biases from baryonic physics, even when the most extreme predictions for AGN feedback or gas cooling are adopted.

The drawback of PCA marginalization, as of any other modelling approaches, is that its success depends on what the underlying true model for baryons is. Hence, it is crucial that this model is contained among the range of realistic predictions used for testing the method \cite{Mohammed18}. In \cite{Eifler15}, the baryonic scenarios considered were obtained from the predictions of several hydrodynamical cosmological simulations (OWLS, \cite{Rudd2008} and another suite\footnote{\url{http://astro.uchicago.edu/~gnedin/WL/}}). The set of simulations was extended in \cite{Huang19} to include Illustris, Horizon-AGN, EAGLE and MassiveBlack-II. This addition resulted in a reduction of the cosmological parameter bias, although the constraining power was also reduced. \cite{Huang19} also explored variations in the PCA scheme and concluded that it performs better when deviations in the data vectors used to construct the PCs are noise-weighted. 

In both \cite{Eifler15} and \cite{Huang19}, the simulations used different cosmologies, and their predictions were re-scaled assuming the baryonic correction to the power spectrum is independent of cosmology. This further assumption needs to be tested in the future, as it might become important for future surveys \cite{McCarthy18}. As we discuss in the next section, to find conclusive evidence in this regard would require a large suite of simulations spanning a wide range of sub-grid models and cosmologies. This would in turn allow for ``emulation" of the matter power spectrum over a parameter space beyond that of existing dark-matter-only emulators \cite{Lawrence17}.

To date, there is no demonstration of the PCA application in existing weak-lensing data. As mentioned above, weak-lensing surveys are currently adopting one of two options: removal of the scales affected by the impact of baryons \cite[e.g.][]{Troxel18} or marginalization over parametrised models \cite[e.g.][]{Hildebrandt19,hikage19}.

\section{The next decade}
\label{sec:next}

\subsection{Observations} 
\label{sec:obs2}

The feedback processes that impact the matter power spectrum induce observable changes in the gas content and distribution in galaxies and in groups and clusters of galaxies.  Thus, an observational program focused on studies of these populations will enable-- when brought together with state-of-the-art hydrodynamical structure formation simulations -- an accurate calibration of feedback processes and a precise prediction of the matter power spectrum.  In the following sections we describe in more detail how cluster and group observations and then cross-correlation studies of the large scale structure can be used for this purpose. We refer the reader to \cite{Battaglia19} for a discussion of other potential probes of feedback that can be complementary to these studies, such as stacked quasar spectral energy distributions \cite{Crichton16,Soergel17} or fast radio bursts \cite{McQuinn14}. 

\subsubsection{Impact of Feedback on Galaxy Clusters and Groups}

There has been clear evidence now for more than two decades that feedback processes are impacting the intracluster medium (ICM) content and distribution on cluster and group scales.  Initially, this was noted as departures from self-similar (i.e., gravity-only) expectations in the observable--observable scaling relations involving ICM properties such as the X-ray temperature, X-ray luminosity \cite{david93}, the X-ray isophotal size \cite{mohr97} and the ICM mass \cite{mohr99} integrated within a radius of $r_{500}$, where the enclosed density within the halo reaches 500 times the critical density. In the simplest picture of structure formation, the distribution of matter within haloes similar on all mass scales, and then, for example, a single, mass dependent characteristic scale like $r_{500}$ can be used to describe the matter distribution on all scales.  Approximate virialization within a region of fixed overdensity with respect to the critical density then implies specific mass and redshift trends for scaling relations (for more detailed discussion, see, e.g., \cite{bulbul19}).  

These departures from self-similarity in nearby galaxy clusters could all be explained by a mass-dependent ICM mass fraction such that lower mass clusters and groups exhibit lower gas fractions than higher mass systems. Efforts to explain these scaling relations focused on the concept of the ICM being heated by feedback processes due to galaxy formation or AGN feedback at early times before the gravitational collapse of the cluster scale halo. Because the entropy level in this preheated gas would be a larger fraction of the entropy induced by gravitational collapse on group and low mass cluster scales than on high mass cluster scales, the ICM distribution would be more impacted in low mass haloes, leading to a reduction of the ICM fraction within the group and cluster virial regions \cite{cavaliere98,ponman99}.

More recently, studies have shown that radio mode accretion within massive elliptical galaxies in clusters and groups leads to mechanical feedback through the production of large, low-density cavities in the ICM through radio jets \cite{cavagnolo10}.  This radio AGN feedback is observed to be ongoing today and therefore represents a continuing source of feedback that-- in contrast to the early preheating model originally explored-- could lead to observable departures from self-similar redshift evolution in the ICM properties of clusters. Studies of ICM observable--mass scaling relations over a broad redshift range provide no strong evidence for departures from self-similar redshift evolution \cite{vikhlinin09b,chiu16b,mantz16,chiu18,bulbul19}, although the observational constraints are still weak.  Direct studies of the ICM distribution within cluster haloes also are consistent with self-similar redshift trends \cite{mcdonald14}.  These results are in general agreement with recent hydrodynamical simulations \cite{barnes17} and suggest that the mechanical feedback from radio AGN within cluster haloes may roughly balance the radiative losses due to X-ray emission from the ICM.  

It should be noted, however, that the energy required to {\it unbind} gas from groups and clusters significantly exceeds that required to offset cooling losses. Thus, this late-time radio mode is often referred to as a ``maintenance'' mode, in that it maintains the ICM thermodynamic structure but does not explain the origin of the overall low baryon content of groups and low-/intermediate-mass clusters. Using simulations which include AGN feedback and that reproduce the observed baryon content, \cite{McCarthy11} showed that most of the gas ejection occurs at high redshift in the low-mass progenitors of groups and clusters, where the AGN are in a quasar mode.  Here the energetics of ejection are much more favourable, as the black holes are accreting at a significantly higher rate than they are today and the potential wells of the progenitor systems are considerably more shallow compared to the present-day collapsed systems. 

In terms of characterising the departures of groups and clusters from self-similarity, the early measurements and inferences relied on observable--observable relations using mass proxies such as the ICM emission weighted mean temperature \TX. Over time, new studies were carried out using cluster by cluster hydrostatic mass estimates \cite{pratt09,vikhlinin09b}.  Often, the selection effects and their impact on the scaling relations were not fully appreciated and modeled. But in recent years, a focus on new, large solid angle cluster surveys in the Sunyaev-Zel'dovich (SZ) effect, X-ray and optical has led to the development of larger and better understood cluster samples \cite{bleem15,PlanckSZ16,rykoff16,hilton18,klein18}. The effort to employ these samples for cosmological studies has led to more sophisticated Bayesian analysis techniques to constrain the observable--mass scaling relations. These techniques include corrections for selection effects, model the cosmological sensitivity of the observable and mass measurements, and employ weak gravitational lensing constraints on the cluster halo masses \cite{mantz10b,dehaan16,mantz16,bocquet18}. Importantly, the posterior parameter distributions for the scaling relations that emerge from these analyses include marginalization over the systematic uncertainties in the mass calibration. It is the emergence of these new cluster samples and the more sophisticated analysis toolkits that provide the possibility to use observed cluster scaling relations to robustly constrain feedback models that have been deployed in hydrodynamical simulations.  Crucial is that the hydrodynamical simulations used to explore baryonic effects on the matter power spectrum have enough volume to enable sizeable simulated cluster and group samples so that a direct comparison of observed and simulated scaling relations is possible.

Future cluster and group surveys with, e.g., eROSITA \cite{predehl10,merloni12,grandis18} SPT-3G \cite{benson14}, AdvACTpol \cite{AdvACTpol18} and the Simons Observatory \cite[SO;][]{SO19} survey instruments will produce large new cluster samples extending to redshifts $z>1$ with high quality ICM observables that will push to lower mass clusters and groups. Because the feedback impact is larger in lower mass haloes, these new samples should enhance the sensitivity of cluster scaling relations to feedback processes. Crucial as well will be the weak lensing mass calibration data coming from the future {\it Euclid} and LSST projects, which are expected to produce far better calibrated and much more sensitive weak lensing shear catalogues and photo-z catalogs than are currently available from DES and KiDS.

\subsubsection{Cross-correlations with the large-scale structure}

While scaling relations give us valuable insights into the total gas content and thermodynamic state of haloes, they make no statement about the spatial distribution of gas within haloes in the case where individual systems can be directly constrained with high signal-to-noise.
Cross-correlating tracers of large-scale structure with maps of the thermal (tSZ) and kinetic SZ (kSZ) effects and X-rays allows us to study ensembles of systems on lower group and even galaxy mass scales, where studies of individual systems are prohibitively expensive or even impossible. Thus, these cross-correlation techniques open up new avenues to measure the density, pressure, and temperature profiles of gas in haloes and to understand the effect of galaxy formation on the large-scale gas distribution and restrict the set of viable simulations \cite[e.g., ][]{Mroczkowski2018,Battaglia2019}.

For example, stacking analyses of the tSZ effect on galaxy and group catalogues have measured the pressure profiles down to masses of $\le10^{12}M_\odot$ \cite{Vikram2017,Hill2018} and compared to predictions from hydrodynamical simulations \cite{Tanimura2019}. Similar comparisons to simulations have been made for cross-correlations between lensing and the tSZ effect \cite{Hojjati2017}, where it has been concluded that future high-resolution measurements will place severe constraints on models of feedback. Such cross-correlations with lensing are of particular interest in the context of this paper, due to their direct dependence on the matter power spectrum and independence of other galaxy-formation processes, such as galaxy bias and the stellar-to-halo mass relation. Recently, the cross-correlation of galaxies and tSZ was measured using data from the DES survey \cite{Pandey19}. Combining such correlation with clustering measurements, it was possible to constrain the halo bias-weighted, gas pressure of the Universe as a function of redshift between $0.15<z<0.75$.

The studies mentioned above rely on data from the {\it Planck} satellite and current ground-based CMB experiments. The low angular resolution of these experiments ($\approx 10\;\text{arcmin}$ for {\it Planck}  and $\approx 5\;\text{arcmin}$ for current ground-based experiments) severely limits the ability to resolve the inner regions of haloes. In the near future, the combination of high resolution and high sensitivity of ground-based CMB experiments, such as the Simons Observatory \cite{2019JCAP...02..056A} or CMB Stage IV \cite{2016arXiv161002743A}, will dramatically improve our understanding of the large-scale properties of gas. This will be done through observations of the three main CMB secondary anisotropies: thermal SZ effect, kinetic SZ effect and CMB lensing, which provide information about the gas pressure, gas density and matter density profiles around different structures \cite{2017JCAP...11..040B,Battaglia19}. The large area and frequency coverage of these experiments will be vital to improve the quality and robustness of the resulting component maps in terms of contamination from other Galactic and extragalactic sources.

Although the aim of this paper is to chart a way towards the characterisation and mitigation of the impact of baryons on the matter power spectrum, it should be noted that measurements of the SZ effects hold a great deal of cosmological information themselves \cite[e.g.,][]{Hill13,Hill13b,vanWaerbeke2014,Hill2014,Planck16b,Horowitz2017,Salvati18}. 
Joint analyses of weak lensing, SZ effects, and their cross-correlation have the potential to yield greatly improved constraints on both cosmological and baryonic parameters.

When including these baryonic probes in cosmological analyses, care needs to be taken to avoid using the same data to calibrate the underlying models and simulations as is later used in the inference (see Section \ref{sec:model2}).

\subsection{Simulations} 
\label{sec:sims2}

In Section \ref{ss:calibIC}, we discussed the current calibration strategies for sub-grid physics in cosmological hydrodynamical simulations. Research teams, a handful of them, take different approaches towards calibration. The least aggressive strategy is to use physically motivated values for sub-grid parameters. A minimal calibration can be put in place by comparing the outputs of preliminary simulation runs with the cosmic star formation history of the Universe \cite[e.g., Illustris][]{Vogelsberger14b}. Other teams rely on galaxy observables to different degrees. For example, the Horizon-AGN simulation is tuned to reproduce the relation between black hole mass and galaxy stellar mass and velocity dispersion \cite{Dubois14}, while EAGLE performs a calibration on the GSMF and galaxy sizes \cite{Crain15}. On top of these different strategies, simulations suites use different initial conditions, cosmologies and hydrodynamical solvers, which complicates comparisons across them. Several insights arise from this comparison.
\begin{itemize}
\item The community would benefit from a comparison of cosmological hydrodynamical simulations across teams, but this can only be done with agreement on a cosmological model and initial conditions.
\item In addition, teams should coordinate their calibration strategies, motivated by further studies of which observables are the most constraining, in which regimes, and to what level and how do uncertainties in those observables propagate to the calibration. New probes should be explored for the calibration, examples are the AGN luminosity function, quasar absorption lines, environmental dependencies of BH scaling relations, observations of galactic winds, probes of supernova feedback, etc.
\item For the purpose of constraining the role of baryons in shaping the matter power spectrum, there is a need to perform calibration in the group regime and spanning the range of redshifts needed for upcoming surveys \citep{McCarthy11}. 
\item When calibrating to specific observational data sets, comparing the simulations to the observations in a like-with-like fashion (e.g., via realistic mock catalogs and maps) is important. For example, current estimates of gas fractions of galaxy groups and clusters are typically derived from X-ray-selected samples and the assumption of hydrostatic equilibrium is often applied when deriving the halo mass. Steps to account for these issues have been taken in only a small number of simulation campaigns so far (e.g., cosmo-OWLS and BAHAMAS).
\item With single runs taking up to millions of CPU hours to complete for current typical volumes and resolution, lower resolution experiments tailored to address the impact of baryons on the matter power spectrum should be given precedence. Strategies such as that of BAHAMAS can speed up computation time and allow exploration of accurate calibration strategies.
\item Calibration runs, typically dismissed due to low resolution or failure in reproducing observables, could (already) be used in preliminary studies of degeneracies in sub-grid models. In particular, if appropriately designed, they could seed the construction of hydrodynamical cosmological emulators in the spirit of similar experiments in emulating the non-linear matter power spectrum by \cite{Heitmann09} and \cite{Lawrence17}. 
\item In addition, a blind challenge could be triggered to determine the accuracy of different proposed methods in recovering cosmological parameters while at the same time marginalizing over parameters for the impact of baryons. Such a challenge could be set up in different stages, with a minimal option providing matter power spectra to participants, and a more sophisticated one providing cosmological observables such as angular power spectra or correlation functions in several tomographic bins. Accuracy metrics could be defined in terms of the recovered bias in the cosmological parameters of interest.

\end{itemize}

\subsection{Approximate methods}

The main advantage of approximate methods is that they are fast and therefore able to probe the parameter space of both baryonic physics and cosmology. Furthermore, they provide an alternative parametrisation to hydrodynamical simulations which -- at least for the case of the halo model and the \emph{baryonification} approach -- are based on halo properties instead of feedback effects such as sub-grid black-hole accretion and ejection. As a consequence, approximate methods are particularly well suited to link theory predictions to observations most particularly in the context parameter inference studies. We therefore believe that in the future they will provide a useful complementary tool to hydrodynamical simulations.

\subsubsection{Halo model}

The halo model is a convenient tool for rapid production of the theory power spectra necessary to analyze cosmological data sets. The main disadvantage is the lack of accuracy which primarily arises due to simplifications made in the derivation of the power spectrum equations. Primarily, the assumption of linear halo bias with respect to the underlying linear density field, which affects power spectrum predictions around the transition region between the two- and the one-halo term. Secondarily, the assumption of a smooth halo profile, devoid of substructure, which affects power in the small scales of the one-halo dominated regime\footnote{How different parts of the matter field are responsible for power can be investigated using techniques like those in \cite{vanDaalen2015}.}. These problems can be addressed by using more complicated halo models \cite[\eg][]{Smith2005,Smith2007,Giocoli2010,Smith2011,Valageas2011,Mohammed2014a} but these all come at the expense of increased computational cost, increased number of fitted parameters and loss of simplicity. Some \cite[\eg][]{Mohammed2014a} additionally break the correspondence of parameters in the model with simple functions of the halo profiles.

The \hmcode model of \cite{Mead15} circumvents the above problems using a mix of physically-motivated and ad-hoc prescriptions to alter the basic halo-model calculation and provides power that is accurate at the $5$ per cent level when compared gravity-only simulations and $\sim 10$ per cent compared to hydrodynamic simulations. This is adequate for current data sets but will not be adequate in the future. It remains to be seen if further ad-hoc tuning will be able to provide a fitting function that is capable of providing unbiased cosmological constraints from future data set or if a better (halo) model, with more theoretical justification, will be required. It has also not been demonstrated that the improvements required to get accurate matter power spectra also improve the accuracy of other cross spectra one could compute using the halo model machinery.

The utility of the halo model is not restricted to providing predictions for the matter power spectrum. In general, it can be used to provide predictions for any $n$-point correlation of any combinations of cosmological fields. Examples of relevance include: $X$-ray emission, the Compton-$y$ parameter, galaxy number counts, 21cm emission and the cosmic infrared background emission. The accuracy of the halo model predictions for higher than two-point correlations of matter is not well documented \cite[see][]{Cooray2002} and nor is the accuracy of the model thoroughly investigated for two-point correlations of field combinations other that the auto-correlation of matter or galaxies \cite{Mandelbaum2005,vandenBosch2013,Kwan2015,Mead15}. This is simply a consequence of the lack of large hydrodynamic simulation campaigns that cover the parameter space necessary to make general statements about the accuracy of the model. Different combinations of fields may have halo-model predictions that are better or worse than those the halo model provides for the matter auto spectrum; a consequence of the fields being generated by different halo populations with their own emission profiles. For example, it is known that Compton-$y$ primarily arises from the highest mass haloes \cite[e.g.,][]{Hill13,McCarthy14,Hill2018}, while 21cm emission is restricted to an intermediate halo-mass range \cite[][]{Villaescusa-Navarro2018}. It is probable that the most fruitful avenue for the future to constrain both cosmological and feedback parameters will involve combinations of (at least) two-point functions of fields that probe the matter in different ways.

\subsubsection{Baryonification}

The \emph{baryonification} model is ideally suited for fast production of full-sky maps including different observables such as weak-lensing, X-ray, or the SZ effect. This opens the path to various cross-correlation studies on the map-level which will allow a simultaneous constraint of baryonic and cosmological parameters. Furthermore, the \emph{baryonification} approach will make it possible to go beyond the two-point function, investigating higher order statistics \cite{Semboloni13,Barreira19}, weak-lensing peak counts \cite{Yang13}, halo mass functions \cite{Cui12,Khandai15,Bocquet16}, or void statistics \cite{Paillas17} to name a few potential cosmological probes. A first step in this direction has been taken by Ref.~\cite{Weiss19} who used the baryonification model to forecast the effects of baryons on peak-count statistics for future surveys.

Another potential field of application is map-based cosmological parameter estimates using deep learning methods. Recently, a first analysis of this type has used the baryonification model to estimate potential baryon-induced systematics on the deep-learning pipeline \cite{Fluri2019}. Of course, such an analysis would also be possible with full hydrodynamical simulations but, as stated above, the \emph{baryonification} model is considerably faster and provides an alternative parametrisation based on observable quantities such as gas and stars around haloes.

Currently, the \emph{baryonification} model only models the total cosmological density field. In order to obtain X-ray and SZ maps, individual density (and pressure) fields of the gas component have to be computed as well. This requires an upgrade of the current algorithm. Instead of displacing all particles following equation~(\ref{displfct}), each simulation particle will have to be tagged as a star, gas, or dark matter particle and displaced with respect to a displacement function describing individual components. This is a straightforward improvement of the algorithm which, however, still needs to be implemented and tested. Depending on the application, it will also be necessary to  model temperature and metallicity variations of the gas, leading to an increase of free model parameters.

In the coming years the goal is to fully implement the \emph{baryonification} model into weak-lensing, X-ray, and SZ prediction pipelines. These are built using very large, multi-trillion $N$-body runs, followed by a light-cone construction and the generation of full-sky maps, assuming a large set of different cosmological parameters. As a consequence, the \emph{baryonification} algorithm has to be implemented into an $N$-body code so that it can be run on-the-fly during the execution of the simulation.

Finally, the \emph{baryonification} model is ideally suited to test the cosmology-dependence of baryonic effects. Do changes of baryonic parameters depend on cosmology? Will it be necessary to run new simulations for each new parameter setup? These questions will have to be answered before designing an efficient and accurate pipeline to predict key observables of next-generation surveys. An example of such an application has already been provided in Ref. \cite{Parimbelli19}. In this case, the authors used a fitting function derived from \emph{baryonification} predictions of the impact of baryons on $P(z)$ and integrated it into an MCMC prediction code for cosmological constraints with an {\it Euclid}-like survey. As mentioned in Section \ref{sec:baryonif}, the fitting function adopted in \cite{Schneider15} has not been thoroughly tested for application to upcoming surveys. In the future, the fitting function step could be by-passed by the direct application of \emph{baryonification} to $N$-body simulations or by the use of deep learning methods.

\subsubsection{Deep learning methods}

The recent emergence of powerful generative models in the machine learning literature has opened up new possibilities to bridge the gap between $N$-body and hydrodynamical simulations. Initial attempts to exploit these methods to model structure formation have yielded promising results \cite{Rodriguez2018,He2018} but more work needs to be done to increase the accuracy and capabilities of these models.

An interesting application would be a deep learning analogue of the \emph{baryonification} model, where the deep learning model learns to convert particles in an $N$-body simulation into dark matter, gas, and star particles with their correct distributions and properties. Since the deep learning models are in principle able to learn complicated environmental dependencies and are not restricted to preconceived notations on what gas and density profiles are supposed to look like, they have the potential to provide very accurate and complete, albeit opaque, mappings between $N$-body and hydrodynamical simulations. The main caveat, as with all machine learning methods, is the availability of the necessary training data. Such methods are starting to be applied in cosmology, such as in the prediction of the three-dimensional distribution of galaxies \cite{Zhang19} or tSZ maps \cite{Troester2019} from the underlying dark matter distribution.

\subsection{Methods for data analysis}
\label{sec:model2}

\subsubsection{Principal Component Analysis}

The original proposal of using PCA to mitigate or remove the impact of baryons on cosmic shear observables was due to \cite{Eifler15}. Reference \cite{Mohammed18} tested the performance of the method with respect to outliers in the baryonic model by incorporating a test set of simulated predictions in addition to the training set used to build the PC basis. Their work highlighted the importance of spanning a wide range of parameter space for baryonic models. Reference \cite{Huang19} recently performed such an extension by including predictions for the impact of baryons on the matter power spectrum from Illustris, Horizon-AGN, EAGLE and MassiveBlack-II, which have only recently become available. Nevertheless, the challenge of establishing {\it representative} training and testing sets of predictions remains. It should be pointed out that the inclusion of every baryonic simulation available is clearly not desirable in this context, e.g. simulations that do not model AGN feedback at all should not enter a PCA training set. In contrast we suggest to opt for a coherent set of simulations that aim to model all aspects of baryonic physics and span the range of observational and modelling uncertainties within this concordance model. (A weighing scheme that includes errors on the individual simulations would be helpful in this context.) This will only be possible by the advent of faster simulations suites exploring a range of sub-grid parameter space and in cosmological volumes. The authors of \cite{Huang19} also noticed that PCs can vary with cosmology, and thus suggested that in the future, this method should be implemented via an iterative approach where after a first fit to the cosmology, the PC basis is re-derived.

Further investigation is needed to explore the use of PCA in a combined probes scenario. An analysis of the efficiency of PCA in a 3x2pt analysis would be the natural extension to existing studies. In the future, external data (Section \ref{sec:obs2}) could help identify more efficient PC basis and further improve the performance of the method.

\subsubsection{Baryonic emulator}

Similarly to the deep learning case, currently available power spectrum emulators \cite{Heitmann09,Lawrence17,Knabenhans19} rely on $N$-body simulations and neglect the effect of gas dynamics, feedback and star formation. Under these assumptions, they are successful in predicting the power spectrum for $k \leq 5$ Mpc$^{-1}$ and $0\leq z\leq 2$ with $4\%$ accuracy. A similar approach could be taken towards constructing a hydro-emulator that spans the parameter space of sub-grid physics models at a given resolution. In \cite{Lawrence17}, the number of simulated outputs used for emulation was $36$. If baryonic effects are independent from cosmology (including massive neutrinos), such a number would not be prohibitive. Degeneracies between the two could severely hinder such a prospect. Hence, we recommend that steps are taken to establish the degree to which the two effects are coupled {\it before} construction of a hydro-emulator. 

The parameter space of sub-grid models to be used for emulation could be significantly reduced by relevant observational priors (their uncertainties and degeneracies) both on astrophysics and cosmology. However, care must be taken not to apply the same constraints twice. For example, observations used to calibrate the emulator must not be used in a second instance for a combined probe analysis. Such circularity could bias cosmological results derived from using the hydro-emulator.

\section{Conclusions} 
\label{sec:conclude}

The impact of baryons in cosmological observables, weak lensing surveys in particular, is a challenge for the next era of experiments. In this work, we have summarized the state-of-the-art in the modelling, numerical simulation, mitigation techniques, and observational constraints available from the literature. We have also provided our outlook into how our knowledge of this effect should develop in the coming decade to support the upcoming LSST, {\it Euclid}, and {\it WFIRST} experiments. We conclude that:

\begin{itemize}
    \item  In current cosmic shear analyses, the impact of baryons on cosmological constraints is sub-dominant at the $\sim~0.5\sigma$ level. For optimal extraction of cosmological information from future surveys (quantified in terms of $w_0$, $w_a$ and $\sum m_\nu$), it will be of interest to make use of non-linear scales in the matter power spectrum ($k>0.1$ Mpc$^{-1}$), for which these will need to be modelled accurately.
    \item A currently common tool to make predictions for the impact of baryons on the matter power spectrum are cosmological hydrodynamical numerical simulations. Simulations vary in the hydrodynamical scheme, resolution, sub-grid model, and calibration strategies. As a result of these difference choices, predictions for $P(k)$ at small scales vary widely. The interplay between them should be investigated in dedicated low resolution suites, significantly reducing the computational cost of current suites. Preliminary studies suggest no significant dependence on cosmology (including neutrinos), though further dedicated studies are needed to confirm this trend.
    \item The halo model approach has been effective in capturing the cosmological imprint of baryons in a wide range of simulations via at most two free parameters that can be marginalized over, potentially using priors from numerical simulations. At increased precision, it will need to be modified to capture the effect of gas cooling and star formation at the center of haloes. It also relies on the assumption that haloes are linearly biased with respect to the matter field. \emph{Baryonification} overcomes this assumption by relying on dark-matter-only simulations that are modified to account for baryons through a displacement of particles belonging to each halo. This displacement is constrained from available observations, though subject to current uncertainties in the mass-observable relation (e.g., hydro-static mass bias). We recommend the development of these two methods in parallel, such as to profit from their cross-validation and to assess their speed and residual biases for cosmological applications in the next decade.  
    \item Multiple observations are available that can be useful in establishing priors on the impact of baryons on $P(k)$ (e.g., as for \emph{baryonification}) or in calibrating hydrodynamical simulations (e.g., BAHAMAS). To this end, constraints on gas and stellar mass fractions and profiles are needed down to at least $10^{13}$ M$_\odot$.   Current estimates of gas fractions come from X-ray and SZ observations, but these are generally limited to $10^{14}$ M$_\odot$ and above.  Furthermore, one would like to select the systems in way which is independent of the gas properties (e.g., via their galaxies or lensing signal). The main challenge is establishing an accurate mass-observable relation, though it is also possible to marginalize over free parameters such as the hydrostatic mass bias at the cost of losing precision in the derived cosmology. In the future, eROSITA and the Simons Observatory will deliver key complementary observations to constrain the role of baryons. Moreover, this will be achieved not only via cluster/group identification but also via cross-correlations. Data combinations that include these baryonic probes could help constrain not only dark energy but neutrino mass. Attention should be paid to avoid circularity (e.g., use of the same probe twice) and to accurately account for uncertainties and covariances between probes. 
    \item Dedicated simulation suites that aim to retrieve $P(k)$ predictions without resolving galaxy formation at kpc scales could be run at reduced computational power, thus enabling the construction of baryonic cosmic emulators or the application of deep learning techniques as generative models to bridge $N$-body and baryonic simulations. These in turn could help validate the PCA and other approaches to mitigate the impact of baryons on $P(k)$.
\end{itemize}

The impact of baryons on cosmological observables remains an open problem. We have reviewed the current status of this topic and provided a set of recommendations to the community with the goal of maximizing the scientific return of LSST, {\it Euclid}, and {\it WFIRST} in the next decade.

\section*{Acknowledgements}
We thank Hung-Jin Huang for providing the MassiveBlack-II matter power spectra used in Figure \ref{fig:allpow}. We thank Tim Eifler for useful discussions, and Yohan Dubois and Joop Schaye for feedback that helped improve this manuscript.
NEC is supported by a Royal Astronomical Society Research Fellowship. PGF and SJ acknowledge support from the Beecroft Trust, STFC, and ERC 693024. AJM and TT acknowledge support from the Horizon 2020 research and innovation programme of the European Union under the Marie Sk\l{}odowska-Curie grant agreements No.~702971 and No.~797794, respectively.  IGM acknowledges support from the ERC under the European Union's Horizon 2020 research and innovation programme (grant agreement No 769130). AS is supported by the Swiss National Science Foundation (PZ00P2\_161363). SMA is supported by the Oxford Hintze Centre for Astrophysical Surveys which is funded through generous support from the Hintze Family Charitable Foundation. We are grateful to the European Research Council for funding the workshop ``Modelling baryons for cosmology" held at University of Oxford in November 2018 and from which this work stemmed.


\bibliographystyle{apsrev4-1}
\bibliography{baryonsbib}

\begin{thebibliography}{199}%
\makeatletter
\providecommand \@ifxundefined [1]{%
 \@ifx{#1\undefined}
}%
\providecommand \@ifnum [1]{%
 \ifnum #1\expandafter \@firstoftwo
 \else \expandafter \@secondoftwo
 \fi
}%
\providecommand \@ifx [1]{%
 \ifx #1\expandafter \@firstoftwo
 \else \expandafter \@secondoftwo
 \fi
}%
\providecommand \natexlab [1]{#1}%
\providecommand \enquote  [1]{``#1''}%
\providecommand \bibnamefont  [1]{#1}%
\providecommand \bibfnamefont [1]{#1}%
\providecommand \citenamefont [1]{#1}%
\providecommand \href@noop [0]{\@secondoftwo}%
\providecommand \href [0]{\begingroup \@sanitize@url \@href}%
\providecommand \@href[1]{\@@startlink{#1}\@@href}%
\providecommand \@@href[1]{\endgroup#1\@@endlink}%
\providecommand \@sanitize@url [0]{\catcode `\\12\catcode `\$12\catcode
  `\&12\catcode `\#12\catcode `\^12\catcode `\_12\catcode `\%12\relax}%
\providecommand \@@startlink[1]{}%
\providecommand \@@endlink[0]{}%
\providecommand \url  [0]{\begingroup\@sanitize@url \@url }%
\providecommand \@url [1]{\endgroup\@href {#1}{\urlprefix }}%
\providecommand \urlprefix  [0]{URL }%
\providecommand \Eprint [0]{\href }%
\providecommand \doibase [0]{http://dx.doi.org/}%
\providecommand \selectlanguage [0]{\@gobble}%
\providecommand \bibinfo  [0]{\@secondoftwo}%
\providecommand \bibfield  [0]{\@secondoftwo}%
\providecommand \translation [1]{[#1]}%
\providecommand \BibitemOpen [0]{}%
\providecommand \bibitemStop [0]{}%
\providecommand \bibitemNoStop [0]{.\EOS\space}%
\providecommand \EOS [0]{\spacefactor3000\relax}%
\providecommand \BibitemShut  [1]{\csname bibitem#1\endcsname}%
\let\auto@bib@innerbib\@empty
\bibitem [{\citenamefont {Aghanim}\ \emph {et~al.}(2018)\citenamefont {Aghanim}
  \emph {et~al.}}]{Aghanim:2018eyx}%
  \BibitemOpen
  \bibfield  {author} {\bibinfo {author} {\bibfnamefont {N.}~\bibnamefont
  {Aghanim}} \emph {et~al.} (\bibinfo {collaboration} {Planck}),\ }\href@noop
  {} {\  (\bibinfo {year} {2018})},\ \Eprint {http://arxiv.org/abs/1807.06209}
  {arXiv:1807.06209 [astro-ph.CO]} \BibitemShut {NoStop}%
\bibitem [{\citenamefont {{Weinberg}}\ \emph {et~al.}(2013)\citenamefont
  {{Weinberg}}, \citenamefont {{Mortonson}}, \citenamefont {{Eisenstein}},
  \citenamefont {{Hirata}}, \citenamefont {{Riess}},\ and\ \citenamefont
  {{Rozo}}}]{Weinberg13}%
  \BibitemOpen
  \bibfield  {author} {\bibinfo {author} {\bibfnamefont {D.~H.}\ \bibnamefont
  {{Weinberg}}}, \bibinfo {author} {\bibfnamefont {M.~J.}\ \bibnamefont
  {{Mortonson}}}, \bibinfo {author} {\bibfnamefont {D.~J.}\ \bibnamefont
  {{Eisenstein}}}, \bibinfo {author} {\bibfnamefont {C.}~\bibnamefont
  {{Hirata}}}, \bibinfo {author} {\bibfnamefont {A.~G.}\ \bibnamefont
  {{Riess}}}, \ and\ \bibinfo {author} {\bibfnamefont {E.}~\bibnamefont
  {{Rozo}}},\ }\href {\doibase 10.1016/j.physrep.2013.05.001} {\bibfield
  {journal} {\bibinfo  {journal} {\physrep}\ }\textbf {\bibinfo {volume}
  {530}},\ \bibinfo {pages} {87} (\bibinfo {year} {2013})},\ \Eprint
  {http://arxiv.org/abs/1201.2434} {arXiv:1201.2434} \BibitemShut {NoStop}%
\bibitem [{\citenamefont {{Ivezic}}\ \emph {et~al.}(2008)\citenamefont
  {{Ivezic}}, \citenamefont {{Tyson}}, \citenamefont {{Abel}}, \citenamefont
  {{Acosta}}, \citenamefont {{Allsman}}, \citenamefont {{AlSayyad}},
  \citenamefont {{Anderson}}, \citenamefont {{Andrew}}, \citenamefont
  {{Angel}}, \citenamefont {{Angeli}}, \citenamefont {{Ansari}}, \citenamefont
  {{Antilogus}}, \citenamefont {{Arndt}}, \citenamefont {{Astier}},
  \citenamefont {{Aubourg}}, \citenamefont {{Axelrod}}, \citenamefont {{Bard}},
  \citenamefont {{Barr}}, \citenamefont {{Barrau}}, \citenamefont {{Bartlett}},
  \citenamefont {{Bauman}}, \citenamefont {{Beaumont}}, \citenamefont
  {{Becker}}, \citenamefont {{Becla}}, \citenamefont {{Beldica}}, \citenamefont
  {{Bellavia}}, \citenamefont {{Blanc}}, \citenamefont {{Blandford}},
  \citenamefont {{Bloom}}, \citenamefont {{Bogart}}, \citenamefont {{Borne}},
  \citenamefont {{Bosch}}, \citenamefont {{Boutigny}}, \citenamefont
  {{Brandt}}, \citenamefont {{Brown}}, \citenamefont {{Bullock}}, \citenamefont
  {{Burchat}}, \citenamefont {{Burke}}, \citenamefont {{Cagnoli}},
  \citenamefont {{Calabrese}}, \citenamefont {{Chandrasekharan}}, \citenamefont
  {{Chesley}}, \citenamefont {{Cheu}}, \citenamefont {{Chiang}}, \citenamefont
  {{Claver}}, \citenamefont {{Connolly}}, \citenamefont {{Cook}}, \citenamefont
  {{Cooray}}, \citenamefont {{Covey}}, \citenamefont {{Cribbs}}, \citenamefont
  {{Cui}}, \citenamefont {{Cutri}}, \citenamefont {{Daubard}}, \citenamefont
  {{Daues}}, \citenamefont {{Delgado}}, \citenamefont {{Digel}}, \citenamefont
  {{Doherty}}, \citenamefont {{Dubois}}, \citenamefont {{Dubois-Felsmann}},
  \citenamefont {{Durech}}, \citenamefont {{Eracleous}}, \citenamefont
  {{Ferguson}}, \citenamefont {{Frank}}, \citenamefont {{Freemon}},
  \citenamefont {{Gangler}}, \citenamefont {{Gawiser}}, \citenamefont
  {{Geary}}, \citenamefont {{Gee}}, \citenamefont {{Geha}}, \citenamefont
  {{Gibson}}, \citenamefont {{Gilmore}}, \citenamefont {{Glanzman}},
  \citenamefont {{Goodenow}}, \citenamefont {{Gressler}}, \citenamefont
  {{Gris}}, \citenamefont {{Guyonnet}}, \citenamefont {{Hascall}},
  \citenamefont {{Haupt}}, \citenamefont {{Hernandez}}, \citenamefont
  {{Hogan}}, \citenamefont {{Huang}}, \citenamefont {{Huffer}}, \citenamefont
  {{Innes}}, \citenamefont {{Jacoby}}, \citenamefont {{Jain}}, \citenamefont
  {{Jee}}, \citenamefont {{Jernigan}}, \citenamefont {{Jevremovic}},
  \citenamefont {{Johns}}, \citenamefont {{Jones}}, \citenamefont
  {{Juramy-Gilles}}, \citenamefont {{Juric}}, \citenamefont {{Kahn}},
  \citenamefont {{Kalirai}}, \citenamefont {{Kallivayalil}}, \citenamefont
  {{Kalmbach}}, \citenamefont {{Kantor}}, \citenamefont {{Kasliwal}},
  \citenamefont {{Kessler}}, \citenamefont {{Kirkby}}, \citenamefont {{Knox}},
  \citenamefont {{Kotov}}, \citenamefont {{Krabbendam}}, \citenamefont
  {{Krughoff}}, \citenamefont {{Kubanek}}, \citenamefont {{Kuczewski}},
  \citenamefont {{Kulkarni}}, \citenamefont {{Lambert}}, \citenamefont {{Le
  Guillou}}, \citenamefont {{Levine}}, \citenamefont {{Liang}}, \citenamefont
  {{Lim}}, \citenamefont {{Lintott}}, \citenamefont {{Lupton}}, \citenamefont
  {{Mahabal}}, \citenamefont {{Marshall}}, \citenamefont {{Marshall}},
  \citenamefont {{May}}, \citenamefont {{McKercher}}, \citenamefont
  {{Migliore}}, \citenamefont {{Miller}}, \citenamefont {{Mills}},
  \citenamefont {{Monet}}, \citenamefont {{Moniez}}, \citenamefont {{Neill}},
  \citenamefont {{Nief}}, \citenamefont {{Nomerotski}}, \citenamefont
  {{Nordby}}, \citenamefont {{O'Connor}}, \citenamefont {{Oliver}},
  \citenamefont {{Olivier}}, \citenamefont {{Olsen}}, \citenamefont {{Ortiz}},
  \citenamefont {{Owen}}, \citenamefont {{Pain}}, \citenamefont {{Peterson}},
  \citenamefont {{Petry}}, \citenamefont {{Pierfederici}}, \citenamefont
  {{Pietrowicz}}, \citenamefont {{Pike}}, \citenamefont {{Pinto}},
  \citenamefont {{Plante}}, \citenamefont {{Plate}}, \citenamefont {{Price}},
  \citenamefont {{Prouza}}, \citenamefont {{Radeka}}, \citenamefont
  {{Rajagopal}}, \citenamefont {{Rasmussen}}, \citenamefont {{Regnault}},
  \citenamefont {{Ridgway}}, \citenamefont {{Ritz}}, \citenamefont {{Rosing}},
  \citenamefont {{Roucelle}}, \citenamefont {{Rumore}}, \citenamefont
  {{Russo}}, \citenamefont {{Saha}}, \citenamefont {{Sassolas}}, \citenamefont
  {{Schalk}}, \citenamefont {{Schindler}}, \citenamefont {{Schneider}},
  \citenamefont {{Schumacher}}, \citenamefont {{Sebag}}, \citenamefont
  {{Sembroski}}, \citenamefont {{Seppala}}, \citenamefont {{Shipsey}},
  \citenamefont {{Silvestri}}, \citenamefont {{Smith}}, \citenamefont
  {{Smith}}, \citenamefont {{Strauss}}, \citenamefont {{Stubbs}}, \citenamefont
  {{Sweeney}}, \citenamefont {{Szalay}}, \citenamefont {{Takacs}},
  \citenamefont {{Thaler}}, \citenamefont {{Van Berg}}, \citenamefont {{Vanden
  Berk}}, \citenamefont {{Vetter}}, \citenamefont {{Virieux}}, \citenamefont
  {{Xin}}, \citenamefont {{Walkowicz}}, \citenamefont {{Walter}}, \citenamefont
  {{Wang}}, \citenamefont {{Warner}}, \citenamefont {{Willman}}, \citenamefont
  {{Wittman}}, \citenamefont {{Wolff}}, \citenamefont {{Wood-Vasey}},
  \citenamefont {{Yoachim}}, \citenamefont {{Zhan}},\ and\ \citenamefont {{for
  the LSST Collaboration}}}]{Ivezic08}%
  \BibitemOpen
  \bibfield  {author} {\bibinfo {author} {\bibfnamefont {Z.}~\bibnamefont
  {{Ivezic}}}, \bibinfo {author} {\bibfnamefont {J.~A.}\ \bibnamefont
  {{Tyson}}}, \bibinfo {author} {\bibfnamefont {B.}~\bibnamefont {{Abel}}},
  \bibinfo {author} {\bibfnamefont {E.}~\bibnamefont {{Acosta}}}, \bibinfo
  {author} {\bibfnamefont {R.}~\bibnamefont {{Allsman}}}, \bibinfo {author}
  {\bibfnamefont {Y.}~\bibnamefont {{AlSayyad}}}, \bibinfo {author}
  {\bibfnamefont {S.~F.}\ \bibnamefont {{Anderson}}}, \bibinfo {author}
  {\bibfnamefont {J.}~\bibnamefont {{Andrew}}}, \bibinfo {author}
  {\bibfnamefont {R.}~\bibnamefont {{Angel}}}, \bibinfo {author} {\bibfnamefont
  {G.}~\bibnamefont {{Angeli}}}, \bibinfo {author} {\bibfnamefont
  {R.}~\bibnamefont {{Ansari}}}, \bibinfo {author} {\bibfnamefont
  {P.}~\bibnamefont {{Antilogus}}}, \bibinfo {author} {\bibfnamefont {K.~T.}\
  \bibnamefont {{Arndt}}}, \bibinfo {author} {\bibfnamefont {P.}~\bibnamefont
  {{Astier}}}, \bibinfo {author} {\bibfnamefont {E.}~\bibnamefont {{Aubourg}}},
  \bibinfo {author} {\bibfnamefont {T.}~\bibnamefont {{Axelrod}}}, \bibinfo
  {author} {\bibfnamefont {D.~J.}\ \bibnamefont {{Bard}}}, \bibinfo {author}
  {\bibfnamefont {J.~D.}\ \bibnamefont {{Barr}}}, \bibinfo {author}
  {\bibfnamefont {A.}~\bibnamefont {{Barrau}}}, \bibinfo {author}
  {\bibfnamefont {J.~G.}\ \bibnamefont {{Bartlett}}}, \bibinfo {author}
  {\bibfnamefont {B.~J.}\ \bibnamefont {{Bauman}}}, \bibinfo {author}
  {\bibfnamefont {S.}~\bibnamefont {{Beaumont}}}, \bibinfo {author}
  {\bibfnamefont {A.~C.}\ \bibnamefont {{Becker}}}, \bibinfo {author}
  {\bibfnamefont {J.}~\bibnamefont {{Becla}}}, \bibinfo {author} {\bibfnamefont
  {C.}~\bibnamefont {{Beldica}}}, \bibinfo {author} {\bibfnamefont
  {S.}~\bibnamefont {{Bellavia}}}, \bibinfo {author} {\bibfnamefont
  {G.}~\bibnamefont {{Blanc}}}, \bibinfo {author} {\bibfnamefont {R.~D.}\
  \bibnamefont {{Blandford}}}, \bibinfo {author} {\bibfnamefont {J.~S.}\
  \bibnamefont {{Bloom}}}, \bibinfo {author} {\bibfnamefont {J.}~\bibnamefont
  {{Bogart}}}, \bibinfo {author} {\bibfnamefont {K.}~\bibnamefont {{Borne}}},
  \bibinfo {author} {\bibfnamefont {J.~F.}\ \bibnamefont {{Bosch}}}, \bibinfo
  {author} {\bibfnamefont {D.}~\bibnamefont {{Boutigny}}}, \bibinfo {author}
  {\bibfnamefont {W.~N.}\ \bibnamefont {{Brandt}}}, \bibinfo {author}
  {\bibfnamefont {M.~E.}\ \bibnamefont {{Brown}}}, \bibinfo {author}
  {\bibfnamefont {J.~S.}\ \bibnamefont {{Bullock}}}, \bibinfo {author}
  {\bibfnamefont {P.}~\bibnamefont {{Burchat}}}, \bibinfo {author}
  {\bibfnamefont {D.~L.}\ \bibnamefont {{Burke}}}, \bibinfo {author}
  {\bibfnamefont {G.}~\bibnamefont {{Cagnoli}}}, \bibinfo {author}
  {\bibfnamefont {D.}~\bibnamefont {{Calabrese}}}, \bibinfo {author}
  {\bibfnamefont {S.}~\bibnamefont {{Chandrasekharan}}}, \bibinfo {author}
  {\bibfnamefont {S.}~\bibnamefont {{Chesley}}}, \bibinfo {author}
  {\bibfnamefont {E.~C.}\ \bibnamefont {{Cheu}}}, \bibinfo {author}
  {\bibfnamefont {J.}~\bibnamefont {{Chiang}}}, \bibinfo {author}
  {\bibfnamefont {C.~F.}\ \bibnamefont {{Claver}}}, \bibinfo {author}
  {\bibfnamefont {A.~J.}\ \bibnamefont {{Connolly}}}, \bibinfo {author}
  {\bibfnamefont {K.~H.}\ \bibnamefont {{Cook}}}, \bibinfo {author}
  {\bibfnamefont {A.}~\bibnamefont {{Cooray}}}, \bibinfo {author}
  {\bibfnamefont {K.~R.}\ \bibnamefont {{Covey}}}, \bibinfo {author}
  {\bibfnamefont {C.}~\bibnamefont {{Cribbs}}}, \bibinfo {author}
  {\bibfnamefont {W.}~\bibnamefont {{Cui}}}, \bibinfo {author} {\bibfnamefont
  {R.}~\bibnamefont {{Cutri}}}, \bibinfo {author} {\bibfnamefont
  {G.}~\bibnamefont {{Daubard}}}, \bibinfo {author} {\bibfnamefont
  {G.}~\bibnamefont {{Daues}}}, \bibinfo {author} {\bibfnamefont
  {F.}~\bibnamefont {{Delgado}}}, \bibinfo {author} {\bibfnamefont
  {S.}~\bibnamefont {{Digel}}}, \bibinfo {author} {\bibfnamefont
  {P.}~\bibnamefont {{Doherty}}}, \bibinfo {author} {\bibfnamefont
  {R.}~\bibnamefont {{Dubois}}}, \bibinfo {author} {\bibfnamefont {G.~P.}\
  \bibnamefont {{Dubois-Felsmann}}}, \bibinfo {author} {\bibfnamefont
  {J.}~\bibnamefont {{Durech}}}, \bibinfo {author} {\bibfnamefont
  {M.}~\bibnamefont {{Eracleous}}}, \bibinfo {author} {\bibfnamefont
  {H.}~\bibnamefont {{Ferguson}}}, \bibinfo {author} {\bibfnamefont
  {J.}~\bibnamefont {{Frank}}}, \bibinfo {author} {\bibfnamefont
  {M.}~\bibnamefont {{Freemon}}}, \bibinfo {author} {\bibfnamefont
  {E.}~\bibnamefont {{Gangler}}}, \bibinfo {author} {\bibfnamefont
  {E.}~\bibnamefont {{Gawiser}}}, \bibinfo {author} {\bibfnamefont {J.~C.}\
  \bibnamefont {{Geary}}}, \bibinfo {author} {\bibfnamefont {P.}~\bibnamefont
  {{Gee}}}, \bibinfo {author} {\bibfnamefont {M.}~\bibnamefont {{Geha}}},
  \bibinfo {author} {\bibfnamefont {R.~R.}\ \bibnamefont {{Gibson}}}, \bibinfo
  {author} {\bibfnamefont {D.~K.}\ \bibnamefont {{Gilmore}}}, \bibinfo {author}
  {\bibfnamefont {T.}~\bibnamefont {{Glanzman}}}, \bibinfo {author}
  {\bibfnamefont {I.}~\bibnamefont {{Goodenow}}}, \bibinfo {author}
  {\bibfnamefont {W.~J.}\ \bibnamefont {{Gressler}}}, \bibinfo {author}
  {\bibfnamefont {P.}~\bibnamefont {{Gris}}}, \bibinfo {author} {\bibfnamefont
  {A.}~\bibnamefont {{Guyonnet}}}, \bibinfo {author} {\bibfnamefont {P.~A.}\
  \bibnamefont {{Hascall}}}, \bibinfo {author} {\bibfnamefont {J.}~\bibnamefont
  {{Haupt}}}, \bibinfo {author} {\bibfnamefont {F.}~\bibnamefont
  {{Hernandez}}}, \bibinfo {author} {\bibfnamefont {C.}~\bibnamefont
  {{Hogan}}}, \bibinfo {author} {\bibfnamefont {D.}~\bibnamefont {{Huang}}},
  \bibinfo {author} {\bibfnamefont {M.~E.}\ \bibnamefont {{Huffer}}}, \bibinfo
  {author} {\bibfnamefont {W.~R.}\ \bibnamefont {{Innes}}}, \bibinfo {author}
  {\bibfnamefont {S.~H.}\ \bibnamefont {{Jacoby}}}, \bibinfo {author}
  {\bibfnamefont {B.}~\bibnamefont {{Jain}}}, \bibinfo {author} {\bibfnamefont
  {J.}~\bibnamefont {{Jee}}}, \bibinfo {author} {\bibfnamefont {J.~G.}\
  \bibnamefont {{Jernigan}}}, \bibinfo {author} {\bibfnamefont
  {D.}~\bibnamefont {{Jevremovic}}}, \bibinfo {author} {\bibfnamefont
  {K.}~\bibnamefont {{Johns}}}, \bibinfo {author} {\bibfnamefont {R.~L.}\
  \bibnamefont {{Jones}}}, \bibinfo {author} {\bibfnamefont {C.}~\bibnamefont
  {{Juramy-Gilles}}}, \bibinfo {author} {\bibfnamefont {M.}~\bibnamefont
  {{Juric}}}, \bibinfo {author} {\bibfnamefont {S.~M.}\ \bibnamefont {{Kahn}}},
  \bibinfo {author} {\bibfnamefont {J.~S.}\ \bibnamefont {{Kalirai}}}, \bibinfo
  {author} {\bibfnamefont {N.}~\bibnamefont {{Kallivayalil}}}, \bibinfo
  {author} {\bibfnamefont {B.}~\bibnamefont {{Kalmbach}}}, \bibinfo {author}
  {\bibfnamefont {J.~P.}\ \bibnamefont {{Kantor}}}, \bibinfo {author}
  {\bibfnamefont {M.~M.}\ \bibnamefont {{Kasliwal}}}, \bibinfo {author}
  {\bibfnamefont {R.}~\bibnamefont {{Kessler}}}, \bibinfo {author}
  {\bibfnamefont {D.}~\bibnamefont {{Kirkby}}}, \bibinfo {author}
  {\bibfnamefont {L.}~\bibnamefont {{Knox}}}, \bibinfo {author} {\bibfnamefont
  {I.}~\bibnamefont {{Kotov}}}, \bibinfo {author} {\bibfnamefont {V.~L.}\
  \bibnamefont {{Krabbendam}}}, \bibinfo {author} {\bibfnamefont
  {S.}~\bibnamefont {{Krughoff}}}, \bibinfo {author} {\bibfnamefont
  {P.}~\bibnamefont {{Kubanek}}}, \bibinfo {author} {\bibfnamefont
  {J.}~\bibnamefont {{Kuczewski}}}, \bibinfo {author} {\bibfnamefont
  {S.}~\bibnamefont {{Kulkarni}}}, \bibinfo {author} {\bibfnamefont
  {R.}~\bibnamefont {{Lambert}}}, \bibinfo {author} {\bibfnamefont
  {L.}~\bibnamefont {{Le Guillou}}}, \bibinfo {author} {\bibfnamefont
  {D.}~\bibnamefont {{Levine}}}, \bibinfo {author} {\bibfnamefont
  {M.}~\bibnamefont {{Liang}}}, \bibinfo {author} {\bibfnamefont
  {K.}~\bibnamefont {{Lim}}}, \bibinfo {author} {\bibfnamefont
  {C.}~\bibnamefont {{Lintott}}}, \bibinfo {author} {\bibfnamefont {R.~H.}\
  \bibnamefont {{Lupton}}}, \bibinfo {author} {\bibfnamefont {A.}~\bibnamefont
  {{Mahabal}}}, \bibinfo {author} {\bibfnamefont {P.}~\bibnamefont
  {{Marshall}}}, \bibinfo {author} {\bibfnamefont {S.}~\bibnamefont
  {{Marshall}}}, \bibinfo {author} {\bibfnamefont {M.}~\bibnamefont {{May}}},
  \bibinfo {author} {\bibfnamefont {R.}~\bibnamefont {{McKercher}}}, \bibinfo
  {author} {\bibfnamefont {M.}~\bibnamefont {{Migliore}}}, \bibinfo {author}
  {\bibfnamefont {M.}~\bibnamefont {{Miller}}}, \bibinfo {author}
  {\bibfnamefont {D.~J.}\ \bibnamefont {{Mills}}}, \bibinfo {author}
  {\bibfnamefont {D.~G.}\ \bibnamefont {{Monet}}}, \bibinfo {author}
  {\bibfnamefont {M.}~\bibnamefont {{Moniez}}}, \bibinfo {author}
  {\bibfnamefont {D.~R.}\ \bibnamefont {{Neill}}}, \bibinfo {author}
  {\bibfnamefont {J.}~\bibnamefont {{Nief}}}, \bibinfo {author} {\bibfnamefont
  {A.}~\bibnamefont {{Nomerotski}}}, \bibinfo {author} {\bibfnamefont
  {M.}~\bibnamefont {{Nordby}}}, \bibinfo {author} {\bibfnamefont
  {P.}~\bibnamefont {{O'Connor}}}, \bibinfo {author} {\bibfnamefont
  {J.}~\bibnamefont {{Oliver}}}, \bibinfo {author} {\bibfnamefont {S.~S.}\
  \bibnamefont {{Olivier}}}, \bibinfo {author} {\bibfnamefont {K.}~\bibnamefont
  {{Olsen}}}, \bibinfo {author} {\bibfnamefont {S.}~\bibnamefont {{Ortiz}}},
  \bibinfo {author} {\bibfnamefont {R.~E.}\ \bibnamefont {{Owen}}}, \bibinfo
  {author} {\bibfnamefont {R.}~\bibnamefont {{Pain}}}, \bibinfo {author}
  {\bibfnamefont {J.~R.}\ \bibnamefont {{Peterson}}}, \bibinfo {author}
  {\bibfnamefont {C.~E.}\ \bibnamefont {{Petry}}}, \bibinfo {author}
  {\bibfnamefont {F.}~\bibnamefont {{Pierfederici}}}, \bibinfo {author}
  {\bibfnamefont {S.}~\bibnamefont {{Pietrowicz}}}, \bibinfo {author}
  {\bibfnamefont {R.}~\bibnamefont {{Pike}}}, \bibinfo {author} {\bibfnamefont
  {P.~A.}\ \bibnamefont {{Pinto}}}, \bibinfo {author} {\bibfnamefont
  {R.}~\bibnamefont {{Plante}}}, \bibinfo {author} {\bibfnamefont
  {S.}~\bibnamefont {{Plate}}}, \bibinfo {author} {\bibfnamefont {P.~A.}\
  \bibnamefont {{Price}}}, \bibinfo {author} {\bibfnamefont {M.}~\bibnamefont
  {{Prouza}}}, \bibinfo {author} {\bibfnamefont {V.}~\bibnamefont {{Radeka}}},
  \bibinfo {author} {\bibfnamefont {J.}~\bibnamefont {{Rajagopal}}}, \bibinfo
  {author} {\bibfnamefont {A.}~\bibnamefont {{Rasmussen}}}, \bibinfo {author}
  {\bibfnamefont {N.}~\bibnamefont {{Regnault}}}, \bibinfo {author}
  {\bibfnamefont {S.~T.}\ \bibnamefont {{Ridgway}}}, \bibinfo {author}
  {\bibfnamefont {S.}~\bibnamefont {{Ritz}}}, \bibinfo {author} {\bibfnamefont
  {W.}~\bibnamefont {{Rosing}}}, \bibinfo {author} {\bibfnamefont
  {C.}~\bibnamefont {{Roucelle}}}, \bibinfo {author} {\bibfnamefont {M.~R.}\
  \bibnamefont {{Rumore}}}, \bibinfo {author} {\bibfnamefont {S.}~\bibnamefont
  {{Russo}}}, \bibinfo {author} {\bibfnamefont {A.}~\bibnamefont {{Saha}}},
  \bibinfo {author} {\bibfnamefont {B.}~\bibnamefont {{Sassolas}}}, \bibinfo
  {author} {\bibfnamefont {T.~L.}\ \bibnamefont {{Schalk}}}, \bibinfo {author}
  {\bibfnamefont {R.~H.}\ \bibnamefont {{Schindler}}}, \bibinfo {author}
  {\bibfnamefont {D.~P.}\ \bibnamefont {{Schneider}}}, \bibinfo {author}
  {\bibfnamefont {G.}~\bibnamefont {{Schumacher}}}, \bibinfo {author}
  {\bibfnamefont {J.}~\bibnamefont {{Sebag}}}, \bibinfo {author} {\bibfnamefont
  {G.~H.}\ \bibnamefont {{Sembroski}}}, \bibinfo {author} {\bibfnamefont
  {L.~G.}\ \bibnamefont {{Seppala}}}, \bibinfo {author} {\bibfnamefont
  {I.}~\bibnamefont {{Shipsey}}}, \bibinfo {author} {\bibfnamefont
  {N.}~\bibnamefont {{Silvestri}}}, \bibinfo {author} {\bibfnamefont {J.~A.}\
  \bibnamefont {{Smith}}}, \bibinfo {author} {\bibfnamefont {R.~C.}\
  \bibnamefont {{Smith}}}, \bibinfo {author} {\bibfnamefont {M.~A.}\
  \bibnamefont {{Strauss}}}, \bibinfo {author} {\bibfnamefont {C.~W.}\
  \bibnamefont {{Stubbs}}}, \bibinfo {author} {\bibfnamefont {D.}~\bibnamefont
  {{Sweeney}}}, \bibinfo {author} {\bibfnamefont {A.}~\bibnamefont {{Szalay}}},
  \bibinfo {author} {\bibfnamefont {P.}~\bibnamefont {{Takacs}}}, \bibinfo
  {author} {\bibfnamefont {J.~J.}\ \bibnamefont {{Thaler}}}, \bibinfo {author}
  {\bibfnamefont {R.}~\bibnamefont {{Van Berg}}}, \bibinfo {author}
  {\bibfnamefont {D.}~\bibnamefont {{Vanden Berk}}}, \bibinfo {author}
  {\bibfnamefont {K.}~\bibnamefont {{Vetter}}}, \bibinfo {author}
  {\bibfnamefont {F.}~\bibnamefont {{Virieux}}}, \bibinfo {author}
  {\bibfnamefont {B.}~\bibnamefont {{Xin}}}, \bibinfo {author} {\bibfnamefont
  {L.}~\bibnamefont {{Walkowicz}}}, \bibinfo {author} {\bibfnamefont {C.~W.}\
  \bibnamefont {{Walter}}}, \bibinfo {author} {\bibfnamefont {D.~L.}\
  \bibnamefont {{Wang}}}, \bibinfo {author} {\bibfnamefont {M.}~\bibnamefont
  {{Warner}}}, \bibinfo {author} {\bibfnamefont {B.}~\bibnamefont {{Willman}}},
  \bibinfo {author} {\bibfnamefont {D.}~\bibnamefont {{Wittman}}}, \bibinfo
  {author} {\bibfnamefont {S.~C.}\ \bibnamefont {{Wolff}}}, \bibinfo {author}
  {\bibfnamefont {W.~M.}\ \bibnamefont {{Wood-Vasey}}}, \bibinfo {author}
  {\bibfnamefont {P.}~\bibnamefont {{Yoachim}}}, \bibinfo {author}
  {\bibfnamefont {H.}~\bibnamefont {{Zhan}}}, \ and\ \bibinfo {author}
  {\bibnamefont {{for the LSST Collaboration}}},\ }\href@noop {} {\bibfield
  {journal} {\bibinfo  {journal} {ArXiv e-prints}\ } (\bibinfo {year}
  {2008})},\ \Eprint {http://arxiv.org/abs/0805.2366} {arXiv:0805.2366}
  \BibitemShut {NoStop}%
\bibitem [{\citenamefont {{Laureijs}}\ \emph {et~al.}(2011)\citenamefont
  {{Laureijs}}, \citenamefont {{Amiaux}}, \citenamefont {{Arduini}},
  \citenamefont {{Augu{\`e}res}}, \citenamefont {{Cole}}, \citenamefont
  {{Cropper}}, \citenamefont {{Dabin}}, \citenamefont {{Duvet}}, \citenamefont
  {{Ealet\ }},\ and\ \citenamefont {et~al.}}]{Laureijs11}%
  \BibitemOpen
  \bibfield  {author} {\bibinfo {author} {\bibfnamefont {R.}~\bibnamefont
  {{Laureijs}}}, \bibinfo {author} {\bibfnamefont {J.}~\bibnamefont
  {{Amiaux}}}, \bibinfo {author} {\bibfnamefont {S.}~\bibnamefont {{Arduini}}},
  \bibinfo {author} {\bibfnamefont {J.}~\bibnamefont {{Augu{\`e}res}},
  \bibfnamefont {J.~. a\ nd~{Brinchmann}}}, \bibinfo {author} {\bibfnamefont
  {R.}~\bibnamefont {{Cole}}}, \bibinfo {author} {\bibfnamefont
  {M.}~\bibnamefont {{Cropper}}}, \bibinfo {author} {\bibfnamefont
  {C.}~\bibnamefont {{Dabin}}}, \bibinfo {author} {\bibfnamefont
  {L.}~\bibnamefont {{Duvet}}}, \bibinfo {author} {\bibfnamefont
  {A.}~\bibnamefont {{Ealet\ }}}, \ and\ \bibinfo {author} {\bibnamefont
  {et~al.}},\ }\href@noop {} {\bibfield  {journal} {\bibinfo  {journal} {ArXiv
  e-prints}\ } (\bibinfo {year} {2011})},\ \Eprint
  {http://arxiv.org/abs/1110.3193} {arXiv:1110.3193 [astro-ph.CO]} \BibitemShut
  {NoStop}%
\bibitem [{\citenamefont {{Spergel}}\ \emph {et~al.}(2013)\citenamefont
  {{Spergel}}, \citenamefont {{Gehrels}}, \citenamefont {{Breckinridge}},
  \citenamefont {{Donahue}}, \citenamefont {{Dressler}}, \citenamefont
  {{Gaudi}}, \citenamefont {{Greene}}, \citenamefont {{Guyon}}, \citenamefont
  {{Hirata}}, \citenamefont {{Kalirai}}, \citenamefont {{Kasdin}},
  \citenamefont {{Moos}}, \citenamefont {{Perlmutter}}, \citenamefont
  {{Postman}}, \citenamefont {{Rauscher}}, \citenamefont {{Rhodes}},
  \citenamefont {{Wang}}, \citenamefont {{Weinberg}}, \citenamefont
  {{Centrella}}, \citenamefont {{Traub}}, \citenamefont {{Baltay}},
  \citenamefont {{Colbert}}, \citenamefont {{Bennett}}, \citenamefont
  {{Kiessling}}, \citenamefont {{Macintosh}}, \citenamefont {{Merten}},
  \citenamefont {{Mortonson}}, \citenamefont {{Penny}}, \citenamefont {{Rozo}},
  \citenamefont {{Savransky}}, \citenamefont {{Stapelfeldt}}, \citenamefont
  {{Zu}}, \citenamefont {{Baker}}, \citenamefont {{Cheng}}, \citenamefont
  {{Content}}, \citenamefont {{Dooley}}, \citenamefont {{Foote}}, \citenamefont
  {{Goullioud}}, \citenamefont {{Grady}}, \citenamefont {{Jackson}},
  \citenamefont {{Kruk}}, \citenamefont {{Levine}}, \citenamefont {{Melton}},
  \citenamefont {{Peddie}}, \citenamefont {{Ruffa}},\ and\ \citenamefont
  {{Shaklan}}}]{Spergel13}%
  \BibitemOpen
  \bibfield  {author} {\bibinfo {author} {\bibfnamefont {D.}~\bibnamefont
  {{Spergel}}}, \bibinfo {author} {\bibfnamefont {N.}~\bibnamefont
  {{Gehrels}}}, \bibinfo {author} {\bibfnamefont {J.}~\bibnamefont
  {{Breckinridge}}}, \bibinfo {author} {\bibfnamefont {M.}~\bibnamefont
  {{Donahue}}}, \bibinfo {author} {\bibfnamefont {A.}~\bibnamefont
  {{Dressler}}}, \bibinfo {author} {\bibfnamefont {B.~S.}\ \bibnamefont
  {{Gaudi}}}, \bibinfo {author} {\bibfnamefont {T.}~\bibnamefont {{Greene}}},
  \bibinfo {author} {\bibfnamefont {O.}~\bibnamefont {{Guyon}}}, \bibinfo
  {author} {\bibfnamefont {C.}~\bibnamefont {{Hirata}}}, \bibinfo {author}
  {\bibfnamefont {J.}~\bibnamefont {{Kalirai}}}, \bibinfo {author}
  {\bibfnamefont {N.~J.}\ \bibnamefont {{Kasdin}}}, \bibinfo {author}
  {\bibfnamefont {W.}~\bibnamefont {{Moos}}}, \bibinfo {author} {\bibfnamefont
  {S.}~\bibnamefont {{Perlmutter}}}, \bibinfo {author} {\bibfnamefont
  {M.}~\bibnamefont {{Postman}}}, \bibinfo {author} {\bibfnamefont
  {B.}~\bibnamefont {{Rauscher}}}, \bibinfo {author} {\bibfnamefont
  {J.}~\bibnamefont {{Rhodes}}}, \bibinfo {author} {\bibfnamefont
  {Y.}~\bibnamefont {{Wang}}}, \bibinfo {author} {\bibfnamefont
  {D.}~\bibnamefont {{Weinberg}}}, \bibinfo {author} {\bibfnamefont
  {J.}~\bibnamefont {{Centrella}}}, \bibinfo {author} {\bibfnamefont
  {W.}~\bibnamefont {{Traub}}}, \bibinfo {author} {\bibfnamefont
  {C.}~\bibnamefont {{Baltay}}}, \bibinfo {author} {\bibfnamefont
  {J.}~\bibnamefont {{Colbert}}}, \bibinfo {author} {\bibfnamefont
  {D.}~\bibnamefont {{Bennett}}}, \bibinfo {author} {\bibfnamefont
  {A.}~\bibnamefont {{Kiessling}}}, \bibinfo {author} {\bibfnamefont
  {B.}~\bibnamefont {{Macintosh}}}, \bibinfo {author} {\bibfnamefont
  {J.}~\bibnamefont {{Merten}}}, \bibinfo {author} {\bibfnamefont
  {M.}~\bibnamefont {{Mortonson}}}, \bibinfo {author} {\bibfnamefont
  {M.}~\bibnamefont {{Penny}}}, \bibinfo {author} {\bibfnamefont
  {E.}~\bibnamefont {{Rozo}}}, \bibinfo {author} {\bibfnamefont
  {D.}~\bibnamefont {{Savransky}}}, \bibinfo {author} {\bibfnamefont
  {K.}~\bibnamefont {{Stapelfeldt}}}, \bibinfo {author} {\bibfnamefont
  {Y.}~\bibnamefont {{Zu}}}, \bibinfo {author} {\bibfnamefont {C.}~\bibnamefont
  {{Baker}}}, \bibinfo {author} {\bibfnamefont {E.}~\bibnamefont {{Cheng}}},
  \bibinfo {author} {\bibfnamefont {D.}~\bibnamefont {{Content}}}, \bibinfo
  {author} {\bibfnamefont {J.}~\bibnamefont {{Dooley}}}, \bibinfo {author}
  {\bibfnamefont {M.}~\bibnamefont {{Foote}}}, \bibinfo {author} {\bibfnamefont
  {R.}~\bibnamefont {{Goullioud}}}, \bibinfo {author} {\bibfnamefont
  {K.}~\bibnamefont {{Grady}}}, \bibinfo {author} {\bibfnamefont
  {C.}~\bibnamefont {{Jackson}}}, \bibinfo {author} {\bibfnamefont
  {J.}~\bibnamefont {{Kruk}}}, \bibinfo {author} {\bibfnamefont
  {M.}~\bibnamefont {{Levine}}}, \bibinfo {author} {\bibfnamefont
  {M.}~\bibnamefont {{Melton}}}, \bibinfo {author} {\bibfnamefont
  {C.}~\bibnamefont {{Peddie}}}, \bibinfo {author} {\bibfnamefont
  {J.}~\bibnamefont {{Ruffa}}}, \ and\ \bibinfo {author} {\bibfnamefont
  {S.}~\bibnamefont {{Shaklan}}},\ }\href@noop {} {\bibfield  {journal}
  {\bibinfo  {journal} {arXiv e-prints}\ } (\bibinfo {year} {2013})},\ \Eprint
  {http://arxiv.org/abs/1305.5425} {arXiv:1305.5425 [astro-ph.IM]} \BibitemShut
  {NoStop}%
\bibitem [{\citenamefont {Bernardeau}\ \emph {et~al.}(2002)\citenamefont
  {Bernardeau}, \citenamefont {Colombi}, \citenamefont {Gaztanaga},\ and\
  \citenamefont {Scoccimarro}}]{Bernardeau:2001qr}%
  \BibitemOpen
  \bibfield  {author} {\bibinfo {author} {\bibfnamefont {F.}~\bibnamefont
  {Bernardeau}}, \bibinfo {author} {\bibfnamefont {S.}~\bibnamefont {Colombi}},
  \bibinfo {author} {\bibfnamefont {E.}~\bibnamefont {Gaztanaga}}, \ and\
  \bibinfo {author} {\bibfnamefont {R.}~\bibnamefont {Scoccimarro}},\ }\href
  {\doibase 10.1016/S0370-1573(02)00135-7} {\bibfield  {journal} {\bibinfo
  {journal} {Phys. Rept.}\ }\textbf {\bibinfo {volume} {367}},\ \bibinfo
  {pages} {1} (\bibinfo {year} {2002})},\ \Eprint
  {http://arxiv.org/abs/astro-ph/0112551} {arXiv:astro-ph/0112551 [astro-ph]}
  \BibitemShut {NoStop}%
\bibitem [{\citenamefont {Cooray}\ and\ \citenamefont
  {Sheth}(2002{\natexlab{a}})}]{Cooray:2002dia}%
  \BibitemOpen
  \bibfield  {author} {\bibinfo {author} {\bibfnamefont {A.}~\bibnamefont
  {Cooray}}\ and\ \bibinfo {author} {\bibfnamefont {R.~K.}\ \bibnamefont
  {Sheth}},\ }\href {\doibase 10.1016/S0370-1573(02)00276-4} {\bibfield
  {journal} {\bibinfo  {journal} {Phys. Rept.}\ }\textbf {\bibinfo {volume}
  {372}},\ \bibinfo {pages} {1} (\bibinfo {year} {2002}{\natexlab{a}})},\
  \Eprint {http://arxiv.org/abs/astro-ph/0206508} {arXiv:astro-ph/0206508
  [astro-ph]} \BibitemShut {NoStop}%
\bibitem [{\citenamefont {Baumann}\ \emph {et~al.}(2012)\citenamefont
  {Baumann}, \citenamefont {Nicolis}, \citenamefont {Senatore},\ and\
  \citenamefont {Zaldarriaga}}]{Baumann:2010tm}%
  \BibitemOpen
  \bibfield  {author} {\bibinfo {author} {\bibfnamefont {D.}~\bibnamefont
  {Baumann}}, \bibinfo {author} {\bibfnamefont {A.}~\bibnamefont {Nicolis}},
  \bibinfo {author} {\bibfnamefont {L.}~\bibnamefont {Senatore}}, \ and\
  \bibinfo {author} {\bibfnamefont {M.}~\bibnamefont {Zaldarriaga}},\ }\href
  {\doibase 10.1088/1475-7516/2012/07/051} {\bibfield  {journal} {\bibinfo
  {journal} {JCAP}\ }\textbf {\bibinfo {volume} {1207}},\ \bibinfo {pages}
  {051} (\bibinfo {year} {2012})},\ \Eprint {http://arxiv.org/abs/1004.2488}
  {arXiv:1004.2488 [astro-ph.CO]} \BibitemShut {NoStop}%
\bibitem [{\citenamefont {Agertz}\ \emph {et~al.}(2007)\citenamefont {Agertz}
  \emph {et~al.}}]{Agertz:2006qb}%
  \BibitemOpen
  \bibfield  {author} {\bibinfo {author} {\bibfnamefont {O.}~\bibnamefont
  {Agertz}} \emph {et~al.},\ }\href {\doibase 10.1111/j.1365-2966.2007.12183.x}
  {\bibfield  {journal} {\bibinfo  {journal} {Mon. Not. Roy. Astron. Soc.}\
  }\textbf {\bibinfo {volume} {380}},\ \bibinfo {pages} {963} (\bibinfo {year}
  {2007})},\ \Eprint {http://arxiv.org/abs/astro-ph/0610051}
  {arXiv:astro-ph/0610051 [astro-ph]} \BibitemShut {NoStop}%
\bibitem [{\citenamefont {{van Daalen}}\ \emph {et~al.}(2011)\citenamefont
  {{van Daalen}}, \citenamefont {{Schaye}}, \citenamefont {{Booth}},\ and\
  \citenamefont {{Dalla Vecchia}}}]{vanDaalen11}%
  \BibitemOpen
  \bibfield  {author} {\bibinfo {author} {\bibfnamefont {M.~P.}\ \bibnamefont
  {{van Daalen}}}, \bibinfo {author} {\bibfnamefont {J.}~\bibnamefont
  {{Schaye}}}, \bibinfo {author} {\bibfnamefont {C.~M.}\ \bibnamefont
  {{Booth}}}, \ and\ \bibinfo {author} {\bibfnamefont {.~C.}\ \bibnamefont
  {{Dalla Vecchia}}},\ }\href {\doibase 10.1111/j.1365-2966.2011.18981.x}
  {\bibfield  {journal} {\bibinfo  {journal} {\mnras}\ }\textbf {\bibinfo
  {volume} {415}},\ \bibinfo {pages} {3649} (\bibinfo {year} {2011})},\ \Eprint
  {http://arxiv.org/abs/1104.1174} {arXiv:1104.1174 [astro-ph.CO]} \BibitemShut
  {NoStop}%
\bibitem [{\citenamefont {{Slosar}}\ \emph {et~al.}(2019)\citenamefont
  {{Slosar}}, \citenamefont {{Davis}}, \citenamefont {{Eisenstein}},
  \citenamefont {{Hlo{\v{z}}ek}}, \citenamefont {{Ishak-Boushaki}},
  \citenamefont {{Mand elbaum}}, \citenamefont {{Marshall}}, \citenamefont
  {{Sakstein}},\ and\ \citenamefont {{White}}}]{Slosar19}%
  \BibitemOpen
  \bibfield  {author} {\bibinfo {author} {\bibfnamefont {A.}~\bibnamefont
  {{Slosar}}}, \bibinfo {author} {\bibfnamefont {T.}~\bibnamefont {{Davis}}},
  \bibinfo {author} {\bibfnamefont {D.}~\bibnamefont {{Eisenstein}}}, \bibinfo
  {author} {\bibfnamefont {R.}~\bibnamefont {{Hlo{\v{z}}ek}}}, \bibinfo
  {author} {\bibfnamefont {M.}~\bibnamefont {{Ishak-Boushaki}}}, \bibinfo
  {author} {\bibfnamefont {R.}~\bibnamefont {{Mand elbaum}}}, \bibinfo {author}
  {\bibfnamefont {P.}~\bibnamefont {{Marshall}}}, \bibinfo {author}
  {\bibfnamefont {J.}~\bibnamefont {{Sakstein}}}, \ and\ \bibinfo {author}
  {\bibfnamefont {M.}~\bibnamefont {{White}}},\ }\href@noop {} {\bibfield
  {journal} {\bibinfo  {journal} {arXiv e-prints}\ ,\ \bibinfo {eid}
  {arXiv:1903.12016}} (\bibinfo {year} {2019})},\ \Eprint
  {http://arxiv.org/abs/1903.12016} {arXiv:1903.12016 [astro-ph.CO]}
  \BibitemShut {NoStop}%
\bibitem [{\citenamefont {{Allison}}\ \emph {et~al.}(2015)\citenamefont
  {{Allison}}, \citenamefont {{Caucal}}, \citenamefont {{Calabrese}},
  \citenamefont {{Dunkley}},\ and\ \citenamefont {{Louis}}}]{Allison15}%
  \BibitemOpen
  \bibfield  {author} {\bibinfo {author} {\bibfnamefont {R.}~\bibnamefont
  {{Allison}}}, \bibinfo {author} {\bibfnamefont {P.}~\bibnamefont {{Caucal}}},
  \bibinfo {author} {\bibfnamefont {E.}~\bibnamefont {{Calabrese}}}, \bibinfo
  {author} {\bibfnamefont {J.}~\bibnamefont {{Dunkley}}}, \ and\ \bibinfo
  {author} {\bibfnamefont {T.}~\bibnamefont {{Louis}}},\ }\href {\doibase
  10.1103/PhysRevD.92.123535} {\bibfield  {journal} {\bibinfo  {journal}
  {\prd}\ }\textbf {\bibinfo {volume} {92}},\ \bibinfo {eid} {123535} (\bibinfo
  {year} {2015})},\ \Eprint {http://arxiv.org/abs/1509.07471} {arXiv:1509.07471
  [astro-ph.CO]} \BibitemShut {NoStop}%
\bibitem [{\citenamefont {{Alvarez}}\ \emph {et~al.}(2014)\citenamefont
  {{Alvarez}}, \citenamefont {{Baldauf}}, \citenamefont {{Bond}}, \citenamefont
  {{Dalal}}, \citenamefont {{de Putter}}, \citenamefont {{Dor{\'e}}},
  \citenamefont {{Green}}, \citenamefont {{Hirata}}, \citenamefont {{Huang}},
  \citenamefont {{Huterer}}, \citenamefont {{Jeong}}, \citenamefont
  {{Johnson}}, \citenamefont {{Krause}}, \citenamefont {{Loverde}},
  \citenamefont {{Meyers}}, \citenamefont {{Meerburg}}, \citenamefont
  {{Senatore}}, \citenamefont {{Shandera}}, \citenamefont {{Silverstein}},
  \citenamefont {{Slosar}}, \citenamefont {{Smith}}, \citenamefont
  {{Zaldarriaga}}, \citenamefont {{Assassi}}, \citenamefont {{Braden}},
  \citenamefont {{Hajian}}, \citenamefont {{Kobayashi}}, \citenamefont
  {{Stein}},\ and\ \citenamefont {{van Engelen}}}]{Alvarez14}%
  \BibitemOpen
  \bibfield  {author} {\bibinfo {author} {\bibfnamefont {M.}~\bibnamefont
  {{Alvarez}}}, \bibinfo {author} {\bibfnamefont {T.}~\bibnamefont
  {{Baldauf}}}, \bibinfo {author} {\bibfnamefont {J.~R.}\ \bibnamefont
  {{Bond}}}, \bibinfo {author} {\bibfnamefont {N.}~\bibnamefont {{Dalal}}},
  \bibinfo {author} {\bibfnamefont {R.}~\bibnamefont {{de Putter}}}, \bibinfo
  {author} {\bibfnamefont {O.}~\bibnamefont {{Dor{\'e}}}}, \bibinfo {author}
  {\bibfnamefont {D.}~\bibnamefont {{Green}}}, \bibinfo {author} {\bibfnamefont
  {C.}~\bibnamefont {{Hirata}}}, \bibinfo {author} {\bibfnamefont
  {Z.}~\bibnamefont {{Huang}}}, \bibinfo {author} {\bibfnamefont
  {D.}~\bibnamefont {{Huterer}}}, \bibinfo {author} {\bibfnamefont
  {D.}~\bibnamefont {{Jeong}}}, \bibinfo {author} {\bibfnamefont {M.~C.}\
  \bibnamefont {{Johnson}}}, \bibinfo {author} {\bibfnamefont {E.}~\bibnamefont
  {{Krause}}}, \bibinfo {author} {\bibfnamefont {M.}~\bibnamefont {{Loverde}}},
  \bibinfo {author} {\bibfnamefont {J.}~\bibnamefont {{Meyers}}}, \bibinfo
  {author} {\bibfnamefont {P.~D.}\ \bibnamefont {{Meerburg}}}, \bibinfo
  {author} {\bibfnamefont {L.}~\bibnamefont {{Senatore}}}, \bibinfo {author}
  {\bibfnamefont {S.}~\bibnamefont {{Shandera}}}, \bibinfo {author}
  {\bibfnamefont {E.}~\bibnamefont {{Silverstein}}}, \bibinfo {author}
  {\bibfnamefont {A.}~\bibnamefont {{Slosar}}}, \bibinfo {author}
  {\bibfnamefont {K.}~\bibnamefont {{Smith}}}, \bibinfo {author} {\bibfnamefont
  {M.}~\bibnamefont {{Zaldarriaga}}}, \bibinfo {author} {\bibfnamefont
  {V.}~\bibnamefont {{Assassi}}}, \bibinfo {author} {\bibfnamefont
  {J.}~\bibnamefont {{Braden}}}, \bibinfo {author} {\bibfnamefont
  {A.}~\bibnamefont {{Hajian}}}, \bibinfo {author} {\bibfnamefont
  {T.}~\bibnamefont {{Kobayashi}}}, \bibinfo {author} {\bibfnamefont
  {G.}~\bibnamefont {{Stein}}}, \ and\ \bibinfo {author} {\bibfnamefont
  {A.}~\bibnamefont {{van Engelen}}},\ }\href@noop {} {\bibfield  {journal}
  {\bibinfo  {journal} {arXiv e-prints}\ ,\ \bibinfo {eid} {arXiv:1412.4671}}
  (\bibinfo {year} {2014})},\ \Eprint {http://arxiv.org/abs/1412.4671}
  {arXiv:1412.4671 [astro-ph.CO]} \BibitemShut {NoStop}%
\bibitem [{\citenamefont {{Meerburg}}\ \emph {et~al.}(2019)\citenamefont
  {{Meerburg}}, \citenamefont {{Green}}, \citenamefont {{Abidi}}, \citenamefont
  {{Amin}}, \citenamefont {{Adshead}}, \citenamefont {{Ahmed}}, \citenamefont
  {{Alonso}}, \citenamefont {{Ansarinejad}}, \citenamefont {{Armstrong}},
  \citenamefont {{Avila}}, \citenamefont {{Baccigalupi}}, \citenamefont
  {{Baldauf}}, \citenamefont {{Ballardini}}, \citenamefont {{Bandura}},
  \citenamefont {{Bartolo}}, \citenamefont {{Battaglia}}, \citenamefont
  {{Baumann}}, \citenamefont {{Bavdhankar}}, \citenamefont {{Bernal}},
  \citenamefont {{Beutler}}, \citenamefont {{Biagetti}}, \citenamefont
  {{Bischoff}}, \citenamefont {{Blazek}}, \citenamefont {{Bond}}, \citenamefont
  {{Borrill}}, \citenamefont {{Bouchet}}, \citenamefont {{Bull}}, \citenamefont
  {{Burgess}}, \citenamefont {{Byrnes}}, \citenamefont {{Calabrese}},
  \citenamefont {{Carlstrom}}, \citenamefont {{Castorina}}, \citenamefont
  {{Challinor}}, \citenamefont {{Chang}}, \citenamefont {{Chaves-Montero}},
  \citenamefont {{Chen}}, \citenamefont {{Yeche}}, \citenamefont {{Cooray}},
  \citenamefont {{Coulton}}, \citenamefont {{Crawford}}, \citenamefont
  {{Chisari}}, \citenamefont {{Cyr-Racine}}, \citenamefont {{D'Amico}},
  \citenamefont {{de Bernardis}}, \citenamefont {{de la Macorra}},
  \citenamefont {{Dor{\'e}}}, \citenamefont {{Duivenvoorden}}, \citenamefont
  {{Dunkley}}, \citenamefont {{Dvorkin}}, \citenamefont {{Eggemeier}},
  \citenamefont {{Escoffier}}, \citenamefont {{Essinger-Hileman}},
  \citenamefont {{Fasiello}}, \citenamefont {{Ferraro}}, \citenamefont
  {{Flauger}}, \citenamefont {{Font-Ribera}}, \citenamefont {{Foreman}},
  \citenamefont {{Friedrich}}, \citenamefont {{Garcia-Bellido}}, \citenamefont
  {{Gerbino}}, \citenamefont {{Gluscevic}}, \citenamefont {{Goon}},
  \citenamefont {{Gorski}}, \citenamefont {{Gudmundsson}}, \citenamefont
  {{Gupta}}, \citenamefont {{Hanany}}, \citenamefont {{Hand ley}},
  \citenamefont {{Hawken}}, \citenamefont {{Hill}}, \citenamefont {{Hirata}},
  \citenamefont {{Hlo{\v{z}}ek}}, \citenamefont {{Holder}}, \citenamefont
  {{Huterer}}, \citenamefont {{Kamionkowski}}, \citenamefont {{Karkare}},
  \citenamefont {{Keeley}}, \citenamefont {{Kinney}}, \citenamefont {{Kisner}},
  \citenamefont {{Kneib}}, \citenamefont {{Knox}}, \citenamefont
  {{Koushiappas}}, \citenamefont {{Kovetz}}, \citenamefont {{Koyama}},
  \citenamefont {{L'Huillier}}, \citenamefont {{Lahav}}, \citenamefont
  {{Lattanzi}}, \citenamefont {{Lee}}, \citenamefont {{Liguori}}, \citenamefont
  {{Loverde}}, \citenamefont {{Madhavacheril}}, \citenamefont {{Maldacena}},
  \citenamefont {{Marsh}}, \citenamefont {{Masui}}, \citenamefont
  {{Matarrese}}, \citenamefont {{McAllister}}, \citenamefont {{McMahon}},
  \citenamefont {{McQuinn}}, \citenamefont {{Meyers}}, \citenamefont
  {{Mirbabayi}}, \citenamefont {{Moradinezhad Dizgah}}, \citenamefont
  {{Motloch}}, \citenamefont {{Mukherjee}}, \citenamefont {{Mu{\~n}oz}},
  \citenamefont {{Myers}}, \citenamefont {{Nagy}}, \citenamefont {{Naselsky}},
  \citenamefont {{Nati}}, \citenamefont {{Newburgh}}, \citenamefont
  {{Nicolis}}, \citenamefont {{Niemack}}, \citenamefont {{Niz}}, \citenamefont
  {{Nomerotski}}, \citenamefont {{Page}}, \citenamefont {{Pajer}},
  \citenamefont {{Padmanabhan}}, \citenamefont {{Palma}}, \citenamefont
  {{Peiris}}, \citenamefont {{Percival}}, \citenamefont {{Piacentni}},
  \citenamefont {{Pimentel}}, \citenamefont {{Pogosian}}, \citenamefont
  {{Prescod-Weinstein}}, \citenamefont {{Pryke}}, \citenamefont {{Puglisi}},
  \citenamefont {{Racine}}, \citenamefont {{Stompor}}, \citenamefont
  {{Raveri}}, \citenamefont {{Remazeilles}}, \citenamefont {{Rocha}},
  \citenamefont {{Ross}}, \citenamefont {{Rossi}}, \citenamefont {{Ruhl}},
  \citenamefont {{Sasaki}}, \citenamefont {{Schaan}}, \citenamefont
  {{Schillaci}}, \citenamefont {{Schmittfull}}, \citenamefont {{Sehgal}},
  \citenamefont {{Senatore}}, \citenamefont {{Seo}}, \citenamefont {{Shan}},
  \citenamefont {{Shandera}}, \citenamefont {{Sherwin}}, \citenamefont
  {{Silverstein}}, \citenamefont {{Simon}}, \citenamefont {{Slosar}},
  \citenamefont {{Staggs}}, \citenamefont {{Starkman}}, \citenamefont
  {{Stebbins}}, \citenamefont {{Suzuki}}, \citenamefont {{Switzer}},
  \citenamefont {{Timbie}}, \citenamefont {{Tolley}}, \citenamefont {{Tomasi}},
  \citenamefont {{Tristram}}, \citenamefont {{Trodden}}, \citenamefont
  {{Tsai}}, \citenamefont {{Uhlemann}}, \citenamefont {{Umilta}}, \citenamefont
  {{van Engelen}}, \citenamefont {{Vargas-Maga{\~n}a}}, \citenamefont
  {{Vieregg}}, \citenamefont {{Wallisch}}, \citenamefont {{Wands}},
  \citenamefont {{Wandelt}}, \citenamefont {{Wang}}, \citenamefont {{Watson}},
  \citenamefont {{Wise}}, \citenamefont {{Wu}}, \citenamefont {{Xianyu}},
  \citenamefont {{Xu}}, \citenamefont {{Yasini}}, \citenamefont {{Young}},
  \citenamefont {{Yutong}}, \citenamefont {{Zaldarriaga}}, \citenamefont
  {{Zemcov}}, \citenamefont {{Zhao}}, \citenamefont {{Zheng}},\ and\
  \citenamefont {{Zhu}}}]{Meerburg19}%
  \BibitemOpen
  \bibfield  {author} {\bibinfo {author} {\bibfnamefont {P.~D.}\ \bibnamefont
  {{Meerburg}}}, \bibinfo {author} {\bibfnamefont {D.}~\bibnamefont {{Green}}},
  \bibinfo {author} {\bibfnamefont {M.}~\bibnamefont {{Abidi}}}, \bibinfo
  {author} {\bibfnamefont {M.~A.}\ \bibnamefont {{Amin}}}, \bibinfo {author}
  {\bibfnamefont {P.}~\bibnamefont {{Adshead}}}, \bibinfo {author}
  {\bibfnamefont {Z.}~\bibnamefont {{Ahmed}}}, \bibinfo {author} {\bibfnamefont
  {D.}~\bibnamefont {{Alonso}}}, \bibinfo {author} {\bibfnamefont
  {B.}~\bibnamefont {{Ansarinejad}}}, \bibinfo {author} {\bibfnamefont
  {R.}~\bibnamefont {{Armstrong}}}, \bibinfo {author} {\bibfnamefont
  {S.}~\bibnamefont {{Avila}}}, \bibinfo {author} {\bibfnamefont
  {C.}~\bibnamefont {{Baccigalupi}}}, \bibinfo {author} {\bibfnamefont
  {T.}~\bibnamefont {{Baldauf}}}, \bibinfo {author} {\bibfnamefont
  {M.}~\bibnamefont {{Ballardini}}}, \bibinfo {author} {\bibfnamefont
  {K.}~\bibnamefont {{Bandura}}}, \bibinfo {author} {\bibfnamefont
  {N.}~\bibnamefont {{Bartolo}}}, \bibinfo {author} {\bibfnamefont
  {N.}~\bibnamefont {{Battaglia}}}, \bibinfo {author} {\bibfnamefont
  {D.}~\bibnamefont {{Baumann}}}, \bibinfo {author} {\bibfnamefont
  {C.}~\bibnamefont {{Bavdhankar}}}, \bibinfo {author} {\bibfnamefont {J.~L.}\
  \bibnamefont {{Bernal}}}, \bibinfo {author} {\bibfnamefont {F.}~\bibnamefont
  {{Beutler}}}, \bibinfo {author} {\bibfnamefont {M.}~\bibnamefont
  {{Biagetti}}}, \bibinfo {author} {\bibfnamefont {C.}~\bibnamefont
  {{Bischoff}}}, \bibinfo {author} {\bibfnamefont {J.}~\bibnamefont
  {{Blazek}}}, \bibinfo {author} {\bibfnamefont {J.~R.}\ \bibnamefont
  {{Bond}}}, \bibinfo {author} {\bibfnamefont {J.}~\bibnamefont {{Borrill}}},
  \bibinfo {author} {\bibfnamefont {F.~R.}\ \bibnamefont {{Bouchet}}}, \bibinfo
  {author} {\bibfnamefont {P.}~\bibnamefont {{Bull}}}, \bibinfo {author}
  {\bibfnamefont {C.}~\bibnamefont {{Burgess}}}, \bibinfo {author}
  {\bibfnamefont {C.}~\bibnamefont {{Byrnes}}}, \bibinfo {author}
  {\bibfnamefont {E.}~\bibnamefont {{Calabrese}}}, \bibinfo {author}
  {\bibfnamefont {J.~E.}\ \bibnamefont {{Carlstrom}}}, \bibinfo {author}
  {\bibfnamefont {E.}~\bibnamefont {{Castorina}}}, \bibinfo {author}
  {\bibfnamefont {A.}~\bibnamefont {{Challinor}}}, \bibinfo {author}
  {\bibfnamefont {T.-C.}\ \bibnamefont {{Chang}}}, \bibinfo {author}
  {\bibfnamefont {J.}~\bibnamefont {{Chaves-Montero}}}, \bibinfo {author}
  {\bibfnamefont {X.}~\bibnamefont {{Chen}}}, \bibinfo {author} {\bibfnamefont
  {C.}~\bibnamefont {{Yeche}}}, \bibinfo {author} {\bibfnamefont
  {A.}~\bibnamefont {{Cooray}}}, \bibinfo {author} {\bibfnamefont
  {W.}~\bibnamefont {{Coulton}}}, \bibinfo {author} {\bibfnamefont
  {T.}~\bibnamefont {{Crawford}}}, \bibinfo {author} {\bibfnamefont
  {E.}~\bibnamefont {{Chisari}}}, \bibinfo {author} {\bibfnamefont {F.-Y.}\
  \bibnamefont {{Cyr-Racine}}}, \bibinfo {author} {\bibfnamefont
  {G.}~\bibnamefont {{D'Amico}}}, \bibinfo {author} {\bibfnamefont
  {P.}~\bibnamefont {{de Bernardis}}}, \bibinfo {author} {\bibfnamefont
  {A.}~\bibnamefont {{de la Macorra}}}, \bibinfo {author} {\bibfnamefont
  {O.}~\bibnamefont {{Dor{\'e}}}}, \bibinfo {author} {\bibfnamefont
  {A.}~\bibnamefont {{Duivenvoorden}}}, \bibinfo {author} {\bibfnamefont
  {J.}~\bibnamefont {{Dunkley}}}, \bibinfo {author} {\bibfnamefont
  {C.}~\bibnamefont {{Dvorkin}}}, \bibinfo {author} {\bibfnamefont
  {A.}~\bibnamefont {{Eggemeier}}}, \bibinfo {author} {\bibfnamefont
  {S.}~\bibnamefont {{Escoffier}}}, \bibinfo {author} {\bibfnamefont
  {T.}~\bibnamefont {{Essinger-Hileman}}}, \bibinfo {author} {\bibfnamefont
  {M.}~\bibnamefont {{Fasiello}}}, \bibinfo {author} {\bibfnamefont
  {S.}~\bibnamefont {{Ferraro}}}, \bibinfo {author} {\bibfnamefont
  {R.}~\bibnamefont {{Flauger}}}, \bibinfo {author} {\bibfnamefont
  {A.}~\bibnamefont {{Font-Ribera}}}, \bibinfo {author} {\bibfnamefont
  {S.}~\bibnamefont {{Foreman}}}, \bibinfo {author} {\bibfnamefont
  {O.}~\bibnamefont {{Friedrich}}}, \bibinfo {author} {\bibfnamefont
  {J.}~\bibnamefont {{Garcia-Bellido}}}, \bibinfo {author} {\bibfnamefont
  {M.}~\bibnamefont {{Gerbino}}}, \bibinfo {author} {\bibfnamefont
  {V.}~\bibnamefont {{Gluscevic}}}, \bibinfo {author} {\bibfnamefont
  {G.}~\bibnamefont {{Goon}}}, \bibinfo {author} {\bibfnamefont {K.~M.}\
  \bibnamefont {{Gorski}}}, \bibinfo {author} {\bibfnamefont {J.~E.}\
  \bibnamefont {{Gudmundsson}}}, \bibinfo {author} {\bibfnamefont
  {N.}~\bibnamefont {{Gupta}}}, \bibinfo {author} {\bibfnamefont
  {S.}~\bibnamefont {{Hanany}}}, \bibinfo {author} {\bibfnamefont
  {W.}~\bibnamefont {{Hand ley}}}, \bibinfo {author} {\bibfnamefont {A.~J.}\
  \bibnamefont {{Hawken}}}, \bibinfo {author} {\bibfnamefont {J.~C.}\
  \bibnamefont {{Hill}}}, \bibinfo {author} {\bibfnamefont {C.~M.}\
  \bibnamefont {{Hirata}}}, \bibinfo {author} {\bibfnamefont {R.}~\bibnamefont
  {{Hlo{\v{z}}ek}}}, \bibinfo {author} {\bibfnamefont {G.}~\bibnamefont
  {{Holder}}}, \bibinfo {author} {\bibfnamefont {D.}~\bibnamefont {{Huterer}}},
  \bibinfo {author} {\bibfnamefont {M.}~\bibnamefont {{Kamionkowski}}},
  \bibinfo {author} {\bibfnamefont {K.~S.}\ \bibnamefont {{Karkare}}}, \bibinfo
  {author} {\bibfnamefont {R.~E.}\ \bibnamefont {{Keeley}}}, \bibinfo {author}
  {\bibfnamefont {W.}~\bibnamefont {{Kinney}}}, \bibinfo {author}
  {\bibfnamefont {T.}~\bibnamefont {{Kisner}}}, \bibinfo {author}
  {\bibfnamefont {J.-P.}\ \bibnamefont {{Kneib}}}, \bibinfo {author}
  {\bibfnamefont {L.}~\bibnamefont {{Knox}}}, \bibinfo {author} {\bibfnamefont
  {S.~M.}\ \bibnamefont {{Koushiappas}}}, \bibinfo {author} {\bibfnamefont
  {E.~D.}\ \bibnamefont {{Kovetz}}}, \bibinfo {author} {\bibfnamefont
  {K.}~\bibnamefont {{Koyama}}}, \bibinfo {author} {\bibfnamefont
  {B.}~\bibnamefont {{L'Huillier}}}, \bibinfo {author} {\bibfnamefont
  {O.}~\bibnamefont {{Lahav}}}, \bibinfo {author} {\bibfnamefont
  {M.}~\bibnamefont {{Lattanzi}}}, \bibinfo {author} {\bibfnamefont
  {H.}~\bibnamefont {{Lee}}}, \bibinfo {author} {\bibfnamefont
  {M.}~\bibnamefont {{Liguori}}}, \bibinfo {author} {\bibfnamefont
  {M.}~\bibnamefont {{Loverde}}}, \bibinfo {author} {\bibfnamefont
  {M.}~\bibnamefont {{Madhavacheril}}}, \bibinfo {author} {\bibfnamefont
  {J.}~\bibnamefont {{Maldacena}}}, \bibinfo {author} {\bibfnamefont
  {M.~C.~D.}\ \bibnamefont {{Marsh}}}, \bibinfo {author} {\bibfnamefont
  {K.}~\bibnamefont {{Masui}}}, \bibinfo {author} {\bibfnamefont
  {S.}~\bibnamefont {{Matarrese}}}, \bibinfo {author} {\bibfnamefont
  {L.}~\bibnamefont {{McAllister}}}, \bibinfo {author} {\bibfnamefont
  {J.}~\bibnamefont {{McMahon}}}, \bibinfo {author} {\bibfnamefont
  {M.}~\bibnamefont {{McQuinn}}}, \bibinfo {author} {\bibfnamefont
  {J.}~\bibnamefont {{Meyers}}}, \bibinfo {author} {\bibfnamefont
  {M.}~\bibnamefont {{Mirbabayi}}}, \bibinfo {author} {\bibfnamefont
  {A.}~\bibnamefont {{Moradinezhad Dizgah}}}, \bibinfo {author} {\bibfnamefont
  {P.}~\bibnamefont {{Motloch}}}, \bibinfo {author} {\bibfnamefont
  {S.}~\bibnamefont {{Mukherjee}}}, \bibinfo {author} {\bibfnamefont {J.~B.}\
  \bibnamefont {{Mu{\~n}oz}}}, \bibinfo {author} {\bibfnamefont {A.~D.}\
  \bibnamefont {{Myers}}}, \bibinfo {author} {\bibfnamefont {J.}~\bibnamefont
  {{Nagy}}}, \bibinfo {author} {\bibfnamefont {P.}~\bibnamefont {{Naselsky}}},
  \bibinfo {author} {\bibfnamefont {F.}~\bibnamefont {{Nati}}}, \bibinfo
  {author} {\bibnamefont {{Newburgh}}}, \bibinfo {author} {\bibfnamefont
  {A.}~\bibnamefont {{Nicolis}}}, \bibinfo {author} {\bibfnamefont {M.~D.}\
  \bibnamefont {{Niemack}}}, \bibinfo {author} {\bibfnamefont {G.}~\bibnamefont
  {{Niz}}}, \bibinfo {author} {\bibfnamefont {A.}~\bibnamefont {{Nomerotski}}},
  \bibinfo {author} {\bibfnamefont {L.}~\bibnamefont {{Page}}}, \bibinfo
  {author} {\bibfnamefont {E.}~\bibnamefont {{Pajer}}}, \bibinfo {author}
  {\bibfnamefont {H.}~\bibnamefont {{Padmanabhan}}}, \bibinfo {author}
  {\bibfnamefont {G.~A.}\ \bibnamefont {{Palma}}}, \bibinfo {author}
  {\bibfnamefont {H.~V.}\ \bibnamefont {{Peiris}}}, \bibinfo {author}
  {\bibfnamefont {W.~J.}\ \bibnamefont {{Percival}}}, \bibinfo {author}
  {\bibfnamefont {F.}~\bibnamefont {{Piacentni}}}, \bibinfo {author}
  {\bibfnamefont {G.~L.}\ \bibnamefont {{Pimentel}}}, \bibinfo {author}
  {\bibfnamefont {L.}~\bibnamefont {{Pogosian}}}, \bibinfo {author}
  {\bibfnamefont {C.}~\bibnamefont {{Prescod-Weinstein}}}, \bibinfo {author}
  {\bibfnamefont {C.}~\bibnamefont {{Pryke}}}, \bibinfo {author} {\bibfnamefont
  {G.}~\bibnamefont {{Puglisi}}}, \bibinfo {author} {\bibfnamefont
  {B.}~\bibnamefont {{Racine}}}, \bibinfo {author} {\bibfnamefont
  {R.}~\bibnamefont {{Stompor}}}, \bibinfo {author} {\bibfnamefont
  {M.}~\bibnamefont {{Raveri}}}, \bibinfo {author} {\bibfnamefont
  {M.}~\bibnamefont {{Remazeilles}}}, \bibinfo {author} {\bibfnamefont
  {G.}~\bibnamefont {{Rocha}}}, \bibinfo {author} {\bibfnamefont {A.~J.}\
  \bibnamefont {{Ross}}}, \bibinfo {author} {\bibfnamefont {G.}~\bibnamefont
  {{Rossi}}}, \bibinfo {author} {\bibfnamefont {J.}~\bibnamefont {{Ruhl}}},
  \bibinfo {author} {\bibfnamefont {M.}~\bibnamefont {{Sasaki}}}, \bibinfo
  {author} {\bibfnamefont {E.}~\bibnamefont {{Schaan}}}, \bibinfo {author}
  {\bibfnamefont {A.}~\bibnamefont {{Schillaci}}}, \bibinfo {author}
  {\bibfnamefont {M.}~\bibnamefont {{Schmittfull}}}, \bibinfo {author}
  {\bibfnamefont {N.}~\bibnamefont {{Sehgal}}}, \bibinfo {author}
  {\bibfnamefont {L.}~\bibnamefont {{Senatore}}}, \bibinfo {author}
  {\bibfnamefont {H.-J.}\ \bibnamefont {{Seo}}}, \bibinfo {author}
  {\bibfnamefont {H.}~\bibnamefont {{Shan}}}, \bibinfo {author} {\bibfnamefont
  {S.}~\bibnamefont {{Shandera}}}, \bibinfo {author} {\bibfnamefont {B.~D.}\
  \bibnamefont {{Sherwin}}}, \bibinfo {author} {\bibfnamefont {E.}~\bibnamefont
  {{Silverstein}}}, \bibinfo {author} {\bibfnamefont {S.}~\bibnamefont
  {{Simon}}}, \bibinfo {author} {\bibfnamefont {A.}~\bibnamefont {{Slosar}}},
  \bibinfo {author} {\bibfnamefont {S.}~\bibnamefont {{Staggs}}}, \bibinfo
  {author} {\bibfnamefont {G.}~\bibnamefont {{Starkman}}}, \bibinfo {author}
  {\bibfnamefont {A.}~\bibnamefont {{Stebbins}}}, \bibinfo {author}
  {\bibfnamefont {A.}~\bibnamefont {{Suzuki}}}, \bibinfo {author}
  {\bibfnamefont {E.~R.}\ \bibnamefont {{Switzer}}}, \bibinfo {author}
  {\bibfnamefont {P.}~\bibnamefont {{Timbie}}}, \bibinfo {author}
  {\bibfnamefont {A.~J.}\ \bibnamefont {{Tolley}}}, \bibinfo {author}
  {\bibfnamefont {M.}~\bibnamefont {{Tomasi}}}, \bibinfo {author}
  {\bibfnamefont {M.}~\bibnamefont {{Tristram}}}, \bibinfo {author}
  {\bibfnamefont {M.}~\bibnamefont {{Trodden}}}, \bibinfo {author}
  {\bibfnamefont {Y.-D.}\ \bibnamefont {{Tsai}}}, \bibinfo {author}
  {\bibfnamefont {C.}~\bibnamefont {{Uhlemann}}}, \bibinfo {author}
  {\bibfnamefont {C.}~\bibnamefont {{Umilta}}}, \bibinfo {author}
  {\bibfnamefont {A.}~\bibnamefont {{van Engelen}}}, \bibinfo {author}
  {\bibfnamefont {M.}~\bibnamefont {{Vargas-Maga{\~n}a}}}, \bibinfo {author}
  {\bibfnamefont {A.}~\bibnamefont {{Vieregg}}}, \bibinfo {author}
  {\bibfnamefont {B.}~\bibnamefont {{Wallisch}}}, \bibinfo {author}
  {\bibfnamefont {D.}~\bibnamefont {{Wands}}}, \bibinfo {author} {\bibfnamefont
  {B.}~\bibnamefont {{Wandelt}}}, \bibinfo {author} {\bibfnamefont
  {Y.}~\bibnamefont {{Wang}}}, \bibinfo {author} {\bibfnamefont
  {S.}~\bibnamefont {{Watson}}}, \bibinfo {author} {\bibfnamefont
  {M.}~\bibnamefont {{Wise}}}, \bibinfo {author} {\bibfnamefont {W.~L.~K.}\
  \bibnamefont {{Wu}}}, \bibinfo {author} {\bibfnamefont {Z.-Z.}\ \bibnamefont
  {{Xianyu}}}, \bibinfo {author} {\bibfnamefont {W.}~\bibnamefont {{Xu}}},
  \bibinfo {author} {\bibfnamefont {S.}~\bibnamefont {{Yasini}}}, \bibinfo
  {author} {\bibfnamefont {S.}~\bibnamefont {{Young}}}, \bibinfo {author}
  {\bibfnamefont {D.}~\bibnamefont {{Yutong}}}, \bibinfo {author}
  {\bibfnamefont {M.}~\bibnamefont {{Zaldarriaga}}}, \bibinfo {author}
  {\bibfnamefont {M.}~\bibnamefont {{Zemcov}}}, \bibinfo {author}
  {\bibfnamefont {G.-B.}\ \bibnamefont {{Zhao}}}, \bibinfo {author}
  {\bibfnamefont {Y.}~\bibnamefont {{Zheng}}}, \ and\ \bibinfo {author}
  {\bibfnamefont {N.}~\bibnamefont {{Zhu}}},\ }\href@noop {} {\bibfield
  {journal} {\bibinfo  {journal} {arXiv e-prints}\ ,\ \bibinfo {eid}
  {arXiv:1903.04409}} (\bibinfo {year} {2019})},\ \Eprint
  {http://arxiv.org/abs/1903.04409} {arXiv:1903.04409 [astro-ph.CO]}
  \BibitemShut {NoStop}%
\bibitem [{\citenamefont {{Huterer}}\ and\ \citenamefont
  {{Takada}}(2005)}]{Huterer05}%
  \BibitemOpen
  \bibfield  {author} {\bibinfo {author} {\bibfnamefont {D.}~\bibnamefont
  {{Huterer}}}\ and\ \bibinfo {author} {\bibfnamefont {M.}~\bibnamefont
  {{Takada}}},\ }\href {\doibase 10.1016/j.astropartphys.2005.02.006}
  {\bibfield  {journal} {\bibinfo  {journal} {Astroparticle Physics}\ }\textbf
  {\bibinfo {volume} {23}},\ \bibinfo {pages} {369} (\bibinfo {year} {2005})},\
  \Eprint {http://arxiv.org/abs/astro-ph/0412142} {astro-ph/0412142}
  \BibitemShut {NoStop}%
\bibitem [{\citenamefont {{Hearin}}\ \emph {et~al.}(2012)\citenamefont
  {{Hearin}}, \citenamefont {{Zentner}},\ and\ \citenamefont
  {{Ma}}}]{Hearin12}%
  \BibitemOpen
  \bibfield  {author} {\bibinfo {author} {\bibfnamefont {A.~P.}\ \bibnamefont
  {{Hearin}}}, \bibinfo {author} {\bibfnamefont {A.~R.}\ \bibnamefont
  {{Zentner}}}, \ and\ \bibinfo {author} {\bibfnamefont {Z.}~\bibnamefont
  {{Ma}}},\ }\href {\doibase 10.1088/1475-7516/2012/04/034} {\bibfield
  {journal} {\bibinfo  {journal} {\jcap}\ }\textbf {\bibinfo {volume} {4}},\
  \bibinfo {eid} {034} (\bibinfo {year} {2012})},\ \Eprint
  {http://arxiv.org/abs/1111.0052} {arXiv:1111.0052} \BibitemShut {NoStop}%
\bibitem [{\citenamefont {{Chisari}}\ \emph {et~al.}(2018)\citenamefont
  {{Chisari}}, \citenamefont {{Richardson}}, \citenamefont {{Devriendt}},
  \citenamefont {{Dubois}}, \citenamefont {{Schneider}}, \citenamefont {{Le
  Brun}}, \citenamefont {{Beckmann}}, \citenamefont {{Peirani}}, \citenamefont
  {{Slyz}},\ and\ \citenamefont {{Pichon}}}]{Chisari18}%
  \BibitemOpen
  \bibfield  {author} {\bibinfo {author} {\bibfnamefont {N.~E.}\ \bibnamefont
  {{Chisari}}}, \bibinfo {author} {\bibfnamefont {M.~L.~A.}\ \bibnamefont
  {{Richardson}}}, \bibinfo {author} {\bibfnamefont {J.}~\bibnamefont
  {{Devriendt}}}, \bibinfo {author} {\bibfnamefont {Y.}~\bibnamefont
  {{Dubois}}}, \bibinfo {author} {\bibfnamefont {A.}~\bibnamefont
  {{Schneider}}}, \bibinfo {author} {\bibfnamefont {A.~M.~C.}\ \bibnamefont
  {{Le Brun}}}, \bibinfo {author} {\bibfnamefont {R.~S.}\ \bibnamefont
  {{Beckmann}}}, \bibinfo {author} {\bibfnamefont {S.}~\bibnamefont
  {{Peirani}}}, \bibinfo {author} {\bibfnamefont {A.}~\bibnamefont {{Slyz}}}, \
  and\ \bibinfo {author} {\bibfnamefont {C.}~\bibnamefont {{Pichon}}},\ }\href
  {\doibase 10.1093/mnras/sty2093} {\bibfield  {journal} {\bibinfo  {journal}
  {\mnras}\ }\textbf {\bibinfo {volume} {480}},\ \bibinfo {pages} {3962}
  (\bibinfo {year} {2018})},\ \Eprint {http://arxiv.org/abs/1801.08559}
  {arXiv:1801.08559} \BibitemShut {NoStop}%
\bibitem [{\citenamefont {{Semboloni}}\ \emph {et~al.}(2011)\citenamefont
  {{Semboloni}}, \citenamefont {{Hoekstra}}, \citenamefont {{Schaye}},\ and\
  \citenamefont {{van Daalen}}}]{Semboloni11}%
  \BibitemOpen
  \bibfield  {author} {\bibinfo {author} {\bibfnamefont {E.}~\bibnamefont
  {{Semboloni}}}, \bibinfo {author} {\bibfnamefont {H.}~\bibnamefont
  {{Hoekstra}}}, \bibinfo {author} {\bibfnamefont {J.}~\bibnamefont
  {{Schaye}}}, \ and\ \bibinfo {author} {\bibfnamefont {I.~G.}\ \bibnamefont
  {{van Daalen}}, \bibfnamefont {M.~P. a\ nd~{McCarthy}}},\ }\href {\doibase
  10.1111/j.1365-2966.2011.19385.x} {\bibfield  {journal} {\bibinfo  {journal}
  {\mnras}\ }\textbf {\bibinfo {volume} {417}},\ \bibinfo {pages} {2020}
  (\bibinfo {year} {2011})},\ \Eprint {http://arxiv.org/abs/1105.1075}
  {arXiv:1105.1075} \BibitemShut {NoStop}%
\bibitem [{\citenamefont {{Eifler}}\ \emph {et~al.}(2015)\citenamefont
  {{Eifler}}, \citenamefont {{Krause}}, \citenamefont {{Dodelson}},
  \citenamefont {{Zentner}}, \citenamefont {{He\ arin}},\ and\ \citenamefont
  {{Gnedin}}}]{Eifler15}%
  \BibitemOpen
  \bibfield  {author} {\bibinfo {author} {\bibfnamefont {T.}~\bibnamefont
  {{Eifler}}}, \bibinfo {author} {\bibfnamefont {E.}~\bibnamefont {{Krause}}},
  \bibinfo {author} {\bibfnamefont {S.}~\bibnamefont {{Dodelson}}}, \bibinfo
  {author} {\bibfnamefont {A.~R.}\ \bibnamefont {{Zentner}}}, \bibinfo {author}
  {\bibfnamefont {A.~P.}\ \bibnamefont {{He\ arin}}}, \ and\ \bibinfo {author}
  {\bibfnamefont {N.~Y.}\ \bibnamefont {{Gnedin}}},\ }\href {\doibase
  10.1093/mnras/stv2000} {\bibfield  {journal} {\bibinfo  {journal} {\mnras}\
  }\textbf {\bibinfo {volume} {454}},\ \bibinfo {pages} {2451} (\bibinfo {year}
  {2015})},\ \Eprint {http://arxiv.org/abs/1405.7423} {arXiv:1405.7423}
  \BibitemShut {NoStop}%
\bibitem [{\citenamefont {{Huang}}\ \emph {et~al.}(2018)\citenamefont
  {{Huang}}, \citenamefont {{Eifler}}, \citenamefont {{Mandelbaum}},\ and\
  \citenamefont {{Dodelson}}}]{Huang19}%
  \BibitemOpen
  \bibfield  {author} {\bibinfo {author} {\bibfnamefont {H.-J.}\ \bibnamefont
  {{Huang}}}, \bibinfo {author} {\bibfnamefont {T.}~\bibnamefont {{Eifler}}},
  \bibinfo {author} {\bibfnamefont {R.}~\bibnamefont {{Mandelbaum}}}, \ and\
  \bibinfo {author} {\bibfnamefont {S.}~\bibnamefont {{Dodelson}}},\
  }\href@noop {} {\bibfield  {journal} {\bibinfo  {journal} {arXiv e-prints}\
  ,\ \bibinfo {eid} {arXiv:1809.01146}} (\bibinfo {year} {2018})},\ \Eprint
  {http://arxiv.org/abs/1809.01146} {arXiv:1809.01146 [astro-ph.CO]}
  \BibitemShut {NoStop}%
\bibitem [{\citenamefont {{Chevallier}}\ and\ \citenamefont
  {{Polarski}}(2001)}]{Chevallier01}%
  \BibitemOpen
  \bibfield  {author} {\bibinfo {author} {\bibfnamefont {M.}~\bibnamefont
  {{Chevallier}}}\ and\ \bibinfo {author} {\bibfnamefont {D.}~\bibnamefont
  {{Polarski}}},\ }\href {\doibase 10.1142/S0218271801000822} {\bibfield
  {journal} {\bibinfo  {journal} {International Journal of Modern Physics D}\
  }\textbf {\bibinfo {volume} {10}},\ \bibinfo {pages} {213} (\bibinfo {year}
  {2001})},\ \Eprint {http://arxiv.org/abs/gr-qc/0009008} {gr-qc/0009008}
  \BibitemShut {NoStop}%
\bibitem [{\citenamefont {{Linder}}(2003)}]{Linder03}%
  \BibitemOpen
  \bibfield  {author} {\bibinfo {author} {\bibfnamefont {E.~V.}\ \bibnamefont
  {{Linder}}},\ }\href {\doibase 10.1103/PhysRevLett.90.091301} {\bibfield
  {journal} {\bibinfo  {journal} {Physical Review Letters}\ }\textbf {\bibinfo
  {volume} {90}},\ \bibinfo {eid} {091301} (\bibinfo {year} {2003})},\ \Eprint
  {http://arxiv.org/abs/astro-ph/0208512} {astro-ph/0208512} \BibitemShut
  {NoStop}%
\bibitem [{\citenamefont {{Mead}}\ \emph {et~al.}(2015)\citenamefont {{Mead}},
  \citenamefont {{Peacock}}, \citenamefont {{Heymans}}, \citenamefont
  {{Joudaki}},\ and\ \citenamefont {{H\ eavens}}}]{Mead15}%
  \BibitemOpen
  \bibfield  {author} {\bibinfo {author} {\bibfnamefont {A.~J.}\ \bibnamefont
  {{Mead}}}, \bibinfo {author} {\bibfnamefont {J.~A.}\ \bibnamefont
  {{Peacock}}}, \bibinfo {author} {\bibfnamefont {C.}~\bibnamefont
  {{Heymans}}}, \bibinfo {author} {\bibfnamefont {S.}~\bibnamefont
  {{Joudaki}}}, \ and\ \bibinfo {author} {\bibfnamefont {A.~F.}\ \bibnamefont
  {{H\ eavens}}},\ }\href {\doibase 10.1093/mnras/stv2036} {\bibfield
  {journal} {\bibinfo  {journal} {\mnras}\ }\textbf {\bibinfo {volume} {454}},\
  \bibinfo {pages} {1958} (\bibinfo {year} {2015})},\ \Eprint
  {http://arxiv.org/abs/1505.07833} {arXiv:1505.07833} \BibitemShut {NoStop}%
\bibitem [{\citenamefont {{Harnois-D{\'e}raps}}\ \emph
  {et~al.}(2015)\citenamefont {{Harnois-D{\'e}raps}}, \citenamefont {{van
  Waerbeke}}, \citenamefont {{Viola}},\ and\ \citenamefont
  {{Heymans}}}]{Harnois15}%
  \BibitemOpen
  \bibfield  {author} {\bibinfo {author} {\bibfnamefont {J.}~\bibnamefont
  {{Harnois-D{\'e}raps}}}, \bibinfo {author} {\bibfnamefont {L.}~\bibnamefont
  {{van Waerbeke}}}, \bibinfo {author} {\bibfnamefont {M.}~\bibnamefont
  {{Viola}}}, \ and\ \bibinfo {author} {\bibfnamefont {C.}~\bibnamefont
  {{Heymans}}},\ }\href {\doibase 10.1093/mnras/stv646} {\bibfield  {journal}
  {\bibinfo  {journal} {\mnras}\ }\textbf {\bibinfo {volume} {450}},\ \bibinfo
  {pages} {1212} (\bibinfo {year} {2015})},\ \Eprint
  {http://arxiv.org/abs/1407.4301} {arXiv:1407.4301} \BibitemShut {NoStop}%
\bibitem [{\citenamefont {{Schneider}}\ and\ \citenamefont
  {{Teyssier}}(2015)}]{Schneider15}%
  \BibitemOpen
  \bibfield  {author} {\bibinfo {author} {\bibfnamefont {A.}~\bibnamefont
  {{Schneider}}}\ and\ \bibinfo {author} {\bibfnamefont {R.}~\bibnamefont
  {{Teyssier}}},\ }\href {\doibase 10.1088/1475-7516/2015/12/049} {\bibfield
  {journal} {\bibinfo  {journal} {\jcap}\ }\textbf {\bibinfo {volume} {12}},\
  \bibinfo {eid} {049} (\bibinfo {year} {2015})},\ \Eprint
  {http://arxiv.org/abs/1510.06034} {arXiv:1510.06034} \BibitemShut {NoStop}%
\bibitem [{\citenamefont {{Hildebrandt}}\ \emph {et~al.}(2018)\citenamefont
  {{Hildebrandt}}, \citenamefont {{K{\"o}hlinger}}, \citenamefont {{van den
  Busch}}, \citenamefont {{Joachimi}}, \citenamefont {{Heymans}}, \citenamefont
  {{Kannawadi}}, \citenamefont {{Wright}}, \citenamefont {{Asgari}},
  \citenamefont {{Blake}}, \citenamefont {{Hoekstra}}, \citenamefont
  {{Joudaki}}, \citenamefont {{Kuijken}}, \citenamefont {{Miller}},
  \citenamefont {{Morrison}}, \citenamefont {{Tr{\"o}ster}}, \citenamefont
  {{Amon}}, \citenamefont {{Archidiacono}}, \citenamefont {{Brieden}},
  \citenamefont {{Choi}}, \citenamefont {{de Jong}}, \citenamefont {{Erben}},
  \citenamefont {{Giblin}}, \citenamefont {{Mead}}, \citenamefont {{Peacock}},
  \citenamefont {{Radovich}}, \citenamefont {{Schneider}}, \citenamefont
  {{Sif{\'o}n}},\ and\ \citenamefont {{Tewes}}}]{Hildebrandt19}%
  \BibitemOpen
  \bibfield  {author} {\bibinfo {author} {\bibfnamefont {H.}~\bibnamefont
  {{Hildebrandt}}}, \bibinfo {author} {\bibfnamefont {F.}~\bibnamefont
  {{K{\"o}hlinger}}}, \bibinfo {author} {\bibfnamefont {J.~L.}\ \bibnamefont
  {{van den Busch}}}, \bibinfo {author} {\bibfnamefont {B.}~\bibnamefont
  {{Joachimi}}}, \bibinfo {author} {\bibfnamefont {C.}~\bibnamefont
  {{Heymans}}}, \bibinfo {author} {\bibfnamefont {A.}~\bibnamefont
  {{Kannawadi}}}, \bibinfo {author} {\bibfnamefont {A.~H.}\ \bibnamefont
  {{Wright}}}, \bibinfo {author} {\bibfnamefont {M.}~\bibnamefont {{Asgari}}},
  \bibinfo {author} {\bibfnamefont {C.}~\bibnamefont {{Blake}}}, \bibinfo
  {author} {\bibfnamefont {H.}~\bibnamefont {{Hoekstra}}}, \bibinfo {author}
  {\bibfnamefont {S.}~\bibnamefont {{Joudaki}}}, \bibinfo {author}
  {\bibfnamefont {K.}~\bibnamefont {{Kuijken}}}, \bibinfo {author}
  {\bibfnamefont {L.}~\bibnamefont {{Miller}}}, \bibinfo {author}
  {\bibfnamefont {C.~B.}\ \bibnamefont {{Morrison}}}, \bibinfo {author}
  {\bibfnamefont {T.}~\bibnamefont {{Tr{\"o}ster}}}, \bibinfo {author}
  {\bibfnamefont {A.}~\bibnamefont {{Amon}}}, \bibinfo {author} {\bibfnamefont
  {M.}~\bibnamefont {{Archidiacono}}}, \bibinfo {author} {\bibfnamefont
  {S.}~\bibnamefont {{Brieden}}}, \bibinfo {author} {\bibfnamefont
  {A.}~\bibnamefont {{Choi}}}, \bibinfo {author} {\bibfnamefont {J.~T.~A.}\
  \bibnamefont {{de Jong}}}, \bibinfo {author} {\bibfnamefont {T.}~\bibnamefont
  {{Erben}}}, \bibinfo {author} {\bibfnamefont {B.}~\bibnamefont {{Giblin}}},
  \bibinfo {author} {\bibfnamefont {A.}~\bibnamefont {{Mead}}}, \bibinfo
  {author} {\bibfnamefont {J.~A.}\ \bibnamefont {{Peacock}}}, \bibinfo {author}
  {\bibfnamefont {M.}~\bibnamefont {{Radovich}}}, \bibinfo {author}
  {\bibfnamefont {P.}~\bibnamefont {{Schneider}}}, \bibinfo {author}
  {\bibfnamefont {C.}~\bibnamefont {{Sif{\'o}n}}}, \ and\ \bibinfo {author}
  {\bibfnamefont {M.}~\bibnamefont {{Tewes}}},\ }\href@noop {} {\bibfield
  {journal} {\bibinfo  {journal} {arXiv e-prints}\ ,\ \bibinfo {eid}
  {arXiv:1812.06076}} (\bibinfo {year} {2018})},\ \Eprint
  {http://arxiv.org/abs/1812.06076} {arXiv:1812.06076 [astro-ph.CO]}
  \BibitemShut {NoStop}%
\bibitem [{\citenamefont {{Semboloni}}\ \emph {et~al.}(2013)\citenamefont
  {{Semboloni}}, \citenamefont {{Hoekstra}},\ and\ \citenamefont
  {{Schaye}}}]{Semboloni13}%
  \BibitemOpen
  \bibfield  {author} {\bibinfo {author} {\bibfnamefont {E.}~\bibnamefont
  {{Semboloni}}}, \bibinfo {author} {\bibfnamefont {H.}~\bibnamefont
  {{Hoekstra}}}, \ and\ \bibinfo {author} {\bibfnamefont {J.}~\bibnamefont
  {{Schaye}}},\ }\href {\doibase 10.1093/mnras/stt1013} {\bibfield  {journal}
  {\bibinfo  {journal} {\mnras}\ }\textbf {\bibinfo {volume} {434}},\ \bibinfo
  {pages} {148} (\bibinfo {year} {2013})},\ \Eprint
  {http://arxiv.org/abs/1210.7303} {arXiv:1210.7303} \BibitemShut {NoStop}%
\bibitem [{\citenamefont {{Fedeli}}(2014)}]{Fedeli2014a}%
  \BibitemOpen
  \bibfield  {author} {\bibinfo {author} {\bibfnamefont {C.}~\bibnamefont
  {{Fedeli}}},\ }\href {\doibase 10.1088/1475-7516/2014/04/028} {\bibfield
  {journal} {\bibinfo  {journal} {\jcap}\ }\textbf {\bibinfo {volume} {4}},\
  \bibinfo {eid} {028} (\bibinfo {year} {2014})},\ \Eprint
  {http://arxiv.org/abs/1401.2997} {arXiv:1401.2997} \BibitemShut {NoStop}%
\bibitem [{\citenamefont {{Schneider}}\ \emph {et~al.}(2019)\citenamefont
  {{Schneider}}, \citenamefont {{Teyssier}}, \citenamefont {{Stadel}},
  \citenamefont {{Chisari}}, \citenamefont {{Le Brun}}, \citenamefont
  {{Amara}},\ and\ \citenamefont {{Refregier}}}]{Schneider18}%
  \BibitemOpen
  \bibfield  {author} {\bibinfo {author} {\bibfnamefont {A.}~\bibnamefont
  {{Schneider}}}, \bibinfo {author} {\bibfnamefont {R.}~\bibnamefont
  {{Teyssier}}}, \bibinfo {author} {\bibfnamefont {J.}~\bibnamefont
  {{Stadel}}}, \bibinfo {author} {\bibfnamefont {N.~E.}\ \bibnamefont
  {{Chisari}}}, \bibinfo {author} {\bibfnamefont {A.~M.~C.}\ \bibnamefont {{Le
  Brun}}}, \bibinfo {author} {\bibfnamefont {A.}~\bibnamefont {{Amara}}}, \
  and\ \bibinfo {author} {\bibfnamefont {A.}~\bibnamefont {{Refregier}}},\
  }\href {\doibase 10.1088/1475-7516/2019/03/020} {\bibfield  {journal}
  {\bibinfo  {journal} {\jcap}\ }\textbf {\bibinfo {volume} {3}},\ \bibinfo
  {eid} {020} (\bibinfo {year} {2019})},\ \Eprint
  {http://arxiv.org/abs/1810.08629} {arXiv:1810.08629} \BibitemShut {NoStop}%
\bibitem [{\citenamefont {{Heymans}}\ \emph {et~al.}(2013)\citenamefont
  {{Heymans}}, \citenamefont {{Grocutt}}, \citenamefont {{Heavens}},
  \citenamefont {{Kilbinger}}, \citenamefont {{Kitching}}, \citenamefont
  {{Simpson}}, \citenamefont {{Benjamin}}, \citenamefont {{Erben}},
  \citenamefont {{Hildebrandt}}, \citenamefont {{Hoekstra}}, \citenamefont
  {{Mellier}}, \citenamefont {{Miller}}, \citenamefont {{Van Waerbeke}},
  \citenamefont {{Brown}}, \citenamefont {{Coupon}}, \citenamefont {{Fu}},
  \citenamefont {{Harnois-D{\'e}raps}}, \citenamefont {{Hudson}}, \citenamefont
  {{Kuijken}}, \citenamefont {{Rowe}}, \citenamefont {{Schrabback}},
  \citenamefont {{Semboloni}}, \citenamefont {{Vafaei}},\ and\ \citenamefont
  {{Velander}}}]{Heymans13}%
  \BibitemOpen
  \bibfield  {author} {\bibinfo {author} {\bibfnamefont {C.}~\bibnamefont
  {{Heymans}}}, \bibinfo {author} {\bibfnamefont {E.}~\bibnamefont
  {{Grocutt}}}, \bibinfo {author} {\bibfnamefont {A.}~\bibnamefont
  {{Heavens}}}, \bibinfo {author} {\bibfnamefont {M.}~\bibnamefont
  {{Kilbinger}}}, \bibinfo {author} {\bibfnamefont {T.~D.}\ \bibnamefont
  {{Kitching}}}, \bibinfo {author} {\bibfnamefont {F.}~\bibnamefont
  {{Simpson}}}, \bibinfo {author} {\bibfnamefont {J.}~\bibnamefont
  {{Benjamin}}}, \bibinfo {author} {\bibfnamefont {T.}~\bibnamefont {{Erben}}},
  \bibinfo {author} {\bibfnamefont {H.}~\bibnamefont {{Hildebrandt}}}, \bibinfo
  {author} {\bibfnamefont {H.}~\bibnamefont {{Hoekstra}}}, \bibinfo {author}
  {\bibfnamefont {Y.}~\bibnamefont {{Mellier}}}, \bibinfo {author}
  {\bibfnamefont {L.}~\bibnamefont {{Miller}}}, \bibinfo {author}
  {\bibfnamefont {L.}~\bibnamefont {{Van Waerbeke}}}, \bibinfo {author}
  {\bibfnamefont {M.~L.}\ \bibnamefont {{Brown}}}, \bibinfo {author}
  {\bibfnamefont {J.}~\bibnamefont {{Coupon}}}, \bibinfo {author}
  {\bibfnamefont {L.}~\bibnamefont {{Fu}}}, \bibinfo {author} {\bibfnamefont
  {J.}~\bibnamefont {{Harnois-D{\'e}raps}}}, \bibinfo {author} {\bibfnamefont
  {M.~J.}\ \bibnamefont {{Hudson}}}, \bibinfo {author} {\bibfnamefont
  {K.}~\bibnamefont {{Kuijken}}}, \bibinfo {author} {\bibfnamefont
  {B.}~\bibnamefont {{Rowe}}}, \bibinfo {author} {\bibfnamefont
  {T.}~\bibnamefont {{Schrabback}}}, \bibinfo {author} {\bibfnamefont
  {E.}~\bibnamefont {{Semboloni}}}, \bibinfo {author} {\bibfnamefont
  {S.}~\bibnamefont {{Vafaei}}}, \ and\ \bibinfo {author} {\bibfnamefont
  {M.}~\bibnamefont {{Velander}}},\ }\href {\doibase 10.1093/mnras/stt601}
  {\bibfield  {journal} {\bibinfo  {journal} {\mnras}\ }\textbf {\bibinfo
  {volume} {432}},\ \bibinfo {pages} {2433} (\bibinfo {year} {2013})},\ \Eprint
  {http://arxiv.org/abs/1303.1808} {arXiv:1303.1808 [astro-ph.CO]} \BibitemShut
  {NoStop}%
\bibitem [{\citenamefont {{Planck Collaboration}}\ \emph
  {et~al.}(2016{\natexlab{a}})\citenamefont {{Planck Collaboration}},
  \citenamefont {{Ade}}, \citenamefont {{Aghanim}}, \citenamefont {{Arnaud}},
  \citenamefont {{Ashdown}}, \citenamefont {{Aumont}}, \citenamefont
  {{Baccigalupi}}, \citenamefont {{Banday}}, \citenamefont {{Barreiro}},
  \citenamefont {{Bartlett}}, \citenamefont {{Bartolo}}, \citenamefont
  {{Battaner}}, \citenamefont {{Battye}}, \citenamefont {{Benabed}},
  \citenamefont {{Beno{\^\i}t}}, \citenamefont {{Benoit-L{\'e}vy}},
  \citenamefont {{Bernard}}, \citenamefont {{Bersanelli}}, \citenamefont
  {{Bielewicz}}, \citenamefont {{Bock}}, \citenamefont {{Bonaldi}},
  \citenamefont {{Bonavera}}, \citenamefont {{Bond}}, \citenamefont
  {{Borrill}}, \citenamefont {{Bouchet}}, \citenamefont {{Boulanger}},
  \citenamefont {{Bucher}}, \citenamefont {{Burigana}}, \citenamefont
  {{Butler}}, \citenamefont {{Calabrese}}, \citenamefont {{Cardoso}},
  \citenamefont {{Catalano}}, \citenamefont {{Challinor}}, \citenamefont
  {{Chamballu}}, \citenamefont {{Chary}}, \citenamefont {{Chiang}},
  \citenamefont {{Chluba}}, \citenamefont {{Christensen}}, \citenamefont
  {{Church}}, \citenamefont {{Clements}}, \citenamefont {{Colombi}},
  \citenamefont {{Colombo}}, \citenamefont {{Combet}}, \citenamefont
  {{Coulais}}, \citenamefont {{Crill}}, \citenamefont {{Curto}}, \citenamefont
  {{Cuttaia}}, \citenamefont {{Danese}}, \citenamefont {{Davies}},
  \citenamefont {{Davis}}, \citenamefont {{de Bernardis}}, \citenamefont {{de
  Rosa}}, \citenamefont {{de Zotti}}, \citenamefont {{Delabrouille}},
  \citenamefont {{D{\'e}sert}}, \citenamefont {{Di Valentino}}, \citenamefont
  {{Dickinson}}, \citenamefont {{Diego}}, \citenamefont {{Dolag}},
  \citenamefont {{Dole}}, \citenamefont {{Donzelli}}, \citenamefont
  {{Dor{\'e}}}, \citenamefont {{Douspis}}, \citenamefont {{Ducout}},
  \citenamefont {{Dunkley}}, \citenamefont {{Dupac}}, \citenamefont
  {{Efstathiou}}, \citenamefont {{Elsner}}, \citenamefont {{En{\ss}lin}},
  \citenamefont {{Eriksen}}, \citenamefont {{Farhang}}, \citenamefont
  {{Fergusson}}, \citenamefont {{Finelli}}, \citenamefont {{Forni}},
  \citenamefont {{Frailis}}, \citenamefont {{Fraisse}}, \citenamefont
  {{Franceschi}}, \citenamefont {{Frejsel}}, \citenamefont {{Galeotta}},
  \citenamefont {{Galli}}, \citenamefont {{Ganga}}, \citenamefont {{Gauthier}},
  \citenamefont {{Gerbino}}, \citenamefont {{Ghosh}}, \citenamefont {{Giard}},
  \citenamefont {{Giraud-H{\'e}raud}}, \citenamefont {{Giusarma}},
  \citenamefont {{Gjerl{\o}w}}, \citenamefont {{Gonz{\'a}lez-Nuevo}},
  \citenamefont {{G{\'o}rski}}, \citenamefont {{Gratton}}, \citenamefont
  {{Gregorio}}, \citenamefont {{Gruppuso}}, \citenamefont {{Gudmundsson}},
  \citenamefont {{Hamann}}, \citenamefont {{Hansen}}, \citenamefont {{Hanson}},
  \citenamefont {{Harrison}}, \citenamefont {{Helou}}, \citenamefont
  {{Henrot-Versill{\'e}}}, \citenamefont {{Hern{\'a}ndez-Monteagudo}},
  \citenamefont {{Herranz}}, \citenamefont {{Hildebrand t}}, \citenamefont
  {{Hivon}}, \citenamefont {{Hobson}}, \citenamefont {{Holmes}}, \citenamefont
  {{Hornstrup}}, \citenamefont {{Hovest}}, \citenamefont {{Huang}},
  \citenamefont {{Huffenberger}}, \citenamefont {{Hurier}}, \citenamefont
  {{Jaffe}}, \citenamefont {{Jaffe}}, \citenamefont {{Jones}}, \citenamefont
  {{Juvela}}, \citenamefont {{Keih{\"a}nen}}, \citenamefont {{Keskitalo}},
  \citenamefont {{Kisner}}, \citenamefont {{Kneissl}}, \citenamefont
  {{Knoche}}, \citenamefont {{Knox}}, \citenamefont {{Kunz}}, \citenamefont
  {{Kurki-Suonio}}, \citenamefont {{Lagache}}, \citenamefont
  {{L{\"a}hteenm{\"a}ki}}, \citenamefont {{Lamarre}}, \citenamefont
  {{Lasenby}}, \citenamefont {{Lattanzi}}, \citenamefont {{Lawrence}},
  \citenamefont {{Leahy}}, \citenamefont {{Leonardi}}, \citenamefont
  {{Lesgourgues}}, \citenamefont {{Levrier}}, \citenamefont {{Lewis}},
  \citenamefont {{Liguori}}, \citenamefont {{Lilje}}, \citenamefont
  {{Linden-V{\o}rnle}}, \citenamefont {{L{\'o}pez-Caniego}}, \citenamefont
  {{Lubin}}, \citenamefont {{Mac{\'\i}as-P{\'e}rez}}, \citenamefont {{Maggio}},
  \citenamefont {{Maino}}, \citenamefont {{Mandolesi}}, \citenamefont
  {{Mangilli}}, \citenamefont {{Marchini}}, \citenamefont {{Maris}},
  \citenamefont {{Martin}}, \citenamefont {{Martinelli}}, \citenamefont
  {{Mart{\'\i}nez-Gonz{\'a}lez}}, \citenamefont {{Masi}}, \citenamefont
  {{Matarrese}}, \citenamefont {{McGehee}}, \citenamefont {{Meinhold}},
  \citenamefont {{Melchiorri}}, \citenamefont {{Melin}}, \citenamefont
  {{Mendes}}, \citenamefont {{Mennella}}, \citenamefont {{Migliaccio}},
  \citenamefont {{Millea}}, \citenamefont {{Mitra}}, \citenamefont
  {{Miville-Desch{\^e}nes}}, \citenamefont {{Moneti}}, \citenamefont
  {{Montier}}, \citenamefont {{Morgante}}, \citenamefont {{Mortlock}},
  \citenamefont {{Moss}}, \citenamefont {{Munshi}}, \citenamefont {{Murphy}},
  \citenamefont {{Naselsky}}, \citenamefont {{Nati}}, \citenamefont {{Natoli}},
  \citenamefont {{Netterfield}}, \citenamefont {{N{\o}rgaard-Nielsen}},
  \citenamefont {{Noviello}}, \citenamefont {{Novikov}}, \citenamefont
  {{Novikov}}, \citenamefont {{Oxborrow}}, \citenamefont {{Paci}},
  \citenamefont {{Pagano}}, \citenamefont {{Pajot}}, \citenamefont
  {{Paladini}}, \citenamefont {{Paoletti}}, \citenamefont {{Partridge}},
  \citenamefont {{Pasian}}, \citenamefont {{Patanchon}}, \citenamefont
  {{Pearson}}, \citenamefont {{Perdereau}}, \citenamefont {{Perotto}},
  \citenamefont {{Perrotta}}, \citenamefont {{Pettorino}}, \citenamefont
  {{Piacentini}}, \citenamefont {{Piat}}, \citenamefont {{Pierpaoli}},
  \citenamefont {{Pietrobon}}, \citenamefont {{Plaszczynski}}, \citenamefont
  {{Pointecouteau}}, \citenamefont {{Polenta}}, \citenamefont {{Popa}},
  \citenamefont {{Pratt}}, \citenamefont {{Pr{\'e}zeau}}, \citenamefont
  {{Prunet}}, \citenamefont {{Puget}}, \citenamefont {{Rachen}}, \citenamefont
  {{Reach}}, \citenamefont {{Rebolo}}, \citenamefont {{Reinecke}},
  \citenamefont {{Remazeilles}}, \citenamefont {{Renault}}, \citenamefont
  {{Renzi}}, \citenamefont {{Ristorcelli}}, \citenamefont {{Rocha}},
  \citenamefont {{Rosset}}, \citenamefont {{Rossetti}}, \citenamefont
  {{Roudier}}, \citenamefont {{Rouill{\'e} d'Orfeuil}}, \citenamefont
  {{Rowan-Robinson}}, \citenamefont {{Rubi{\~n}o-Mart{\'\i}n}}, \citenamefont
  {{Rusholme}}, \citenamefont {{Said}}, \citenamefont {{Salvatelli}},
  \citenamefont {{Salvati}}, \citenamefont {{Sandri}}, \citenamefont
  {{Santos}}, \citenamefont {{Savelainen}}, \citenamefont {{Savini}},
  \citenamefont {{Scott}}, \citenamefont {{Seiffert}}, \citenamefont {{Serra}},
  \citenamefont {{Shellard}}, \citenamefont {{Spencer}}, \citenamefont
  {{Spinelli}}, \citenamefont {{Stolyarov}}, \citenamefont {{Stompor}},
  \citenamefont {{Sudiwala}}, \citenamefont {{Sunyaev}}, \citenamefont
  {{Sutton}}, \citenamefont {{Suur-Uski}}, \citenamefont {{Sygnet}},
  \citenamefont {{Tauber}}, \citenamefont {{Terenzi}}, \citenamefont
  {{Toffolatti}}, \citenamefont {{Tomasi}}, \citenamefont {{Tristram}},
  \citenamefont {{Trombetti}}, \citenamefont {{Tucci}}, \citenamefont
  {{Tuovinen}}, \citenamefont {{T{\"u}rler}}, \citenamefont {{Umana}},
  \citenamefont {{Valenziano}}, \citenamefont {{Valiviita}}, \citenamefont
  {{Van Tent}}, \citenamefont {{Vielva}}, \citenamefont {{Villa}},
  \citenamefont {{Wade}}, \citenamefont {{Wandelt}}, \citenamefont {{Wehus}},
  \citenamefont {{White}}, \citenamefont {{White}}, \citenamefont
  {{Wilkinson}}, \citenamefont {{Yvon}}, \citenamefont {{Zacchei}},\ and\
  \citenamefont {{Zonca}}}]{Planck15}%
  \BibitemOpen
  \bibfield  {author} {\bibinfo {author} {\bibnamefont {{Planck
  Collaboration}}}, \bibinfo {author} {\bibfnamefont {P.~A.~R.}\ \bibnamefont
  {{Ade}}}, \bibinfo {author} {\bibfnamefont {N.}~\bibnamefont {{Aghanim}}},
  \bibinfo {author} {\bibfnamefont {M.}~\bibnamefont {{Arnaud}}}, \bibinfo
  {author} {\bibfnamefont {M.}~\bibnamefont {{Ashdown}}}, \bibinfo {author}
  {\bibfnamefont {J.}~\bibnamefont {{Aumont}}}, \bibinfo {author}
  {\bibfnamefont {C.}~\bibnamefont {{Baccigalupi}}}, \bibinfo {author}
  {\bibfnamefont {A.~J.}\ \bibnamefont {{Banday}}}, \bibinfo {author}
  {\bibfnamefont {R.~B.}\ \bibnamefont {{Barreiro}}}, \bibinfo {author}
  {\bibfnamefont {J.~G.}\ \bibnamefont {{Bartlett}}}, \bibinfo {author}
  {\bibfnamefont {N.}~\bibnamefont {{Bartolo}}}, \bibinfo {author}
  {\bibfnamefont {E.}~\bibnamefont {{Battaner}}}, \bibinfo {author}
  {\bibfnamefont {R.}~\bibnamefont {{Battye}}}, \bibinfo {author}
  {\bibfnamefont {K.}~\bibnamefont {{Benabed}}}, \bibinfo {author}
  {\bibfnamefont {A.}~\bibnamefont {{Beno{\^\i}t}}}, \bibinfo {author}
  {\bibfnamefont {A.}~\bibnamefont {{Benoit-L{\'e}vy}}}, \bibinfo {author}
  {\bibfnamefont {J.~P.}\ \bibnamefont {{Bernard}}}, \bibinfo {author}
  {\bibfnamefont {M.}~\bibnamefont {{Bersanelli}}}, \bibinfo {author}
  {\bibfnamefont {P.}~\bibnamefont {{Bielewicz}}}, \bibinfo {author}
  {\bibfnamefont {J.~J.}\ \bibnamefont {{Bock}}}, \bibinfo {author}
  {\bibfnamefont {A.}~\bibnamefont {{Bonaldi}}}, \bibinfo {author}
  {\bibfnamefont {L.}~\bibnamefont {{Bonavera}}}, \bibinfo {author}
  {\bibfnamefont {J.~R.}\ \bibnamefont {{Bond}}}, \bibinfo {author}
  {\bibfnamefont {J.}~\bibnamefont {{Borrill}}}, \bibinfo {author}
  {\bibfnamefont {F.~R.}\ \bibnamefont {{Bouchet}}}, \bibinfo {author}
  {\bibfnamefont {F.}~\bibnamefont {{Boulanger}}}, \bibinfo {author}
  {\bibfnamefont {M.}~\bibnamefont {{Bucher}}}, \bibinfo {author}
  {\bibfnamefont {C.}~\bibnamefont {{Burigana}}}, \bibinfo {author}
  {\bibfnamefont {R.~C.}\ \bibnamefont {{Butler}}}, \bibinfo {author}
  {\bibfnamefont {E.}~\bibnamefont {{Calabrese}}}, \bibinfo {author}
  {\bibfnamefont {J.~F.}\ \bibnamefont {{Cardoso}}}, \bibinfo {author}
  {\bibfnamefont {A.}~\bibnamefont {{Catalano}}}, \bibinfo {author}
  {\bibfnamefont {A.}~\bibnamefont {{Challinor}}}, \bibinfo {author}
  {\bibfnamefont {A.}~\bibnamefont {{Chamballu}}}, \bibinfo {author}
  {\bibfnamefont {R.~R.}\ \bibnamefont {{Chary}}}, \bibinfo {author}
  {\bibfnamefont {H.~C.}\ \bibnamefont {{Chiang}}}, \bibinfo {author}
  {\bibfnamefont {J.}~\bibnamefont {{Chluba}}}, \bibinfo {author}
  {\bibfnamefont {P.~R.}\ \bibnamefont {{Christensen}}}, \bibinfo {author}
  {\bibfnamefont {S.}~\bibnamefont {{Church}}}, \bibinfo {author}
  {\bibfnamefont {D.~L.}\ \bibnamefont {{Clements}}}, \bibinfo {author}
  {\bibfnamefont {S.}~\bibnamefont {{Colombi}}}, \bibinfo {author}
  {\bibfnamefont {L.~P.~L.}\ \bibnamefont {{Colombo}}}, \bibinfo {author}
  {\bibfnamefont {C.}~\bibnamefont {{Combet}}}, \bibinfo {author}
  {\bibfnamefont {A.}~\bibnamefont {{Coulais}}}, \bibinfo {author}
  {\bibfnamefont {B.~P.}\ \bibnamefont {{Crill}}}, \bibinfo {author}
  {\bibfnamefont {A.}~\bibnamefont {{Curto}}}, \bibinfo {author} {\bibfnamefont
  {F.}~\bibnamefont {{Cuttaia}}}, \bibinfo {author} {\bibfnamefont
  {L.}~\bibnamefont {{Danese}}}, \bibinfo {author} {\bibfnamefont {R.~D.}\
  \bibnamefont {{Davies}}}, \bibinfo {author} {\bibfnamefont {R.~J.}\
  \bibnamefont {{Davis}}}, \bibinfo {author} {\bibfnamefont {P.}~\bibnamefont
  {{de Bernardis}}}, \bibinfo {author} {\bibfnamefont {A.}~\bibnamefont {{de
  Rosa}}}, \bibinfo {author} {\bibfnamefont {G.}~\bibnamefont {{de Zotti}}},
  \bibinfo {author} {\bibfnamefont {J.}~\bibnamefont {{Delabrouille}}},
  \bibinfo {author} {\bibfnamefont {F.~X.}\ \bibnamefont {{D{\'e}sert}}},
  \bibinfo {author} {\bibfnamefont {E.}~\bibnamefont {{Di Valentino}}},
  \bibinfo {author} {\bibfnamefont {C.}~\bibnamefont {{Dickinson}}}, \bibinfo
  {author} {\bibfnamefont {J.~M.}\ \bibnamefont {{Diego}}}, \bibinfo {author}
  {\bibfnamefont {K.}~\bibnamefont {{Dolag}}}, \bibinfo {author} {\bibfnamefont
  {H.}~\bibnamefont {{Dole}}}, \bibinfo {author} {\bibfnamefont
  {S.}~\bibnamefont {{Donzelli}}}, \bibinfo {author} {\bibfnamefont
  {O.}~\bibnamefont {{Dor{\'e}}}}, \bibinfo {author} {\bibfnamefont
  {M.}~\bibnamefont {{Douspis}}}, \bibinfo {author} {\bibfnamefont
  {A.}~\bibnamefont {{Ducout}}}, \bibinfo {author} {\bibfnamefont
  {J.}~\bibnamefont {{Dunkley}}}, \bibinfo {author} {\bibfnamefont
  {X.}~\bibnamefont {{Dupac}}}, \bibinfo {author} {\bibfnamefont
  {G.}~\bibnamefont {{Efstathiou}}}, \bibinfo {author} {\bibfnamefont
  {F.}~\bibnamefont {{Elsner}}}, \bibinfo {author} {\bibfnamefont {T.~A.}\
  \bibnamefont {{En{\ss}lin}}}, \bibinfo {author} {\bibfnamefont {H.~K.}\
  \bibnamefont {{Eriksen}}}, \bibinfo {author} {\bibfnamefont {M.}~\bibnamefont
  {{Farhang}}}, \bibinfo {author} {\bibfnamefont {J.}~\bibnamefont
  {{Fergusson}}}, \bibinfo {author} {\bibfnamefont {F.}~\bibnamefont
  {{Finelli}}}, \bibinfo {author} {\bibfnamefont {O.}~\bibnamefont {{Forni}}},
  \bibinfo {author} {\bibfnamefont {M.}~\bibnamefont {{Frailis}}}, \bibinfo
  {author} {\bibfnamefont {A.~A.}\ \bibnamefont {{Fraisse}}}, \bibinfo {author}
  {\bibfnamefont {E.}~\bibnamefont {{Franceschi}}}, \bibinfo {author}
  {\bibfnamefont {A.}~\bibnamefont {{Frejsel}}}, \bibinfo {author}
  {\bibfnamefont {S.}~\bibnamefont {{Galeotta}}}, \bibinfo {author}
  {\bibfnamefont {S.}~\bibnamefont {{Galli}}}, \bibinfo {author} {\bibfnamefont
  {K.}~\bibnamefont {{Ganga}}}, \bibinfo {author} {\bibfnamefont
  {C.}~\bibnamefont {{Gauthier}}}, \bibinfo {author} {\bibfnamefont
  {M.}~\bibnamefont {{Gerbino}}}, \bibinfo {author} {\bibfnamefont
  {T.}~\bibnamefont {{Ghosh}}}, \bibinfo {author} {\bibfnamefont
  {M.}~\bibnamefont {{Giard}}}, \bibinfo {author} {\bibfnamefont
  {Y.}~\bibnamefont {{Giraud-H{\'e}raud}}}, \bibinfo {author} {\bibfnamefont
  {E.}~\bibnamefont {{Giusarma}}}, \bibinfo {author} {\bibfnamefont
  {E.}~\bibnamefont {{Gjerl{\o}w}}}, \bibinfo {author} {\bibfnamefont
  {J.}~\bibnamefont {{Gonz{\'a}lez-Nuevo}}}, \bibinfo {author} {\bibfnamefont
  {K.~M.}\ \bibnamefont {{G{\'o}rski}}}, \bibinfo {author} {\bibfnamefont
  {S.}~\bibnamefont {{Gratton}}}, \bibinfo {author} {\bibfnamefont
  {A.}~\bibnamefont {{Gregorio}}}, \bibinfo {author} {\bibfnamefont
  {A.}~\bibnamefont {{Gruppuso}}}, \bibinfo {author} {\bibfnamefont {J.~E.}\
  \bibnamefont {{Gudmundsson}}}, \bibinfo {author} {\bibfnamefont
  {J.}~\bibnamefont {{Hamann}}}, \bibinfo {author} {\bibfnamefont {F.~K.}\
  \bibnamefont {{Hansen}}}, \bibinfo {author} {\bibfnamefont {D.}~\bibnamefont
  {{Hanson}}}, \bibinfo {author} {\bibfnamefont {D.~L.}\ \bibnamefont
  {{Harrison}}}, \bibinfo {author} {\bibfnamefont {G.}~\bibnamefont {{Helou}}},
  \bibinfo {author} {\bibfnamefont {S.}~\bibnamefont {{Henrot-Versill{\'e}}}},
  \bibinfo {author} {\bibfnamefont {C.}~\bibnamefont
  {{Hern{\'a}ndez-Monteagudo}}}, \bibinfo {author} {\bibfnamefont
  {D.}~\bibnamefont {{Herranz}}}, \bibinfo {author} {\bibfnamefont {S.~R.}\
  \bibnamefont {{Hildebrand t}}}, \bibinfo {author} {\bibfnamefont
  {E.}~\bibnamefont {{Hivon}}}, \bibinfo {author} {\bibfnamefont
  {M.}~\bibnamefont {{Hobson}}}, \bibinfo {author} {\bibfnamefont {W.~A.}\
  \bibnamefont {{Holmes}}}, \bibinfo {author} {\bibfnamefont {A.}~\bibnamefont
  {{Hornstrup}}}, \bibinfo {author} {\bibfnamefont {W.}~\bibnamefont
  {{Hovest}}}, \bibinfo {author} {\bibfnamefont {Z.}~\bibnamefont {{Huang}}},
  \bibinfo {author} {\bibfnamefont {K.~M.}\ \bibnamefont {{Huffenberger}}},
  \bibinfo {author} {\bibfnamefont {G.}~\bibnamefont {{Hurier}}}, \bibinfo
  {author} {\bibfnamefont {A.~H.}\ \bibnamefont {{Jaffe}}}, \bibinfo {author}
  {\bibfnamefont {T.~R.}\ \bibnamefont {{Jaffe}}}, \bibinfo {author}
  {\bibfnamefont {W.~C.}\ \bibnamefont {{Jones}}}, \bibinfo {author}
  {\bibfnamefont {M.}~\bibnamefont {{Juvela}}}, \bibinfo {author}
  {\bibfnamefont {E.}~\bibnamefont {{Keih{\"a}nen}}}, \bibinfo {author}
  {\bibfnamefont {R.}~\bibnamefont {{Keskitalo}}}, \bibinfo {author}
  {\bibfnamefont {T.~S.}\ \bibnamefont {{Kisner}}}, \bibinfo {author}
  {\bibfnamefont {R.}~\bibnamefont {{Kneissl}}}, \bibinfo {author}
  {\bibfnamefont {J.}~\bibnamefont {{Knoche}}}, \bibinfo {author}
  {\bibfnamefont {L.}~\bibnamefont {{Knox}}}, \bibinfo {author} {\bibfnamefont
  {M.}~\bibnamefont {{Kunz}}}, \bibinfo {author} {\bibfnamefont
  {H.}~\bibnamefont {{Kurki-Suonio}}}, \bibinfo {author} {\bibfnamefont
  {G.}~\bibnamefont {{Lagache}}}, \bibinfo {author} {\bibfnamefont
  {A.}~\bibnamefont {{L{\"a}hteenm{\"a}ki}}}, \bibinfo {author} {\bibfnamefont
  {J.~M.}\ \bibnamefont {{Lamarre}}}, \bibinfo {author} {\bibfnamefont
  {A.}~\bibnamefont {{Lasenby}}}, \bibinfo {author} {\bibfnamefont
  {M.}~\bibnamefont {{Lattanzi}}}, \bibinfo {author} {\bibfnamefont {C.~R.}\
  \bibnamefont {{Lawrence}}}, \bibinfo {author} {\bibfnamefont {J.~P.}\
  \bibnamefont {{Leahy}}}, \bibinfo {author} {\bibfnamefont {R.}~\bibnamefont
  {{Leonardi}}}, \bibinfo {author} {\bibfnamefont {J.}~\bibnamefont
  {{Lesgourgues}}}, \bibinfo {author} {\bibfnamefont {F.}~\bibnamefont
  {{Levrier}}}, \bibinfo {author} {\bibfnamefont {A.}~\bibnamefont {{Lewis}}},
  \bibinfo {author} {\bibfnamefont {M.}~\bibnamefont {{Liguori}}}, \bibinfo
  {author} {\bibfnamefont {P.~B.}\ \bibnamefont {{Lilje}}}, \bibinfo {author}
  {\bibfnamefont {M.}~\bibnamefont {{Linden-V{\o}rnle}}}, \bibinfo {author}
  {\bibfnamefont {M.}~\bibnamefont {{L{\'o}pez-Caniego}}}, \bibinfo {author}
  {\bibfnamefont {P.~M.}\ \bibnamefont {{Lubin}}}, \bibinfo {author}
  {\bibfnamefont {J.~F.}\ \bibnamefont {{Mac{\'\i}as-P{\'e}rez}}}, \bibinfo
  {author} {\bibfnamefont {G.}~\bibnamefont {{Maggio}}}, \bibinfo {author}
  {\bibfnamefont {D.}~\bibnamefont {{Maino}}}, \bibinfo {author} {\bibfnamefont
  {N.}~\bibnamefont {{Mandolesi}}}, \bibinfo {author} {\bibfnamefont
  {A.}~\bibnamefont {{Mangilli}}}, \bibinfo {author} {\bibfnamefont
  {A.}~\bibnamefont {{Marchini}}}, \bibinfo {author} {\bibfnamefont
  {M.}~\bibnamefont {{Maris}}}, \bibinfo {author} {\bibfnamefont {P.~G.}\
  \bibnamefont {{Martin}}}, \bibinfo {author} {\bibfnamefont {M.}~\bibnamefont
  {{Martinelli}}}, \bibinfo {author} {\bibfnamefont {E.}~\bibnamefont
  {{Mart{\'\i}nez-Gonz{\'a}lez}}}, \bibinfo {author} {\bibfnamefont
  {S.}~\bibnamefont {{Masi}}}, \bibinfo {author} {\bibfnamefont
  {S.}~\bibnamefont {{Matarrese}}}, \bibinfo {author} {\bibfnamefont
  {P.}~\bibnamefont {{McGehee}}}, \bibinfo {author} {\bibfnamefont {P.~R.}\
  \bibnamefont {{Meinhold}}}, \bibinfo {author} {\bibfnamefont
  {A.}~\bibnamefont {{Melchiorri}}}, \bibinfo {author} {\bibfnamefont {J.~B.}\
  \bibnamefont {{Melin}}}, \bibinfo {author} {\bibfnamefont {L.}~\bibnamefont
  {{Mendes}}}, \bibinfo {author} {\bibfnamefont {A.}~\bibnamefont
  {{Mennella}}}, \bibinfo {author} {\bibfnamefont {M.}~\bibnamefont
  {{Migliaccio}}}, \bibinfo {author} {\bibfnamefont {M.}~\bibnamefont
  {{Millea}}}, \bibinfo {author} {\bibfnamefont {S.}~\bibnamefont {{Mitra}}},
  \bibinfo {author} {\bibfnamefont {M.~A.}\ \bibnamefont
  {{Miville-Desch{\^e}nes}}}, \bibinfo {author} {\bibfnamefont
  {A.}~\bibnamefont {{Moneti}}}, \bibinfo {author} {\bibfnamefont
  {L.}~\bibnamefont {{Montier}}}, \bibinfo {author} {\bibfnamefont
  {G.}~\bibnamefont {{Morgante}}}, \bibinfo {author} {\bibfnamefont
  {D.}~\bibnamefont {{Mortlock}}}, \bibinfo {author} {\bibfnamefont
  {A.}~\bibnamefont {{Moss}}}, \bibinfo {author} {\bibfnamefont
  {D.}~\bibnamefont {{Munshi}}}, \bibinfo {author} {\bibfnamefont {J.~A.}\
  \bibnamefont {{Murphy}}}, \bibinfo {author} {\bibfnamefont {P.}~\bibnamefont
  {{Naselsky}}}, \bibinfo {author} {\bibfnamefont {F.}~\bibnamefont {{Nati}}},
  \bibinfo {author} {\bibfnamefont {P.}~\bibnamefont {{Natoli}}}, \bibinfo
  {author} {\bibfnamefont {C.~B.}\ \bibnamefont {{Netterfield}}}, \bibinfo
  {author} {\bibfnamefont {H.~U.}\ \bibnamefont {{N{\o}rgaard-Nielsen}}},
  \bibinfo {author} {\bibfnamefont {F.}~\bibnamefont {{Noviello}}}, \bibinfo
  {author} {\bibfnamefont {D.}~\bibnamefont {{Novikov}}}, \bibinfo {author}
  {\bibfnamefont {I.}~\bibnamefont {{Novikov}}}, \bibinfo {author}
  {\bibfnamefont {C.~A.}\ \bibnamefont {{Oxborrow}}}, \bibinfo {author}
  {\bibfnamefont {F.}~\bibnamefont {{Paci}}}, \bibinfo {author} {\bibfnamefont
  {L.}~\bibnamefont {{Pagano}}}, \bibinfo {author} {\bibfnamefont
  {F.}~\bibnamefont {{Pajot}}}, \bibinfo {author} {\bibfnamefont
  {R.}~\bibnamefont {{Paladini}}}, \bibinfo {author} {\bibfnamefont
  {D.}~\bibnamefont {{Paoletti}}}, \bibinfo {author} {\bibfnamefont
  {B.}~\bibnamefont {{Partridge}}}, \bibinfo {author} {\bibfnamefont
  {F.}~\bibnamefont {{Pasian}}}, \bibinfo {author} {\bibfnamefont
  {G.}~\bibnamefont {{Patanchon}}}, \bibinfo {author} {\bibfnamefont {T.~J.}\
  \bibnamefont {{Pearson}}}, \bibinfo {author} {\bibfnamefont {O.}~\bibnamefont
  {{Perdereau}}}, \bibinfo {author} {\bibfnamefont {L.}~\bibnamefont
  {{Perotto}}}, \bibinfo {author} {\bibfnamefont {F.}~\bibnamefont
  {{Perrotta}}}, \bibinfo {author} {\bibfnamefont {V.}~\bibnamefont
  {{Pettorino}}}, \bibinfo {author} {\bibfnamefont {F.}~\bibnamefont
  {{Piacentini}}}, \bibinfo {author} {\bibfnamefont {M.}~\bibnamefont
  {{Piat}}}, \bibinfo {author} {\bibfnamefont {E.}~\bibnamefont {{Pierpaoli}}},
  \bibinfo {author} {\bibfnamefont {D.}~\bibnamefont {{Pietrobon}}}, \bibinfo
  {author} {\bibfnamefont {S.}~\bibnamefont {{Plaszczynski}}}, \bibinfo
  {author} {\bibfnamefont {E.}~\bibnamefont {{Pointecouteau}}}, \bibinfo
  {author} {\bibfnamefont {G.}~\bibnamefont {{Polenta}}}, \bibinfo {author}
  {\bibfnamefont {L.}~\bibnamefont {{Popa}}}, \bibinfo {author} {\bibfnamefont
  {G.~W.}\ \bibnamefont {{Pratt}}}, \bibinfo {author} {\bibfnamefont
  {G.}~\bibnamefont {{Pr{\'e}zeau}}}, \bibinfo {author} {\bibfnamefont
  {S.}~\bibnamefont {{Prunet}}}, \bibinfo {author} {\bibfnamefont {J.~L.}\
  \bibnamefont {{Puget}}}, \bibinfo {author} {\bibfnamefont {J.~P.}\
  \bibnamefont {{Rachen}}}, \bibinfo {author} {\bibfnamefont {W.~T.}\
  \bibnamefont {{Reach}}}, \bibinfo {author} {\bibfnamefont {R.}~\bibnamefont
  {{Rebolo}}}, \bibinfo {author} {\bibfnamefont {M.}~\bibnamefont
  {{Reinecke}}}, \bibinfo {author} {\bibfnamefont {M.}~\bibnamefont
  {{Remazeilles}}}, \bibinfo {author} {\bibfnamefont {C.}~\bibnamefont
  {{Renault}}}, \bibinfo {author} {\bibfnamefont {A.}~\bibnamefont {{Renzi}}},
  \bibinfo {author} {\bibfnamefont {I.}~\bibnamefont {{Ristorcelli}}}, \bibinfo
  {author} {\bibfnamefont {G.}~\bibnamefont {{Rocha}}}, \bibinfo {author}
  {\bibfnamefont {C.}~\bibnamefont {{Rosset}}}, \bibinfo {author}
  {\bibfnamefont {M.}~\bibnamefont {{Rossetti}}}, \bibinfo {author}
  {\bibfnamefont {G.}~\bibnamefont {{Roudier}}}, \bibinfo {author}
  {\bibfnamefont {B.}~\bibnamefont {{Rouill{\'e} d'Orfeuil}}}, \bibinfo
  {author} {\bibfnamefont {M.}~\bibnamefont {{Rowan-Robinson}}}, \bibinfo
  {author} {\bibfnamefont {J.~A.}\ \bibnamefont {{Rubi{\~n}o-Mart{\'\i}n}}},
  \bibinfo {author} {\bibfnamefont {B.}~\bibnamefont {{Rusholme}}}, \bibinfo
  {author} {\bibfnamefont {N.}~\bibnamefont {{Said}}}, \bibinfo {author}
  {\bibfnamefont {V.}~\bibnamefont {{Salvatelli}}}, \bibinfo {author}
  {\bibfnamefont {L.}~\bibnamefont {{Salvati}}}, \bibinfo {author}
  {\bibfnamefont {M.}~\bibnamefont {{Sandri}}}, \bibinfo {author}
  {\bibfnamefont {D.}~\bibnamefont {{Santos}}}, \bibinfo {author}
  {\bibfnamefont {M.}~\bibnamefont {{Savelainen}}}, \bibinfo {author}
  {\bibfnamefont {G.}~\bibnamefont {{Savini}}}, \bibinfo {author}
  {\bibfnamefont {D.}~\bibnamefont {{Scott}}}, \bibinfo {author} {\bibfnamefont
  {M.~D.}\ \bibnamefont {{Seiffert}}}, \bibinfo {author} {\bibfnamefont
  {P.}~\bibnamefont {{Serra}}}, \bibinfo {author} {\bibfnamefont {E.~P.~S.}\
  \bibnamefont {{Shellard}}}, \bibinfo {author} {\bibfnamefont {L.~D.}\
  \bibnamefont {{Spencer}}}, \bibinfo {author} {\bibfnamefont {M.}~\bibnamefont
  {{Spinelli}}}, \bibinfo {author} {\bibfnamefont {V.}~\bibnamefont
  {{Stolyarov}}}, \bibinfo {author} {\bibfnamefont {R.}~\bibnamefont
  {{Stompor}}}, \bibinfo {author} {\bibfnamefont {R.}~\bibnamefont
  {{Sudiwala}}}, \bibinfo {author} {\bibfnamefont {R.}~\bibnamefont
  {{Sunyaev}}}, \bibinfo {author} {\bibfnamefont {D.}~\bibnamefont {{Sutton}}},
  \bibinfo {author} {\bibfnamefont {A.~S.}\ \bibnamefont {{Suur-Uski}}},
  \bibinfo {author} {\bibfnamefont {J.~F.}\ \bibnamefont {{Sygnet}}}, \bibinfo
  {author} {\bibfnamefont {J.~A.}\ \bibnamefont {{Tauber}}}, \bibinfo {author}
  {\bibfnamefont {L.}~\bibnamefont {{Terenzi}}}, \bibinfo {author}
  {\bibfnamefont {L.}~\bibnamefont {{Toffolatti}}}, \bibinfo {author}
  {\bibfnamefont {M.}~\bibnamefont {{Tomasi}}}, \bibinfo {author}
  {\bibfnamefont {M.}~\bibnamefont {{Tristram}}}, \bibinfo {author}
  {\bibfnamefont {T.}~\bibnamefont {{Trombetti}}}, \bibinfo {author}
  {\bibfnamefont {M.}~\bibnamefont {{Tucci}}}, \bibinfo {author} {\bibfnamefont
  {J.}~\bibnamefont {{Tuovinen}}}, \bibinfo {author} {\bibfnamefont
  {M.}~\bibnamefont {{T{\"u}rler}}}, \bibinfo {author} {\bibfnamefont
  {G.}~\bibnamefont {{Umana}}}, \bibinfo {author} {\bibfnamefont
  {L.}~\bibnamefont {{Valenziano}}}, \bibinfo {author} {\bibfnamefont
  {J.}~\bibnamefont {{Valiviita}}}, \bibinfo {author} {\bibfnamefont
  {F.}~\bibnamefont {{Van Tent}}}, \bibinfo {author} {\bibfnamefont
  {P.}~\bibnamefont {{Vielva}}}, \bibinfo {author} {\bibfnamefont
  {F.}~\bibnamefont {{Villa}}}, \bibinfo {author} {\bibfnamefont {L.~A.}\
  \bibnamefont {{Wade}}}, \bibinfo {author} {\bibfnamefont {B.~D.}\
  \bibnamefont {{Wandelt}}}, \bibinfo {author} {\bibfnamefont {I.~K.}\
  \bibnamefont {{Wehus}}}, \bibinfo {author} {\bibfnamefont {M.}~\bibnamefont
  {{White}}}, \bibinfo {author} {\bibfnamefont {S.~D.~M.}\ \bibnamefont
  {{White}}}, \bibinfo {author} {\bibfnamefont {A.}~\bibnamefont
  {{Wilkinson}}}, \bibinfo {author} {\bibfnamefont {D.}~\bibnamefont {{Yvon}}},
  \bibinfo {author} {\bibfnamefont {A.}~\bibnamefont {{Zacchei}}}, \ and\
  \bibinfo {author} {\bibfnamefont {A.}~\bibnamefont {{Zonca}}},\ }\href
  {\doibase 10.1051/0004-6361/201525830} {\bibfield  {journal} {\bibinfo
  {journal} {\aap}\ }\textbf {\bibinfo {volume} {594}},\ \bibinfo {eid} {A13}
  (\bibinfo {year} {2016}{\natexlab{a}})},\ \Eprint
  {http://arxiv.org/abs/1502.01589} {arXiv:1502.01589 [astro-ph.CO]}
  \BibitemShut {NoStop}%
\bibitem [{\citenamefont {{Planck Collaboration}}\ \emph
  {et~al.}(2016{\natexlab{b}})\citenamefont {{Planck Collaboration}},
  \citenamefont {{Ade}}, \citenamefont {{Aghanim}}, \citenamefont {{Arnaud}},
  \citenamefont {{Ashdown}}, \citenamefont {{Aumont}}, \citenamefont
  {{Baccigalupi}}, \citenamefont {{Banday}}, \citenamefont {{Barreiro}},
  \citenamefont {{Bartolo}}, \citenamefont {{Battaner}}, \citenamefont
  {{Battye}}, \citenamefont {{Benabed}}, \citenamefont {{Beno{\^\i}t}},
  \citenamefont {{Benoit-L{\'e}vy}}, \citenamefont {{Bernard}}, \citenamefont
  {{Bersanelli}}, \citenamefont {{Bielewicz}}, \citenamefont {{Bock}},
  \citenamefont {{Bonaldi}}, \citenamefont {{Bonavera}}, \citenamefont
  {{Bond}}, \citenamefont {{Borrill}}, \citenamefont {{Bouchet}}, \citenamefont
  {{Bucher}}, \citenamefont {{Burigana}}, \citenamefont {{Butler}},
  \citenamefont {{Calabrese}}, \citenamefont {{Cardoso}}, \citenamefont
  {{Catalano}}, \citenamefont {{Challinor}}, \citenamefont {{Chamballu}},
  \citenamefont {{Chiang}}, \citenamefont {{Christensen}}, \citenamefont
  {{Church}}, \citenamefont {{Clements}}, \citenamefont {{Colombi}},
  \citenamefont {{Colombo}}, \citenamefont {{Combet}}, \citenamefont
  {{Couchot}}, \citenamefont {{Coulais}}, \citenamefont {{Crill}},
  \citenamefont {{Curto}}, \citenamefont {{Cuttaia}}, \citenamefont {{Danese}},
  \citenamefont {{Davies}}, \citenamefont {{Davis}}, \citenamefont {{de
  Bernardis}}, \citenamefont {{de Rosa}}, \citenamefont {{de Zotti}},
  \citenamefont {{Delabrouille}}, \citenamefont {{D{\'e}sert}}, \citenamefont
  {{Diego}}, \citenamefont {{Dole}}, \citenamefont {{Donzelli}}, \citenamefont
  {{Dor{\'e}}}, \citenamefont {{Douspis}}, \citenamefont {{Ducout}},
  \citenamefont {{Dupac}}, \citenamefont {{Efstathiou}}, \citenamefont
  {{Elsner}}, \citenamefont {{En{\ss}lin}}, \citenamefont {{Eriksen}},
  \citenamefont {{Fergusson}}, \citenamefont {{Finelli}}, \citenamefont
  {{Forni}}, \citenamefont {{Frailis}}, \citenamefont {{Fraisse}},
  \citenamefont {{Franceschi}}, \citenamefont {{Frejsel}}, \citenamefont
  {{Galeotta}}, \citenamefont {{Galli}}, \citenamefont {{Ganga}}, \citenamefont
  {{Giard}}, \citenamefont {{Giraud-H{\'e}raud}}, \citenamefont {{Gjerl{\o}w}},
  \citenamefont {{Gonz{\'a}lez-Nuevo}}, \citenamefont {{G{\'o}rski}},
  \citenamefont {{Gratton}}, \citenamefont {{Gregorio}}, \citenamefont
  {{Gruppuso}}, \citenamefont {{Gudmundsson}}, \citenamefont {{Hansen}},
  \citenamefont {{Hanson}}, \citenamefont {{Harrison}}, \citenamefont
  {{Heavens}}, \citenamefont {{Helou}}, \citenamefont {{Henrot-Versill{\'e}}},
  \citenamefont {{Hern{\'a}ndez-Monteagudo}}, \citenamefont {{Herranz}},
  \citenamefont {{Hildebrand t}}, \citenamefont {{Hivon}}, \citenamefont
  {{Hobson}}, \citenamefont {{Holmes}}, \citenamefont {{Hornstrup}},
  \citenamefont {{Hovest}}, \citenamefont {{Huang}}, \citenamefont
  {{Huffenberger}}, \citenamefont {{Hurier}}, \citenamefont {{Jaffe}},
  \citenamefont {{Jaffe}}, \citenamefont {{Jones}}, \citenamefont {{Juvela}},
  \citenamefont {{Keih{\"a}nen}}, \citenamefont {{Keskitalo}}, \citenamefont
  {{Kisner}}, \citenamefont {{Knoche}}, \citenamefont {{Kunz}}, \citenamefont
  {{Kurki-Suonio}}, \citenamefont {{Lagache}}, \citenamefont
  {{L{\"a}hteenm{\"a}ki}}, \citenamefont {{Lamarre}}, \citenamefont
  {{Lasenby}}, \citenamefont {{Lattanzi}}, \citenamefont {{Lawrence}},
  \citenamefont {{Leonardi}}, \citenamefont {{Lesgourgues}}, \citenamefont
  {{Levrier}}, \citenamefont {{Lewis}}, \citenamefont {{Liguori}},
  \citenamefont {{Lilje}}, \citenamefont {{Linden-V{\o}rnle}}, \citenamefont
  {{L{\'o}pez-Caniego}}, \citenamefont {{Lubin}}, \citenamefont {{Ma}},
  \citenamefont {{Mac{\'\i}as-P{\'e}rez}}, \citenamefont {{Maggio}},
  \citenamefont {{Maino}}, \citenamefont {{Mandolesi}}, \citenamefont
  {{Mangilli}}, \citenamefont {{Marchini}}, \citenamefont {{Maris}},
  \citenamefont {{Martin}}, \citenamefont {{Martinelli}}, \citenamefont
  {{Mart{\'\i}nez-Gonz{\'a}lez}}, \citenamefont {{Masi}}, \citenamefont
  {{Matarrese}}, \citenamefont {{McGehee}}, \citenamefont {{Meinhold}},
  \citenamefont {{Melchiorri}}, \citenamefont {{Mendes}}, \citenamefont
  {{Mennella}}, \citenamefont {{Migliaccio}}, \citenamefont {{Mitra}},
  \citenamefont {{Miville-Desch{\^e}nes}}, \citenamefont {{Moneti}},
  \citenamefont {{Montier}}, \citenamefont {{Morgante}}, \citenamefont
  {{Mortlock}}, \citenamefont {{Moss}}, \citenamefont {{Munshi}}, \citenamefont
  {{Murphy}}, \citenamefont {{Narimani}}, \citenamefont {{Naselsky}},
  \citenamefont {{Nati}}, \citenamefont {{Natoli}}, \citenamefont
  {{Netterfield}}, \citenamefont {{N{\o}rgaard-Nielsen}}, \citenamefont
  {{Noviello}}, \citenamefont {{Novikov}}, \citenamefont {{Novikov}},
  \citenamefont {{Oxborrow}}, \citenamefont {{Paci}}, \citenamefont {{Pagano}},
  \citenamefont {{Pajot}}, \citenamefont {{Paoletti}}, \citenamefont
  {{Pasian}}, \citenamefont {{Patanchon}}, \citenamefont {{Pearson}},
  \citenamefont {{Perdereau}}, \citenamefont {{Perotto}}, \citenamefont
  {{Perrotta}}, \citenamefont {{Pettorino}}, \citenamefont {{Piacentini}},
  \citenamefont {{Piat}}, \citenamefont {{Pierpaoli}}, \citenamefont
  {{Pietrobon}}, \citenamefont {{Plaszczynski}}, \citenamefont
  {{Pointecouteau}}, \citenamefont {{Polenta}}, \citenamefont {{Popa}},
  \citenamefont {{Pratt}}, \citenamefont {{Pr{\'e}zeau}}, \citenamefont
  {{Prunet}}, \citenamefont {{Puget}}, \citenamefont {{Rachen}}, \citenamefont
  {{Reach}}, \citenamefont {{Rebolo}}, \citenamefont {{Reinecke}},
  \citenamefont {{Remazeilles}}, \citenamefont {{Renault}}, \citenamefont
  {{Renzi}}, \citenamefont {{Ristorcelli}}, \citenamefont {{Rocha}},
  \citenamefont {{Rosset}}, \citenamefont {{Rossetti}}, \citenamefont
  {{Roudier}}, \citenamefont {{Rowan-Robinson}}, \citenamefont
  {{Rubi{\~n}o-Mart{\'\i}n}}, \citenamefont {{Rusholme}}, \citenamefont
  {{Salvatelli}}, \citenamefont {{Sandri}}, \citenamefont {{Santos}},
  \citenamefont {{Savelainen}}, \citenamefont {{Savini}}, \citenamefont
  {{Schaefer}}, \citenamefont {{Scott}}, \citenamefont {{Seiffert}},
  \citenamefont {{Shellard}}, \citenamefont {{Spencer}}, \citenamefont
  {{Stolyarov}}, \citenamefont {{Stompor}}, \citenamefont {{Sudiwala}},
  \citenamefont {{Sunyaev}}, \citenamefont {{Sutton}}, \citenamefont
  {{Suur-Uski}}, \citenamefont {{Sygnet}}, \citenamefont {{Tauber}},
  \citenamefont {{Terenzi}}, \citenamefont {{Toffolatti}}, \citenamefont
  {{Tomasi}}, \citenamefont {{Tristram}}, \citenamefont {{Tucci}},
  \citenamefont {{Tuovinen}}, \citenamefont {{Valenziano}}, \citenamefont
  {{Valiviita}}, \citenamefont {{Van Tent}}, \citenamefont {{Viel}},
  \citenamefont {{Vielva}}, \citenamefont {{Villa}}, \citenamefont {{Wade}},
  \citenamefont {{Wandelt}}, \citenamefont {{Wehus}}, \citenamefont {{White}},
  \citenamefont {{Yvon}}, \citenamefont {{Zacchei}},\ and\ \citenamefont
  {{Zonca}}}]{Planck15mg}%
  \BibitemOpen
  \bibfield  {author} {\bibinfo {author} {\bibnamefont {{Planck
  Collaboration}}}, \bibinfo {author} {\bibfnamefont {P.~A.~R.}\ \bibnamefont
  {{Ade}}}, \bibinfo {author} {\bibfnamefont {N.}~\bibnamefont {{Aghanim}}},
  \bibinfo {author} {\bibfnamefont {M.}~\bibnamefont {{Arnaud}}}, \bibinfo
  {author} {\bibfnamefont {M.}~\bibnamefont {{Ashdown}}}, \bibinfo {author}
  {\bibfnamefont {J.}~\bibnamefont {{Aumont}}}, \bibinfo {author}
  {\bibfnamefont {C.}~\bibnamefont {{Baccigalupi}}}, \bibinfo {author}
  {\bibfnamefont {A.~J.}\ \bibnamefont {{Banday}}}, \bibinfo {author}
  {\bibfnamefont {R.~B.}\ \bibnamefont {{Barreiro}}}, \bibinfo {author}
  {\bibfnamefont {N.}~\bibnamefont {{Bartolo}}}, \bibinfo {author}
  {\bibfnamefont {E.}~\bibnamefont {{Battaner}}}, \bibinfo {author}
  {\bibfnamefont {R.}~\bibnamefont {{Battye}}}, \bibinfo {author}
  {\bibfnamefont {K.}~\bibnamefont {{Benabed}}}, \bibinfo {author}
  {\bibfnamefont {A.}~\bibnamefont {{Beno{\^\i}t}}}, \bibinfo {author}
  {\bibfnamefont {A.}~\bibnamefont {{Benoit-L{\'e}vy}}}, \bibinfo {author}
  {\bibfnamefont {J.~P.}\ \bibnamefont {{Bernard}}}, \bibinfo {author}
  {\bibfnamefont {M.}~\bibnamefont {{Bersanelli}}}, \bibinfo {author}
  {\bibfnamefont {P.}~\bibnamefont {{Bielewicz}}}, \bibinfo {author}
  {\bibfnamefont {J.~J.}\ \bibnamefont {{Bock}}}, \bibinfo {author}
  {\bibfnamefont {A.}~\bibnamefont {{Bonaldi}}}, \bibinfo {author}
  {\bibfnamefont {L.}~\bibnamefont {{Bonavera}}}, \bibinfo {author}
  {\bibfnamefont {J.~R.}\ \bibnamefont {{Bond}}}, \bibinfo {author}
  {\bibfnamefont {J.}~\bibnamefont {{Borrill}}}, \bibinfo {author}
  {\bibfnamefont {F.~R.}\ \bibnamefont {{Bouchet}}}, \bibinfo {author}
  {\bibfnamefont {M.}~\bibnamefont {{Bucher}}}, \bibinfo {author}
  {\bibfnamefont {C.}~\bibnamefont {{Burigana}}}, \bibinfo {author}
  {\bibfnamefont {R.~C.}\ \bibnamefont {{Butler}}}, \bibinfo {author}
  {\bibfnamefont {E.}~\bibnamefont {{Calabrese}}}, \bibinfo {author}
  {\bibfnamefont {J.~F.}\ \bibnamefont {{Cardoso}}}, \bibinfo {author}
  {\bibfnamefont {A.}~\bibnamefont {{Catalano}}}, \bibinfo {author}
  {\bibfnamefont {A.}~\bibnamefont {{Challinor}}}, \bibinfo {author}
  {\bibfnamefont {A.}~\bibnamefont {{Chamballu}}}, \bibinfo {author}
  {\bibfnamefont {H.~C.}\ \bibnamefont {{Chiang}}}, \bibinfo {author}
  {\bibfnamefont {P.~R.}\ \bibnamefont {{Christensen}}}, \bibinfo {author}
  {\bibfnamefont {S.}~\bibnamefont {{Church}}}, \bibinfo {author}
  {\bibfnamefont {D.~L.}\ \bibnamefont {{Clements}}}, \bibinfo {author}
  {\bibfnamefont {S.}~\bibnamefont {{Colombi}}}, \bibinfo {author}
  {\bibfnamefont {L.~P.~L.}\ \bibnamefont {{Colombo}}}, \bibinfo {author}
  {\bibfnamefont {C.}~\bibnamefont {{Combet}}}, \bibinfo {author}
  {\bibfnamefont {F.}~\bibnamefont {{Couchot}}}, \bibinfo {author}
  {\bibfnamefont {A.}~\bibnamefont {{Coulais}}}, \bibinfo {author}
  {\bibfnamefont {B.~P.}\ \bibnamefont {{Crill}}}, \bibinfo {author}
  {\bibfnamefont {A.}~\bibnamefont {{Curto}}}, \bibinfo {author} {\bibfnamefont
  {F.}~\bibnamefont {{Cuttaia}}}, \bibinfo {author} {\bibfnamefont
  {L.}~\bibnamefont {{Danese}}}, \bibinfo {author} {\bibfnamefont {R.~D.}\
  \bibnamefont {{Davies}}}, \bibinfo {author} {\bibfnamefont {R.~J.}\
  \bibnamefont {{Davis}}}, \bibinfo {author} {\bibfnamefont {P.}~\bibnamefont
  {{de Bernardis}}}, \bibinfo {author} {\bibfnamefont {A.}~\bibnamefont {{de
  Rosa}}}, \bibinfo {author} {\bibfnamefont {G.}~\bibnamefont {{de Zotti}}},
  \bibinfo {author} {\bibfnamefont {J.}~\bibnamefont {{Delabrouille}}},
  \bibinfo {author} {\bibfnamefont {F.~X.}\ \bibnamefont {{D{\'e}sert}}},
  \bibinfo {author} {\bibfnamefont {J.~M.}\ \bibnamefont {{Diego}}}, \bibinfo
  {author} {\bibfnamefont {H.}~\bibnamefont {{Dole}}}, \bibinfo {author}
  {\bibfnamefont {S.}~\bibnamefont {{Donzelli}}}, \bibinfo {author}
  {\bibfnamefont {O.}~\bibnamefont {{Dor{\'e}}}}, \bibinfo {author}
  {\bibfnamefont {M.}~\bibnamefont {{Douspis}}}, \bibinfo {author}
  {\bibfnamefont {A.}~\bibnamefont {{Ducout}}}, \bibinfo {author}
  {\bibfnamefont {X.}~\bibnamefont {{Dupac}}}, \bibinfo {author} {\bibfnamefont
  {G.}~\bibnamefont {{Efstathiou}}}, \bibinfo {author} {\bibfnamefont
  {F.}~\bibnamefont {{Elsner}}}, \bibinfo {author} {\bibfnamefont {T.~A.}\
  \bibnamefont {{En{\ss}lin}}}, \bibinfo {author} {\bibfnamefont {H.~K.}\
  \bibnamefont {{Eriksen}}}, \bibinfo {author} {\bibfnamefont {J.}~\bibnamefont
  {{Fergusson}}}, \bibinfo {author} {\bibfnamefont {F.}~\bibnamefont
  {{Finelli}}}, \bibinfo {author} {\bibfnamefont {O.}~\bibnamefont {{Forni}}},
  \bibinfo {author} {\bibfnamefont {M.}~\bibnamefont {{Frailis}}}, \bibinfo
  {author} {\bibfnamefont {A.~A.}\ \bibnamefont {{Fraisse}}}, \bibinfo {author}
  {\bibfnamefont {E.}~\bibnamefont {{Franceschi}}}, \bibinfo {author}
  {\bibfnamefont {A.}~\bibnamefont {{Frejsel}}}, \bibinfo {author}
  {\bibfnamefont {S.}~\bibnamefont {{Galeotta}}}, \bibinfo {author}
  {\bibfnamefont {S.}~\bibnamefont {{Galli}}}, \bibinfo {author} {\bibfnamefont
  {K.}~\bibnamefont {{Ganga}}}, \bibinfo {author} {\bibfnamefont
  {M.}~\bibnamefont {{Giard}}}, \bibinfo {author} {\bibfnamefont
  {Y.}~\bibnamefont {{Giraud-H{\'e}raud}}}, \bibinfo {author} {\bibfnamefont
  {E.}~\bibnamefont {{Gjerl{\o}w}}}, \bibinfo {author} {\bibfnamefont
  {J.}~\bibnamefont {{Gonz{\'a}lez-Nuevo}}}, \bibinfo {author} {\bibfnamefont
  {K.~M.}\ \bibnamefont {{G{\'o}rski}}}, \bibinfo {author} {\bibfnamefont
  {S.}~\bibnamefont {{Gratton}}}, \bibinfo {author} {\bibfnamefont
  {A.}~\bibnamefont {{Gregorio}}}, \bibinfo {author} {\bibfnamefont
  {A.}~\bibnamefont {{Gruppuso}}}, \bibinfo {author} {\bibfnamefont {J.~E.}\
  \bibnamefont {{Gudmundsson}}}, \bibinfo {author} {\bibfnamefont {F.~K.}\
  \bibnamefont {{Hansen}}}, \bibinfo {author} {\bibfnamefont {D.}~\bibnamefont
  {{Hanson}}}, \bibinfo {author} {\bibfnamefont {D.~L.}\ \bibnamefont
  {{Harrison}}}, \bibinfo {author} {\bibfnamefont {A.}~\bibnamefont
  {{Heavens}}}, \bibinfo {author} {\bibfnamefont {G.}~\bibnamefont {{Helou}}},
  \bibinfo {author} {\bibfnamefont {S.}~\bibnamefont {{Henrot-Versill{\'e}}}},
  \bibinfo {author} {\bibfnamefont {C.}~\bibnamefont
  {{Hern{\'a}ndez-Monteagudo}}}, \bibinfo {author} {\bibfnamefont
  {D.}~\bibnamefont {{Herranz}}}, \bibinfo {author} {\bibfnamefont {S.~R.}\
  \bibnamefont {{Hildebrand t}}}, \bibinfo {author} {\bibfnamefont
  {E.}~\bibnamefont {{Hivon}}}, \bibinfo {author} {\bibfnamefont
  {M.}~\bibnamefont {{Hobson}}}, \bibinfo {author} {\bibfnamefont {W.~A.}\
  \bibnamefont {{Holmes}}}, \bibinfo {author} {\bibfnamefont {A.}~\bibnamefont
  {{Hornstrup}}}, \bibinfo {author} {\bibfnamefont {W.}~\bibnamefont
  {{Hovest}}}, \bibinfo {author} {\bibfnamefont {Z.}~\bibnamefont {{Huang}}},
  \bibinfo {author} {\bibfnamefont {K.~M.}\ \bibnamefont {{Huffenberger}}},
  \bibinfo {author} {\bibfnamefont {G.}~\bibnamefont {{Hurier}}}, \bibinfo
  {author} {\bibfnamefont {A.~H.}\ \bibnamefont {{Jaffe}}}, \bibinfo {author}
  {\bibfnamefont {T.~R.}\ \bibnamefont {{Jaffe}}}, \bibinfo {author}
  {\bibfnamefont {W.~C.}\ \bibnamefont {{Jones}}}, \bibinfo {author}
  {\bibfnamefont {M.}~\bibnamefont {{Juvela}}}, \bibinfo {author}
  {\bibfnamefont {E.}~\bibnamefont {{Keih{\"a}nen}}}, \bibinfo {author}
  {\bibfnamefont {R.}~\bibnamefont {{Keskitalo}}}, \bibinfo {author}
  {\bibfnamefont {T.~S.}\ \bibnamefont {{Kisner}}}, \bibinfo {author}
  {\bibfnamefont {J.}~\bibnamefont {{Knoche}}}, \bibinfo {author}
  {\bibfnamefont {M.}~\bibnamefont {{Kunz}}}, \bibinfo {author} {\bibfnamefont
  {H.}~\bibnamefont {{Kurki-Suonio}}}, \bibinfo {author} {\bibfnamefont
  {G.}~\bibnamefont {{Lagache}}}, \bibinfo {author} {\bibfnamefont
  {A.}~\bibnamefont {{L{\"a}hteenm{\"a}ki}}}, \bibinfo {author} {\bibfnamefont
  {J.~M.}\ \bibnamefont {{Lamarre}}}, \bibinfo {author} {\bibfnamefont
  {A.}~\bibnamefont {{Lasenby}}}, \bibinfo {author} {\bibfnamefont
  {M.}~\bibnamefont {{Lattanzi}}}, \bibinfo {author} {\bibfnamefont {C.~R.}\
  \bibnamefont {{Lawrence}}}, \bibinfo {author} {\bibfnamefont
  {R.}~\bibnamefont {{Leonardi}}}, \bibinfo {author} {\bibfnamefont
  {J.}~\bibnamefont {{Lesgourgues}}}, \bibinfo {author} {\bibfnamefont
  {F.}~\bibnamefont {{Levrier}}}, \bibinfo {author} {\bibfnamefont
  {A.}~\bibnamefont {{Lewis}}}, \bibinfo {author} {\bibfnamefont
  {M.}~\bibnamefont {{Liguori}}}, \bibinfo {author} {\bibfnamefont {P.~B.}\
  \bibnamefont {{Lilje}}}, \bibinfo {author} {\bibfnamefont {M.}~\bibnamefont
  {{Linden-V{\o}rnle}}}, \bibinfo {author} {\bibfnamefont {M.}~\bibnamefont
  {{L{\'o}pez-Caniego}}}, \bibinfo {author} {\bibfnamefont {P.~M.}\
  \bibnamefont {{Lubin}}}, \bibinfo {author} {\bibfnamefont {Y.~Z.}\
  \bibnamefont {{Ma}}}, \bibinfo {author} {\bibfnamefont {J.~F.}\ \bibnamefont
  {{Mac{\'\i}as-P{\'e}rez}}}, \bibinfo {author} {\bibfnamefont
  {G.}~\bibnamefont {{Maggio}}}, \bibinfo {author} {\bibfnamefont
  {D.}~\bibnamefont {{Maino}}}, \bibinfo {author} {\bibfnamefont
  {N.}~\bibnamefont {{Mandolesi}}}, \bibinfo {author} {\bibfnamefont
  {A.}~\bibnamefont {{Mangilli}}}, \bibinfo {author} {\bibfnamefont
  {A.}~\bibnamefont {{Marchini}}}, \bibinfo {author} {\bibfnamefont
  {M.}~\bibnamefont {{Maris}}}, \bibinfo {author} {\bibfnamefont {P.~G.}\
  \bibnamefont {{Martin}}}, \bibinfo {author} {\bibfnamefont {M.}~\bibnamefont
  {{Martinelli}}}, \bibinfo {author} {\bibfnamefont {E.}~\bibnamefont
  {{Mart{\'\i}nez-Gonz{\'a}lez}}}, \bibinfo {author} {\bibfnamefont
  {S.}~\bibnamefont {{Masi}}}, \bibinfo {author} {\bibfnamefont
  {S.}~\bibnamefont {{Matarrese}}}, \bibinfo {author} {\bibfnamefont
  {P.}~\bibnamefont {{McGehee}}}, \bibinfo {author} {\bibfnamefont {P.~R.}\
  \bibnamefont {{Meinhold}}}, \bibinfo {author} {\bibfnamefont
  {A.}~\bibnamefont {{Melchiorri}}}, \bibinfo {author} {\bibfnamefont
  {L.}~\bibnamefont {{Mendes}}}, \bibinfo {author} {\bibfnamefont
  {A.}~\bibnamefont {{Mennella}}}, \bibinfo {author} {\bibfnamefont
  {M.}~\bibnamefont {{Migliaccio}}}, \bibinfo {author} {\bibfnamefont
  {S.}~\bibnamefont {{Mitra}}}, \bibinfo {author} {\bibfnamefont {M.~A.}\
  \bibnamefont {{Miville-Desch{\^e}nes}}}, \bibinfo {author} {\bibfnamefont
  {A.}~\bibnamefont {{Moneti}}}, \bibinfo {author} {\bibfnamefont
  {L.}~\bibnamefont {{Montier}}}, \bibinfo {author} {\bibfnamefont
  {G.}~\bibnamefont {{Morgante}}}, \bibinfo {author} {\bibfnamefont
  {D.}~\bibnamefont {{Mortlock}}}, \bibinfo {author} {\bibfnamefont
  {A.}~\bibnamefont {{Moss}}}, \bibinfo {author} {\bibfnamefont
  {D.}~\bibnamefont {{Munshi}}}, \bibinfo {author} {\bibfnamefont {J.~A.}\
  \bibnamefont {{Murphy}}}, \bibinfo {author} {\bibfnamefont {A.}~\bibnamefont
  {{Narimani}}}, \bibinfo {author} {\bibfnamefont {P.}~\bibnamefont
  {{Naselsky}}}, \bibinfo {author} {\bibfnamefont {F.}~\bibnamefont {{Nati}}},
  \bibinfo {author} {\bibfnamefont {P.}~\bibnamefont {{Natoli}}}, \bibinfo
  {author} {\bibfnamefont {C.~B.}\ \bibnamefont {{Netterfield}}}, \bibinfo
  {author} {\bibfnamefont {H.~U.}\ \bibnamefont {{N{\o}rgaard-Nielsen}}},
  \bibinfo {author} {\bibfnamefont {F.}~\bibnamefont {{Noviello}}}, \bibinfo
  {author} {\bibfnamefont {D.}~\bibnamefont {{Novikov}}}, \bibinfo {author}
  {\bibfnamefont {I.}~\bibnamefont {{Novikov}}}, \bibinfo {author}
  {\bibfnamefont {C.~A.}\ \bibnamefont {{Oxborrow}}}, \bibinfo {author}
  {\bibfnamefont {F.}~\bibnamefont {{Paci}}}, \bibinfo {author} {\bibfnamefont
  {L.}~\bibnamefont {{Pagano}}}, \bibinfo {author} {\bibfnamefont
  {F.}~\bibnamefont {{Pajot}}}, \bibinfo {author} {\bibfnamefont
  {D.}~\bibnamefont {{Paoletti}}}, \bibinfo {author} {\bibfnamefont
  {F.}~\bibnamefont {{Pasian}}}, \bibinfo {author} {\bibfnamefont
  {G.}~\bibnamefont {{Patanchon}}}, \bibinfo {author} {\bibfnamefont {T.~J.}\
  \bibnamefont {{Pearson}}}, \bibinfo {author} {\bibfnamefont {O.}~\bibnamefont
  {{Perdereau}}}, \bibinfo {author} {\bibfnamefont {L.}~\bibnamefont
  {{Perotto}}}, \bibinfo {author} {\bibfnamefont {F.}~\bibnamefont
  {{Perrotta}}}, \bibinfo {author} {\bibfnamefont {V.}~\bibnamefont
  {{Pettorino}}}, \bibinfo {author} {\bibfnamefont {F.}~\bibnamefont
  {{Piacentini}}}, \bibinfo {author} {\bibfnamefont {M.}~\bibnamefont
  {{Piat}}}, \bibinfo {author} {\bibfnamefont {E.}~\bibnamefont {{Pierpaoli}}},
  \bibinfo {author} {\bibfnamefont {D.}~\bibnamefont {{Pietrobon}}}, \bibinfo
  {author} {\bibfnamefont {S.}~\bibnamefont {{Plaszczynski}}}, \bibinfo
  {author} {\bibfnamefont {E.}~\bibnamefont {{Pointecouteau}}}, \bibinfo
  {author} {\bibfnamefont {G.}~\bibnamefont {{Polenta}}}, \bibinfo {author}
  {\bibfnamefont {L.}~\bibnamefont {{Popa}}}, \bibinfo {author} {\bibfnamefont
  {G.~W.}\ \bibnamefont {{Pratt}}}, \bibinfo {author} {\bibfnamefont
  {G.}~\bibnamefont {{Pr{\'e}zeau}}}, \bibinfo {author} {\bibfnamefont
  {S.}~\bibnamefont {{Prunet}}}, \bibinfo {author} {\bibfnamefont {J.~L.}\
  \bibnamefont {{Puget}}}, \bibinfo {author} {\bibfnamefont {J.~P.}\
  \bibnamefont {{Rachen}}}, \bibinfo {author} {\bibfnamefont {W.~T.}\
  \bibnamefont {{Reach}}}, \bibinfo {author} {\bibfnamefont {R.}~\bibnamefont
  {{Rebolo}}}, \bibinfo {author} {\bibfnamefont {M.}~\bibnamefont
  {{Reinecke}}}, \bibinfo {author} {\bibfnamefont {M.}~\bibnamefont
  {{Remazeilles}}}, \bibinfo {author} {\bibfnamefont {C.}~\bibnamefont
  {{Renault}}}, \bibinfo {author} {\bibfnamefont {A.}~\bibnamefont {{Renzi}}},
  \bibinfo {author} {\bibfnamefont {I.}~\bibnamefont {{Ristorcelli}}}, \bibinfo
  {author} {\bibfnamefont {G.}~\bibnamefont {{Rocha}}}, \bibinfo {author}
  {\bibfnamefont {C.}~\bibnamefont {{Rosset}}}, \bibinfo {author}
  {\bibfnamefont {M.}~\bibnamefont {{Rossetti}}}, \bibinfo {author}
  {\bibfnamefont {G.}~\bibnamefont {{Roudier}}}, \bibinfo {author}
  {\bibfnamefont {M.}~\bibnamefont {{Rowan-Robinson}}}, \bibinfo {author}
  {\bibfnamefont {J.~A.}\ \bibnamefont {{Rubi{\~n}o-Mart{\'\i}n}}}, \bibinfo
  {author} {\bibfnamefont {B.}~\bibnamefont {{Rusholme}}}, \bibinfo {author}
  {\bibfnamefont {V.}~\bibnamefont {{Salvatelli}}}, \bibinfo {author}
  {\bibfnamefont {M.}~\bibnamefont {{Sandri}}}, \bibinfo {author}
  {\bibfnamefont {D.}~\bibnamefont {{Santos}}}, \bibinfo {author}
  {\bibfnamefont {M.}~\bibnamefont {{Savelainen}}}, \bibinfo {author}
  {\bibfnamefont {G.}~\bibnamefont {{Savini}}}, \bibinfo {author}
  {\bibfnamefont {B.~M.}\ \bibnamefont {{Schaefer}}}, \bibinfo {author}
  {\bibfnamefont {D.}~\bibnamefont {{Scott}}}, \bibinfo {author} {\bibfnamefont
  {M.~D.}\ \bibnamefont {{Seiffert}}}, \bibinfo {author} {\bibfnamefont
  {E.~P.~S.}\ \bibnamefont {{Shellard}}}, \bibinfo {author} {\bibfnamefont
  {L.~D.}\ \bibnamefont {{Spencer}}}, \bibinfo {author} {\bibfnamefont
  {V.}~\bibnamefont {{Stolyarov}}}, \bibinfo {author} {\bibfnamefont
  {R.}~\bibnamefont {{Stompor}}}, \bibinfo {author} {\bibfnamefont
  {R.}~\bibnamefont {{Sudiwala}}}, \bibinfo {author} {\bibfnamefont
  {R.}~\bibnamefont {{Sunyaev}}}, \bibinfo {author} {\bibfnamefont
  {D.}~\bibnamefont {{Sutton}}}, \bibinfo {author} {\bibfnamefont {A.~S.}\
  \bibnamefont {{Suur-Uski}}}, \bibinfo {author} {\bibfnamefont {J.~F.}\
  \bibnamefont {{Sygnet}}}, \bibinfo {author} {\bibfnamefont {J.~A.}\
  \bibnamefont {{Tauber}}}, \bibinfo {author} {\bibfnamefont {L.}~\bibnamefont
  {{Terenzi}}}, \bibinfo {author} {\bibfnamefont {L.}~\bibnamefont
  {{Toffolatti}}}, \bibinfo {author} {\bibfnamefont {M.}~\bibnamefont
  {{Tomasi}}}, \bibinfo {author} {\bibfnamefont {M.}~\bibnamefont
  {{Tristram}}}, \bibinfo {author} {\bibfnamefont {M.}~\bibnamefont {{Tucci}}},
  \bibinfo {author} {\bibfnamefont {J.}~\bibnamefont {{Tuovinen}}}, \bibinfo
  {author} {\bibfnamefont {L.}~\bibnamefont {{Valenziano}}}, \bibinfo {author}
  {\bibfnamefont {J.}~\bibnamefont {{Valiviita}}}, \bibinfo {author}
  {\bibfnamefont {B.}~\bibnamefont {{Van Tent}}}, \bibinfo {author}
  {\bibfnamefont {M.}~\bibnamefont {{Viel}}}, \bibinfo {author} {\bibfnamefont
  {P.}~\bibnamefont {{Vielva}}}, \bibinfo {author} {\bibfnamefont
  {F.}~\bibnamefont {{Villa}}}, \bibinfo {author} {\bibfnamefont {L.~A.}\
  \bibnamefont {{Wade}}}, \bibinfo {author} {\bibfnamefont {B.~D.}\
  \bibnamefont {{Wandelt}}}, \bibinfo {author} {\bibfnamefont {I.~K.}\
  \bibnamefont {{Wehus}}}, \bibinfo {author} {\bibfnamefont {M.}~\bibnamefont
  {{White}}}, \bibinfo {author} {\bibfnamefont {D.}~\bibnamefont {{Yvon}}},
  \bibinfo {author} {\bibfnamefont {A.}~\bibnamefont {{Zacchei}}}, \ and\
  \bibinfo {author} {\bibfnamefont {A.}~\bibnamefont {{Zonca}}},\ }\href
  {\doibase 10.1051/0004-6361/201525814} {\bibfield  {journal} {\bibinfo
  {journal} {\aap}\ }\textbf {\bibinfo {volume} {594}},\ \bibinfo {eid} {A14}
  (\bibinfo {year} {2016}{\natexlab{b}})},\ \Eprint
  {http://arxiv.org/abs/1502.01590} {arXiv:1502.01590 [astro-ph.CO]}
  \BibitemShut {NoStop}%
\bibitem [{\citenamefont {{Troxel}}\ \emph {et~al.}(2018)\citenamefont
  {{Troxel}}, \citenamefont {{MacCrann}}, \citenamefont {{Zuntz}},
  \citenamefont {{Eifler}}, \citenamefont {{Krause}}, \citenamefont
  {{Dodelson}}, \citenamefont {{Gruen}}, \citenamefont {{Blazek}},
  \citenamefont {{Friedrich}}, \citenamefont {{Samuroff}}, \citenamefont
  {{Prat}}, \citenamefont {{Secco}}, \citenamefont {{Davis}}, \citenamefont
  {{Fert{\'e}}}, \citenamefont {{DeRose}}, \citenamefont {{Alarcon}},
  \citenamefont {{Amara}}, \citenamefont {{Baxter}}, \citenamefont {{Becker}},
  \citenamefont {{Bernstein}}, \citenamefont {{Bridle}}, \citenamefont
  {{Cawthon}}, \citenamefont {{Chang}}, \citenamefont {{Choi}}, \citenamefont
  {{De Vicente}}, \citenamefont {{Drlica-Wagner}}, \citenamefont
  {{Elvin-Poole}}, \citenamefont {{Frieman}}, \citenamefont {{Gatti}},
  \citenamefont {{Hartley}}, \citenamefont {{Honscheid}}, \citenamefont
  {{Hoyle}}, \citenamefont {{Huff}}, \citenamefont {{Huterer}}, \citenamefont
  {{Jain}}, \citenamefont {{Jarvis}}, \citenamefont {{Kacprzak}}, \citenamefont
  {{Kirk}}, \citenamefont {{Kokron}}, \citenamefont {{Krawiec}}, \citenamefont
  {{Lahav}}, \citenamefont {{Liddle}}, \citenamefont {{Peacock}}, \citenamefont
  {{Rau}}, \citenamefont {{Refregier}}, \citenamefont {{Rollins}},
  \citenamefont {{Rozo}}, \citenamefont {{Rykoff}}, \citenamefont
  {{S{\'a}nchez}}, \citenamefont {{Sevilla-Noarbe}}, \citenamefont {{Sheldon}},
  \citenamefont {{Stebbins}}, \citenamefont {{Varga}}, \citenamefont
  {{Vielzeuf}}, \citenamefont {{Wang}}, \citenamefont {{Wechsler}},
  \citenamefont {{Yanny}}, \citenamefont {{Abbott}}, \citenamefont {{Abdalla}},
  \citenamefont {{Allam}}, \citenamefont {{Annis}}, \citenamefont {{Bechtol}},
  \citenamefont {{Benoit-L{\'e}vy}}, \citenamefont {{Bertin}}, \citenamefont
  {{Brooks}}, \citenamefont {{Buckley-Geer}}, \citenamefont {{Burke}},
  \citenamefont {{Carnero Rosell}}, \citenamefont {{Carrasco Kind}},
  \citenamefont {{Carretero}}, \citenamefont {{Castander}}, \citenamefont
  {{Crocce}}, \citenamefont {{Cunha}}, \citenamefont {{D'Andrea}},
  \citenamefont {{da Costa}}, \citenamefont {{DePoy}}, \citenamefont {{Desai}},
  \citenamefont {{Diehl}}, \citenamefont {{Dietrich}}, \citenamefont {{Doel}},
  \citenamefont {{Fernandez}}, \citenamefont {{Flaugher}}, \citenamefont
  {{Fosalba}}, \citenamefont {{Garc{\'\i}a-Bellido}}, \citenamefont
  {{Gaztanaga}}, \citenamefont {{Gerdes}}, \citenamefont {{Giannantonio}},
  \citenamefont {{Goldstein}}, \citenamefont {{Gruendl}}, \citenamefont
  {{Gschwend}}, \citenamefont {{Gutierrez}}, \citenamefont {{James}},
  \citenamefont {{Jeltema}}, \citenamefont {{Johnson}}, \citenamefont
  {{Johnson}}, \citenamefont {{Kent}}, \citenamefont {{Kuehn}}, \citenamefont
  {{Kuhlmann}}, \citenamefont {{Kuropatkin}}, \citenamefont {{Li}},
  \citenamefont {{Lima}}, \citenamefont {{Lin}}, \citenamefont {{Maia}},
  \citenamefont {{March}}, \citenamefont {{Marshall}}, \citenamefont
  {{Martini}}, \citenamefont {{Melchior}}, \citenamefont {{Menanteau}},
  \citenamefont {{Miquel}}, \citenamefont {{Mohr}}, \citenamefont {{Neilsen}},
  \citenamefont {{Nichol}}, \citenamefont {{Nord}}, \citenamefont
  {{Petravick}}, \citenamefont {{Plazas}}, \citenamefont {{Romer}},
  \citenamefont {{Roodman}}, \citenamefont {{Sako}}, \citenamefont {{Sanchez}},
  \citenamefont {{Scarpine}}, \citenamefont {{Schindler}}, \citenamefont
  {{Schubnell}}, \citenamefont {{Smith}}, \citenamefont {{Smith}},
  \citenamefont {{Soares-Santos}}, \citenamefont {{Sobreira}}, \citenamefont
  {{Suchyta}}, \citenamefont {{Swanson}}, \citenamefont {{Tarle}},
  \citenamefont {{Thomas}}, \citenamefont {{Tucker}}, \citenamefont {{Vikram}},
  \citenamefont {{Walker}}, \citenamefont {{Weller}}, \citenamefont {{Zhang}},\
  and\ \citenamefont {{DES Collaboration}}}]{Troxel18}%
  \BibitemOpen
  \bibfield  {author} {\bibinfo {author} {\bibfnamefont {M.~A.}\ \bibnamefont
  {{Troxel}}}, \bibinfo {author} {\bibfnamefont {N.}~\bibnamefont
  {{MacCrann}}}, \bibinfo {author} {\bibfnamefont {J.}~\bibnamefont {{Zuntz}}},
  \bibinfo {author} {\bibfnamefont {T.~F.}\ \bibnamefont {{Eifler}}}, \bibinfo
  {author} {\bibfnamefont {E.}~\bibnamefont {{Krause}}}, \bibinfo {author}
  {\bibfnamefont {S.}~\bibnamefont {{Dodelson}}}, \bibinfo {author}
  {\bibfnamefont {D.}~\bibnamefont {{Gruen}}}, \bibinfo {author} {\bibfnamefont
  {J.}~\bibnamefont {{Blazek}}}, \bibinfo {author} {\bibfnamefont
  {O.}~\bibnamefont {{Friedrich}}}, \bibinfo {author} {\bibfnamefont
  {S.}~\bibnamefont {{Samuroff}}}, \bibinfo {author} {\bibfnamefont
  {J.}~\bibnamefont {{Prat}}}, \bibinfo {author} {\bibfnamefont {L.~F.}\
  \bibnamefont {{Secco}}}, \bibinfo {author} {\bibfnamefont {C.}~\bibnamefont
  {{Davis}}}, \bibinfo {author} {\bibfnamefont {A.}~\bibnamefont
  {{Fert{\'e}}}}, \bibinfo {author} {\bibfnamefont {J.}~\bibnamefont
  {{DeRose}}}, \bibinfo {author} {\bibfnamefont {A.}~\bibnamefont {{Alarcon}}},
  \bibinfo {author} {\bibfnamefont {A.}~\bibnamefont {{Amara}}}, \bibinfo
  {author} {\bibfnamefont {E.}~\bibnamefont {{Baxter}}}, \bibinfo {author}
  {\bibfnamefont {M.~R.}\ \bibnamefont {{Becker}}}, \bibinfo {author}
  {\bibfnamefont {G.~M.}\ \bibnamefont {{Bernstein}}}, \bibinfo {author}
  {\bibfnamefont {S.~L.}\ \bibnamefont {{Bridle}}}, \bibinfo {author}
  {\bibfnamefont {R.}~\bibnamefont {{Cawthon}}}, \bibinfo {author}
  {\bibfnamefont {C.}~\bibnamefont {{Chang}}}, \bibinfo {author} {\bibfnamefont
  {A.}~\bibnamefont {{Choi}}}, \bibinfo {author} {\bibfnamefont
  {J.}~\bibnamefont {{De Vicente}}}, \bibinfo {author} {\bibfnamefont
  {A.}~\bibnamefont {{Drlica-Wagner}}}, \bibinfo {author} {\bibfnamefont
  {J.}~\bibnamefont {{Elvin-Poole}}}, \bibinfo {author} {\bibfnamefont
  {J.}~\bibnamefont {{Frieman}}}, \bibinfo {author} {\bibfnamefont
  {M.}~\bibnamefont {{Gatti}}}, \bibinfo {author} {\bibfnamefont {W.~G.}\
  \bibnamefont {{Hartley}}}, \bibinfo {author} {\bibfnamefont {K.}~\bibnamefont
  {{Honscheid}}}, \bibinfo {author} {\bibfnamefont {B.}~\bibnamefont
  {{Hoyle}}}, \bibinfo {author} {\bibfnamefont {E.~M.}\ \bibnamefont {{Huff}}},
  \bibinfo {author} {\bibfnamefont {D.}~\bibnamefont {{Huterer}}}, \bibinfo
  {author} {\bibfnamefont {B.}~\bibnamefont {{Jain}}}, \bibinfo {author}
  {\bibfnamefont {M.}~\bibnamefont {{Jarvis}}}, \bibinfo {author}
  {\bibfnamefont {T.}~\bibnamefont {{Kacprzak}}}, \bibinfo {author}
  {\bibfnamefont {D.}~\bibnamefont {{Kirk}}}, \bibinfo {author} {\bibfnamefont
  {N.}~\bibnamefont {{Kokron}}}, \bibinfo {author} {\bibfnamefont
  {C.}~\bibnamefont {{Krawiec}}}, \bibinfo {author} {\bibfnamefont
  {O.}~\bibnamefont {{Lahav}}}, \bibinfo {author} {\bibfnamefont {A.~R.}\
  \bibnamefont {{Liddle}}}, \bibinfo {author} {\bibfnamefont {J.}~\bibnamefont
  {{Peacock}}}, \bibinfo {author} {\bibfnamefont {M.~M.}\ \bibnamefont
  {{Rau}}}, \bibinfo {author} {\bibfnamefont {A.}~\bibnamefont {{Refregier}}},
  \bibinfo {author} {\bibfnamefont {R.~P.}\ \bibnamefont {{Rollins}}}, \bibinfo
  {author} {\bibfnamefont {E.}~\bibnamefont {{Rozo}}}, \bibinfo {author}
  {\bibfnamefont {E.~S.}\ \bibnamefont {{Rykoff}}}, \bibinfo {author}
  {\bibfnamefont {C.}~\bibnamefont {{S{\'a}nchez}}}, \bibinfo {author}
  {\bibfnamefont {I.}~\bibnamefont {{Sevilla-Noarbe}}}, \bibinfo {author}
  {\bibfnamefont {E.}~\bibnamefont {{Sheldon}}}, \bibinfo {author}
  {\bibfnamefont {A.}~\bibnamefont {{Stebbins}}}, \bibinfo {author}
  {\bibfnamefont {T.~N.}\ \bibnamefont {{Varga}}}, \bibinfo {author}
  {\bibfnamefont {P.}~\bibnamefont {{Vielzeuf}}}, \bibinfo {author}
  {\bibfnamefont {M.}~\bibnamefont {{Wang}}}, \bibinfo {author} {\bibfnamefont
  {R.~H.}\ \bibnamefont {{Wechsler}}}, \bibinfo {author} {\bibfnamefont
  {B.}~\bibnamefont {{Yanny}}}, \bibinfo {author} {\bibfnamefont {T.~M.~C.}\
  \bibnamefont {{Abbott}}}, \bibinfo {author} {\bibfnamefont {F.~B.}\
  \bibnamefont {{Abdalla}}}, \bibinfo {author} {\bibfnamefont {S.}~\bibnamefont
  {{Allam}}}, \bibinfo {author} {\bibfnamefont {J.}~\bibnamefont {{Annis}}},
  \bibinfo {author} {\bibfnamefont {K.}~\bibnamefont {{Bechtol}}}, \bibinfo
  {author} {\bibfnamefont {A.}~\bibnamefont {{Benoit-L{\'e}vy}}}, \bibinfo
  {author} {\bibfnamefont {E.}~\bibnamefont {{Bertin}}}, \bibinfo {author}
  {\bibfnamefont {D.}~\bibnamefont {{Brooks}}}, \bibinfo {author}
  {\bibfnamefont {E.}~\bibnamefont {{Buckley-Geer}}}, \bibinfo {author}
  {\bibfnamefont {D.~L.}\ \bibnamefont {{Burke}}}, \bibinfo {author}
  {\bibfnamefont {A.}~\bibnamefont {{Carnero Rosell}}}, \bibinfo {author}
  {\bibfnamefont {M.}~\bibnamefont {{Carrasco Kind}}}, \bibinfo {author}
  {\bibfnamefont {J.}~\bibnamefont {{Carretero}}}, \bibinfo {author}
  {\bibfnamefont {F.~J.}\ \bibnamefont {{Castander}}}, \bibinfo {author}
  {\bibfnamefont {M.}~\bibnamefont {{Crocce}}}, \bibinfo {author}
  {\bibfnamefont {C.~E.}\ \bibnamefont {{Cunha}}}, \bibinfo {author}
  {\bibfnamefont {C.~B.}\ \bibnamefont {{D'Andrea}}}, \bibinfo {author}
  {\bibfnamefont {L.~N.}\ \bibnamefont {{da Costa}}}, \bibinfo {author}
  {\bibfnamefont {D.~L.}\ \bibnamefont {{DePoy}}}, \bibinfo {author}
  {\bibfnamefont {S.}~\bibnamefont {{Desai}}}, \bibinfo {author} {\bibfnamefont
  {H.~T.}\ \bibnamefont {{Diehl}}}, \bibinfo {author} {\bibfnamefont {J.~P.}\
  \bibnamefont {{Dietrich}}}, \bibinfo {author} {\bibfnamefont
  {P.}~\bibnamefont {{Doel}}}, \bibinfo {author} {\bibfnamefont
  {E.}~\bibnamefont {{Fernandez}}}, \bibinfo {author} {\bibfnamefont
  {B.}~\bibnamefont {{Flaugher}}}, \bibinfo {author} {\bibfnamefont
  {P.}~\bibnamefont {{Fosalba}}}, \bibinfo {author} {\bibfnamefont
  {J.}~\bibnamefont {{Garc{\'\i}a-Bellido}}}, \bibinfo {author} {\bibfnamefont
  {E.}~\bibnamefont {{Gaztanaga}}}, \bibinfo {author} {\bibfnamefont {D.~W.}\
  \bibnamefont {{Gerdes}}}, \bibinfo {author} {\bibfnamefont {T.}~\bibnamefont
  {{Giannantonio}}}, \bibinfo {author} {\bibfnamefont {D.~A.}\ \bibnamefont
  {{Goldstein}}}, \bibinfo {author} {\bibfnamefont {R.~A.}\ \bibnamefont
  {{Gruendl}}}, \bibinfo {author} {\bibfnamefont {J.}~\bibnamefont
  {{Gschwend}}}, \bibinfo {author} {\bibfnamefont {G.}~\bibnamefont
  {{Gutierrez}}}, \bibinfo {author} {\bibfnamefont {D.~J.}\ \bibnamefont
  {{James}}}, \bibinfo {author} {\bibfnamefont {T.}~\bibnamefont {{Jeltema}}},
  \bibinfo {author} {\bibfnamefont {M.~W.~G.}\ \bibnamefont {{Johnson}}},
  \bibinfo {author} {\bibfnamefont {M.~D.}\ \bibnamefont {{Johnson}}}, \bibinfo
  {author} {\bibfnamefont {S.}~\bibnamefont {{Kent}}}, \bibinfo {author}
  {\bibfnamefont {K.}~\bibnamefont {{Kuehn}}}, \bibinfo {author} {\bibfnamefont
  {S.}~\bibnamefont {{Kuhlmann}}}, \bibinfo {author} {\bibfnamefont
  {N.}~\bibnamefont {{Kuropatkin}}}, \bibinfo {author} {\bibfnamefont {T.~S.}\
  \bibnamefont {{Li}}}, \bibinfo {author} {\bibfnamefont {M.}~\bibnamefont
  {{Lima}}}, \bibinfo {author} {\bibfnamefont {H.}~\bibnamefont {{Lin}}},
  \bibinfo {author} {\bibfnamefont {M.~A.~G.}\ \bibnamefont {{Maia}}}, \bibinfo
  {author} {\bibfnamefont {M.}~\bibnamefont {{March}}}, \bibinfo {author}
  {\bibfnamefont {J.~L.}\ \bibnamefont {{Marshall}}}, \bibinfo {author}
  {\bibfnamefont {P.}~\bibnamefont {{Martini}}}, \bibinfo {author}
  {\bibfnamefont {P.}~\bibnamefont {{Melchior}}}, \bibinfo {author}
  {\bibfnamefont {F.}~\bibnamefont {{Menanteau}}}, \bibinfo {author}
  {\bibfnamefont {R.}~\bibnamefont {{Miquel}}}, \bibinfo {author}
  {\bibfnamefont {J.~J.}\ \bibnamefont {{Mohr}}}, \bibinfo {author}
  {\bibfnamefont {E.}~\bibnamefont {{Neilsen}}}, \bibinfo {author}
  {\bibfnamefont {R.~C.}\ \bibnamefont {{Nichol}}}, \bibinfo {author}
  {\bibfnamefont {B.}~\bibnamefont {{Nord}}}, \bibinfo {author} {\bibfnamefont
  {D.}~\bibnamefont {{Petravick}}}, \bibinfo {author} {\bibfnamefont {A.~A.}\
  \bibnamefont {{Plazas}}}, \bibinfo {author} {\bibfnamefont {A.~K.}\
  \bibnamefont {{Romer}}}, \bibinfo {author} {\bibfnamefont {A.}~\bibnamefont
  {{Roodman}}}, \bibinfo {author} {\bibfnamefont {M.}~\bibnamefont {{Sako}}},
  \bibinfo {author} {\bibfnamefont {E.}~\bibnamefont {{Sanchez}}}, \bibinfo
  {author} {\bibfnamefont {V.}~\bibnamefont {{Scarpine}}}, \bibinfo {author}
  {\bibfnamefont {R.}~\bibnamefont {{Schindler}}}, \bibinfo {author}
  {\bibfnamefont {M.}~\bibnamefont {{Schubnell}}}, \bibinfo {author}
  {\bibfnamefont {M.}~\bibnamefont {{Smith}}}, \bibinfo {author} {\bibfnamefont
  {R.~C.}\ \bibnamefont {{Smith}}}, \bibinfo {author} {\bibfnamefont
  {M.}~\bibnamefont {{Soares-Santos}}}, \bibinfo {author} {\bibfnamefont
  {F.}~\bibnamefont {{Sobreira}}}, \bibinfo {author} {\bibfnamefont
  {E.}~\bibnamefont {{Suchyta}}}, \bibinfo {author} {\bibfnamefont {M.~E.~C.}\
  \bibnamefont {{Swanson}}}, \bibinfo {author} {\bibfnamefont {G.}~\bibnamefont
  {{Tarle}}}, \bibinfo {author} {\bibfnamefont {D.}~\bibnamefont {{Thomas}}},
  \bibinfo {author} {\bibfnamefont {D.~L.}\ \bibnamefont {{Tucker}}}, \bibinfo
  {author} {\bibfnamefont {V.}~\bibnamefont {{Vikram}}}, \bibinfo {author}
  {\bibfnamefont {A.~R.}\ \bibnamefont {{Walker}}}, \bibinfo {author}
  {\bibfnamefont {J.}~\bibnamefont {{Weller}}}, \bibinfo {author}
  {\bibfnamefont {Y.}~\bibnamefont {{Zhang}}}, \ and\ \bibinfo {author}
  {\bibnamefont {{DES Collaboration}}},\ }\href {\doibase
  10.1103/PhysRevD.98.043528} {\bibfield  {journal} {\bibinfo  {journal}
  {\prd}\ }\textbf {\bibinfo {volume} {98}},\ \bibinfo {eid} {043528} (\bibinfo
  {year} {2018})},\ \Eprint {http://arxiv.org/abs/1708.01538} {arXiv:1708.01538
  [astro-ph.CO]} \BibitemShut {NoStop}%
\bibitem [{\citenamefont {{MacCrann}}\ \emph {et~al.}(2015)\citenamefont
  {{MacCrann}}, \citenamefont {{Zuntz}}, \citenamefont {{Bridle}},
  \citenamefont {{Jain}},\ and\ \citenamefont {{Becker}}}]{MacCrann15}%
  \BibitemOpen
  \bibfield  {author} {\bibinfo {author} {\bibfnamefont {N.}~\bibnamefont
  {{MacCrann}}}, \bibinfo {author} {\bibfnamefont {J.}~\bibnamefont {{Zuntz}}},
  \bibinfo {author} {\bibfnamefont {S.}~\bibnamefont {{Bridle}}}, \bibinfo
  {author} {\bibfnamefont {B.}~\bibnamefont {{Jain}}}, \ and\ \bibinfo {author}
  {\bibfnamefont {M.~R.}\ \bibnamefont {{Becker}}},\ }\href {\doibase
  10.1093/mnras/stv1154} {\bibfield  {journal} {\bibinfo  {journal} {\mnras}\
  }\textbf {\bibinfo {volume} {451}},\ \bibinfo {pages} {2877} (\bibinfo {year}
  {2015})},\ \Eprint {http://arxiv.org/abs/1408.4742} {arXiv:1408.4742
  [astro-ph.CO]} \BibitemShut {NoStop}%
\bibitem [{\citenamefont {{Joudaki}}\ \emph
  {et~al.}(2017{\natexlab{a}})\citenamefont {{Joudaki}}, \citenamefont
  {{Blake}}, \citenamefont {{Heymans}}, \citenamefont {{Choi}}, \citenamefont
  {{Harnois-Deraps}}, \citenamefont {{Hildebrandt}}, \citenamefont
  {{Joachimi}}, \citenamefont {{Johnson}}, \citenamefont {{Mead}},
  \citenamefont {{Parkinson}}, \citenamefont {{Viola}},\ and\ \citenamefont
  {{van Waerbeke}}}]{Joudaki17a}%
  \BibitemOpen
  \bibfield  {author} {\bibinfo {author} {\bibfnamefont {S.}~\bibnamefont
  {{Joudaki}}}, \bibinfo {author} {\bibfnamefont {C.}~\bibnamefont {{Blake}}},
  \bibinfo {author} {\bibfnamefont {C.}~\bibnamefont {{Heymans}}}, \bibinfo
  {author} {\bibfnamefont {A.}~\bibnamefont {{Choi}}}, \bibinfo {author}
  {\bibfnamefont {J.}~\bibnamefont {{Harnois-Deraps}}}, \bibinfo {author}
  {\bibfnamefont {H.}~\bibnamefont {{Hildebrandt}}}, \bibinfo {author}
  {\bibfnamefont {B.}~\bibnamefont {{Joachimi}}}, \bibinfo {author}
  {\bibfnamefont {A.}~\bibnamefont {{Johnson}}}, \bibinfo {author}
  {\bibfnamefont {A.}~\bibnamefont {{Mead}}}, \bibinfo {author} {\bibfnamefont
  {D.}~\bibnamefont {{Parkinson}}}, \bibinfo {author} {\bibfnamefont
  {M.}~\bibnamefont {{Viola}}}, \ and\ \bibinfo {author} {\bibfnamefont
  {L.}~\bibnamefont {{van Waerbeke}}},\ }\href {\doibase 10.1093/mnras/stw2665}
  {\bibfield  {journal} {\bibinfo  {journal} {\mnras}\ }\textbf {\bibinfo
  {volume} {465}},\ \bibinfo {pages} {2033} (\bibinfo {year}
  {2017}{\natexlab{a}})},\ \Eprint {http://arxiv.org/abs/1601.05786}
  {arXiv:1601.05786 [astro-ph.CO]} \BibitemShut {NoStop}%
\bibitem [{\citenamefont {{Schaye}}\ \emph {et~al.}(2010)\citenamefont
  {{Schaye}}, \citenamefont {{Dalla Vecchia}}, \citenamefont {{Booth}},
  \citenamefont {{Wiersma}}, \citenamefont {{Theuns}}, \citenamefont {{Haas}},
  \citenamefont {{Bertone}}, \citenamefont {{Duffy}}, \citenamefont
  {{McCarthy}},\ and\ \citenamefont {{van de Voort}}}]{Schaye10}%
  \BibitemOpen
  \bibfield  {author} {\bibinfo {author} {\bibfnamefont {J.}~\bibnamefont
  {{Schaye}}}, \bibinfo {author} {\bibfnamefont {C.}~\bibnamefont {{Dalla
  Vecchia}}}, \bibinfo {author} {\bibfnamefont {C.~M.}\ \bibnamefont
  {{Booth}}}, \bibinfo {author} {\bibfnamefont {R.~P.~C.}\ \bibnamefont
  {{Wiersma}}}, \bibinfo {author} {\bibfnamefont {T.}~\bibnamefont {{Theuns}}},
  \bibinfo {author} {\bibfnamefont {M.~R.}\ \bibnamefont {{Haas}}}, \bibinfo
  {author} {\bibfnamefont {S.}~\bibnamefont {{Bertone}}}, \bibinfo {author}
  {\bibfnamefont {A.~R.}\ \bibnamefont {{Duffy}}}, \bibinfo {author}
  {\bibfnamefont {I.~G.}\ \bibnamefont {{McCarthy}}}, \ and\ \bibinfo {author}
  {\bibfnamefont {F.}~\bibnamefont {{van de Voort}}},\ }\href {\doibase
  10.1111/j.1365-2966.2009.16029.x} {\bibfield  {journal} {\bibinfo  {journal}
  {\mnras}\ }\textbf {\bibinfo {volume} {402}},\ \bibinfo {pages} {1536}
  (\bibinfo {year} {2010})},\ \Eprint {http://arxiv.org/abs/0909.5196}
  {arXiv:0909.5196 [astro-ph.CO]} \BibitemShut {NoStop}%
\bibitem [{\citenamefont {{Hildebrandt}}\ \emph {et~al.}(2017)\citenamefont
  {{Hildebrandt}}, \citenamefont {{Viola}}, \citenamefont {{Heymans}},
  \citenamefont {{Joudaki}}, \citenamefont {{Kuijken}}, \citenamefont
  {{Blake}}, \citenamefont {{Erben}}, \citenamefont {{Joachimi}}, \citenamefont
  {{Klaes}}, \citenamefont {{Miller}}, \citenamefont {{Morrison}},
  \citenamefont {{Nakajima}}, \citenamefont {{Verdoes Kleijn}}, \citenamefont
  {{Amon}}, \citenamefont {{Choi}}, \citenamefont {{Covone}}, \citenamefont
  {{de Jong}}, \citenamefont {{Dvornik}}, \citenamefont {{Fenech Conti}},
  \citenamefont {{Grado}}, \citenamefont {{Harnois-D{\'e}raps}}, \citenamefont
  {{Herbonnet}}, \citenamefont {{Hoekstra}}, \citenamefont {{K{\"o}hlinger}},
  \citenamefont {{McFarland}}, \citenamefont {{Mead}}, \citenamefont
  {{Merten}}, \citenamefont {{Napolitano}}, \citenamefont {{Peacock}},
  \citenamefont {{Radovich}}, \citenamefont {{Schneider}}, \citenamefont
  {{Simon}}, \citenamefont {{Valentijn}}, \citenamefont {{van den Busch}},
  \citenamefont {{van Uitert}},\ and\ \citenamefont {{Van
  Waerbeke}}}]{hildebrandt17}%
  \BibitemOpen
  \bibfield  {author} {\bibinfo {author} {\bibfnamefont {H.}~\bibnamefont
  {{Hildebrandt}}}, \bibinfo {author} {\bibfnamefont {M.}~\bibnamefont
  {{Viola}}}, \bibinfo {author} {\bibfnamefont {C.}~\bibnamefont {{Heymans}}},
  \bibinfo {author} {\bibfnamefont {S.}~\bibnamefont {{Joudaki}}}, \bibinfo
  {author} {\bibfnamefont {K.}~\bibnamefont {{Kuijken}}}, \bibinfo {author}
  {\bibfnamefont {C.}~\bibnamefont {{Blake}}}, \bibinfo {author} {\bibfnamefont
  {T.}~\bibnamefont {{Erben}}}, \bibinfo {author} {\bibfnamefont
  {B.}~\bibnamefont {{Joachimi}}}, \bibinfo {author} {\bibfnamefont
  {D.}~\bibnamefont {{Klaes}}}, \bibinfo {author} {\bibfnamefont
  {L.}~\bibnamefont {{Miller}}}, \bibinfo {author} {\bibfnamefont {C.~B.}\
  \bibnamefont {{Morrison}}}, \bibinfo {author} {\bibfnamefont
  {R.}~\bibnamefont {{Nakajima}}}, \bibinfo {author} {\bibfnamefont
  {G.}~\bibnamefont {{Verdoes Kleijn}}}, \bibinfo {author} {\bibfnamefont
  {A.}~\bibnamefont {{Amon}}}, \bibinfo {author} {\bibfnamefont
  {A.}~\bibnamefont {{Choi}}}, \bibinfo {author} {\bibfnamefont
  {G.}~\bibnamefont {{Covone}}}, \bibinfo {author} {\bibfnamefont {J.~T.~A.}\
  \bibnamefont {{de Jong}}}, \bibinfo {author} {\bibfnamefont {A.}~\bibnamefont
  {{Dvornik}}}, \bibinfo {author} {\bibfnamefont {I.}~\bibnamefont {{Fenech
  Conti}}}, \bibinfo {author} {\bibfnamefont {A.}~\bibnamefont {{Grado}}},
  \bibinfo {author} {\bibfnamefont {J.}~\bibnamefont {{Harnois-D{\'e}raps}}},
  \bibinfo {author} {\bibfnamefont {R.}~\bibnamefont {{Herbonnet}}}, \bibinfo
  {author} {\bibfnamefont {H.}~\bibnamefont {{Hoekstra}}}, \bibinfo {author}
  {\bibfnamefont {F.}~\bibnamefont {{K{\"o}hlinger}}}, \bibinfo {author}
  {\bibfnamefont {J.}~\bibnamefont {{McFarland}}}, \bibinfo {author}
  {\bibfnamefont {A.}~\bibnamefont {{Mead}}}, \bibinfo {author} {\bibfnamefont
  {J.}~\bibnamefont {{Merten}}}, \bibinfo {author} {\bibfnamefont
  {N.}~\bibnamefont {{Napolitano}}}, \bibinfo {author} {\bibfnamefont {J.~A.}\
  \bibnamefont {{Peacock}}}, \bibinfo {author} {\bibfnamefont {M.}~\bibnamefont
  {{Radovich}}}, \bibinfo {author} {\bibfnamefont {P.}~\bibnamefont
  {{Schneider}}}, \bibinfo {author} {\bibfnamefont {P.}~\bibnamefont
  {{Simon}}}, \bibinfo {author} {\bibfnamefont {E.~A.}\ \bibnamefont
  {{Valentijn}}}, \bibinfo {author} {\bibfnamefont {J.~L.}\ \bibnamefont {{van
  den Busch}}}, \bibinfo {author} {\bibfnamefont {E.}~\bibnamefont {{van
  Uitert}}}, \ and\ \bibinfo {author} {\bibfnamefont {L.}~\bibnamefont {{Van
  Waerbeke}}},\ }\href {\doibase 10.1093/mnras/stw2805} {\bibfield  {journal}
  {\bibinfo  {journal} {\mnras}\ }\textbf {\bibinfo {volume} {465}},\ \bibinfo
  {pages} {1454} (\bibinfo {year} {2017})},\ \Eprint
  {http://arxiv.org/abs/1606.05338} {arXiv:1606.05338} \BibitemShut {NoStop}%
\bibitem [{\citenamefont {{Joudaki}}\ \emph
  {et~al.}(2017{\natexlab{b}})\citenamefont {{Joudaki}}, \citenamefont
  {{Mead}}, \citenamefont {{Blake}}, \citenamefont {{Choi}}, \citenamefont {{de
  Jong}}, \citenamefont {{Erben}}, \citenamefont {{Fenech Conti}},
  \citenamefont {{Herbonnet}}, \citenamefont {{Heymans}}, \citenamefont
  {{Hildebrandt}}, \citenamefont {{Hoekstra}}, \citenamefont {{Joachimi}},
  \citenamefont {{Klaes}}, \citenamefont {{K{\"o}hlinger}}, \citenamefont
  {{Kuijken}}, \citenamefont {{McFarland}}, \citenamefont {{Miller}},
  \citenamefont {{Schneider}},\ and\ \citenamefont {{Viola}}}]{J17b}%
  \BibitemOpen
  \bibfield  {author} {\bibinfo {author} {\bibfnamefont {S.}~\bibnamefont
  {{Joudaki}}}, \bibinfo {author} {\bibfnamefont {A.}~\bibnamefont {{Mead}}},
  \bibinfo {author} {\bibfnamefont {C.}~\bibnamefont {{Blake}}}, \bibinfo
  {author} {\bibfnamefont {A.}~\bibnamefont {{Choi}}}, \bibinfo {author}
  {\bibfnamefont {J.}~\bibnamefont {{de Jong}}}, \bibinfo {author}
  {\bibfnamefont {T.}~\bibnamefont {{Erben}}}, \bibinfo {author} {\bibfnamefont
  {I.}~\bibnamefont {{Fenech Conti}}}, \bibinfo {author} {\bibfnamefont
  {R.}~\bibnamefont {{Herbonnet}}}, \bibinfo {author} {\bibfnamefont
  {C.}~\bibnamefont {{Heymans}}}, \bibinfo {author} {\bibfnamefont
  {H.}~\bibnamefont {{Hildebrandt}}}, \bibinfo {author} {\bibfnamefont
  {H.}~\bibnamefont {{Hoekstra}}}, \bibinfo {author} {\bibfnamefont
  {B.}~\bibnamefont {{Joachimi}}}, \bibinfo {author} {\bibfnamefont
  {D.}~\bibnamefont {{Klaes}}}, \bibinfo {author} {\bibfnamefont
  {F.}~\bibnamefont {{K{\"o}hlinger}}}, \bibinfo {author} {\bibfnamefont
  {K.}~\bibnamefont {{Kuijken}}}, \bibinfo {author} {\bibfnamefont
  {J.}~\bibnamefont {{McFarland}}}, \bibinfo {author} {\bibfnamefont
  {L.}~\bibnamefont {{Miller}}}, \bibinfo {author} {\bibfnamefont
  {P.}~\bibnamefont {{Schneider}}}, \ and\ \bibinfo {author} {\bibfnamefont
  {M.}~\bibnamefont {{Viola}}},\ }\href {\doibase 10.1093/mnras/stx998}
  {\bibfield  {journal} {\bibinfo  {journal} {\mnras}\ }\textbf {\bibinfo
  {volume} {471}},\ \bibinfo {pages} {1259} (\bibinfo {year}
  {2017}{\natexlab{b}})},\ \Eprint {http://arxiv.org/abs/1610.04606}
  {arXiv:1610.04606 [astro-ph.CO]} \BibitemShut {NoStop}%
\bibitem [{\citenamefont {{Joudaki}}\ \emph {et~al.}(2018)\citenamefont
  {{Joudaki}}, \citenamefont {{Blake}}, \citenamefont {{Johnson}},
  \citenamefont {{Amon}}, \citenamefont {{Asgari}}, \citenamefont {{Choi}},
  \citenamefont {{Erben}}, \citenamefont {{Glazebrook}}, \citenamefont
  {{Harnois-D{\'e}raps}}, \citenamefont {{Heymans}}, \citenamefont {{Hildebrand
  t}}, \citenamefont {{Hoekstra}}, \citenamefont {{Klaes}}, \citenamefont
  {{Kuijken}}, \citenamefont {{Lidman}}, \citenamefont {{Mead}}, \citenamefont
  {{Miller}}, \citenamefont {{Parkinson}}, \citenamefont {{Poole}},
  \citenamefont {{Schneider}}, \citenamefont {{Viola}},\ and\ \citenamefont
  {{Wolf}}}]{Joudaki18}%
  \BibitemOpen
  \bibfield  {author} {\bibinfo {author} {\bibfnamefont {S.}~\bibnamefont
  {{Joudaki}}}, \bibinfo {author} {\bibfnamefont {C.}~\bibnamefont {{Blake}}},
  \bibinfo {author} {\bibfnamefont {A.}~\bibnamefont {{Johnson}}}, \bibinfo
  {author} {\bibfnamefont {A.}~\bibnamefont {{Amon}}}, \bibinfo {author}
  {\bibfnamefont {M.}~\bibnamefont {{Asgari}}}, \bibinfo {author}
  {\bibfnamefont {A.}~\bibnamefont {{Choi}}}, \bibinfo {author} {\bibfnamefont
  {T.}~\bibnamefont {{Erben}}}, \bibinfo {author} {\bibfnamefont
  {K.}~\bibnamefont {{Glazebrook}}}, \bibinfo {author} {\bibfnamefont
  {J.}~\bibnamefont {{Harnois-D{\'e}raps}}}, \bibinfo {author} {\bibfnamefont
  {C.}~\bibnamefont {{Heymans}}}, \bibinfo {author} {\bibfnamefont
  {H.}~\bibnamefont {{Hildebrand t}}}, \bibinfo {author} {\bibfnamefont
  {H.}~\bibnamefont {{Hoekstra}}}, \bibinfo {author} {\bibfnamefont
  {D.}~\bibnamefont {{Klaes}}}, \bibinfo {author} {\bibfnamefont
  {K.}~\bibnamefont {{Kuijken}}}, \bibinfo {author} {\bibfnamefont
  {C.}~\bibnamefont {{Lidman}}}, \bibinfo {author} {\bibfnamefont
  {A.}~\bibnamefont {{Mead}}}, \bibinfo {author} {\bibfnamefont
  {L.}~\bibnamefont {{Miller}}}, \bibinfo {author} {\bibfnamefont
  {D.}~\bibnamefont {{Parkinson}}}, \bibinfo {author} {\bibfnamefont {G.~B.}\
  \bibnamefont {{Poole}}}, \bibinfo {author} {\bibfnamefont {P.}~\bibnamefont
  {{Schneider}}}, \bibinfo {author} {\bibfnamefont {M.}~\bibnamefont
  {{Viola}}}, \ and\ \bibinfo {author} {\bibfnamefont {C.}~\bibnamefont
  {{Wolf}}},\ }\href {\doibase 10.1093/mnras/stx2820} {\bibfield  {journal}
  {\bibinfo  {journal} {\mnras}\ }\textbf {\bibinfo {volume} {474}},\ \bibinfo
  {pages} {4894} (\bibinfo {year} {2018})},\ \Eprint
  {http://arxiv.org/abs/1707.06627} {arXiv:1707.06627 [astro-ph.CO]}
  \BibitemShut {NoStop}%
\bibitem [{\citenamefont {{van Uitert}}\ \emph {et~al.}(2018)\citenamefont
  {{van Uitert}}, \citenamefont {{Joachimi}}, \citenamefont {{Joudaki}},
  \citenamefont {{Amon}}, \citenamefont {{Heymans}}, \citenamefont
  {{K{\"o}hlinger}}, \citenamefont {{Asgari}}, \citenamefont {{Blake}},
  \citenamefont {{Choi}}, \citenamefont {{Erben}}, \citenamefont {{Farrow}},
  \citenamefont {{Harnois-D{\'e}raps}}, \citenamefont {{Hildebrandt}},
  \citenamefont {{Hoekstra}}, \citenamefont {{Kitching}}, \citenamefont
  {{Klaes}}, \citenamefont {{Kuijken}}, \citenamefont {{Merten}}, \citenamefont
  {{Miller}}, \citenamefont {{Nakajima}}, \citenamefont {{Schneider}},
  \citenamefont {{Valentijn}},\ and\ \citenamefont {{Viola}}}]{vanUitert18}%
  \BibitemOpen
  \bibfield  {author} {\bibinfo {author} {\bibfnamefont {E.}~\bibnamefont {{van
  Uitert}}}, \bibinfo {author} {\bibfnamefont {B.}~\bibnamefont {{Joachimi}}},
  \bibinfo {author} {\bibfnamefont {S.}~\bibnamefont {{Joudaki}}}, \bibinfo
  {author} {\bibfnamefont {A.}~\bibnamefont {{Amon}}}, \bibinfo {author}
  {\bibfnamefont {C.}~\bibnamefont {{Heymans}}}, \bibinfo {author}
  {\bibfnamefont {F.}~\bibnamefont {{K{\"o}hlinger}}}, \bibinfo {author}
  {\bibfnamefont {M.}~\bibnamefont {{Asgari}}}, \bibinfo {author}
  {\bibfnamefont {C.}~\bibnamefont {{Blake}}}, \bibinfo {author} {\bibfnamefont
  {A.}~\bibnamefont {{Choi}}}, \bibinfo {author} {\bibfnamefont
  {T.}~\bibnamefont {{Erben}}}, \bibinfo {author} {\bibfnamefont {D.~J.}\
  \bibnamefont {{Farrow}}}, \bibinfo {author} {\bibfnamefont {J.}~\bibnamefont
  {{Harnois-D{\'e}raps}}}, \bibinfo {author} {\bibfnamefont {H.}~\bibnamefont
  {{Hildebrandt}}}, \bibinfo {author} {\bibfnamefont {H.}~\bibnamefont
  {{Hoekstra}}}, \bibinfo {author} {\bibfnamefont {T.~D.}\ \bibnamefont
  {{Kitching}}}, \bibinfo {author} {\bibfnamefont {D.}~\bibnamefont {{Klaes}}},
  \bibinfo {author} {\bibfnamefont {K.}~\bibnamefont {{Kuijken}}}, \bibinfo
  {author} {\bibfnamefont {J.}~\bibnamefont {{Merten}}}, \bibinfo {author}
  {\bibfnamefont {L.}~\bibnamefont {{Miller}}}, \bibinfo {author}
  {\bibfnamefont {R.}~\bibnamefont {{Nakajima}}}, \bibinfo {author}
  {\bibfnamefont {P.}~\bibnamefont {{Schneider}}}, \bibinfo {author}
  {\bibfnamefont {E.}~\bibnamefont {{Valentijn}}}, \ and\ \bibinfo {author}
  {\bibfnamefont {M.}~\bibnamefont {{Viola}}},\ }\href {\doibase
  10.1093/mnras/sty551} {\bibfield  {journal} {\bibinfo  {journal} {\mnras}\
  }\textbf {\bibinfo {volume} {476}},\ \bibinfo {pages} {4662} (\bibinfo {year}
  {2018})},\ \Eprint {http://arxiv.org/abs/1706.05004} {arXiv:1706.05004
  [astro-ph.CO]} \BibitemShut {NoStop}%
\bibitem [{\citenamefont {{Yoon}}\ \emph {et~al.}(2019)\citenamefont {{Yoon}},
  \citenamefont {{Jee}}, \citenamefont {{Tyson}}, \citenamefont {{Schmidt}},
  \citenamefont {{Wittman}},\ and\ \citenamefont {{Choi}}}]{Yoon19}%
  \BibitemOpen
  \bibfield  {author} {\bibinfo {author} {\bibfnamefont {M.}~\bibnamefont
  {{Yoon}}}, \bibinfo {author} {\bibfnamefont {M.~J.}\ \bibnamefont {{Jee}}},
  \bibinfo {author} {\bibfnamefont {J.~A.}\ \bibnamefont {{Tyson}}}, \bibinfo
  {author} {\bibfnamefont {S.}~\bibnamefont {{Schmidt}}}, \bibinfo {author}
  {\bibfnamefont {D.}~\bibnamefont {{Wittman}}}, \ and\ \bibinfo {author}
  {\bibfnamefont {A.}~\bibnamefont {{Choi}}},\ }\href {\doibase
  10.3847/1538-4357/aaf3a9} {\bibfield  {journal} {\bibinfo  {journal} {\apj}\
  }\textbf {\bibinfo {volume} {870}},\ \bibinfo {eid} {111} (\bibinfo {year}
  {2019})},\ \Eprint {http://arxiv.org/abs/1807.09195} {arXiv:1807.09195
  [astro-ph.CO]} \BibitemShut {NoStop}%
\bibitem [{\citenamefont {{Hikage}}\ \emph {et~al.}(2019)\citenamefont
  {{Hikage}}, \citenamefont {{Oguri}}, \citenamefont {{Hamana}}, \citenamefont
  {{More}}, \citenamefont {{Mandelbaum}}, \citenamefont {{Takada}},
  \citenamefont {{K{\"o}hlinger}}, \citenamefont {{Miyatake}}, \citenamefont
  {{Nishizawa}}, \citenamefont {{Aihara}}, \citenamefont {{Armstrong}},
  \citenamefont {{Bosch}}, \citenamefont {{Coupon}}, \citenamefont {{Ducout}},
  \citenamefont {{Ho}}, \citenamefont {{Hsieh}}, \citenamefont {{Komiyama}},
  \citenamefont {{Lanusse}}, \citenamefont {{Leauthaud}}, \citenamefont
  {{Lupton}}, \citenamefont {{Medezinski}}, \citenamefont {{Mineo}},
  \citenamefont {{Miyama}}, \citenamefont {{Miyazaki}}, \citenamefont
  {{Murata}}, \citenamefont {{Murayama}}, \citenamefont {{Shirasaki}},
  \citenamefont {{Sif{\'o}n}}, \citenamefont {{Simet}}, \citenamefont
  {{Speagle}}, \citenamefont {{Spergel}}, \citenamefont {{Strauss}},
  \citenamefont {{Sugiyama}}, \citenamefont {{Tanaka}}, \citenamefont
  {{Utsumi}}, \citenamefont {{Wang}},\ and\ \citenamefont
  {{Yamada}}}]{hikage19}%
  \BibitemOpen
  \bibfield  {author} {\bibinfo {author} {\bibfnamefont {C.}~\bibnamefont
  {{Hikage}}}, \bibinfo {author} {\bibfnamefont {M.}~\bibnamefont {{Oguri}}},
  \bibinfo {author} {\bibfnamefont {T.}~\bibnamefont {{Hamana}}}, \bibinfo
  {author} {\bibfnamefont {S.}~\bibnamefont {{More}}}, \bibinfo {author}
  {\bibfnamefont {R.}~\bibnamefont {{Mandelbaum}}}, \bibinfo {author}
  {\bibfnamefont {M.}~\bibnamefont {{Takada}}}, \bibinfo {author}
  {\bibfnamefont {F.}~\bibnamefont {{K{\"o}hlinger}}}, \bibinfo {author}
  {\bibfnamefont {H.}~\bibnamefont {{Miyatake}}}, \bibinfo {author}
  {\bibfnamefont {A.~J.}\ \bibnamefont {{Nishizawa}}}, \bibinfo {author}
  {\bibfnamefont {H.}~\bibnamefont {{Aihara}}}, \bibinfo {author}
  {\bibfnamefont {R.}~\bibnamefont {{Armstrong}}}, \bibinfo {author}
  {\bibfnamefont {J.}~\bibnamefont {{Bosch}}}, \bibinfo {author} {\bibfnamefont
  {J.}~\bibnamefont {{Coupon}}}, \bibinfo {author} {\bibfnamefont
  {A.}~\bibnamefont {{Ducout}}}, \bibinfo {author} {\bibfnamefont
  {P.}~\bibnamefont {{Ho}}}, \bibinfo {author} {\bibfnamefont {B.-C.}\
  \bibnamefont {{Hsieh}}}, \bibinfo {author} {\bibfnamefont {Y.}~\bibnamefont
  {{Komiyama}}}, \bibinfo {author} {\bibfnamefont {F.}~\bibnamefont
  {{Lanusse}}}, \bibinfo {author} {\bibfnamefont {A.}~\bibnamefont
  {{Leauthaud}}}, \bibinfo {author} {\bibfnamefont {R.~H.}\ \bibnamefont
  {{Lupton}}}, \bibinfo {author} {\bibfnamefont {E.}~\bibnamefont
  {{Medezinski}}}, \bibinfo {author} {\bibfnamefont {S.}~\bibnamefont
  {{Mineo}}}, \bibinfo {author} {\bibfnamefont {S.}~\bibnamefont {{Miyama}}},
  \bibinfo {author} {\bibfnamefont {S.}~\bibnamefont {{Miyazaki}}}, \bibinfo
  {author} {\bibfnamefont {R.}~\bibnamefont {{Murata}}}, \bibinfo {author}
  {\bibfnamefont {H.}~\bibnamefont {{Murayama}}}, \bibinfo {author}
  {\bibfnamefont {M.}~\bibnamefont {{Shirasaki}}}, \bibinfo {author}
  {\bibfnamefont {C.}~\bibnamefont {{Sif{\'o}n}}}, \bibinfo {author}
  {\bibfnamefont {M.}~\bibnamefont {{Simet}}}, \bibinfo {author} {\bibfnamefont
  {J.}~\bibnamefont {{Speagle}}}, \bibinfo {author} {\bibfnamefont {D.~N.}\
  \bibnamefont {{Spergel}}}, \bibinfo {author} {\bibfnamefont {M.~A.}\
  \bibnamefont {{Strauss}}}, \bibinfo {author} {\bibfnamefont {N.}~\bibnamefont
  {{Sugiyama}}}, \bibinfo {author} {\bibfnamefont {M.}~\bibnamefont
  {{Tanaka}}}, \bibinfo {author} {\bibfnamefont {Y.}~\bibnamefont {{Utsumi}}},
  \bibinfo {author} {\bibfnamefont {S.-Y.}\ \bibnamefont {{Wang}}}, \ and\
  \bibinfo {author} {\bibfnamefont {Y.}~\bibnamefont {{Yamada}}},\ }\href
  {\doibase 10.1093/pasj/psz010} {\bibfield  {journal} {\bibinfo  {journal}
  {Publications of the Astronomical Society of Japan}\ ,\ \bibinfo {pages}
  {22}} (\bibinfo {year} {2019})},\ \Eprint {http://arxiv.org/abs/1809.09148}
  {arXiv:1809.09148 [astro-ph.CO]} \BibitemShut {NoStop}%
\bibitem [{\citenamefont {{K{\"o}hlinger}}\ \emph {et~al.}(2016)\citenamefont
  {{K{\"o}hlinger}}, \citenamefont {{Viola}}, \citenamefont {{Valkenburg}},
  \citenamefont {{Joachimi}}, \citenamefont {{Hoekstra}},\ and\ \citenamefont
  {{Kuijken}}}]{Kohlinger16}%
  \BibitemOpen
  \bibfield  {author} {\bibinfo {author} {\bibfnamefont {F.}~\bibnamefont
  {{K{\"o}hlinger}}}, \bibinfo {author} {\bibfnamefont {M.}~\bibnamefont
  {{Viola}}}, \bibinfo {author} {\bibfnamefont {W.}~\bibnamefont
  {{Valkenburg}}}, \bibinfo {author} {\bibfnamefont {B.}~\bibnamefont
  {{Joachimi}}}, \bibinfo {author} {\bibfnamefont {H.}~\bibnamefont
  {{Hoekstra}}}, \ and\ \bibinfo {author} {\bibfnamefont {K.}~\bibnamefont
  {{Kuijken}}},\ }\href {\doibase 10.1093/mnras/stv2762} {\bibfield  {journal}
  {\bibinfo  {journal} {\mnras}\ }\textbf {\bibinfo {volume} {456}},\ \bibinfo
  {pages} {1508} (\bibinfo {year} {2016})},\ \Eprint
  {http://arxiv.org/abs/1509.04071} {arXiv:1509.04071 [astro-ph.CO]}
  \BibitemShut {NoStop}%
\bibitem [{\citenamefont {{K{\"o}hlinger}}\ \emph {et~al.}(2017)\citenamefont
  {{K{\"o}hlinger}}, \citenamefont {{Viola}}, \citenamefont {{Joachimi}},
  \citenamefont {{Hoekstra}}, \citenamefont {{van Uitert}}, \citenamefont
  {{Hildebrandt}}, \citenamefont {{Choi}}, \citenamefont {{Erben}},
  \citenamefont {{Heymans}}, \citenamefont {{Joudaki}}, \citenamefont
  {{Klaes}}, \citenamefont {{Kuijken}}, \citenamefont {{Merten}}, \citenamefont
  {{Miller}}, \citenamefont {{Schneider}},\ and\ \citenamefont
  {{Valentijn}}}]{Kohlinger17}%
  \BibitemOpen
  \bibfield  {author} {\bibinfo {author} {\bibfnamefont {F.}~\bibnamefont
  {{K{\"o}hlinger}}}, \bibinfo {author} {\bibfnamefont {M.}~\bibnamefont
  {{Viola}}}, \bibinfo {author} {\bibfnamefont {B.}~\bibnamefont {{Joachimi}}},
  \bibinfo {author} {\bibfnamefont {H.}~\bibnamefont {{Hoekstra}}}, \bibinfo
  {author} {\bibfnamefont {E.}~\bibnamefont {{van Uitert}}}, \bibinfo {author}
  {\bibfnamefont {H.}~\bibnamefont {{Hildebrandt}}}, \bibinfo {author}
  {\bibfnamefont {A.}~\bibnamefont {{Choi}}}, \bibinfo {author} {\bibfnamefont
  {T.}~\bibnamefont {{Erben}}}, \bibinfo {author} {\bibfnamefont
  {C.}~\bibnamefont {{Heymans}}}, \bibinfo {author} {\bibfnamefont
  {S.}~\bibnamefont {{Joudaki}}}, \bibinfo {author} {\bibfnamefont
  {D.}~\bibnamefont {{Klaes}}}, \bibinfo {author} {\bibfnamefont
  {K.}~\bibnamefont {{Kuijken}}}, \bibinfo {author} {\bibfnamefont
  {J.}~\bibnamefont {{Merten}}}, \bibinfo {author} {\bibfnamefont
  {L.}~\bibnamefont {{Miller}}}, \bibinfo {author} {\bibfnamefont
  {P.}~\bibnamefont {{Schneider}}}, \ and\ \bibinfo {author} {\bibfnamefont
  {E.~A.}\ \bibnamefont {{Valentijn}}},\ }\href {\doibase
  10.1093/mnras/stx1820} {\bibfield  {journal} {\bibinfo  {journal} {\mnras}\
  }\textbf {\bibinfo {volume} {471}},\ \bibinfo {pages} {4412} (\bibinfo {year}
  {2017})},\ \Eprint {http://arxiv.org/abs/1706.02892} {arXiv:1706.02892
  [astro-ph.CO]} \BibitemShut {NoStop}%
\bibitem [{\citenamefont {{Foreman}}\ \emph {et~al.}(2016)\citenamefont
  {{Foreman}}, \citenamefont {{Becker}},\ and\ \citenamefont
  {{Wechsler}}}]{Foreman16}%
  \BibitemOpen
  \bibfield  {author} {\bibinfo {author} {\bibfnamefont {S.}~\bibnamefont
  {{Foreman}}}, \bibinfo {author} {\bibfnamefont {M.~R.}\ \bibnamefont
  {{Becker}}}, \ and\ \bibinfo {author} {\bibfnamefont {R.~H.}\ \bibnamefont
  {{Wechsler}}},\ }\href {\doibase 10.1093/mnras/stw2189} {\bibfield  {journal}
  {\bibinfo  {journal} {\mnras}\ }\textbf {\bibinfo {volume} {463}},\ \bibinfo
  {pages} {3326} (\bibinfo {year} {2016})},\ \Eprint
  {http://arxiv.org/abs/1605.09056} {arXiv:1605.09056} \BibitemShut {NoStop}%
\bibitem [{\citenamefont {{Mead}}\ \emph {et~al.}(2016)\citenamefont {{Mead}},
  \citenamefont {{Heymans}}, \citenamefont {{Lombriser}}, \citenamefont
  {{Peacock}}, \citenamefont {{Steele}},\ and\ \citenamefont
  {{Winther}}}]{Mead16}%
  \BibitemOpen
  \bibfield  {author} {\bibinfo {author} {\bibfnamefont {A.~J.}\ \bibnamefont
  {{Mead}}}, \bibinfo {author} {\bibfnamefont {C.}~\bibnamefont {{Heymans}}},
  \bibinfo {author} {\bibfnamefont {L.}~\bibnamefont {{Lombriser}}}, \bibinfo
  {author} {\bibfnamefont {J.~A.}\ \bibnamefont {{Peacock}}}, \bibinfo {author}
  {\bibfnamefont {O.~I.}\ \bibnamefont {{Steele}}}, \ and\ \bibinfo {author}
  {\bibfnamefont {H.~A.}\ \bibnamefont {{Winther}}},\ }\href {\doibase
  10.1093/mnras/stw681} {\bibfield  {journal} {\bibinfo  {journal} {\mnras}\
  }\textbf {\bibinfo {volume} {459}},\ \bibinfo {pages} {1468} (\bibinfo {year}
  {2016})},\ \Eprint {http://arxiv.org/abs/1602.02154} {arXiv:1602.02154
  [astro-ph.CO]} \BibitemShut {NoStop}%
\bibitem [{\citenamefont {{Mummery}}\ \emph {et~al.}(2017)\citenamefont
  {{Mummery}}, \citenamefont {{McCarthy}}, \citenamefont {{Bird}},\ and\
  \citenamefont {{Schaye}}}]{Mummery}%
  \BibitemOpen
  \bibfield  {author} {\bibinfo {author} {\bibfnamefont {B.~O.}\ \bibnamefont
  {{Mummery}}}, \bibinfo {author} {\bibfnamefont {I.~G.}\ \bibnamefont
  {{McCarthy}}}, \bibinfo {author} {\bibfnamefont {S.}~\bibnamefont {{Bird}}},
  \ and\ \bibinfo {author} {\bibfnamefont {J.}~\bibnamefont {{Schaye}}},\
  }\href {\doibase 10.1093/mnras/stx1469} {\bibfield  {journal} {\bibinfo
  {journal} {\mnras}\ }\textbf {\bibinfo {volume} {471}},\ \bibinfo {pages}
  {227} (\bibinfo {year} {2017})},\ \Eprint {http://arxiv.org/abs/1702.02064}
  {arXiv:1702.02064} \BibitemShut {NoStop}%
\bibitem [{\citenamefont {{Teyssier}}(2002)}]{Teyssier02}%
  \BibitemOpen
  \bibfield  {author} {\bibinfo {author} {\bibfnamefont {R.}~\bibnamefont
  {{Teyssier}}},\ }\href {\doibase 10.1051/0004-6361:20011817} {\bibfield
  {journal} {\bibinfo  {journal} {\aap}\ }\textbf {\bibinfo {volume} {385}},\
  \bibinfo {pages} {337} (\bibinfo {year} {2002})},\ \Eprint
  {http://arxiv.org/abs/astro-ph/0111367} {astro-ph/0111367} \BibitemShut
  {NoStop}%
\bibitem [{\citenamefont {{Dubois}}\ \emph {et~al.}(2014)\citenamefont
  {{Dubois}}, \citenamefont {{Pichon}}, \citenamefont {{Welker}}, \citenamefont
  {{Le Borgne}}, \citenamefont {{Devriendt}}, \citenamefont {{Laigle}},
  \citenamefont {{Codis}}, \citenamefont {{Pogosyan}}, \citenamefont
  {{Arnouts}}, \citenamefont {{Benabed}}, \citenamefont {{Bertin}},
  \citenamefont {{Blaizot}}, \citenamefont {{Bouchet}}, \citenamefont
  {{Cardoso}}, \citenamefont {{Colombi}}, \citenamefont {{de Lapparent}},
  \citenamefont {{Desjacques}}, \citenamefont {{Gavazzi}}, \citenamefont
  {{Kassin}}, \citenamefont {{Kimm}}, \citenamefont {{McCracken}},
  \citenamefont {{Milliard}}, \citenamefont {{Peirani}}, \citenamefont
  {{Prunet}}, \citenamefont {{Rouberol}}, \citenamefont {{Silk}}, \citenamefont
  {{Slyz}}, \citenamefont {{Sousbie}}, \citenamefont {{Teyssier}},
  \citenamefont {{Tresse}}, \citenamefont {{Treyer}}, \citenamefont
  {{Vibert}},\ and\ \citenamefont {{Volonteri}}}]{Dubois14}%
  \BibitemOpen
  \bibfield  {author} {\bibinfo {author} {\bibfnamefont {Y.}~\bibnamefont
  {{Dubois}}}, \bibinfo {author} {\bibfnamefont {C.}~\bibnamefont {{Pichon}}},
  \bibinfo {author} {\bibfnamefont {C.}~\bibnamefont {{Welker}}}, \bibinfo
  {author} {\bibfnamefont {D.}~\bibnamefont {{Le Borgne}}}, \bibinfo {author}
  {\bibfnamefont {J.}~\bibnamefont {{Devriendt}}}, \bibinfo {author}
  {\bibfnamefont {C.}~\bibnamefont {{Laigle}}}, \bibinfo {author}
  {\bibfnamefont {S.}~\bibnamefont {{Codis}}}, \bibinfo {author} {\bibfnamefont
  {D.}~\bibnamefont {{Pogosyan}}}, \bibinfo {author} {\bibfnamefont
  {S.}~\bibnamefont {{Arnouts}}}, \bibinfo {author} {\bibfnamefont
  {K.}~\bibnamefont {{Benabed}}}, \bibinfo {author} {\bibfnamefont
  {E.}~\bibnamefont {{Bertin}}}, \bibinfo {author} {\bibfnamefont
  {J.}~\bibnamefont {{Blaizot}}}, \bibinfo {author} {\bibfnamefont
  {F.}~\bibnamefont {{Bouchet}}}, \bibinfo {author} {\bibfnamefont {J.-F.}\
  \bibnamefont {{Cardoso}}}, \bibinfo {author} {\bibfnamefont {S.}~\bibnamefont
  {{Colombi}}}, \bibinfo {author} {\bibfnamefont {V.}~\bibnamefont {{de
  Lapparent}}}, \bibinfo {author} {\bibfnamefont {V.}~\bibnamefont
  {{Desjacques}}}, \bibinfo {author} {\bibfnamefont {R.}~\bibnamefont
  {{Gavazzi}}}, \bibinfo {author} {\bibfnamefont {S.}~\bibnamefont {{Kassin}}},
  \bibinfo {author} {\bibfnamefont {T.}~\bibnamefont {{Kimm}}}, \bibinfo
  {author} {\bibfnamefont {H.}~\bibnamefont {{McCracken}}}, \bibinfo {author}
  {\bibfnamefont {B.}~\bibnamefont {{Milliard}}}, \bibinfo {author}
  {\bibfnamefont {S.}~\bibnamefont {{Peirani}}}, \bibinfo {author}
  {\bibfnamefont {S.}~\bibnamefont {{Prunet}}}, \bibinfo {author}
  {\bibfnamefont {S.}~\bibnamefont {{Rouberol}}}, \bibinfo {author}
  {\bibfnamefont {J.}~\bibnamefont {{Silk}}}, \bibinfo {author} {\bibfnamefont
  {A.}~\bibnamefont {{Slyz}}}, \bibinfo {author} {\bibfnamefont
  {T.}~\bibnamefont {{Sousbie}}}, \bibinfo {author} {\bibfnamefont
  {R.}~\bibnamefont {{Teyssier}}}, \bibinfo {author} {\bibfnamefont
  {L.}~\bibnamefont {{Tresse}}}, \bibinfo {author} {\bibfnamefont
  {M.}~\bibnamefont {{Treyer}}}, \bibinfo {author} {\bibfnamefont
  {D.}~\bibnamefont {{Vibert}}}, \ and\ \bibinfo {author} {\bibfnamefont
  {M.}~\bibnamefont {{Volonteri}}},\ }\href {\doibase 10.1093/mnras/stu1227}
  {\bibfield  {journal} {\bibinfo  {journal} {\mnras}\ }\textbf {\bibinfo
  {volume} {444}},\ \bibinfo {pages} {1453} (\bibinfo {year} {2014})},\ \Eprint
  {http://arxiv.org/abs/1402.1165} {arXiv:1402.1165} \BibitemShut {NoStop}%
\bibitem [{\citenamefont {{Peirani}}\ \emph {et~al.}(2017)\citenamefont
  {{Peirani}}, \citenamefont {{Dubois}}, \citenamefont {{Volonteri}},
  \citenamefont {{Devriendt}}, \citenamefont {{Bundy}}, \citenamefont {{Silk}},
  \citenamefont {{Pichon}}, \citenamefont {{Kaviraj}}, \citenamefont
  {{Gavazzi}},\ and\ \citenamefont {{Habouzit}}}]{Peirani17}%
  \BibitemOpen
  \bibfield  {author} {\bibinfo {author} {\bibfnamefont {S.}~\bibnamefont
  {{Peirani}}}, \bibinfo {author} {\bibfnamefont {Y.}~\bibnamefont {{Dubois}}},
  \bibinfo {author} {\bibfnamefont {M.}~\bibnamefont {{Volonteri}}}, \bibinfo
  {author} {\bibfnamefont {J.}~\bibnamefont {{Devriendt}}}, \bibinfo {author}
  {\bibfnamefont {K.}~\bibnamefont {{Bundy}}}, \bibinfo {author} {\bibfnamefont
  {J.}~\bibnamefont {{Silk}}}, \bibinfo {author} {\bibfnamefont
  {C.}~\bibnamefont {{Pichon}}}, \bibinfo {author} {\bibfnamefont
  {S.}~\bibnamefont {{Kaviraj}}}, \bibinfo {author} {\bibfnamefont
  {R.}~\bibnamefont {{Gavazzi}}}, \ and\ \bibinfo {author} {\bibfnamefont
  {M.}~\bibnamefont {{Habouzit}}},\ }\href {\doibase 10.1093/mnras/stx2099}
  {\bibfield  {journal} {\bibinfo  {journal} {\mnras}\ }\textbf {\bibinfo
  {volume} {472}},\ \bibinfo {pages} {2153} (\bibinfo {year} {2017})},\ \Eprint
  {http://arxiv.org/abs/1611.09922} {arXiv:1611.09922 [astro-ph.GA]}
  \BibitemShut {NoStop}%
\bibitem [{\citenamefont {{Gouin}}\ \emph {et~al.}(2019)\citenamefont
  {{Gouin}}, \citenamefont {{Gavazzi}}, \citenamefont {{Pichon}}, \citenamefont
  {{Dubois}}, \citenamefont {{Laigle}}, \citenamefont {{Chisari}},
  \citenamefont {{Codis}}, \citenamefont {{Devriendt}},\ and\ \citenamefont
  {{Peirani}}}]{Gouin19}%
  \BibitemOpen
  \bibfield  {author} {\bibinfo {author} {\bibfnamefont {C.}~\bibnamefont
  {{Gouin}}}, \bibinfo {author} {\bibfnamefont {R.}~\bibnamefont {{Gavazzi}}},
  \bibinfo {author} {\bibfnamefont {C.}~\bibnamefont {{Pichon}}}, \bibinfo
  {author} {\bibfnamefont {Y.}~\bibnamefont {{Dubois}}}, \bibinfo {author}
  {\bibfnamefont {C.}~\bibnamefont {{Laigle}}}, \bibinfo {author}
  {\bibfnamefont {N.~E.}\ \bibnamefont {{Chisari}}}, \bibinfo {author}
  {\bibfnamefont {S.}~\bibnamefont {{Codis}}}, \bibinfo {author} {\bibfnamefont
  {J.}~\bibnamefont {{Devriendt}}}, \ and\ \bibinfo {author} {\bibfnamefont
  {S.}~\bibnamefont {{Peirani}}},\ }\href@noop {} {\bibfield  {journal}
  {\bibinfo  {journal} {arXiv e-prints}\ } (\bibinfo {year} {2019})},\ \Eprint
  {http://arxiv.org/abs/1904.07905} {arXiv:1904.07905} \BibitemShut {NoStop}%
\bibitem [{\citenamefont {{Beckmann}}\ \emph {et~al.}(2017)\citenamefont
  {{Beckmann}}, \citenamefont {{Devriendt}}, \citenamefont {{Slyz}},
  \citenamefont {{Peirani}}, \citenamefont {{Richardson}}, \citenamefont
  {{Dubois}}, \citenamefont {{Pichon}}, \citenamefont {{Chisari}},
  \citenamefont {{Kaviraj}}, \citenamefont {{Laigle}},\ and\ \citenamefont
  {{Volonteri}}}]{Beckmann17}%
  \BibitemOpen
  \bibfield  {author} {\bibinfo {author} {\bibfnamefont {R.~S.}\ \bibnamefont
  {{Beckmann}}}, \bibinfo {author} {\bibfnamefont {J.}~\bibnamefont
  {{Devriendt}}}, \bibinfo {author} {\bibfnamefont {A.}~\bibnamefont {{Slyz}}},
  \bibinfo {author} {\bibfnamefont {S.}~\bibnamefont {{Peirani}}}, \bibinfo
  {author} {\bibfnamefont {M.~L.~A.}\ \bibnamefont {{Richardson}}}, \bibinfo
  {author} {\bibfnamefont {Y.}~\bibnamefont {{Dubois}}}, \bibinfo {author}
  {\bibfnamefont {C.}~\bibnamefont {{Pichon}}}, \bibinfo {author}
  {\bibfnamefont {N.~E.}\ \bibnamefont {{Chisari}}}, \bibinfo {author}
  {\bibfnamefont {S.}~\bibnamefont {{Kaviraj}}}, \bibinfo {author}
  {\bibfnamefont {C.}~\bibnamefont {{Laigle}}}, \ and\ \bibinfo {author}
  {\bibfnamefont {M.}~\bibnamefont {{Volonteri}}},\ }\href {\doibase
  10.1093/mnras/stx1831} {\bibfield  {journal} {\bibinfo  {journal} {\mnras}\
  }\textbf {\bibinfo {volume} {472}},\ \bibinfo {pages} {949} (\bibinfo {year}
  {2017})},\ \Eprint {http://arxiv.org/abs/1701.07838} {arXiv:1701.07838}
  \BibitemShut {NoStop}%
\bibitem [{\citenamefont {{Kaviraj}}\ \emph {et~al.}(2017)\citenamefont
  {{Kaviraj}}, \citenamefont {{Laigle}}, \citenamefont {{Kimm}}, \citenamefont
  {{Devriendt}}, \citenamefont {{Dubois}}, \citenamefont {{Pichon}},
  \citenamefont {{Slyz}}, \citenamefont {{Chisari}},\ and\ \citenamefont
  {{Peirani}}}]{Kaviraj17}%
  \BibitemOpen
  \bibfield  {author} {\bibinfo {author} {\bibfnamefont {S.}~\bibnamefont
  {{Kaviraj}}}, \bibinfo {author} {\bibfnamefont {C.}~\bibnamefont {{Laigle}}},
  \bibinfo {author} {\bibfnamefont {T.}~\bibnamefont {{Kimm}}}, \bibinfo
  {author} {\bibfnamefont {J.~E.~G.}\ \bibnamefont {{Devriendt}}}, \bibinfo
  {author} {\bibfnamefont {Y.}~\bibnamefont {{Dubois}}}, \bibinfo {author}
  {\bibfnamefont {C.}~\bibnamefont {{Pichon}}}, \bibinfo {author}
  {\bibfnamefont {A.}~\bibnamefont {{Slyz}}}, \bibinfo {author} {\bibfnamefont
  {E.}~\bibnamefont {{Chisari}}}, \ and\ \bibinfo {author} {\bibfnamefont
  {S.}~\bibnamefont {{Peirani}}},\ }\href {\doibase 10.1093/mnras/stx126}
  {\bibfield  {journal} {\bibinfo  {journal} {\mnras}\ }\textbf {\bibinfo
  {volume} {467}},\ \bibinfo {pages} {4739} (\bibinfo {year} {2017})},\ \Eprint
  {http://arxiv.org/abs/1605.09379} {arXiv:1605.09379} \BibitemShut {NoStop}%
\bibitem [{\citenamefont {{Dubois}}\ \emph {et~al.}(2016)\citenamefont
  {{Dubois}}, \citenamefont {{Peirani}}, \citenamefont {{Pichon}},
  \citenamefont {{Devriendt}}, \citenamefont {{Gavazzi}}, \citenamefont
  {{Welker}},\ and\ \citenamefont {{Volonteri}}}]{Dubois16}%
  \BibitemOpen
  \bibfield  {author} {\bibinfo {author} {\bibfnamefont {Y.}~\bibnamefont
  {{Dubois}}}, \bibinfo {author} {\bibfnamefont {S.}~\bibnamefont {{Peirani}}},
  \bibinfo {author} {\bibfnamefont {C.}~\bibnamefont {{Pichon}}}, \bibinfo
  {author} {\bibfnamefont {J.}~\bibnamefont {{Devriendt}}}, \bibinfo {author}
  {\bibfnamefont {R.}~\bibnamefont {{Gavazzi}}}, \bibinfo {author}
  {\bibfnamefont {C.}~\bibnamefont {{Welker}}}, \ and\ \bibinfo {author}
  {\bibfnamefont {M.}~\bibnamefont {{Volonteri}}},\ }\href {\doibase
  10.1093/mnras/stw2265} {\bibfield  {journal} {\bibinfo  {journal} {\mnras}\
  }\textbf {\bibinfo {volume} {463}},\ \bibinfo {pages} {3948} (\bibinfo {year}
  {2016})},\ \Eprint {http://arxiv.org/abs/1606.03086} {arXiv:1606.03086}
  \BibitemShut {NoStop}%
\bibitem [{\citenamefont {{Khandai}}\ \emph {et~al.}(2015)\citenamefont
  {{Khandai}}, \citenamefont {{Di Matteo}}, \citenamefont {{Croft}},
  \citenamefont {{Wilkins}}, \citenamefont {{Feng}}, \citenamefont {{Tucker}},
  \citenamefont {{DeGraf}},\ and\ \citenamefont {{Liu}}}]{Khandai15}%
  \BibitemOpen
  \bibfield  {author} {\bibinfo {author} {\bibfnamefont {N.}~\bibnamefont
  {{Khandai}}}, \bibinfo {author} {\bibfnamefont {T.}~\bibnamefont {{Di
  Matteo}}}, \bibinfo {author} {\bibfnamefont {R.}~\bibnamefont {{Croft}}},
  \bibinfo {author} {\bibfnamefont {S.}~\bibnamefont {{Wilkins}}}, \bibinfo
  {author} {\bibfnamefont {Y.}~\bibnamefont {{Feng}}}, \bibinfo {author}
  {\bibfnamefont {E.}~\bibnamefont {{Tucker}}}, \bibinfo {author}
  {\bibfnamefont {C.}~\bibnamefont {{DeGraf}}}, \ and\ \bibinfo {author}
  {\bibfnamefont {M.-S.}\ \bibnamefont {{Liu}}},\ }\href {\doibase
  10.1093/mnras/stv627} {\bibfield  {journal} {\bibinfo  {journal} {\mnras}\
  }\textbf {\bibinfo {volume} {450}},\ \bibinfo {pages} {1349} (\bibinfo {year}
  {2015})},\ \Eprint {http://arxiv.org/abs/1402.0888} {arXiv:1402.0888
  [astro-ph.CO]} \BibitemShut {NoStop}%
\bibitem [{\citenamefont {{Springel}}(2005)}]{Springel05}%
  \BibitemOpen
  \bibfield  {author} {\bibinfo {author} {\bibfnamefont {V.}~\bibnamefont
  {{Springel}}},\ }\href {\doibase 10.1111/j.1365-2966.2005.09655.x} {\bibfield
   {journal} {\bibinfo  {journal} {\mnras}\ }\textbf {\bibinfo {volume}
  {364}},\ \bibinfo {pages} {1105} (\bibinfo {year} {2005})},\ \Eprint
  {http://arxiv.org/abs/astro-ph/0505010} {astro-ph/0505010} \BibitemShut
  {NoStop}%
\bibitem [{\citenamefont {{Di Matteo}}\ \emph {et~al.}(2012)\citenamefont {{Di
  Matteo}}, \citenamefont {{Khandai}}, \citenamefont {{DeGraf}}, \citenamefont
  {{Feng}}, \citenamefont {{Croft}}, \citenamefont {{Lopez}},\ and\
  \citenamefont {{Springel}}}]{DiMatteo12}%
  \BibitemOpen
  \bibfield  {author} {\bibinfo {author} {\bibfnamefont {T.}~\bibnamefont {{Di
  Matteo}}}, \bibinfo {author} {\bibfnamefont {N.}~\bibnamefont {{Khandai}}},
  \bibinfo {author} {\bibfnamefont {C.}~\bibnamefont {{DeGraf}}}, \bibinfo
  {author} {\bibfnamefont {Y.}~\bibnamefont {{Feng}}}, \bibinfo {author}
  {\bibfnamefont {R.~A.~C.}\ \bibnamefont {{Croft}}}, \bibinfo {author}
  {\bibfnamefont {J.}~\bibnamefont {{Lopez}}}, \ and\ \bibinfo {author}
  {\bibfnamefont {V.}~\bibnamefont {{Springel}}},\ }\href {\doibase
  10.1088/2041-8205/745/2/L29} {\bibfield  {journal} {\bibinfo  {journal}
  {\apjl}\ }\textbf {\bibinfo {volume} {745}},\ \bibinfo {eid} {L29} (\bibinfo
  {year} {2012})},\ \Eprint {http://arxiv.org/abs/1107.1253} {arXiv:1107.1253}
  \BibitemShut {NoStop}%
\bibitem [{\citenamefont {{Tenneti}}\ \emph {et~al.}(2015)\citenamefont
  {{Tenneti}}, \citenamefont {{Mandelbaum}}, \citenamefont {{Di Matteo}},
  \citenamefont {{Kiessling}},\ and\ \citenamefont {{Khandai}}}]{Tenneti15}%
  \BibitemOpen
  \bibfield  {author} {\bibinfo {author} {\bibfnamefont {A.}~\bibnamefont
  {{Tenneti}}}, \bibinfo {author} {\bibfnamefont {R.}~\bibnamefont
  {{Mandelbaum}}}, \bibinfo {author} {\bibfnamefont {T.}~\bibnamefont {{Di
  Matteo}}}, \bibinfo {author} {\bibfnamefont {A.}~\bibnamefont {{Kiessling}}},
  \ and\ \bibinfo {author} {\bibfnamefont {N.}~\bibnamefont {{Khandai}}},\
  }\href {\doibase 10.1093/mnras/stv1625} {\bibfield  {journal} {\bibinfo
  {journal} {\mnras}\ }\textbf {\bibinfo {volume} {453}},\ \bibinfo {pages}
  {469} (\bibinfo {year} {2015})},\ \Eprint {http://arxiv.org/abs/1505.03124}
  {arXiv:1505.03124 [astro-ph.CO]} \BibitemShut {NoStop}%
\bibitem [{\citenamefont {{Le Brun}}\ \emph {et~al.}(2014)\citenamefont {{Le
  Brun}}, \citenamefont {{McCarthy}}, \citenamefont {{Schaye}},\ and\
  \citenamefont {{Ponman}}}]{LeBrun14}%
  \BibitemOpen
  \bibfield  {author} {\bibinfo {author} {\bibfnamefont {A.~M.~C.}\
  \bibnamefont {{Le Brun}}}, \bibinfo {author} {\bibfnamefont {I.~G.}\
  \bibnamefont {{McCarthy}}}, \bibinfo {author} {\bibfnamefont
  {J.}~\bibnamefont {{Schaye}}}, \ and\ \bibinfo {author} {\bibfnamefont
  {T.~J.}\ \bibnamefont {{Ponman}}},\ }\href {\doibase 10.1093/mnras/stu608}
  {\bibfield  {journal} {\bibinfo  {journal} {\mnras}\ }\textbf {\bibinfo
  {volume} {441}},\ \bibinfo {pages} {1270} (\bibinfo {year} {2014})},\ \Eprint
  {http://arxiv.org/abs/1312.5462} {arXiv:1312.5462 [astro-ph.CO]} \BibitemShut
  {NoStop}%
\bibitem [{\citenamefont {{McCarthy}}\ \emph {et~al.}(2014)\citenamefont
  {{McCarthy}}, \citenamefont {{Le Brun}}, \citenamefont {{Schaye}},\ and\
  \citenamefont {{Holder}}}]{McCarthy14}%
  \BibitemOpen
  \bibfield  {author} {\bibinfo {author} {\bibfnamefont {I.~G.}\ \bibnamefont
  {{McCarthy}}}, \bibinfo {author} {\bibfnamefont {A.~M.~C.}\ \bibnamefont {{Le
  Brun}}}, \bibinfo {author} {\bibfnamefont {J.}~\bibnamefont {{Schaye}}}, \
  and\ \bibinfo {author} {\bibfnamefont {G.~P.}\ \bibnamefont {{Holder}}},\
  }\href {\doibase 10.1093/mnras/stu543} {\bibfield  {journal} {\bibinfo
  {journal} {\mnras}\ }\textbf {\bibinfo {volume} {440}},\ \bibinfo {pages}
  {3645} (\bibinfo {year} {2014})},\ \Eprint {http://arxiv.org/abs/1312.5341}
  {arXiv:1312.5341} \BibitemShut {NoStop}%
\bibitem [{\citenamefont {{Schaye}}\ \emph {et~al.}(2015)\citenamefont
  {{Schaye}}, \citenamefont {{Crain}}, \citenamefont {{Bower}}, \citenamefont
  {{Furlong}}, \citenamefont {{Schaller}}, \citenamefont {{Theuns}},
  \citenamefont {{Dalla Vecchia}}, \citenamefont {{Frenk}}, \citenamefont
  {{McCarthy}}, \citenamefont {{Helly}}, \citenamefont {{Jenkins}},
  \citenamefont {{Rosas-Guevara}}, \citenamefont {{White}}, \citenamefont
  {{Baes}}, \citenamefont {{Booth}}, \citenamefont {{Camps}}, \citenamefont
  {{Navarro}}, \citenamefont {{Qu}}, \citenamefont {{Rahmati}}, \citenamefont
  {{Sawala}}, \citenamefont {{Thomas}},\ and\ \citenamefont
  {{Trayford}}}]{Schaye15}%
  \BibitemOpen
  \bibfield  {author} {\bibinfo {author} {\bibfnamefont {J.}~\bibnamefont
  {{Schaye}}}, \bibinfo {author} {\bibfnamefont {R.~A.}\ \bibnamefont
  {{Crain}}}, \bibinfo {author} {\bibfnamefont {R.~G.}\ \bibnamefont
  {{Bower}}}, \bibinfo {author} {\bibfnamefont {M.}~\bibnamefont {{Furlong}}},
  \bibinfo {author} {\bibfnamefont {M.}~\bibnamefont {{Schaller}}}, \bibinfo
  {author} {\bibfnamefont {T.}~\bibnamefont {{Theuns}}}, \bibinfo {author}
  {\bibfnamefont {C.}~\bibnamefont {{Dalla Vecchia}}}, \bibinfo {author}
  {\bibfnamefont {C.~S.}\ \bibnamefont {{Frenk}}}, \bibinfo {author}
  {\bibfnamefont {I.~G.}\ \bibnamefont {{McCarthy}}}, \bibinfo {author}
  {\bibfnamefont {J.~C.}\ \bibnamefont {{Helly}}}, \bibinfo {author}
  {\bibfnamefont {A.}~\bibnamefont {{Jenkins}}}, \bibinfo {author}
  {\bibfnamefont {Y.~M.}\ \bibnamefont {{Rosas-Guevara}}}, \bibinfo {author}
  {\bibfnamefont {S.~D.~M.}\ \bibnamefont {{White}}}, \bibinfo {author}
  {\bibfnamefont {M.}~\bibnamefont {{Baes}}}, \bibinfo {author} {\bibfnamefont
  {C.~M.}\ \bibnamefont {{Booth}}}, \bibinfo {author} {\bibfnamefont
  {P.}~\bibnamefont {{Camps}}}, \bibinfo {author} {\bibfnamefont {J.~F.}\
  \bibnamefont {{Navarro}}}, \bibinfo {author} {\bibfnamefont {Y.}~\bibnamefont
  {{Qu}}}, \bibinfo {author} {\bibfnamefont {A.}~\bibnamefont {{Rahmati}}},
  \bibinfo {author} {\bibfnamefont {T.}~\bibnamefont {{Sawala}}}, \bibinfo
  {author} {\bibfnamefont {P.~A.}\ \bibnamefont {{Thomas}}}, \ and\ \bibinfo
  {author} {\bibfnamefont {J.}~\bibnamefont {{Trayford}}},\ }\href {\doibase
  10.1093/mnras/stu2058} {\bibfield  {journal} {\bibinfo  {journal} {\mnras}\
  }\textbf {\bibinfo {volume} {446}},\ \bibinfo {pages} {521} (\bibinfo {year}
  {2015})},\ \Eprint {http://arxiv.org/abs/1407.7040} {arXiv:1407.7040
  [astro-ph.GA]} \BibitemShut {NoStop}%
\bibitem [{\citenamefont {{McCarthy}}\ \emph {et~al.}(2017)\citenamefont
  {{McCarthy}}, \citenamefont {{Schaye}}, \citenamefont {{Bird}},\ and\
  \citenamefont {{Le Brun}}}]{McCarthy17}%
  \BibitemOpen
  \bibfield  {author} {\bibinfo {author} {\bibfnamefont {I.~G.}\ \bibnamefont
  {{McCarthy}}}, \bibinfo {author} {\bibfnamefont {J.}~\bibnamefont
  {{Schaye}}}, \bibinfo {author} {\bibfnamefont {S.}~\bibnamefont {{Bird}}}, \
  and\ \bibinfo {author} {\bibfnamefont {A.~M.~C.}\ \bibnamefont {{Le Brun}}},\
  }\href {\doibase 10.1093/mnras/stw2792} {\bibfield  {journal} {\bibinfo
  {journal} {\mnras}\ }\textbf {\bibinfo {volume} {465}},\ \bibinfo {pages}
  {2936} (\bibinfo {year} {2017})},\ \Eprint {http://arxiv.org/abs/1603.02702}
  {arXiv:1603.02702} \BibitemShut {NoStop}%
\bibitem [{\citenamefont {{Vogelsberger}}\ \emph
  {et~al.}(2014{\natexlab{a}})\citenamefont {{Vogelsberger}}, \citenamefont
  {{Genel}}, \citenamefont {{Springel}}, \citenamefont {{Torrey}},
  \citenamefont {{Sijacki}}, \citenamefont {{Xu}}, \citenamefont {{Snyder}},
  \citenamefont {{Bird}}, \citenamefont {{Nelson}},\ and\ \citenamefont
  {{Hernquist}}}]{Vogelsberger14}%
  \BibitemOpen
  \bibfield  {author} {\bibinfo {author} {\bibfnamefont {M.}~\bibnamefont
  {{Vogelsberger}}}, \bibinfo {author} {\bibfnamefont {S.}~\bibnamefont
  {{Genel}}}, \bibinfo {author} {\bibfnamefont {V.}~\bibnamefont {{Springel}}},
  \bibinfo {author} {\bibfnamefont {P.}~\bibnamefont {{Torrey}}}, \bibinfo
  {author} {\bibfnamefont {D.}~\bibnamefont {{Sijacki}}}, \bibinfo {author}
  {\bibfnamefont {D.}~\bibnamefont {{Xu}}}, \bibinfo {author} {\bibfnamefont
  {G.}~\bibnamefont {{Snyder}}}, \bibinfo {author} {\bibfnamefont
  {S.}~\bibnamefont {{Bird}}}, \bibinfo {author} {\bibfnamefont
  {D.}~\bibnamefont {{Nelson}}}, \ and\ \bibinfo {author} {\bibfnamefont
  {L.}~\bibnamefont {{Hernquist}}},\ }\href {\doibase 10.1038/nature13316}
  {\bibfield  {journal} {\bibinfo  {journal} {\nat}\ }\textbf {\bibinfo
  {volume} {509}},\ \bibinfo {pages} {177} (\bibinfo {year}
  {2014}{\natexlab{a}})},\ \Eprint {http://arxiv.org/abs/1405.1418}
  {arXiv:1405.1418 [astro-ph.CO]} \BibitemShut {NoStop}%
\bibitem [{\citenamefont {{Vogelsberger}}\ \emph
  {et~al.}(2014{\natexlab{b}})\citenamefont {{Vogelsberger}}, \citenamefont
  {{Genel}}, \citenamefont {{Springel}}, \citenamefont {{Torrey}},
  \citenamefont {{Sijacki}}, \citenamefont {{Xu}}, \citenamefont {{Snyder}},
  \citenamefont {{Nelson}},\ and\ \citenamefont
  {{Hernquist}}}]{Vogelsberger14b}%
  \BibitemOpen
  \bibfield  {author} {\bibinfo {author} {\bibfnamefont {M.}~\bibnamefont
  {{Vogelsberger}}}, \bibinfo {author} {\bibfnamefont {S.}~\bibnamefont
  {{Genel}}}, \bibinfo {author} {\bibfnamefont {V.}~\bibnamefont {{Springel}}},
  \bibinfo {author} {\bibfnamefont {P.}~\bibnamefont {{Torrey}}}, \bibinfo
  {author} {\bibfnamefont {D.}~\bibnamefont {{Sijacki}}}, \bibinfo {author}
  {\bibfnamefont {D.}~\bibnamefont {{Xu}}}, \bibinfo {author} {\bibfnamefont
  {G.}~\bibnamefont {{Snyder}}}, \bibinfo {author} {\bibfnamefont
  {D.}~\bibnamefont {{Nelson}}}, \ and\ \bibinfo {author} {\bibfnamefont
  {L.}~\bibnamefont {{Hernquist}}},\ }\href {\doibase 10.1093/mnras/stu1536}
  {\bibfield  {journal} {\bibinfo  {journal} {\mnras}\ }\textbf {\bibinfo
  {volume} {444}},\ \bibinfo {pages} {1518} (\bibinfo {year}
  {2014}{\natexlab{b}})},\ \Eprint {http://arxiv.org/abs/1405.2921}
  {arXiv:1405.2921 [astro-ph.CO]} \BibitemShut {NoStop}%
\bibitem [{\citenamefont {{Genel}}\ \emph {et~al.}(2014)\citenamefont
  {{Genel}}, \citenamefont {{Vogelsberger}}, \citenamefont {{Springel}},
  \citenamefont {{Sijacki}}, \citenamefont {{Nelson}}, \citenamefont
  {{Snyder}}, \citenamefont {{Rodriguez-Gomez}}, \citenamefont {{Torrey}},\
  and\ \citenamefont {{Hernquist}}}]{Genel14}%
  \BibitemOpen
  \bibfield  {author} {\bibinfo {author} {\bibfnamefont {S.}~\bibnamefont
  {{Genel}}}, \bibinfo {author} {\bibfnamefont {M.}~\bibnamefont
  {{Vogelsberger}}}, \bibinfo {author} {\bibfnamefont {V.}~\bibnamefont
  {{Springel}}}, \bibinfo {author} {\bibfnamefont {D.}~\bibnamefont
  {{Sijacki}}}, \bibinfo {author} {\bibfnamefont {D.}~\bibnamefont {{Nelson}}},
  \bibinfo {author} {\bibfnamefont {G.}~\bibnamefont {{Snyder}}}, \bibinfo
  {author} {\bibfnamefont {V.}~\bibnamefont {{Rodriguez-Gomez}}}, \bibinfo
  {author} {\bibfnamefont {P.}~\bibnamefont {{Torrey}}}, \ and\ \bibinfo
  {author} {\bibfnamefont {L.}~\bibnamefont {{Hernquist}}},\ }\href {\doibase
  10.1093/mnras/stu1654} {\bibfield  {journal} {\bibinfo  {journal} {\mnras}\
  }\textbf {\bibinfo {volume} {445}},\ \bibinfo {pages} {175} (\bibinfo {year}
  {2014})},\ \Eprint {http://arxiv.org/abs/1405.3749} {arXiv:1405.3749}
  \BibitemShut {NoStop}%
\bibitem [{\citenamefont {{Springel}}(2010)}]{Springel10}%
  \BibitemOpen
  \bibfield  {author} {\bibinfo {author} {\bibfnamefont {V.}~\bibnamefont
  {{Springel}}},\ }\href {\doibase 10.1111/j.1365-2966.2009.15715.x} {\bibfield
   {journal} {\bibinfo  {journal} {\mnras}\ }\textbf {\bibinfo {volume}
  {401}},\ \bibinfo {pages} {791} (\bibinfo {year} {2010})},\ \Eprint
  {http://arxiv.org/abs/0901.4107} {arXiv:0901.4107} \BibitemShut {NoStop}%
\bibitem [{\citenamefont {{Vogelsberger}}\ \emph {et~al.}(2013)\citenamefont
  {{Vogelsberger}}, \citenamefont {{Genel}}, \citenamefont {{Sijacki}},
  \citenamefont {{Torrey}}, \citenamefont {{Springel}},\ and\ \citenamefont
  {{Hernquist}}}]{Vogelsberger13}%
  \BibitemOpen
  \bibfield  {author} {\bibinfo {author} {\bibfnamefont {M.}~\bibnamefont
  {{Vogelsberger}}}, \bibinfo {author} {\bibfnamefont {S.}~\bibnamefont
  {{Genel}}}, \bibinfo {author} {\bibfnamefont {D.}~\bibnamefont {{Sijacki}}},
  \bibinfo {author} {\bibfnamefont {P.}~\bibnamefont {{Torrey}}}, \bibinfo
  {author} {\bibfnamefont {V.}~\bibnamefont {{Springel}}}, \ and\ \bibinfo
  {author} {\bibfnamefont {L.}~\bibnamefont {{Hernquist}}},\ }\href {\doibase
  10.1093/mnras/stt1789} {\bibfield  {journal} {\bibinfo  {journal} {\mnras}\
  }\textbf {\bibinfo {volume} {436}},\ \bibinfo {pages} {3031} (\bibinfo {year}
  {2013})},\ \Eprint {http://arxiv.org/abs/1305.2913} {arXiv:1305.2913}
  \BibitemShut {NoStop}%
\bibitem [{\citenamefont {{Weinberger}}\ \emph {et~al.}(2017)\citenamefont
  {{Weinberger}}, \citenamefont {{Springel}}, \citenamefont {{Hernquist}},
  \citenamefont {{Pillepich}}, \citenamefont {{Marinacci}}, \citenamefont
  {{Pakmor}}, \citenamefont {{Nelson}}, \citenamefont {{Genel}}, \citenamefont
  {{Vogelsberger}}, \citenamefont {{Naiman}},\ and\ \citenamefont
  {{Torrey}}}]{Weinberger17}%
  \BibitemOpen
  \bibfield  {author} {\bibinfo {author} {\bibfnamefont {R.}~\bibnamefont
  {{Weinberger}}}, \bibinfo {author} {\bibfnamefont {V.}~\bibnamefont
  {{Springel}}}, \bibinfo {author} {\bibfnamefont {L.}~\bibnamefont
  {{Hernquist}}}, \bibinfo {author} {\bibfnamefont {A.}~\bibnamefont
  {{Pillepich}}}, \bibinfo {author} {\bibfnamefont {F.}~\bibnamefont
  {{Marinacci}}}, \bibinfo {author} {\bibfnamefont {R.}~\bibnamefont
  {{Pakmor}}}, \bibinfo {author} {\bibfnamefont {D.}~\bibnamefont {{Nelson}}},
  \bibinfo {author} {\bibfnamefont {S.}~\bibnamefont {{Genel}}}, \bibinfo
  {author} {\bibfnamefont {M.}~\bibnamefont {{Vogelsberger}}}, \bibinfo
  {author} {\bibfnamefont {J.}~\bibnamefont {{Naiman}}}, \ and\ \bibinfo
  {author} {\bibfnamefont {P.}~\bibnamefont {{Torrey}}},\ }\href {\doibase
  10.1093/mnras/stw2944} {\bibfield  {journal} {\bibinfo  {journal} {\mnras}\
  }\textbf {\bibinfo {volume} {465}},\ \bibinfo {pages} {3291} (\bibinfo {year}
  {2017})},\ \Eprint {http://arxiv.org/abs/1607.03486} {arXiv:1607.03486}
  \BibitemShut {NoStop}%
\bibitem [{\citenamefont {{Pillepich}}\ \emph {et~al.}(2018)\citenamefont
  {{Pillepich}}, \citenamefont {{Springel}}, \citenamefont {{Nelson}},
  \citenamefont {{Genel}}, \citenamefont {{Naiman}}, \citenamefont {{Pakmor}},
  \citenamefont {{Hernquist}}, \citenamefont {{Torrey}}, \citenamefont
  {{Vogelsberger}}, \citenamefont {{Weinberger}},\ and\ \citenamefont
  {{Marinacci}}}]{Pillepich18}%
  \BibitemOpen
  \bibfield  {author} {\bibinfo {author} {\bibfnamefont {A.}~\bibnamefont
  {{Pillepich}}}, \bibinfo {author} {\bibfnamefont {V.}~\bibnamefont
  {{Springel}}}, \bibinfo {author} {\bibfnamefont {D.}~\bibnamefont
  {{Nelson}}}, \bibinfo {author} {\bibfnamefont {S.}~\bibnamefont {{Genel}}},
  \bibinfo {author} {\bibfnamefont {J.}~\bibnamefont {{Naiman}}}, \bibinfo
  {author} {\bibfnamefont {R.}~\bibnamefont {{Pakmor}}}, \bibinfo {author}
  {\bibfnamefont {L.}~\bibnamefont {{Hernquist}}}, \bibinfo {author}
  {\bibfnamefont {P.}~\bibnamefont {{Torrey}}}, \bibinfo {author}
  {\bibfnamefont {M.}~\bibnamefont {{Vogelsberger}}}, \bibinfo {author}
  {\bibfnamefont {R.}~\bibnamefont {{Weinberger}}}, \ and\ \bibinfo {author}
  {\bibfnamefont {F.}~\bibnamefont {{Marinacci}}},\ }\href {\doibase
  10.1093/mnras/stx2656} {\bibfield  {journal} {\bibinfo  {journal} {\mnras}\
  }\textbf {\bibinfo {volume} {473}},\ \bibinfo {pages} {4077} (\bibinfo {year}
  {2018})},\ \Eprint {http://arxiv.org/abs/1703.02970} {arXiv:1703.02970}
  \BibitemShut {NoStop}%
\bibitem [{\citenamefont {{Hellwing}}\ \emph {et~al.}(2016)\citenamefont
  {{Hellwing}}, \citenamefont {{Schaller}}, \citenamefont {{Frenk}},
  \citenamefont {{Theuns}}, \citenamefont {{Schaye}}, \citenamefont {{Bower}},\
  and\ \citenamefont {{Crain}}}]{Hellwing16}%
  \BibitemOpen
  \bibfield  {author} {\bibinfo {author} {\bibfnamefont {W.~A.}\ \bibnamefont
  {{Hellwing}}}, \bibinfo {author} {\bibfnamefont {M.}~\bibnamefont
  {{Schaller}}}, \bibinfo {author} {\bibfnamefont {C.~S.}\ \bibnamefont
  {{Frenk}}}, \bibinfo {author} {\bibfnamefont {T.}~\bibnamefont {{Theuns}}},
  \bibinfo {author} {\bibfnamefont {J.}~\bibnamefont {{Schaye}}}, \bibinfo
  {author} {\bibfnamefont {R.~G.}\ \bibnamefont {{Bower}}}, \ and\ \bibinfo
  {author} {\bibfnamefont {R.~A.}\ \bibnamefont {{Crain}}},\ }\href {\doibase
  10.1093/mnrasl/slw081} {\bibfield  {journal} {\bibinfo  {journal} {\mnras}\
  }\textbf {\bibinfo {volume} {461}},\ \bibinfo {pages} {L11} (\bibinfo {year}
  {2016})},\ \Eprint {http://arxiv.org/abs/1603.03328} {arXiv:1603.03328
  [astro-ph.CO]} \BibitemShut {NoStop}%
\bibitem [{\citenamefont {{Springel}}\ \emph {et~al.}(2018)\citenamefont
  {{Springel}}, \citenamefont {{Pakmor}}, \citenamefont {{Pillepich}},
  \citenamefont {{Weinberger}}, \citenamefont {{Nelson}}, \citenamefont
  {{Hernquist}}, \citenamefont {{Vogelsberger}}, \citenamefont {{Genel}},
  \citenamefont {{Torrey}}, \citenamefont {{Marinacci}},\ and\ \citenamefont
  {{Naiman}}}]{Springel18}%
  \BibitemOpen
  \bibfield  {author} {\bibinfo {author} {\bibfnamefont {V.}~\bibnamefont
  {{Springel}}}, \bibinfo {author} {\bibfnamefont {R.}~\bibnamefont
  {{Pakmor}}}, \bibinfo {author} {\bibfnamefont {A.}~\bibnamefont
  {{Pillepich}}}, \bibinfo {author} {\bibfnamefont {R.}~\bibnamefont
  {{Weinberger}}}, \bibinfo {author} {\bibfnamefont {D.}~\bibnamefont
  {{Nelson}}}, \bibinfo {author} {\bibfnamefont {L.}~\bibnamefont
  {{Hernquist}}}, \bibinfo {author} {\bibfnamefont {M.}~\bibnamefont
  {{Vogelsberger}}}, \bibinfo {author} {\bibfnamefont {S.}~\bibnamefont
  {{Genel}}}, \bibinfo {author} {\bibfnamefont {P.}~\bibnamefont {{Torrey}}},
  \bibinfo {author} {\bibfnamefont {F.}~\bibnamefont {{Marinacci}}}, \ and\
  \bibinfo {author} {\bibfnamefont {J.}~\bibnamefont {{Naiman}}},\ }\href
  {\doibase 10.1093/mnras/stx3304} {\bibfield  {journal} {\bibinfo  {journal}
  {\mnras}\ }\textbf {\bibinfo {volume} {475}},\ \bibinfo {pages} {676}
  (\bibinfo {year} {2018})},\ \Eprint {http://arxiv.org/abs/1707.03397}
  {arXiv:1707.03397} \BibitemShut {NoStop}%
\bibitem [{\citenamefont {{McCarthy}}\ \emph {et~al.}(2018)\citenamefont
  {{McCarthy}}, \citenamefont {{Bird}}, \citenamefont {{Schaye}}, \citenamefont
  {{Harnois-Deraps}}, \citenamefont {{Font}},\ and\ \citenamefont {{van
  Waerbeke}}}]{McCarthy18}%
  \BibitemOpen
  \bibfield  {author} {\bibinfo {author} {\bibfnamefont {I.~G.}\ \bibnamefont
  {{McCarthy}}}, \bibinfo {author} {\bibfnamefont {S.}~\bibnamefont {{Bird}}},
  \bibinfo {author} {\bibfnamefont {J.}~\bibnamefont {{Schaye}}}, \bibinfo
  {author} {\bibfnamefont {J.}~\bibnamefont {{Harnois-Deraps}}}, \bibinfo
  {author} {\bibfnamefont {A.~S.}\ \bibnamefont {{Font}}}, \ and\ \bibinfo
  {author} {\bibfnamefont {L.}~\bibnamefont {{van Waerbeke}}},\ }\href
  {\doibase 10.1093/mnras/sty377} {\bibfield  {journal} {\bibinfo  {journal}
  {\mnras}\ }\textbf {\bibinfo {volume} {476}},\ \bibinfo {pages} {2999}
  (\bibinfo {year} {2018})},\ \Eprint {http://arxiv.org/abs/1712.02411}
  {arXiv:1712.02411} \BibitemShut {NoStop}%
\bibitem [{\citenamefont {{Dav{\'e}}}\ \emph {et~al.}(2016)\citenamefont
  {{Dav{\'e}}}, \citenamefont {{Thompson}},\ and\ \citenamefont
  {{Hopkins}}}]{MUFASA}%
  \BibitemOpen
  \bibfield  {author} {\bibinfo {author} {\bibfnamefont {R.}~\bibnamefont
  {{Dav{\'e}}}}, \bibinfo {author} {\bibfnamefont {R.}~\bibnamefont
  {{Thompson}}}, \ and\ \bibinfo {author} {\bibfnamefont {P.~F.}\ \bibnamefont
  {{Hopkins}}},\ }\href {\doibase 10.1093/mnras/stw1862} {\bibfield  {journal}
  {\bibinfo  {journal} {\mnras}\ }\textbf {\bibinfo {volume} {462}},\ \bibinfo
  {pages} {3265} (\bibinfo {year} {2016})},\ \Eprint
  {http://arxiv.org/abs/1604.01418} {arXiv:1604.01418 [astro-ph.GA]}
  \BibitemShut {NoStop}%
\bibitem [{\citenamefont {{Dav{\'e}}}\ \emph {et~al.}(2019)\citenamefont
  {{Dav{\'e}}}, \citenamefont {{Angl{\'e}s-Alc{\'a}zar}}, \citenamefont
  {{Narayanan}}, \citenamefont {{Li}}, \citenamefont {{Rafieferantsoa}},\ and\
  \citenamefont {{Appleby}}}]{Simba}%
  \BibitemOpen
  \bibfield  {author} {\bibinfo {author} {\bibfnamefont {R.}~\bibnamefont
  {{Dav{\'e}}}}, \bibinfo {author} {\bibfnamefont {D.}~\bibnamefont
  {{Angl{\'e}s-Alc{\'a}zar}}}, \bibinfo {author} {\bibfnamefont
  {D.}~\bibnamefont {{Narayanan}}}, \bibinfo {author} {\bibfnamefont
  {Q.}~\bibnamefont {{Li}}}, \bibinfo {author} {\bibfnamefont {M.~H.}\
  \bibnamefont {{Rafieferantsoa}}}, \ and\ \bibinfo {author} {\bibfnamefont
  {S.}~\bibnamefont {{Appleby}}},\ }\href@noop {} {\bibfield  {journal}
  {\bibinfo  {journal} {arXiv e-prints}\ ,\ \bibinfo {eid} {arXiv:1901.10203}}
  (\bibinfo {year} {2019})},\ \Eprint {http://arxiv.org/abs/1901.10203}
  {arXiv:1901.10203 [astro-ph.GA]} \BibitemShut {NoStop}%
\bibitem [{\citenamefont {{Tremmel}}\ \emph {et~al.}(2017)\citenamefont
  {{Tremmel}}, \citenamefont {{Karcher}}, \citenamefont {{Governato}},
  \citenamefont {{Volonteri}}, \citenamefont {{Quinn}}, \citenamefont
  {{Pontzen}}, \citenamefont {{Anderson}},\ and\ \citenamefont
  {{Bellovary}}}]{Tremmel17}%
  \BibitemOpen
  \bibfield  {author} {\bibinfo {author} {\bibfnamefont {M.}~\bibnamefont
  {{Tremmel}}}, \bibinfo {author} {\bibfnamefont {M.}~\bibnamefont
  {{Karcher}}}, \bibinfo {author} {\bibfnamefont {F.}~\bibnamefont
  {{Governato}}}, \bibinfo {author} {\bibfnamefont {M.}~\bibnamefont
  {{Volonteri}}}, \bibinfo {author} {\bibfnamefont {T.~R.}\ \bibnamefont
  {{Quinn}}}, \bibinfo {author} {\bibfnamefont {A.}~\bibnamefont {{Pontzen}}},
  \bibinfo {author} {\bibfnamefont {L.}~\bibnamefont {{Anderson}}}, \ and\
  \bibinfo {author} {\bibfnamefont {J.}~\bibnamefont {{Bellovary}}},\ }\href
  {\doibase 10.1093/mnras/stx1160} {\bibfield  {journal} {\bibinfo  {journal}
  {\mnras}\ }\textbf {\bibinfo {volume} {470}},\ \bibinfo {pages} {1121}
  (\bibinfo {year} {2017})},\ \Eprint {http://arxiv.org/abs/1607.02151}
  {arXiv:1607.02151 [astro-ph.GA]} \BibitemShut {NoStop}%
\bibitem [{\citenamefont {{Hopkins}}(2015)}]{Hopkins15}%
  \BibitemOpen
  \bibfield  {author} {\bibinfo {author} {\bibfnamefont {P.~F.}\ \bibnamefont
  {{Hopkins}}},\ }\href {\doibase 10.1093/mnras/stv195} {\bibfield  {journal}
  {\bibinfo  {journal} {\mnras}\ }\textbf {\bibinfo {volume} {450}},\ \bibinfo
  {pages} {53} (\bibinfo {year} {2015})},\ \Eprint
  {http://arxiv.org/abs/1409.7395} {arXiv:1409.7395} \BibitemShut {NoStop}%
\bibitem [{\citenamefont {{Gabor}}\ and\ \citenamefont
  {{Dav{\'e}}}(2015)}]{Gabor15}%
  \BibitemOpen
  \bibfield  {author} {\bibinfo {author} {\bibfnamefont {J.~M.}\ \bibnamefont
  {{Gabor}}}\ and\ \bibinfo {author} {\bibfnamefont {R.}~\bibnamefont
  {{Dav{\'e}}}},\ }\href {\doibase 10.1093/mnras/stu2399} {\bibfield  {journal}
  {\bibinfo  {journal} {\mnras}\ }\textbf {\bibinfo {volume} {447}},\ \bibinfo
  {pages} {374} (\bibinfo {year} {2015})},\ \Eprint
  {http://arxiv.org/abs/1405.1043} {arXiv:1405.1043} \BibitemShut {NoStop}%
\bibitem [{\citenamefont {{Hopkins}}\ and\ \citenamefont
  {{Quataert}}(2011)}]{Hopkins11}%
  \BibitemOpen
  \bibfield  {author} {\bibinfo {author} {\bibfnamefont {P.~F.}\ \bibnamefont
  {{Hopkins}}}\ and\ \bibinfo {author} {\bibfnamefont {E.}~\bibnamefont
  {{Quataert}}},\ }\href {\doibase 10.1111/j.1365-2966.2011.18542.x} {\bibfield
   {journal} {\bibinfo  {journal} {\mnras}\ }\textbf {\bibinfo {volume}
  {415}},\ \bibinfo {pages} {1027} (\bibinfo {year} {2011})},\ \Eprint
  {http://arxiv.org/abs/1007.2647} {arXiv:1007.2647 [astro-ph.CO]} \BibitemShut
  {NoStop}%
\bibitem [{\citenamefont {{Angl{\'e}s-Alc{\'a}zar}}\ \emph
  {et~al.}(2015)\citenamefont {{Angl{\'e}s-Alc{\'a}zar}}, \citenamefont
  {{{\"O}zel}}, \citenamefont {{Dav{\'e}}}, \citenamefont {{Katz}},
  \citenamefont {{Kollmeier}},\ and\ \citenamefont {{Oppenheimer}}}]{Angles15}%
  \BibitemOpen
  \bibfield  {author} {\bibinfo {author} {\bibfnamefont {D.}~\bibnamefont
  {{Angl{\'e}s-Alc{\'a}zar}}}, \bibinfo {author} {\bibfnamefont
  {F.}~\bibnamefont {{{\"O}zel}}}, \bibinfo {author} {\bibfnamefont
  {R.}~\bibnamefont {{Dav{\'e}}}}, \bibinfo {author} {\bibfnamefont
  {N.}~\bibnamefont {{Katz}}}, \bibinfo {author} {\bibfnamefont {J.~A.}\
  \bibnamefont {{Kollmeier}}}, \ and\ \bibinfo {author} {\bibfnamefont {B.~D.}\
  \bibnamefont {{Oppenheimer}}},\ }\href {\doibase 10.1088/0004-637X/800/2/127}
  {\bibfield  {journal} {\bibinfo  {journal} {\apj}\ }\textbf {\bibinfo
  {volume} {800}},\ \bibinfo {eid} {127} (\bibinfo {year} {2015})},\ \Eprint
  {http://arxiv.org/abs/1309.5963} {arXiv:1309.5963} \BibitemShut {NoStop}%
\bibitem [{\citenamefont {{Menon}}\ \emph {et~al.}(2015)\citenamefont
  {{Menon}}, \citenamefont {{Wesolowski}}, \citenamefont {{Zheng}},
  \citenamefont {{Jetley}}, \citenamefont {{Kale}}, \citenamefont {{Quinn}},\
  and\ \citenamefont {{Governato}}}]{Menon15}%
  \BibitemOpen
  \bibfield  {author} {\bibinfo {author} {\bibfnamefont {H.}~\bibnamefont
  {{Menon}}}, \bibinfo {author} {\bibfnamefont {L.}~\bibnamefont
  {{Wesolowski}}}, \bibinfo {author} {\bibfnamefont {G.}~\bibnamefont
  {{Zheng}}}, \bibinfo {author} {\bibfnamefont {P.}~\bibnamefont {{Jetley}}},
  \bibinfo {author} {\bibfnamefont {L.}~\bibnamefont {{Kale}}}, \bibinfo
  {author} {\bibfnamefont {T.}~\bibnamefont {{Quinn}}}, \ and\ \bibinfo
  {author} {\bibfnamefont {F.}~\bibnamefont {{Governato}}},\ }\href {\doibase
  10.1186/s40668-015-0007-9} {\bibfield  {journal} {\bibinfo  {journal}
  {Computational Astrophysics and Cosmology}\ }\textbf {\bibinfo {volume}
  {2}},\ \bibinfo {eid} {1} (\bibinfo {year} {2015})},\ \Eprint
  {http://arxiv.org/abs/1409.1929} {arXiv:1409.1929 [astro-ph.IM]} \BibitemShut
  {NoStop}%
\bibitem [{\citenamefont {{Dolag}}\ \emph {et~al.}(2016)\citenamefont
  {{Dolag}}, \citenamefont {{Komatsu}},\ and\ \citenamefont
  {{Sunyaev}}}]{Dolag16}%
  \BibitemOpen
  \bibfield  {author} {\bibinfo {author} {\bibfnamefont {K.}~\bibnamefont
  {{Dolag}}}, \bibinfo {author} {\bibfnamefont {E.}~\bibnamefont {{Komatsu}}},
  \ and\ \bibinfo {author} {\bibfnamefont {R.}~\bibnamefont {{Sunyaev}}},\
  }\href {\doibase 10.1093/mnras/stw2035} {\bibfield  {journal} {\bibinfo
  {journal} {\mnras}\ }\textbf {\bibinfo {volume} {463}},\ \bibinfo {pages}
  {1797} (\bibinfo {year} {2016})},\ \Eprint {http://arxiv.org/abs/1509.05134}
  {arXiv:1509.05134} \BibitemShut {NoStop}%
\bibitem [{Boc()}]{Bocquet16}%
  \BibitemOpen
  \href@noop {} {\ }\BibitemShut {NoStop}%
\bibitem [{\citenamefont {{Castro}}\ \emph {et~al.}(2018)\citenamefont
  {{Castro}}, \citenamefont {{Quartin}}, \citenamefont {{Giocoli}},
  \citenamefont {{Borgani}},\ and\ \citenamefont {{Dolag}}}]{Castro18}%
  \BibitemOpen
  \bibfield  {author} {\bibinfo {author} {\bibfnamefont {T.}~\bibnamefont
  {{Castro}}}, \bibinfo {author} {\bibfnamefont {M.}~\bibnamefont {{Quartin}}},
  \bibinfo {author} {\bibfnamefont {C.}~\bibnamefont {{Giocoli}}}, \bibinfo
  {author} {\bibfnamefont {S.}~\bibnamefont {{Borgani}}}, \ and\ \bibinfo
  {author} {\bibfnamefont {K.}~\bibnamefont {{Dolag}}},\ }\href {\doibase
  10.1093/mnras/sty1117} {\bibfield  {journal} {\bibinfo  {journal} {\mnras}\
  }\textbf {\bibinfo {volume} {478}},\ \bibinfo {pages} {1305} (\bibinfo {year}
  {2018})},\ \Eprint {http://arxiv.org/abs/1711.10017} {arXiv:1711.10017}
  \BibitemShut {NoStop}%
\bibitem [{\citenamefont {Lewis}\ \emph {et~al.}(2000)\citenamefont {Lewis},
  \citenamefont {Challinor},\ and\ \citenamefont {Lasenby}}]{camb1}%
  \BibitemOpen
  \bibfield  {author} {\bibinfo {author} {\bibfnamefont {A.}~\bibnamefont
  {Lewis}}, \bibinfo {author} {\bibfnamefont {A.}~\bibnamefont {Challinor}}, \
  and\ \bibinfo {author} {\bibfnamefont {A.}~\bibnamefont {Lasenby}},\ }\href
  {\doibase 10.1086/309179} {\bibfield  {journal} {\bibinfo  {journal} {\apj}\
  }\textbf {\bibinfo {volume} {538}},\ \bibinfo {pages} {473} (\bibinfo {year}
  {2000})},\ \Eprint {http://arxiv.org/abs/astro-ph/9911177}
  {arXiv:astro-ph/9911177 [astro-ph]} \BibitemShut {NoStop}%
\bibitem [{\citenamefont {Howlett}\ \emph {et~al.}(2012)\citenamefont
  {Howlett}, \citenamefont {Lewis}, \citenamefont {Hall},\ and\ \citenamefont
  {Challinor}}]{camb2}%
  \BibitemOpen
  \bibfield  {author} {\bibinfo {author} {\bibfnamefont {C.}~\bibnamefont
  {Howlett}}, \bibinfo {author} {\bibfnamefont {A.}~\bibnamefont {Lewis}},
  \bibinfo {author} {\bibfnamefont {A.}~\bibnamefont {Hall}}, \ and\ \bibinfo
  {author} {\bibfnamefont {A.}~\bibnamefont {Challinor}},\ }\href {\doibase
  10.1088/1475-7516/2012/04/027} {\bibfield  {journal} {\bibinfo  {journal}
  {\jcap}\ }\textbf {\bibinfo {volume} {1204}},\ \bibinfo {pages} {027}
  (\bibinfo {year} {2012})},\ \Eprint {http://arxiv.org/abs/1201.3654}
  {arXiv:1201.3654 [astro-ph.CO]} \BibitemShut {NoStop}%
\bibitem [{\citenamefont {{Blas}}\ \emph {et~al.}(2011)\citenamefont {{Blas}},
  \citenamefont {{Lesgourgues}},\ and\ \citenamefont {{Tram}}}]{CLASS}%
  \BibitemOpen
  \bibfield  {author} {\bibinfo {author} {\bibfnamefont {D.}~\bibnamefont
  {{Blas}}}, \bibinfo {author} {\bibfnamefont {J.}~\bibnamefont
  {{Lesgourgues}}}, \ and\ \bibinfo {author} {\bibfnamefont {T.}~\bibnamefont
  {{Tram}}},\ }\href {\doibase 10.1088/1475-7516/2011/07/034} {\bibfield
  {journal} {\bibinfo  {journal} {Journal of Cosmology and Astro-Particle
  Physics}\ }\textbf {\bibinfo {volume} {2011}},\ \bibinfo {eid} {034}
  (\bibinfo {year} {2011})},\ \Eprint {http://arxiv.org/abs/1104.2933}
  {arXiv:1104.2933 [astro-ph.CO]} \BibitemShut {NoStop}%
\bibitem [{\citenamefont {{Planck Collaboration}}\ \emph
  {et~al.}(2018)\citenamefont {{Planck Collaboration}}, \citenamefont
  {{Aghanim}}, \citenamefont {{Akrami}}, \citenamefont {{Ashdown}},
  \citenamefont {{Aumont}}, \citenamefont {{Baccigalupi}}, \citenamefont
  {{Ballardini}}, \citenamefont {{Banday}}, \citenamefont {{Barreiro}},
  \citenamefont {{Bartolo}}, \citenamefont {{Basak}}, \citenamefont {{Battye}},
  \citenamefont {{Benabed}}, \citenamefont {{Bernard}}, \citenamefont
  {{Bersanelli}}, \citenamefont {{Bielewicz}}, \citenamefont {{Bock}},
  \citenamefont {{Bond}}, \citenamefont {{Borrill}}, \citenamefont {{Bouchet}},
  \citenamefont {{Boulanger}}, \citenamefont {{Bucher}}, \citenamefont
  {{Burigana}}, \citenamefont {{Butler}}, \citenamefont {{Calabrese}},
  \citenamefont {{Cardoso}}, \citenamefont {{Carron}}, \citenamefont
  {{Challinor}}, \citenamefont {{Chiang}}, \citenamefont {{Chluba}},
  \citenamefont {{Colombo}}, \citenamefont {{Combet}}, \citenamefont
  {{Contreras}}, \citenamefont {{Crill}}, \citenamefont {{Cuttaia}},
  \citenamefont {{de Bernardis}}, \citenamefont {{de Zotti}}, \citenamefont
  {{Delabrouille}}, \citenamefont {{Delouis}}, \citenamefont {{Di Valentino}},
  \citenamefont {{Diego}}, \citenamefont {{Dor{\'e}}}, \citenamefont
  {{Douspis}}, \citenamefont {{Ducout}}, \citenamefont {{Dupac}}, \citenamefont
  {{Dusini}}, \citenamefont {{Efstathiou}}, \citenamefont {{Elsner}},
  \citenamefont {{En{\ss}lin}}, \citenamefont {{Eriksen}}, \citenamefont
  {{Fantaye}}, \citenamefont {{Farhang}}, \citenamefont {{Fergusson}},
  \citenamefont {{Fernandez-Cobos}}, \citenamefont {{Finelli}}, \citenamefont
  {{Forastieri}}, \citenamefont {{Frailis}}, \citenamefont {{Franceschi}},
  \citenamefont {{Frolov}}, \citenamefont {{Galeotta}}, \citenamefont
  {{Galli}}, \citenamefont {{Ganga}}, \citenamefont {{G{\'e}nova-Santos}},
  \citenamefont {{Gerbino}}, \citenamefont {{Ghosh}}, \citenamefont
  {{Gonz{\'a}lez-Nuevo}}, \citenamefont {{G{\'o}rski}}, \citenamefont
  {{Gratton}}, \citenamefont {{Gruppuso}}, \citenamefont {{Gudmundsson}},
  \citenamefont {{Hamann}}, \citenamefont {{Handley}}, \citenamefont
  {{Herranz}}, \citenamefont {{Hivon}}, \citenamefont {{Huang}}, \citenamefont
  {{Jaffe}}, \citenamefont {{Jones}}, \citenamefont {{Karakci}}, \citenamefont
  {{Keih{\"a}nen}}, \citenamefont {{Keskitalo}}, \citenamefont {{Kiiveri}},
  \citenamefont {{Kim}}, \citenamefont {{Kisner}}, \citenamefont {{Knox}},
  \citenamefont {{Krachmalnicoff}}, \citenamefont {{Kunz}}, \citenamefont
  {{Kurki-Suonio}}, \citenamefont {{Lagache}}, \citenamefont {{Lamarre}},
  \citenamefont {{Lasenby}}, \citenamefont {{Lattanzi}}, \citenamefont
  {{Lawrence}}, \citenamefont {{Le Jeune}}, \citenamefont {{Lemos}},
  \citenamefont {{Lesgourgues}}, \citenamefont {{Levrier}}, \citenamefont
  {{Lewis}}, \citenamefont {{Liguori}}, \citenamefont {{Lilje}}, \citenamefont
  {{Lilley}}, \citenamefont {{Lindholm}}, \citenamefont {{L{\'o}pez-Caniego}},
  \citenamefont {{Lubin}}, \citenamefont {{Ma}}, \citenamefont
  {{Mac{\'\i}as-P{\'e}rez}}, \citenamefont {{Maggio}}, \citenamefont {{Maino}},
  \citenamefont {{Mandolesi}}, \citenamefont {{Mangilli}}, \citenamefont
  {{Marcos-Caballero}}, \citenamefont {{Maris}}, \citenamefont {{Martin}},
  \citenamefont {{Martinelli}}, \citenamefont {{Mart{\'\i}nez- Gonz{\'a}lez}},
  \citenamefont {{Matarrese}}, \citenamefont {{Mauri}}, \citenamefont
  {{McEwen}}, \citenamefont {{Meinhold}}, \citenamefont {{Melchiorri}},
  \citenamefont {{Mennella}}, \citenamefont {{Migliaccio}}, \citenamefont
  {{Millea}}, \citenamefont {{Mitra}}, \citenamefont {{Miville-Desch{\^e}nes}},
  \citenamefont {{Molinari}}, \citenamefont {{Montier}}, \citenamefont
  {{Morgante}}, \citenamefont {{Moss}}, \citenamefont {{Natoli}}, \citenamefont
  {{N{\o}rgaard-Nielsen}}, \citenamefont {{Pagano}}, \citenamefont
  {{Paoletti}}, \citenamefont {{Partridge}}, \citenamefont {{Patanchon}},
  \citenamefont {{Peiris}}, \citenamefont {{Perrotta}}, \citenamefont
  {{Pettorino}}, \citenamefont {{Piacentini}}, \citenamefont {{Polastri}},
  \citenamefont {{Polenta}}, \citenamefont {{Puget}}, \citenamefont {{Rachen}},
  \citenamefont {{Reinecke}}, \citenamefont {{Remazeilles}}, \citenamefont
  {{Renzi}}, \citenamefont {{Rocha}}, \citenamefont {{Rosset}}, \citenamefont
  {{Roudier}}, \citenamefont {{Rubi{\~n}o-Mart{\'\i}n}}, \citenamefont
  {{Ruiz-Granados}}, \citenamefont {{Salvati}}, \citenamefont {{Sandri}},
  \citenamefont {{Savelainen}}, \citenamefont {{Scott}}, \citenamefont
  {{Shellard}}, \citenamefont {{Sirignano}}, \citenamefont {{Sirri}},
  \citenamefont {{Spencer}}, \citenamefont {{Sunyaev}}, \citenamefont
  {{Suur-Uski}}, \citenamefont {{Tauber}}, \citenamefont {{Tavagnacco}},
  \citenamefont {{Tenti}}, \citenamefont {{Toffolatti}}, \citenamefont
  {{Tomasi}}, \citenamefont {{Trombetti}}, \citenamefont {{Valenziano}},
  \citenamefont {{Valiviita}}, \citenamefont {{Van Tent}}, \citenamefont
  {{Vibert}}, \citenamefont {{Vielva}}, \citenamefont {{Villa}}, \citenamefont
  {{Vittorio}}, \citenamefont {{Wandelt}}, \citenamefont {{Wehus}},
  \citenamefont {{White}}, \citenamefont {{White}}, \citenamefont {{Zacchei}},\
  and\ \citenamefont {{Zonca}}}]{Planck18}%
  \BibitemOpen
  \bibfield  {author} {\bibinfo {author} {\bibnamefont {{Planck
  Collaboration}}}, \bibinfo {author} {\bibfnamefont {N.}~\bibnamefont
  {{Aghanim}}}, \bibinfo {author} {\bibfnamefont {Y.}~\bibnamefont {{Akrami}}},
  \bibinfo {author} {\bibfnamefont {M.}~\bibnamefont {{Ashdown}}}, \bibinfo
  {author} {\bibfnamefont {J.}~\bibnamefont {{Aumont}}}, \bibinfo {author}
  {\bibfnamefont {C.}~\bibnamefont {{Baccigalupi}}}, \bibinfo {author}
  {\bibfnamefont {M.}~\bibnamefont {{Ballardini}}}, \bibinfo {author}
  {\bibfnamefont {A.~J.}\ \bibnamefont {{Banday}}}, \bibinfo {author}
  {\bibfnamefont {R.~B.}\ \bibnamefont {{Barreiro}}}, \bibinfo {author}
  {\bibfnamefont {N.}~\bibnamefont {{Bartolo}}}, \bibinfo {author}
  {\bibfnamefont {S.}~\bibnamefont {{Basak}}}, \bibinfo {author} {\bibfnamefont
  {R.}~\bibnamefont {{Battye}}}, \bibinfo {author} {\bibfnamefont
  {K.}~\bibnamefont {{Benabed}}}, \bibinfo {author} {\bibfnamefont {J.~P.}\
  \bibnamefont {{Bernard}}}, \bibinfo {author} {\bibfnamefont {M.}~\bibnamefont
  {{Bersanelli}}}, \bibinfo {author} {\bibfnamefont {P.}~\bibnamefont
  {{Bielewicz}}}, \bibinfo {author} {\bibfnamefont {J.~J.}\ \bibnamefont
  {{Bock}}}, \bibinfo {author} {\bibfnamefont {J.~R.}\ \bibnamefont {{Bond}}},
  \bibinfo {author} {\bibfnamefont {J.}~\bibnamefont {{Borrill}}}, \bibinfo
  {author} {\bibfnamefont {F.~R.}\ \bibnamefont {{Bouchet}}}, \bibinfo {author}
  {\bibfnamefont {F.}~\bibnamefont {{Boulanger}}}, \bibinfo {author}
  {\bibfnamefont {M.}~\bibnamefont {{Bucher}}}, \bibinfo {author}
  {\bibfnamefont {C.}~\bibnamefont {{Burigana}}}, \bibinfo {author}
  {\bibfnamefont {R.~C.}\ \bibnamefont {{Butler}}}, \bibinfo {author}
  {\bibfnamefont {E.}~\bibnamefont {{Calabrese}}}, \bibinfo {author}
  {\bibfnamefont {J.~F.}\ \bibnamefont {{Cardoso}}}, \bibinfo {author}
  {\bibfnamefont {J.}~\bibnamefont {{Carron}}}, \bibinfo {author}
  {\bibfnamefont {A.}~\bibnamefont {{Challinor}}}, \bibinfo {author}
  {\bibfnamefont {H.~C.}\ \bibnamefont {{Chiang}}}, \bibinfo {author}
  {\bibfnamefont {J.}~\bibnamefont {{Chluba}}}, \bibinfo {author}
  {\bibfnamefont {L.~P.~L.}\ \bibnamefont {{Colombo}}}, \bibinfo {author}
  {\bibfnamefont {C.}~\bibnamefont {{Combet}}}, \bibinfo {author}
  {\bibfnamefont {D.}~\bibnamefont {{Contreras}}}, \bibinfo {author}
  {\bibfnamefont {B.~P.}\ \bibnamefont {{Crill}}}, \bibinfo {author}
  {\bibfnamefont {F.}~\bibnamefont {{Cuttaia}}}, \bibinfo {author}
  {\bibfnamefont {P.}~\bibnamefont {{de Bernardis}}}, \bibinfo {author}
  {\bibfnamefont {G.}~\bibnamefont {{de Zotti}}}, \bibinfo {author}
  {\bibfnamefont {J.}~\bibnamefont {{Delabrouille}}}, \bibinfo {author}
  {\bibfnamefont {J.~M.}\ \bibnamefont {{Delouis}}}, \bibinfo {author}
  {\bibfnamefont {E.}~\bibnamefont {{Di Valentino}}}, \bibinfo {author}
  {\bibfnamefont {J.~M.}\ \bibnamefont {{Diego}}}, \bibinfo {author}
  {\bibfnamefont {O.}~\bibnamefont {{Dor{\'e}}}}, \bibinfo {author}
  {\bibfnamefont {M.}~\bibnamefont {{Douspis}}}, \bibinfo {author}
  {\bibfnamefont {A.}~\bibnamefont {{Ducout}}}, \bibinfo {author}
  {\bibfnamefont {X.}~\bibnamefont {{Dupac}}}, \bibinfo {author} {\bibfnamefont
  {S.}~\bibnamefont {{Dusini}}}, \bibinfo {author} {\bibfnamefont
  {G.}~\bibnamefont {{Efstathiou}}}, \bibinfo {author} {\bibfnamefont
  {F.}~\bibnamefont {{Elsner}}}, \bibinfo {author} {\bibfnamefont {T.~A.}\
  \bibnamefont {{En{\ss}lin}}}, \bibinfo {author} {\bibfnamefont {H.~K.}\
  \bibnamefont {{Eriksen}}}, \bibinfo {author} {\bibfnamefont {Y.}~\bibnamefont
  {{Fantaye}}}, \bibinfo {author} {\bibfnamefont {M.}~\bibnamefont
  {{Farhang}}}, \bibinfo {author} {\bibfnamefont {J.}~\bibnamefont
  {{Fergusson}}}, \bibinfo {author} {\bibfnamefont {R.}~\bibnamefont
  {{Fernandez-Cobos}}}, \bibinfo {author} {\bibfnamefont {F.}~\bibnamefont
  {{Finelli}}}, \bibinfo {author} {\bibfnamefont {F.}~\bibnamefont
  {{Forastieri}}}, \bibinfo {author} {\bibfnamefont {M.}~\bibnamefont
  {{Frailis}}}, \bibinfo {author} {\bibfnamefont {E.}~\bibnamefont
  {{Franceschi}}}, \bibinfo {author} {\bibfnamefont {A.}~\bibnamefont
  {{Frolov}}}, \bibinfo {author} {\bibfnamefont {S.}~\bibnamefont
  {{Galeotta}}}, \bibinfo {author} {\bibfnamefont {S.}~\bibnamefont {{Galli}}},
  \bibinfo {author} {\bibfnamefont {K.}~\bibnamefont {{Ganga}}}, \bibinfo
  {author} {\bibfnamefont {R.~T.}\ \bibnamefont {{G{\'e}nova-Santos}}},
  \bibinfo {author} {\bibfnamefont {M.}~\bibnamefont {{Gerbino}}}, \bibinfo
  {author} {\bibfnamefont {T.}~\bibnamefont {{Ghosh}}}, \bibinfo {author}
  {\bibfnamefont {J.}~\bibnamefont {{Gonz{\'a}lez-Nuevo}}}, \bibinfo {author}
  {\bibfnamefont {K.~M.}\ \bibnamefont {{G{\'o}rski}}}, \bibinfo {author}
  {\bibfnamefont {S.}~\bibnamefont {{Gratton}}}, \bibinfo {author}
  {\bibfnamefont {A.}~\bibnamefont {{Gruppuso}}}, \bibinfo {author}
  {\bibfnamefont {J.~E.}\ \bibnamefont {{Gudmundsson}}}, \bibinfo {author}
  {\bibfnamefont {J.}~\bibnamefont {{Hamann}}}, \bibinfo {author}
  {\bibfnamefont {W.}~\bibnamefont {{Handley}}}, \bibinfo {author}
  {\bibfnamefont {D.}~\bibnamefont {{Herranz}}}, \bibinfo {author}
  {\bibfnamefont {E.}~\bibnamefont {{Hivon}}}, \bibinfo {author} {\bibfnamefont
  {Z.}~\bibnamefont {{Huang}}}, \bibinfo {author} {\bibfnamefont {A.~H.}\
  \bibnamefont {{Jaffe}}}, \bibinfo {author} {\bibfnamefont {W.~C.}\
  \bibnamefont {{Jones}}}, \bibinfo {author} {\bibfnamefont {A.}~\bibnamefont
  {{Karakci}}}, \bibinfo {author} {\bibfnamefont {E.}~\bibnamefont
  {{Keih{\"a}nen}}}, \bibinfo {author} {\bibfnamefont {R.}~\bibnamefont
  {{Keskitalo}}}, \bibinfo {author} {\bibfnamefont {K.}~\bibnamefont
  {{Kiiveri}}}, \bibinfo {author} {\bibfnamefont {J.}~\bibnamefont {{Kim}}},
  \bibinfo {author} {\bibfnamefont {T.~S.}\ \bibnamefont {{Kisner}}}, \bibinfo
  {author} {\bibfnamefont {L.}~\bibnamefont {{Knox}}}, \bibinfo {author}
  {\bibfnamefont {N.}~\bibnamefont {{Krachmalnicoff}}}, \bibinfo {author}
  {\bibfnamefont {M.}~\bibnamefont {{Kunz}}}, \bibinfo {author} {\bibfnamefont
  {H.}~\bibnamefont {{Kurki-Suonio}}}, \bibinfo {author} {\bibfnamefont
  {G.}~\bibnamefont {{Lagache}}}, \bibinfo {author} {\bibfnamefont {J.~M.}\
  \bibnamefont {{Lamarre}}}, \bibinfo {author} {\bibfnamefont {A.}~\bibnamefont
  {{Lasenby}}}, \bibinfo {author} {\bibfnamefont {M.}~\bibnamefont
  {{Lattanzi}}}, \bibinfo {author} {\bibfnamefont {C.~R.}\ \bibnamefont
  {{Lawrence}}}, \bibinfo {author} {\bibfnamefont {M.}~\bibnamefont {{Le
  Jeune}}}, \bibinfo {author} {\bibfnamefont {P.}~\bibnamefont {{Lemos}}},
  \bibinfo {author} {\bibfnamefont {J.}~\bibnamefont {{Lesgourgues}}}, \bibinfo
  {author} {\bibfnamefont {F.}~\bibnamefont {{Levrier}}}, \bibinfo {author}
  {\bibfnamefont {A.}~\bibnamefont {{Lewis}}}, \bibinfo {author} {\bibfnamefont
  {M.}~\bibnamefont {{Liguori}}}, \bibinfo {author} {\bibfnamefont {P.~B.}\
  \bibnamefont {{Lilje}}}, \bibinfo {author} {\bibfnamefont {M.}~\bibnamefont
  {{Lilley}}}, \bibinfo {author} {\bibfnamefont {V.}~\bibnamefont
  {{Lindholm}}}, \bibinfo {author} {\bibfnamefont {M.}~\bibnamefont
  {{L{\'o}pez-Caniego}}}, \bibinfo {author} {\bibfnamefont {P.~M.}\
  \bibnamefont {{Lubin}}}, \bibinfo {author} {\bibfnamefont {Y.~Z.}\
  \bibnamefont {{Ma}}}, \bibinfo {author} {\bibfnamefont {J.~F.}\ \bibnamefont
  {{Mac{\'\i}as-P{\'e}rez}}}, \bibinfo {author} {\bibfnamefont
  {G.}~\bibnamefont {{Maggio}}}, \bibinfo {author} {\bibfnamefont
  {D.}~\bibnamefont {{Maino}}}, \bibinfo {author} {\bibfnamefont
  {N.}~\bibnamefont {{Mandolesi}}}, \bibinfo {author} {\bibfnamefont
  {A.}~\bibnamefont {{Mangilli}}}, \bibinfo {author} {\bibfnamefont
  {A.}~\bibnamefont {{Marcos-Caballero}}}, \bibinfo {author} {\bibfnamefont
  {M.}~\bibnamefont {{Maris}}}, \bibinfo {author} {\bibfnamefont {P.~G.}\
  \bibnamefont {{Martin}}}, \bibinfo {author} {\bibfnamefont {M.}~\bibnamefont
  {{Martinelli}}}, \bibinfo {author} {\bibfnamefont {E.}~\bibnamefont
  {{Mart{\'\i}nez- Gonz{\'a}lez}}}, \bibinfo {author} {\bibfnamefont
  {S.}~\bibnamefont {{Matarrese}}}, \bibinfo {author} {\bibfnamefont
  {N.}~\bibnamefont {{Mauri}}}, \bibinfo {author} {\bibfnamefont {J.~D.}\
  \bibnamefont {{McEwen}}}, \bibinfo {author} {\bibfnamefont {P.~R.}\
  \bibnamefont {{Meinhold}}}, \bibinfo {author} {\bibfnamefont
  {A.}~\bibnamefont {{Melchiorri}}}, \bibinfo {author} {\bibfnamefont
  {A.}~\bibnamefont {{Mennella}}}, \bibinfo {author} {\bibfnamefont
  {M.}~\bibnamefont {{Migliaccio}}}, \bibinfo {author} {\bibfnamefont
  {M.}~\bibnamefont {{Millea}}}, \bibinfo {author} {\bibfnamefont
  {S.}~\bibnamefont {{Mitra}}}, \bibinfo {author} {\bibfnamefont {M.~A.}\
  \bibnamefont {{Miville-Desch{\^e}nes}}}, \bibinfo {author} {\bibfnamefont
  {D.}~\bibnamefont {{Molinari}}}, \bibinfo {author} {\bibfnamefont
  {L.}~\bibnamefont {{Montier}}}, \bibinfo {author} {\bibfnamefont
  {G.}~\bibnamefont {{Morgante}}}, \bibinfo {author} {\bibfnamefont
  {A.}~\bibnamefont {{Moss}}}, \bibinfo {author} {\bibfnamefont
  {P.}~\bibnamefont {{Natoli}}}, \bibinfo {author} {\bibfnamefont {H.~U.}\
  \bibnamefont {{N{\o}rgaard-Nielsen}}}, \bibinfo {author} {\bibfnamefont
  {L.}~\bibnamefont {{Pagano}}}, \bibinfo {author} {\bibfnamefont
  {D.}~\bibnamefont {{Paoletti}}}, \bibinfo {author} {\bibfnamefont
  {B.}~\bibnamefont {{Partridge}}}, \bibinfo {author} {\bibfnamefont
  {G.}~\bibnamefont {{Patanchon}}}, \bibinfo {author} {\bibfnamefont {H.~V.}\
  \bibnamefont {{Peiris}}}, \bibinfo {author} {\bibfnamefont {F.}~\bibnamefont
  {{Perrotta}}}, \bibinfo {author} {\bibfnamefont {V.}~\bibnamefont
  {{Pettorino}}}, \bibinfo {author} {\bibfnamefont {F.}~\bibnamefont
  {{Piacentini}}}, \bibinfo {author} {\bibfnamefont {L.}~\bibnamefont
  {{Polastri}}}, \bibinfo {author} {\bibfnamefont {G.}~\bibnamefont
  {{Polenta}}}, \bibinfo {author} {\bibfnamefont {J.~L.}\ \bibnamefont
  {{Puget}}}, \bibinfo {author} {\bibfnamefont {J.~P.}\ \bibnamefont
  {{Rachen}}}, \bibinfo {author} {\bibfnamefont {M.}~\bibnamefont
  {{Reinecke}}}, \bibinfo {author} {\bibfnamefont {M.}~\bibnamefont
  {{Remazeilles}}}, \bibinfo {author} {\bibfnamefont {A.}~\bibnamefont
  {{Renzi}}}, \bibinfo {author} {\bibfnamefont {G.}~\bibnamefont {{Rocha}}},
  \bibinfo {author} {\bibfnamefont {C.}~\bibnamefont {{Rosset}}}, \bibinfo
  {author} {\bibfnamefont {G.}~\bibnamefont {{Roudier}}}, \bibinfo {author}
  {\bibfnamefont {J.~A.}\ \bibnamefont {{Rubi{\~n}o-Mart{\'\i}n}}}, \bibinfo
  {author} {\bibfnamefont {B.}~\bibnamefont {{Ruiz-Granados}}}, \bibinfo
  {author} {\bibfnamefont {L.}~\bibnamefont {{Salvati}}}, \bibinfo {author}
  {\bibfnamefont {M.}~\bibnamefont {{Sandri}}}, \bibinfo {author}
  {\bibfnamefont {M.}~\bibnamefont {{Savelainen}}}, \bibinfo {author}
  {\bibfnamefont {D.}~\bibnamefont {{Scott}}}, \bibinfo {author} {\bibfnamefont
  {E.~P.~S.}\ \bibnamefont {{Shellard}}}, \bibinfo {author} {\bibfnamefont
  {C.}~\bibnamefont {{Sirignano}}}, \bibinfo {author} {\bibfnamefont
  {G.}~\bibnamefont {{Sirri}}}, \bibinfo {author} {\bibfnamefont {L.~D.}\
  \bibnamefont {{Spencer}}}, \bibinfo {author} {\bibfnamefont {R.}~\bibnamefont
  {{Sunyaev}}}, \bibinfo {author} {\bibfnamefont {A.~S.}\ \bibnamefont
  {{Suur-Uski}}}, \bibinfo {author} {\bibfnamefont {J.~A.}\ \bibnamefont
  {{Tauber}}}, \bibinfo {author} {\bibfnamefont {D.}~\bibnamefont
  {{Tavagnacco}}}, \bibinfo {author} {\bibfnamefont {M.}~\bibnamefont
  {{Tenti}}}, \bibinfo {author} {\bibfnamefont {L.}~\bibnamefont
  {{Toffolatti}}}, \bibinfo {author} {\bibfnamefont {M.}~\bibnamefont
  {{Tomasi}}}, \bibinfo {author} {\bibfnamefont {T.}~\bibnamefont
  {{Trombetti}}}, \bibinfo {author} {\bibfnamefont {L.}~\bibnamefont
  {{Valenziano}}}, \bibinfo {author} {\bibfnamefont {J.}~\bibnamefont
  {{Valiviita}}}, \bibinfo {author} {\bibfnamefont {B.}~\bibnamefont {{Van
  Tent}}}, \bibinfo {author} {\bibfnamefont {L.}~\bibnamefont {{Vibert}}},
  \bibinfo {author} {\bibfnamefont {P.}~\bibnamefont {{Vielva}}}, \bibinfo
  {author} {\bibfnamefont {F.}~\bibnamefont {{Villa}}}, \bibinfo {author}
  {\bibfnamefont {N.}~\bibnamefont {{Vittorio}}}, \bibinfo {author}
  {\bibfnamefont {B.~D.}\ \bibnamefont {{Wandelt}}}, \bibinfo {author}
  {\bibfnamefont {I.~K.}\ \bibnamefont {{Wehus}}}, \bibinfo {author}
  {\bibfnamefont {M.}~\bibnamefont {{White}}}, \bibinfo {author} {\bibfnamefont
  {S.~D.~M.}\ \bibnamefont {{White}}}, \bibinfo {author} {\bibfnamefont
  {A.}~\bibnamefont {{Zacchei}}}, \ and\ \bibinfo {author} {\bibfnamefont
  {A.}~\bibnamefont {{Zonca}}},\ }\href@noop {} {\bibfield  {journal} {\bibinfo
   {journal} {arXiv e-prints}\ ,\ \bibinfo {eid} {arXiv:1807.06209}} (\bibinfo
  {year} {2018})},\ \Eprint {http://arxiv.org/abs/1807.06209} {arXiv:1807.06209
  [astro-ph.CO]} \BibitemShut {NoStop}%
\bibitem [{\citenamefont {{Angulo}}\ \emph {et~al.}(2013)\citenamefont
  {{Angulo}}, \citenamefont {{Hahn}},\ and\ \citenamefont {{Abel}}}]{Angulo13}%
  \BibitemOpen
  \bibfield  {author} {\bibinfo {author} {\bibfnamefont {R.~E.}\ \bibnamefont
  {{Angulo}}}, \bibinfo {author} {\bibfnamefont {O.}~\bibnamefont {{Hahn}}}, \
  and\ \bibinfo {author} {\bibfnamefont {T.}~\bibnamefont {{Abel}}},\ }\href
  {\doibase 10.1093/mnras/stt1135} {\bibfield  {journal} {\bibinfo  {journal}
  {\mnras}\ }\textbf {\bibinfo {volume} {434}},\ \bibinfo {pages} {1756}
  (\bibinfo {year} {2013})},\ \Eprint {http://arxiv.org/abs/1301.7426}
  {arXiv:1301.7426 [astro-ph.CO]} \BibitemShut {NoStop}%
\bibitem [{\citenamefont {{Valkenburg}}\ and\ \citenamefont
  {{Villaescusa-Navarro}}(2017)}]{Valkenburg17}%
  \BibitemOpen
  \bibfield  {author} {\bibinfo {author} {\bibfnamefont {W.}~\bibnamefont
  {{Valkenburg}}}\ and\ \bibinfo {author} {\bibfnamefont {F.}~\bibnamefont
  {{Villaescusa-Navarro}}},\ }\href {\doibase 10.1093/mnras/stx376} {\bibfield
  {journal} {\bibinfo  {journal} {\mnras}\ }\textbf {\bibinfo {volume} {467}},\
  \bibinfo {pages} {4401} (\bibinfo {year} {2017})},\ \Eprint
  {http://arxiv.org/abs/1610.08501} {arXiv:1610.08501 [astro-ph.CO]}
  \BibitemShut {NoStop}%
\bibitem [{\citenamefont {{Schneider}}\ \emph {et~al.}(2016)\citenamefont
  {{Schneider}}, \citenamefont {{Teyssier}}, \citenamefont {{Potter}},
  \citenamefont {{Stadel}}, \citenamefont {{Onions}}, \citenamefont {{Reed}},
  \citenamefont {{Smith}}, \citenamefont {{Springel}}, \citenamefont
  {{Pearce}},\ and\ \citenamefont {{Scoccimarro}}}]{Schneider16}%
  \BibitemOpen
  \bibfield  {author} {\bibinfo {author} {\bibfnamefont {A.}~\bibnamefont
  {{Schneider}}}, \bibinfo {author} {\bibfnamefont {R.}~\bibnamefont
  {{Teyssier}}}, \bibinfo {author} {\bibfnamefont {D.}~\bibnamefont
  {{Potter}}}, \bibinfo {author} {\bibfnamefont {J.}~\bibnamefont {{Stadel}}},
  \bibinfo {author} {\bibfnamefont {J.}~\bibnamefont {{Onions}}}, \bibinfo
  {author} {\bibfnamefont {D.~S.}\ \bibnamefont {{Reed}}}, \bibinfo {author}
  {\bibfnamefont {R.~E.}\ \bibnamefont {{Smith}}}, \bibinfo {author}
  {\bibfnamefont {V.}~\bibnamefont {{Springel}}}, \bibinfo {author}
  {\bibfnamefont {F.~R.}\ \bibnamefont {{Pearce}}}, \ and\ \bibinfo {author}
  {\bibfnamefont {R.}~\bibnamefont {{Scoccimarro}}},\ }\href {\doibase
  10.1088/1475-7516/2016/04/047} {\bibfield  {journal} {\bibinfo  {journal}
  {Journal of Cosmology and Astro-Particle Physics}\ }\textbf {\bibinfo
  {volume} {2016}},\ \bibinfo {eid} {047} (\bibinfo {year} {2016})},\ \Eprint
  {http://arxiv.org/abs/1503.05920} {arXiv:1503.05920 [astro-ph.CO]}
  \BibitemShut {NoStop}%
\bibitem [{\citenamefont {{Kim}}\ \emph {et~al.}(1996)\citenamefont {{Kim}},
  \citenamefont {{Olinto}},\ and\ \citenamefont {{Rosner}}}]{Kim96}%
  \BibitemOpen
  \bibfield  {author} {\bibinfo {author} {\bibfnamefont {E.-J.}\ \bibnamefont
  {{Kim}}}, \bibinfo {author} {\bibfnamefont {A.~V.}\ \bibnamefont {{Olinto}}},
  \ and\ \bibinfo {author} {\bibfnamefont {R.}~\bibnamefont {{Rosner}}},\
  }\href {\doibase 10.1086/177667} {\bibfield  {journal} {\bibinfo  {journal}
  {\apj}\ }\textbf {\bibinfo {volume} {468}},\ \bibinfo {pages} {28} (\bibinfo
  {year} {1996})},\ \Eprint {http://arxiv.org/abs/astro-ph/9412070}
  {astro-ph/9412070} \BibitemShut {NoStop}%
\bibitem [{\citenamefont {{Tsagas}}\ and\ \citenamefont
  {{Maartens}}(2000)}]{Tsagas00}%
  \BibitemOpen
  \bibfield  {author} {\bibinfo {author} {\bibfnamefont {C.~G.}\ \bibnamefont
  {{Tsagas}}}\ and\ \bibinfo {author} {\bibfnamefont {R.}~\bibnamefont
  {{Maartens}}},\ }\href {\doibase 10.1103/PhysRevD.61.083519} {\bibfield
  {journal} {\bibinfo  {journal} {\prd}\ }\textbf {\bibinfo {volume} {61}},\
  \bibinfo {eid} {083519} (\bibinfo {year} {2000})},\ \Eprint
  {http://arxiv.org/abs/astro-ph/9904390} {astro-ph/9904390} \BibitemShut
  {NoStop}%
\bibitem [{\citenamefont {{Kandus}}\ \emph {et~al.}(2011)\citenamefont
  {{Kandus}}, \citenamefont {{Kunze}},\ and\ \citenamefont
  {{Tsagas}}}]{Kandus11}%
  \BibitemOpen
  \bibfield  {author} {\bibinfo {author} {\bibfnamefont {A.}~\bibnamefont
  {{Kandus}}}, \bibinfo {author} {\bibfnamefont {K.~E.}\ \bibnamefont
  {{Kunze}}}, \ and\ \bibinfo {author} {\bibfnamefont {C.~G.}\ \bibnamefont
  {{Tsagas}}},\ }\href {\doibase 10.1016/j.physrep.2011.03.001} {\bibfield
  {journal} {\bibinfo  {journal} {\physrep}\ }\textbf {\bibinfo {volume}
  {505}},\ \bibinfo {pages} {1} (\bibinfo {year} {2011})},\ \Eprint
  {http://arxiv.org/abs/1007.3891} {arXiv:1007.3891} \BibitemShut {NoStop}%
\bibitem [{\citenamefont {{Planck Collaboration}}\ \emph
  {et~al.}(2016{\natexlab{c}})\citenamefont {{Planck Collaboration}},
  \citenamefont {{Ade}}, \citenamefont {{Aghanim}}, \citenamefont {{Arnaud}},
  \citenamefont {{Arroja}}, \citenamefont {{Ashdown}}, \citenamefont
  {{Aumont}}, \citenamefont {{Baccigalupi}}, \citenamefont {{Ballardini}},
  \citenamefont {{Banday}},\ and\ \citenamefont {et~al.}}]{Planck16}%
  \BibitemOpen
  \bibfield  {author} {\bibinfo {author} {\bibnamefont {{Planck
  Collaboration}}}, \bibinfo {author} {\bibfnamefont {P.~A.~R.}\ \bibnamefont
  {{Ade}}}, \bibinfo {author} {\bibfnamefont {N.}~\bibnamefont {{Aghanim}}},
  \bibinfo {author} {\bibfnamefont {M.}~\bibnamefont {{Arnaud}}}, \bibinfo
  {author} {\bibfnamefont {F.}~\bibnamefont {{Arroja}}}, \bibinfo {author}
  {\bibfnamefont {M.}~\bibnamefont {{Ashdown}}}, \bibinfo {author}
  {\bibfnamefont {J.}~\bibnamefont {{Aumont}}}, \bibinfo {author}
  {\bibfnamefont {C.}~\bibnamefont {{Baccigalupi}}}, \bibinfo {author}
  {\bibfnamefont {M.}~\bibnamefont {{Ballardini}}}, \bibinfo {author}
  {\bibfnamefont {A.~J.}\ \bibnamefont {{Banday}}}, \ and\ \bibinfo {author}
  {\bibnamefont {et~al.}},\ }\href {\doibase 10.1051/0004-6361/201525821}
  {\bibfield  {journal} {\bibinfo  {journal} {\aap}\ }\textbf {\bibinfo
  {volume} {594}},\ \bibinfo {eid} {A19} (\bibinfo {year}
  {2016}{\natexlab{c}})},\ \Eprint {http://arxiv.org/abs/1502.01594}
  {arXiv:1502.01594} \BibitemShut {NoStop}%
\bibitem [{\citenamefont {{Ryu}}\ \emph {et~al.}(2012)\citenamefont {{Ryu}},
  \citenamefont {{Schleicher}}, \citenamefont {{Treumann}}, \citenamefont
  {{Tsagas}},\ and\ \citenamefont {{Widrow}}}]{Ryu12}%
  \BibitemOpen
  \bibfield  {author} {\bibinfo {author} {\bibfnamefont {D.}~\bibnamefont
  {{Ryu}}}, \bibinfo {author} {\bibfnamefont {D.~R.~G.}\ \bibnamefont
  {{Schleicher}}}, \bibinfo {author} {\bibfnamefont {R.~A.}\ \bibnamefont
  {{Treumann}}}, \bibinfo {author} {\bibfnamefont {C.~G.}\ \bibnamefont
  {{Tsagas}}}, \ and\ \bibinfo {author} {\bibfnamefont {L.~M.}\ \bibnamefont
  {{Widrow}}},\ }\href {\doibase 10.1007/s11214-011-9839-z} {\bibfield
  {journal} {\bibinfo  {journal} {\ssr}\ }\textbf {\bibinfo {volume} {166}},\
  \bibinfo {pages} {1} (\bibinfo {year} {2012})},\ \Eprint
  {http://arxiv.org/abs/1109.4055} {arXiv:1109.4055 [astro-ph.CO]} \BibitemShut
  {NoStop}%
\bibitem [{\citenamefont {{Vazza}}\ \emph {et~al.}(2014)\citenamefont
  {{Vazza}}, \citenamefont {{Gheller}},\ and\ \citenamefont
  {{Br{\"u}ggen}}}]{Vazza14}%
  \BibitemOpen
  \bibfield  {author} {\bibinfo {author} {\bibfnamefont {F.}~\bibnamefont
  {{Vazza}}}, \bibinfo {author} {\bibfnamefont {C.}~\bibnamefont {{Gheller}}},
  \ and\ \bibinfo {author} {\bibfnamefont {M.}~\bibnamefont {{Br{\"u}ggen}}},\
  }\href {\doibase 10.1093/mnras/stu126} {\bibfield  {journal} {\bibinfo
  {journal} {\mnras}\ }\textbf {\bibinfo {volume} {439}},\ \bibinfo {pages}
  {2662} (\bibinfo {year} {2014})},\ \Eprint {http://arxiv.org/abs/1401.4454}
  {arXiv:1401.4454 [astro-ph.CO]} \BibitemShut {NoStop}%
\bibitem [{\citenamefont {Marinacci}\ and\ \citenamefont
  {Vogelsberger}(2016)}]{Marinacci16}%
  \BibitemOpen
  \bibfield  {author} {\bibinfo {author} {\bibfnamefont {F.}~\bibnamefont
  {Marinacci}}\ and\ \bibinfo {author} {\bibfnamefont {M.}~\bibnamefont
  {Vogelsberger}},\ }\href {\doibase 10.1093/mnrasl/slv176} {\bibfield
  {journal} {\bibinfo  {journal} {Monthly Notices of the Royal Astronomical
  Society: Letters}\ }\textbf {\bibinfo {volume} {456}},\ \bibinfo {pages}
  {L69} (\bibinfo {year} {2016})}\BibitemShut {NoStop}%
\bibitem [{\citenamefont {{Salem}}\ and\ \citenamefont
  {{Bryan}}(2014)}]{Salem14}%
  \BibitemOpen
  \bibfield  {author} {\bibinfo {author} {\bibfnamefont {M.}~\bibnamefont
  {{Salem}}}\ and\ \bibinfo {author} {\bibfnamefont {G.~L.}\ \bibnamefont
  {{Bryan}}},\ }\href {\doibase 10.1093/mnras/stt2121} {\bibfield  {journal}
  {\bibinfo  {journal} {\mnras}\ }\textbf {\bibinfo {volume} {437}},\ \bibinfo
  {pages} {3312} (\bibinfo {year} {2014})},\ \Eprint
  {http://arxiv.org/abs/1307.6215} {arXiv:1307.6215} \BibitemShut {NoStop}%
\bibitem [{\citenamefont {{Jacob}}\ \emph {et~al.}(2018)\citenamefont
  {{Jacob}}, \citenamefont {{Pakmor}}, \citenamefont {{Simpson}}, \citenamefont
  {{Springel}},\ and\ \citenamefont {{Pfrommer}}}]{Jacob18}%
  \BibitemOpen
  \bibfield  {author} {\bibinfo {author} {\bibfnamefont {S.}~\bibnamefont
  {{Jacob}}}, \bibinfo {author} {\bibfnamefont {R.}~\bibnamefont {{Pakmor}}},
  \bibinfo {author} {\bibfnamefont {C.~M.}\ \bibnamefont {{Simpson}}}, \bibinfo
  {author} {\bibfnamefont {V.}~\bibnamefont {{Springel}}}, \ and\ \bibinfo
  {author} {\bibfnamefont {C.}~\bibnamefont {{Pfrommer}}},\ }\href {\doibase
  10.1093/mnras/stx3221} {\bibfield  {journal} {\bibinfo  {journal} {\mnras}\
  }\textbf {\bibinfo {volume} {475}},\ \bibinfo {pages} {570} (\bibinfo {year}
  {2018})},\ \Eprint {http://arxiv.org/abs/1712.04947} {arXiv:1712.04947
  [astro-ph.GA]} \BibitemShut {NoStop}%
\bibitem [{\citenamefont {{Dubois}}\ \emph {et~al.}(2012)\citenamefont
  {{Dubois}}, \citenamefont {{Devriendt}}, \citenamefont {{Slyz}},\ and\
  \citenamefont {{Teyssier}}}]{Dubois12}%
  \BibitemOpen
  \bibfield  {author} {\bibinfo {author} {\bibfnamefont {Y.}~\bibnamefont
  {{Dubois}}}, \bibinfo {author} {\bibfnamefont {J.}~\bibnamefont
  {{Devriendt}}}, \bibinfo {author} {\bibfnamefont {A.}~\bibnamefont {{Slyz}}},
  \ and\ \bibinfo {author} {\bibfnamefont {R.}~\bibnamefont {{Teyssier}}},\
  }\href {\doibase 10.1111/j.1365-2966.2011.20236.x} {\bibfield  {journal}
  {\bibinfo  {journal} {\mnras}\ }\textbf {\bibinfo {volume} {420}},\ \bibinfo
  {pages} {2662} (\bibinfo {year} {2012})},\ \Eprint
  {http://arxiv.org/abs/1108.0110} {arXiv:1108.0110 [astro-ph.CO]} \BibitemShut
  {NoStop}%
\bibitem [{\citenamefont {{Booth}}\ and\ \citenamefont
  {{Schaye}}(2009)}]{Booth09}%
  \BibitemOpen
  \bibfield  {author} {\bibinfo {author} {\bibfnamefont {C.~M.}\ \bibnamefont
  {{Booth}}}\ and\ \bibinfo {author} {\bibfnamefont {J.}~\bibnamefont
  {{Schaye}}},\ }\href {\doibase 10.1111/j.1365-2966.2009.15043.x} {\bibfield
  {journal} {\bibinfo  {journal} {\mnras}\ }\textbf {\bibinfo {volume} {398}},\
  \bibinfo {pages} {53} (\bibinfo {year} {2009})},\ \Eprint
  {http://arxiv.org/abs/0904.2572} {arXiv:0904.2572 [astro-ph.CO]} \BibitemShut
  {NoStop}%
\bibitem [{\citenamefont {{Di Matteo}}\ \emph {et~al.}(2008)\citenamefont {{Di
  Matteo}}, \citenamefont {{Colberg}}, \citenamefont {{Springel}},
  \citenamefont {{Hernquist}},\ and\ \citenamefont {{Sijacki}}}]{DiMatteo08}%
  \BibitemOpen
  \bibfield  {author} {\bibinfo {author} {\bibfnamefont {T.}~\bibnamefont {{Di
  Matteo}}}, \bibinfo {author} {\bibfnamefont {J.}~\bibnamefont {{Colberg}}},
  \bibinfo {author} {\bibfnamefont {V.}~\bibnamefont {{Springel}}}, \bibinfo
  {author} {\bibfnamefont {L.}~\bibnamefont {{Hernquist}}}, \ and\ \bibinfo
  {author} {\bibfnamefont {D.}~\bibnamefont {{Sijacki}}},\ }\href {\doibase
  10.1086/524921} {\bibfield  {journal} {\bibinfo  {journal} {\apj}\ }\textbf
  {\bibinfo {volume} {676}},\ \bibinfo {pages} {33} (\bibinfo {year} {2008})},\
  \Eprint {http://arxiv.org/abs/0705.2269} {arXiv:0705.2269} \BibitemShut
  {NoStop}%
\bibitem [{\citenamefont {{Croft}}\ \emph {et~al.}(2009)\citenamefont
  {{Croft}}, \citenamefont {{Di Matteo}}, \citenamefont {{Springel}},\ and\
  \citenamefont {{Hernquist}}}]{Croft09}%
  \BibitemOpen
  \bibfield  {author} {\bibinfo {author} {\bibfnamefont {R.~A.~C.}\
  \bibnamefont {{Croft}}}, \bibinfo {author} {\bibfnamefont {T.}~\bibnamefont
  {{Di Matteo}}}, \bibinfo {author} {\bibfnamefont {V.}~\bibnamefont
  {{Springel}}}, \ and\ \bibinfo {author} {\bibfnamefont {L.}~\bibnamefont
  {{Hernquist}}},\ }\href {\doibase 10.1111/j.1365-2966.2009.15446.x}
  {\bibfield  {journal} {\bibinfo  {journal} {\mnras}\ }\textbf {\bibinfo
  {volume} {400}},\ \bibinfo {pages} {43} (\bibinfo {year} {2009})},\ \Eprint
  {http://arxiv.org/abs/0803.4003} {arXiv:0803.4003} \BibitemShut {NoStop}%
\bibitem [{\citenamefont {{Degraf}}\ \emph {et~al.}(2010)\citenamefont
  {{Degraf}}, \citenamefont {{Di Matteo}},\ and\ \citenamefont
  {{Springel}}}]{Degraf10}%
  \BibitemOpen
  \bibfield  {author} {\bibinfo {author} {\bibfnamefont {C.}~\bibnamefont
  {{Degraf}}}, \bibinfo {author} {\bibfnamefont {T.}~\bibnamefont {{Di
  Matteo}}}, \ and\ \bibinfo {author} {\bibfnamefont {V.}~\bibnamefont
  {{Springel}}},\ }\href {\doibase 10.1111/j.1365-2966.2009.16018.x} {\bibfield
   {journal} {\bibinfo  {journal} {\mnras}\ }\textbf {\bibinfo {volume}
  {402}},\ \bibinfo {pages} {1927} (\bibinfo {year} {2010})},\ \Eprint
  {http://arxiv.org/abs/0910.1843} {arXiv:0910.1843} \BibitemShut {NoStop}%
\bibitem [{\citenamefont {{McCarthy}}\ \emph {et~al.}(2010)\citenamefont
  {{McCarthy}}, \citenamefont {{Schaye}}, \citenamefont {{Ponman}},
  \citenamefont {{Bower}}, \citenamefont {{Booth}}, \citenamefont {{Dalla
  Vecchia}}, \citenamefont {{Crain}}, \citenamefont {{Springel}}, \citenamefont
  {{Theuns}},\ and\ \citenamefont {{Wiersma}}}]{McCarthy10}%
  \BibitemOpen
  \bibfield  {author} {\bibinfo {author} {\bibfnamefont {I.~G.}\ \bibnamefont
  {{McCarthy}}}, \bibinfo {author} {\bibfnamefont {J.}~\bibnamefont
  {{Schaye}}}, \bibinfo {author} {\bibfnamefont {T.~J.}\ \bibnamefont
  {{Ponman}}}, \bibinfo {author} {\bibfnamefont {R.~G.}\ \bibnamefont
  {{Bower}}}, \bibinfo {author} {\bibfnamefont {C.~M.}\ \bibnamefont
  {{Booth}}}, \bibinfo {author} {\bibfnamefont {C.}~\bibnamefont {{Dalla
  Vecchia}}}, \bibinfo {author} {\bibfnamefont {R.~A.}\ \bibnamefont
  {{Crain}}}, \bibinfo {author} {\bibfnamefont {V.}~\bibnamefont {{Springel}}},
  \bibinfo {author} {\bibfnamefont {T.}~\bibnamefont {{Theuns}}}, \ and\
  \bibinfo {author} {\bibfnamefont {R.~P.~C.}\ \bibnamefont {{Wiersma}}},\
  }\href {\doibase 10.1111/j.1365-2966.2010.16750.x} {\bibfield  {journal}
  {\bibinfo  {journal} {\mnras}\ }\textbf {\bibinfo {volume} {406}},\ \bibinfo
  {pages} {822} (\bibinfo {year} {2010})},\ \Eprint
  {http://arxiv.org/abs/0911.2641} {arXiv:0911.2641 [astro-ph.CO]} \BibitemShut
  {NoStop}%
\bibitem [{\citenamefont {{McCarthy}}\ \emph {et~al.}(2011)\citenamefont
  {{McCarthy}}, \citenamefont {{Schaye}}, \citenamefont {{Bower}},
  \citenamefont {{Ponman}}, \citenamefont {{Booth}}, \citenamefont {{Dalla
  Vecchia}},\ and\ \citenamefont {{Springel}}}]{McCarthy11}%
  \BibitemOpen
  \bibfield  {author} {\bibinfo {author} {\bibfnamefont {I.~G.}\ \bibnamefont
  {{McCarthy}}}, \bibinfo {author} {\bibfnamefont {J.}~\bibnamefont
  {{Schaye}}}, \bibinfo {author} {\bibfnamefont {R.~G.}\ \bibnamefont
  {{Bower}}}, \bibinfo {author} {\bibfnamefont {T.~J.}\ \bibnamefont
  {{Ponman}}}, \bibinfo {author} {\bibfnamefont {C.~M.}\ \bibnamefont
  {{Booth}}}, \bibinfo {author} {\bibfnamefont {C.}~\bibnamefont {{Dalla
  Vecchia}}}, \ and\ \bibinfo {author} {\bibfnamefont {V.}~\bibnamefont
  {{Springel}}},\ }\href {\doibase 10.1111/j.1365-2966.2010.18033.x} {\bibfield
   {journal} {\bibinfo  {journal} {\mnras}\ }\textbf {\bibinfo {volume}
  {412}},\ \bibinfo {pages} {1965} (\bibinfo {year} {2011})},\ \Eprint
  {http://arxiv.org/abs/1008.4799} {arXiv:1008.4799 [astro-ph.CO]} \BibitemShut
  {NoStop}%
\bibitem [{\citenamefont {{Crain}}\ \emph {et~al.}(2015)\citenamefont
  {{Crain}}, \citenamefont {{Schaye}}, \citenamefont {{Bower}}, \citenamefont
  {{Furlong}}, \citenamefont {{Schaller}}, \citenamefont {{Theuns}},
  \citenamefont {{Dalla Vecchia}}, \citenamefont {{Frenk}}, \citenamefont
  {{McCarthy}}, \citenamefont {{Helly}}, \citenamefont {{Jenkins}},
  \citenamefont {{Rosas-Guevara}}, \citenamefont {{White}},\ and\ \citenamefont
  {{Trayford}}}]{Crain15}%
  \BibitemOpen
  \bibfield  {author} {\bibinfo {author} {\bibfnamefont {R.~A.}\ \bibnamefont
  {{Crain}}}, \bibinfo {author} {\bibfnamefont {J.}~\bibnamefont {{Schaye}}},
  \bibinfo {author} {\bibfnamefont {R.~G.}\ \bibnamefont {{Bower}}}, \bibinfo
  {author} {\bibfnamefont {M.}~\bibnamefont {{Furlong}}}, \bibinfo {author}
  {\bibfnamefont {M.}~\bibnamefont {{Schaller}}}, \bibinfo {author}
  {\bibfnamefont {T.}~\bibnamefont {{Theuns}}}, \bibinfo {author}
  {\bibfnamefont {C.}~\bibnamefont {{Dalla Vecchia}}}, \bibinfo {author}
  {\bibfnamefont {C.~S.}\ \bibnamefont {{Frenk}}}, \bibinfo {author}
  {\bibfnamefont {I.~G.}\ \bibnamefont {{McCarthy}}}, \bibinfo {author}
  {\bibfnamefont {J.~C.}\ \bibnamefont {{Helly}}}, \bibinfo {author}
  {\bibfnamefont {A.}~\bibnamefont {{Jenkins}}}, \bibinfo {author}
  {\bibfnamefont {Y.~M.}\ \bibnamefont {{Rosas-Guevara}}}, \bibinfo {author}
  {\bibfnamefont {S.~D.~M.}\ \bibnamefont {{White}}}, \ and\ \bibinfo {author}
  {\bibfnamefont {J.~W.}\ \bibnamefont {{Trayford}}},\ }\href {\doibase
  10.1093/mnras/stv725} {\bibfield  {journal} {\bibinfo  {journal} {\mnras}\
  }\textbf {\bibinfo {volume} {450}},\ \bibinfo {pages} {1937} (\bibinfo {year}
  {2015})},\ \Eprint {http://arxiv.org/abs/1501.01311} {arXiv:1501.01311}
  \BibitemShut {NoStop}%
\bibitem [{\citenamefont {{Pakmor}}\ \emph {et~al.}(2011)\citenamefont
  {{Pakmor}}, \citenamefont {{Bauer}},\ and\ \citenamefont
  {{Springel}}}]{Pakmor11}%
  \BibitemOpen
  \bibfield  {author} {\bibinfo {author} {\bibfnamefont {R.}~\bibnamefont
  {{Pakmor}}}, \bibinfo {author} {\bibfnamefont {A.}~\bibnamefont {{Bauer}}}, \
  and\ \bibinfo {author} {\bibfnamefont {V.}~\bibnamefont {{Springel}}},\
  }\href {\doibase 10.1111/j.1365-2966.2011.19591.x} {\bibfield  {journal}
  {\bibinfo  {journal} {\mnras}\ }\textbf {\bibinfo {volume} {418}},\ \bibinfo
  {pages} {1392} (\bibinfo {year} {2011})},\ \Eprint
  {http://arxiv.org/abs/1108.1792} {arXiv:1108.1792 [astro-ph.IM]} \BibitemShut
  {NoStop}%
\bibitem [{\citenamefont {{Seljak}}(2000)}]{Seljak2000}%
  \BibitemOpen
  \bibfield  {author} {\bibinfo {author} {\bibfnamefont {U.}~\bibnamefont
  {{Seljak}}},\ }\href {\doibase 10.1046/j.1365-8711.2000.03715.x} {\bibfield
  {journal} {\bibinfo  {journal} {\mnras}\ }\textbf {\bibinfo {volume} {318}},\
  \bibinfo {pages} {203} (\bibinfo {year} {2000})},\ \Eprint
  {http://arxiv.org/abs/arXiv:astro-ph/0001493} {arXiv:astro-ph/0001493}
  \BibitemShut {NoStop}%
\bibitem [{\citenamefont {{Peacock}}\ and\ \citenamefont
  {{Smith}}(2000)}]{Peacock2000}%
  \BibitemOpen
  \bibfield  {author} {\bibinfo {author} {\bibfnamefont {J.~A.}\ \bibnamefont
  {{Peacock}}}\ and\ \bibinfo {author} {\bibfnamefont {R.~E.}\ \bibnamefont
  {{Smith}}},\ }\href {\doibase 10.1046/j.1365-8711.2000.03779.x} {\bibfield
  {journal} {\bibinfo  {journal} {\mnras}\ }\textbf {\bibinfo {volume} {318}},\
  \bibinfo {pages} {1144} (\bibinfo {year} {2000})},\ \Eprint
  {http://arxiv.org/abs/astro-ph/0005010} {astro-ph/0005010} \BibitemShut
  {NoStop}%
\bibitem [{\citenamefont {Cooray}\ and\ \citenamefont
  {Sheth}(2002{\natexlab{b}})}]{Cooray2002}%
  \BibitemOpen
  \bibfield  {author} {\bibinfo {author} {\bibfnamefont {A.}~\bibnamefont
  {Cooray}}\ and\ \bibinfo {author} {\bibfnamefont {R.}~\bibnamefont {Sheth}},\
  }\href {\doibase 10.1016/S0370-1573(02)00276-4} {\bibfield  {journal}
  {\bibinfo  {journal} {Physics Reports}\ }\textbf {\bibinfo {volume} {372}},\
  \bibinfo {pages} {1} (\bibinfo {year} {2002}{\natexlab{b}})}\BibitemShut
  {NoStop}%
\bibitem [{\citenamefont {Navarro}\ \emph {et~al.}(1997)\citenamefont
  {Navarro}, \citenamefont {Frenk},\ and\ \citenamefont {White}}]{Navarro1997}%
  \BibitemOpen
  \bibfield  {author} {\bibinfo {author} {\bibfnamefont {J.~F.}\ \bibnamefont
  {Navarro}}, \bibinfo {author} {\bibfnamefont {C.~S.}\ \bibnamefont {Frenk}},
  \ and\ \bibinfo {author} {\bibfnamefont {S.~D.~M.}\ \bibnamefont {White}},\
  }\href {\doibase 10.1086/304888} {\bibfield  {journal} {\bibinfo  {journal}
  {\apj}\ }\textbf {\bibinfo {volume} {490}},\ \bibinfo {pages} {493} (\bibinfo
  {year} {1997})}\BibitemShut {NoStop}%
\bibitem [{\citenamefont {{White}}(2004)}]{White2004}%
  \BibitemOpen
  \bibfield  {author} {\bibinfo {author} {\bibfnamefont {M.}~\bibnamefont
  {{White}}},\ }\href {\doibase 10.1016/j.astropartphys.2004.06.001} {\bibfield
   {journal} {\bibinfo  {journal} {Astroparticle Physics}\ }\textbf {\bibinfo
  {volume} {22}},\ \bibinfo {pages} {211} (\bibinfo {year} {2004})},\ \Eprint
  {http://arxiv.org/abs/astro-ph/0405593} {astro-ph/0405593} \BibitemShut
  {NoStop}%
\bibitem [{\citenamefont {{Jing}}\ \emph {et~al.}(2006)\citenamefont {{Jing}},
  \citenamefont {{Zhang}}, \citenamefont {{Lin}}, \citenamefont {{Gao}},\ and\
  \citenamefont {{Springel}}}]{Jing2006}%
  \BibitemOpen
  \bibfield  {author} {\bibinfo {author} {\bibfnamefont {Y.~P.}\ \bibnamefont
  {{Jing}}}, \bibinfo {author} {\bibfnamefont {P.}~\bibnamefont {{Zhang}}},
  \bibinfo {author} {\bibfnamefont {W.~P.}\ \bibnamefont {{Lin}}}, \bibinfo
  {author} {\bibfnamefont {L.}~\bibnamefont {{Gao}}}, \ and\ \bibinfo {author}
  {\bibfnamefont {V.}~\bibnamefont {{Springel}}},\ }\href {\doibase
  10.1086/503547} {\bibfield  {journal} {\bibinfo  {journal} {\apjl}\ }\textbf
  {\bibinfo {volume} {640}},\ \bibinfo {pages} {L119} (\bibinfo {year}
  {2006})},\ \Eprint {http://arxiv.org/abs/astro-ph/0512426} {astro-ph/0512426}
  \BibitemShut {NoStop}%
\bibitem [{\citenamefont {{Rudd}}\ \emph {et~al.}(2008)\citenamefont {{Rudd}},
  \citenamefont {{Zentner}},\ and\ \citenamefont {{Kravtsov}}}]{Rudd2008}%
  \BibitemOpen
  \bibfield  {author} {\bibinfo {author} {\bibfnamefont {D.~H.}\ \bibnamefont
  {{Rudd}}}, \bibinfo {author} {\bibfnamefont {A.~R.}\ \bibnamefont
  {{Zentner}}}, \ and\ \bibinfo {author} {\bibfnamefont {A.~V.}\ \bibnamefont
  {{Kravtsov}}},\ }\href {\doibase 10.1086/523836} {\bibfield  {journal}
  {\bibinfo  {journal} {\apj}\ }\textbf {\bibinfo {volume} {672}},\ \bibinfo
  {pages} {19} (\bibinfo {year} {2008})},\ \Eprint
  {http://arxiv.org/abs/astro-ph/0703741} {astro-ph/0703741} \BibitemShut
  {NoStop}%
\bibitem [{\citenamefont {{Mohammed}}\ \emph {et~al.}(2014)\citenamefont
  {{Mohammed}}, \citenamefont {{Martizzi}}, \citenamefont {{Teyssier}},\ and\
  \citenamefont {{Amara}}}]{Mohammed2014b}%
  \BibitemOpen
  \bibfield  {author} {\bibinfo {author} {\bibfnamefont {I.}~\bibnamefont
  {{Mohammed}}}, \bibinfo {author} {\bibfnamefont {D.}~\bibnamefont
  {{Martizzi}}}, \bibinfo {author} {\bibfnamefont {R.}~\bibnamefont
  {{Teyssier}}}, \ and\ \bibinfo {author} {\bibfnamefont {A.}~\bibnamefont
  {{Amara}}},\ }\href@noop {} {\bibfield  {journal} {\bibinfo  {journal}
  {preprint (arXiv:e-prints 1410.6826)}\ } (\bibinfo {year} {2014})},\ \Eprint
  {http://arxiv.org/abs/1410.6826} {arXiv:1410.6826} \BibitemShut {NoStop}%
\bibitem [{\citenamefont {{Osmond}}\ and\ \citenamefont
  {{Ponman}}(2004)}]{Osmond2004}%
  \BibitemOpen
  \bibfield  {author} {\bibinfo {author} {\bibfnamefont {J.~P.~F.}\
  \bibnamefont {{Osmond}}}\ and\ \bibinfo {author} {\bibfnamefont {T.~J.}\
  \bibnamefont {{Ponman}}},\ }\href {\doibase 10.1111/j.1365-2966.2004.07742.x}
  {\bibfield  {journal} {\bibinfo  {journal} {\mnras}\ }\textbf {\bibinfo
  {volume} {350}},\ \bibinfo {pages} {1511} (\bibinfo {year} {2004})},\ \Eprint
  {http://arxiv.org/abs/astro-ph/0402439} {astro-ph/0402439} \BibitemShut
  {NoStop}%
\bibitem [{\citenamefont {{Fedeli}}\ \emph {et~al.}(2014)\citenamefont
  {{Fedeli}}, \citenamefont {{Semboloni}}, \citenamefont {{Velliscig}},
  \citenamefont {{Van Daalen}}, \citenamefont {{Schaye}},\ and\ \citenamefont
  {{Hoekstra}}}]{Fedeli2014b}%
  \BibitemOpen
  \bibfield  {author} {\bibinfo {author} {\bibfnamefont {C.}~\bibnamefont
  {{Fedeli}}}, \bibinfo {author} {\bibfnamefont {E.}~\bibnamefont
  {{Semboloni}}}, \bibinfo {author} {\bibfnamefont {M.}~\bibnamefont
  {{Velliscig}}}, \bibinfo {author} {\bibfnamefont {M.}~\bibnamefont {{Van
  Daalen}}}, \bibinfo {author} {\bibfnamefont {J.}~\bibnamefont {{Schaye}}}, \
  and\ \bibinfo {author} {\bibfnamefont {H.}~\bibnamefont {{Hoekstra}}},\
  }\href {\doibase 10.1088/1475-7516/2014/08/028} {\bibfield  {journal}
  {\bibinfo  {journal} {\jcap}\ }\textbf {\bibinfo {volume} {8}},\ \bibinfo
  {eid} {028} (\bibinfo {year} {2014})},\ \Eprint
  {http://arxiv.org/abs/1406.5013} {arXiv:1406.5013} \BibitemShut {NoStop}%
\bibitem [{\citenamefont {Bullock}\ \emph {et~al.}(2001)\citenamefont
  {Bullock}, \citenamefont {Kolatt}, \citenamefont {Sigad}, \citenamefont
  {Somerville}, \citenamefont {Kravtsov}, \citenamefont {Klypin}, \citenamefont
  {Primack},\ and\ \citenamefont {Dekel}}]{Bullock2001}%
  \BibitemOpen
  \bibfield  {author} {\bibinfo {author} {\bibfnamefont {J.~S.}\ \bibnamefont
  {Bullock}}, \bibinfo {author} {\bibfnamefont {T.~S.}\ \bibnamefont {Kolatt}},
  \bibinfo {author} {\bibfnamefont {Y.}~\bibnamefont {Sigad}}, \bibinfo
  {author} {\bibfnamefont {R.~S.}\ \bibnamefont {Somerville}}, \bibinfo
  {author} {\bibfnamefont {A.~V.}\ \bibnamefont {Kravtsov}}, \bibinfo {author}
  {\bibfnamefont {A.~A.}\ \bibnamefont {Klypin}}, \bibinfo {author}
  {\bibfnamefont {J.~R.}\ \bibnamefont {Primack}}, \ and\ \bibinfo {author}
  {\bibfnamefont {A.}~\bibnamefont {Dekel}},\ }\href {\doibase
  10.1046/j.1365-8711.2001.04068.x} {\bibfield  {journal} {\bibinfo  {journal}
  {\mnras}\ }\textbf {\bibinfo {volume} {321}},\ \bibinfo {pages} {559}
  (\bibinfo {year} {2001})}\BibitemShut {NoStop}%
\bibitem [{\citenamefont {{Lawrence}}\ \emph {et~al.}(2010)\citenamefont
  {{Lawrence}}, \citenamefont {{Heitmann}}, \citenamefont {{White}},
  \citenamefont {{Higdon}}, \citenamefont {{Wagner}}, \citenamefont {{Habib}},\
  and\ \citenamefont {{Williams}}}]{Lawrence10}%
  \BibitemOpen
  \bibfield  {author} {\bibinfo {author} {\bibfnamefont {E.}~\bibnamefont
  {{Lawrence}}}, \bibinfo {author} {\bibfnamefont {K.}~\bibnamefont
  {{Heitmann}}}, \bibinfo {author} {\bibfnamefont {M.}~\bibnamefont {{White}}},
  \bibinfo {author} {\bibfnamefont {D.}~\bibnamefont {{Higdon}}}, \bibinfo
  {author} {\bibfnamefont {C.}~\bibnamefont {{Wagner}}}, \bibinfo {author}
  {\bibfnamefont {S.}~\bibnamefont {{Habib}}}, \ and\ \bibinfo {author}
  {\bibfnamefont {B.}~\bibnamefont {{Williams}}},\ }\href {\doibase
  10.1088/0004-637X/713/2/1322} {\bibfield  {journal} {\bibinfo  {journal}
  {\apj}\ }\textbf {\bibinfo {volume} {713}},\ \bibinfo {pages} {1322}
  (\bibinfo {year} {2010})},\ \Eprint {http://arxiv.org/abs/0912.4490}
  {arXiv:0912.4490 [astro-ph.CO]} \BibitemShut {NoStop}%
\bibitem [{\citenamefont {{Copeland}}\ \emph {et~al.}(2019)\citenamefont
  {{Copeland}}, \citenamefont {{Taylor}},\ and\ \citenamefont
  {{Hall}}}]{Copeland2019}%
  \BibitemOpen
  \bibfield  {author} {\bibinfo {author} {\bibfnamefont {D.}~\bibnamefont
  {{Copeland}}}, \bibinfo {author} {\bibfnamefont {A.}~\bibnamefont
  {{Taylor}}}, \ and\ \bibinfo {author} {\bibfnamefont {A.}~\bibnamefont
  {{Hall}}},\ }\href@noop {} {\bibfield  {journal} {\bibinfo  {journal} {arXiv
  e-prints}\ ,\ \bibinfo {eid} {arXiv:1905.08754}} (\bibinfo {year} {2019})},\
  \Eprint {http://arxiv.org/abs/1905.08754} {arXiv:1905.08754 [astro-ph.CO]}
  \BibitemShut {NoStop}%
\bibitem [{\citenamefont {{Copeland}}\ \emph {et~al.}(2018)\citenamefont
  {{Copeland}}, \citenamefont {{Taylor}},\ and\ \citenamefont
  {{Hall}}}]{Copeland2018}%
  \BibitemOpen
  \bibfield  {author} {\bibinfo {author} {\bibfnamefont {D.}~\bibnamefont
  {{Copeland}}}, \bibinfo {author} {\bibfnamefont {A.}~\bibnamefont
  {{Taylor}}}, \ and\ \bibinfo {author} {\bibfnamefont {A.}~\bibnamefont
  {{Hall}}},\ }\href {\doibase 10.1093/mnras/sty2001} {\bibfield  {journal}
  {\bibinfo  {journal} {\mnras}\ }\textbf {\bibinfo {volume} {480}},\ \bibinfo
  {pages} {2247} (\bibinfo {year} {2018})},\ \Eprint
  {http://arxiv.org/abs/1712.07112} {arXiv:1712.07112} \BibitemShut {NoStop}%
\bibitem [{\citenamefont {{MacCrann}}\ \emph {et~al.}(2017)\citenamefont
  {{MacCrann}}, \citenamefont {{Aleksi{\'c}}}, \citenamefont {{Amara}},
  \citenamefont {{Bridle}}, \citenamefont {{Bruderer}}, \citenamefont
  {{Chang}}, \citenamefont {{Dodelson}}, \citenamefont {{Eifler}},
  \citenamefont {{Huff}}, \citenamefont {{Huterer}}, \citenamefont
  {{Kacprzak}}, \citenamefont {{Refregier}}, \citenamefont {{Suchyta}},
  \citenamefont {{Wechsler}}, \citenamefont {{Zuntz}}, \citenamefont
  {{Abbott}}, \citenamefont {{Allam}}, \citenamefont {{Annis}}, \citenamefont
  {{Armstrong}}, \citenamefont {{Benoit-L{\'e}vy}}, \citenamefont {{Brooks}},
  \citenamefont {{Burke}}, \citenamefont {{Carnero Rosell}}, \citenamefont
  {{Carrasco Kind}}, \citenamefont {{Carretero}}, \citenamefont {{Castander}},
  \citenamefont {{Crocce}}, \citenamefont {{Cunha}}, \citenamefont {{da
  Costa}}, \citenamefont {{Desai}}, \citenamefont {{Diehl}}, \citenamefont
  {{Dietrich}}, \citenamefont {{Doel}}, \citenamefont {{Evrard}}, \citenamefont
  {{Flaugher}}, \citenamefont {{Fosalba}}, \citenamefont {{Gerdes}},
  \citenamefont {{Goldstein}}, \citenamefont {{Gruen}}, \citenamefont
  {{Gruendl}}, \citenamefont {{Gutierrez}}, \citenamefont {{Honscheid}},
  \citenamefont {{James}}, \citenamefont {{Jarvis}}, \citenamefont {{Krause}},
  \citenamefont {{Kuehn}}, \citenamefont {{Kuropatkin}}, \citenamefont
  {{Lima}}, \citenamefont {{Marshall}}, \citenamefont {{Melchior}},
  \citenamefont {{Menanteau}}, \citenamefont {{Miquel}}, \citenamefont
  {{Plazas}}, \citenamefont {{Romer}}, \citenamefont {{Rykoff}}, \citenamefont
  {{Sanchez}}, \citenamefont {{Scarpine}}, \citenamefont {{Sevilla-Noarbe}},
  \citenamefont {{Sheldon}}, \citenamefont {{Soares-Santos}}, \citenamefont
  {{Swanson}}, \citenamefont {{Tarle}}, \citenamefont {{Thomas}},\ and\
  \citenamefont {{Vikram}}}]{MacCrann2017}%
  \BibitemOpen
  \bibfield  {author} {\bibinfo {author} {\bibfnamefont {N.}~\bibnamefont
  {{MacCrann}}}, \bibinfo {author} {\bibfnamefont {J.}~\bibnamefont
  {{Aleksi{\'c}}}}, \bibinfo {author} {\bibfnamefont {A.}~\bibnamefont
  {{Amara}}}, \bibinfo {author} {\bibfnamefont {S.~L.}\ \bibnamefont
  {{Bridle}}}, \bibinfo {author} {\bibfnamefont {C.}~\bibnamefont
  {{Bruderer}}}, \bibinfo {author} {\bibfnamefont {C.}~\bibnamefont {{Chang}}},
  \bibinfo {author} {\bibfnamefont {S.}~\bibnamefont {{Dodelson}}}, \bibinfo
  {author} {\bibfnamefont {T.~F.}\ \bibnamefont {{Eifler}}}, \bibinfo {author}
  {\bibfnamefont {E.~M.}\ \bibnamefont {{Huff}}}, \bibinfo {author}
  {\bibfnamefont {D.}~\bibnamefont {{Huterer}}}, \bibinfo {author}
  {\bibfnamefont {T.}~\bibnamefont {{Kacprzak}}}, \bibinfo {author}
  {\bibfnamefont {A.}~\bibnamefont {{Refregier}}}, \bibinfo {author}
  {\bibfnamefont {E.}~\bibnamefont {{Suchyta}}}, \bibinfo {author}
  {\bibfnamefont {R.~H.}\ \bibnamefont {{Wechsler}}}, \bibinfo {author}
  {\bibfnamefont {J.}~\bibnamefont {{Zuntz}}}, \bibinfo {author} {\bibfnamefont
  {T.~M.~C.}\ \bibnamefont {{Abbott}}}, \bibinfo {author} {\bibfnamefont
  {S.}~\bibnamefont {{Allam}}}, \bibinfo {author} {\bibfnamefont
  {J.}~\bibnamefont {{Annis}}}, \bibinfo {author} {\bibfnamefont
  {R.}~\bibnamefont {{Armstrong}}}, \bibinfo {author} {\bibfnamefont
  {A.}~\bibnamefont {{Benoit-L{\'e}vy}}}, \bibinfo {author} {\bibfnamefont
  {D.}~\bibnamefont {{Brooks}}}, \bibinfo {author} {\bibfnamefont {D.~L.}\
  \bibnamefont {{Burke}}}, \bibinfo {author} {\bibfnamefont {A.}~\bibnamefont
  {{Carnero Rosell}}}, \bibinfo {author} {\bibfnamefont {M.}~\bibnamefont
  {{Carrasco Kind}}}, \bibinfo {author} {\bibfnamefont {J.}~\bibnamefont
  {{Carretero}}}, \bibinfo {author} {\bibfnamefont {F.~J.}\ \bibnamefont
  {{Castander}}}, \bibinfo {author} {\bibfnamefont {M.}~\bibnamefont
  {{Crocce}}}, \bibinfo {author} {\bibfnamefont {C.~E.}\ \bibnamefont
  {{Cunha}}}, \bibinfo {author} {\bibfnamefont {L.~N.}\ \bibnamefont {{da
  Costa}}}, \bibinfo {author} {\bibfnamefont {S.}~\bibnamefont {{Desai}}},
  \bibinfo {author} {\bibfnamefont {H.~T.}\ \bibnamefont {{Diehl}}}, \bibinfo
  {author} {\bibfnamefont {J.~P.}\ \bibnamefont {{Dietrich}}}, \bibinfo
  {author} {\bibfnamefont {P.}~\bibnamefont {{Doel}}}, \bibinfo {author}
  {\bibfnamefont {A.~E.}\ \bibnamefont {{Evrard}}}, \bibinfo {author}
  {\bibfnamefont {B.}~\bibnamefont {{Flaugher}}}, \bibinfo {author}
  {\bibfnamefont {P.}~\bibnamefont {{Fosalba}}}, \bibinfo {author}
  {\bibfnamefont {D.~W.}\ \bibnamefont {{Gerdes}}}, \bibinfo {author}
  {\bibfnamefont {D.~A.}\ \bibnamefont {{Goldstein}}}, \bibinfo {author}
  {\bibfnamefont {D.}~\bibnamefont {{Gruen}}}, \bibinfo {author} {\bibfnamefont
  {R.~A.}\ \bibnamefont {{Gruendl}}}, \bibinfo {author} {\bibfnamefont
  {G.}~\bibnamefont {{Gutierrez}}}, \bibinfo {author} {\bibfnamefont
  {K.}~\bibnamefont {{Honscheid}}}, \bibinfo {author} {\bibfnamefont {D.~J.}\
  \bibnamefont {{James}}}, \bibinfo {author} {\bibfnamefont {M.}~\bibnamefont
  {{Jarvis}}}, \bibinfo {author} {\bibfnamefont {E.}~\bibnamefont {{Krause}}},
  \bibinfo {author} {\bibfnamefont {K.}~\bibnamefont {{Kuehn}}}, \bibinfo
  {author} {\bibfnamefont {N.}~\bibnamefont {{Kuropatkin}}}, \bibinfo {author}
  {\bibfnamefont {M.}~\bibnamefont {{Lima}}}, \bibinfo {author} {\bibfnamefont
  {J.~L.}\ \bibnamefont {{Marshall}}}, \bibinfo {author} {\bibfnamefont
  {P.}~\bibnamefont {{Melchior}}}, \bibinfo {author} {\bibfnamefont
  {F.}~\bibnamefont {{Menanteau}}}, \bibinfo {author} {\bibfnamefont
  {R.}~\bibnamefont {{Miquel}}}, \bibinfo {author} {\bibfnamefont {A.~A.}\
  \bibnamefont {{Plazas}}}, \bibinfo {author} {\bibfnamefont {A.~K.}\
  \bibnamefont {{Romer}}}, \bibinfo {author} {\bibfnamefont {E.~S.}\
  \bibnamefont {{Rykoff}}}, \bibinfo {author} {\bibfnamefont {E.}~\bibnamefont
  {{Sanchez}}}, \bibinfo {author} {\bibfnamefont {V.}~\bibnamefont
  {{Scarpine}}}, \bibinfo {author} {\bibfnamefont {I.}~\bibnamefont
  {{Sevilla-Noarbe}}}, \bibinfo {author} {\bibfnamefont {E.}~\bibnamefont
  {{Sheldon}}}, \bibinfo {author} {\bibfnamefont {M.}~\bibnamefont
  {{Soares-Santos}}}, \bibinfo {author} {\bibfnamefont {M.~E.~C.}\ \bibnamefont
  {{Swanson}}}, \bibinfo {author} {\bibfnamefont {G.}~\bibnamefont {{Tarle}}},
  \bibinfo {author} {\bibfnamefont {D.}~\bibnamefont {{Thomas}}}, \ and\
  \bibinfo {author} {\bibfnamefont {V.}~\bibnamefont {{Vikram}}},\ }\href
  {\doibase 10.1093/mnras/stw2849} {\bibfield  {journal} {\bibinfo  {journal}
  {\mnras}\ }\textbf {\bibinfo {volume} {465}},\ \bibinfo {pages} {2567}
  (\bibinfo {year} {2017})},\ \Eprint {http://arxiv.org/abs/1608.01838}
  {arXiv:1608.01838} \BibitemShut {NoStop}%
\bibitem [{\citenamefont {{van Daalen}}\ \emph {et~al.}(2014)\citenamefont
  {{van Daalen}}, \citenamefont {{Schaye}}, \citenamefont {{McCarthy}},
  \citenamefont {{Booth}},\ and\ \citenamefont {{Dalla
  Vecchia}}}]{vanDaalen14}%
  \BibitemOpen
  \bibfield  {author} {\bibinfo {author} {\bibfnamefont {M.~P.}\ \bibnamefont
  {{van Daalen}}}, \bibinfo {author} {\bibfnamefont {J.}~\bibnamefont
  {{Schaye}}}, \bibinfo {author} {\bibfnamefont {I.~G.}\ \bibnamefont
  {{McCarthy}}}, \bibinfo {author} {\bibfnamefont {C.~M.}\ \bibnamefont
  {{Booth}}}, \ and\ \bibinfo {author} {\bibfnamefont {C.}~\bibnamefont {{Dalla
  Vecchia}}},\ }\href {\doibase 10.1093/mnras/stu482} {\bibfield  {journal}
  {\bibinfo  {journal} {\mnras}\ }\textbf {\bibinfo {volume} {440}},\ \bibinfo
  {pages} {2997} (\bibinfo {year} {2014})},\ \Eprint
  {http://arxiv.org/abs/1310.7571} {arXiv:1310.7571 [astro-ph.CO]} \BibitemShut
  {NoStop}%
\bibitem [{\citenamefont {{Parimbelli}}\ \emph {et~al.}(2019)\citenamefont
  {{Parimbelli}}, \citenamefont {{Viel}},\ and\ \citenamefont
  {{Sefusatti}}}]{Parimbelli19}%
  \BibitemOpen
  \bibfield  {author} {\bibinfo {author} {\bibfnamefont {G.}~\bibnamefont
  {{Parimbelli}}}, \bibinfo {author} {\bibfnamefont {M.}~\bibnamefont
  {{Viel}}}, \ and\ \bibinfo {author} {\bibfnamefont {E.}~\bibnamefont
  {{Sefusatti}}},\ }\href {\doibase 10.1088/1475-7516/2019/01/010} {\bibfield
  {journal} {\bibinfo  {journal} {Journal of Cosmology and Astro-Particle
  Physics}\ }\textbf {\bibinfo {volume} {2019}},\ \bibinfo {eid} {010}
  (\bibinfo {year} {2019})},\ \Eprint {http://arxiv.org/abs/1809.06634}
  {arXiv:1809.06634 [astro-ph.CO]} \BibitemShut {NoStop}%
\bibitem [{\citenamefont {{Mohammed}}\ and\ \citenamefont
  {{Gnedin}}(2018)}]{Mohammed18}%
  \BibitemOpen
  \bibfield  {author} {\bibinfo {author} {\bibfnamefont {I.}~\bibnamefont
  {{Mohammed}}}\ and\ \bibinfo {author} {\bibfnamefont {N.~Y.}\ \bibnamefont
  {{Gnedin}}},\ }\href {\doibase 10.3847/1538-4357/aad3b1} {\bibfield
  {journal} {\bibinfo  {journal} {\apj}\ }\textbf {\bibinfo {volume} {863}},\
  \bibinfo {eid} {173} (\bibinfo {year} {2018})},\ \Eprint
  {http://arxiv.org/abs/1707.02332} {arXiv:1707.02332 [astro-ph.CO]}
  \BibitemShut {NoStop}%
\bibitem [{\citenamefont {{Lawrence}}\ \emph {et~al.}(2017)\citenamefont
  {{Lawrence}}, \citenamefont {{Heitmann}}, \citenamefont {{Kwan}},
  \citenamefont {{Upadhye}}, \citenamefont {{Bingham}}, \citenamefont
  {{Habib}}, \citenamefont {{Higdon}}, \citenamefont {{Pope}}, \citenamefont
  {{Finkel}},\ and\ \citenamefont {{Frontiere}}}]{Lawrence17}%
  \BibitemOpen
  \bibfield  {author} {\bibinfo {author} {\bibfnamefont {E.}~\bibnamefont
  {{Lawrence}}}, \bibinfo {author} {\bibfnamefont {K.}~\bibnamefont
  {{Heitmann}}}, \bibinfo {author} {\bibfnamefont {J.}~\bibnamefont {{Kwan}}},
  \bibinfo {author} {\bibfnamefont {A.}~\bibnamefont {{Upadhye}}}, \bibinfo
  {author} {\bibfnamefont {D.}~\bibnamefont {{Bingham}}}, \bibinfo {author}
  {\bibfnamefont {S.}~\bibnamefont {{Habib}}}, \bibinfo {author} {\bibfnamefont
  {D.}~\bibnamefont {{Higdon}}}, \bibinfo {author} {\bibfnamefont
  {A.}~\bibnamefont {{Pope}}}, \bibinfo {author} {\bibfnamefont
  {H.}~\bibnamefont {{Finkel}}}, \ and\ \bibinfo {author} {\bibfnamefont
  {N.}~\bibnamefont {{Frontiere}}},\ }\href {\doibase 10.3847/1538-4357/aa86a9}
  {\bibfield  {journal} {\bibinfo  {journal} {\apj}\ }\textbf {\bibinfo
  {volume} {847}},\ \bibinfo {eid} {50} (\bibinfo {year} {2017})},\ \Eprint
  {http://arxiv.org/abs/1705.03388} {arXiv:1705.03388} \BibitemShut {NoStop}%
\bibitem [{\citenamefont {{Battaglia}}\ \emph
  {et~al.}(2019{\natexlab{a}})\citenamefont {{Battaglia}}, \citenamefont
  {{Hill}}, \citenamefont {{Amodeo}}, \citenamefont {{Bartlett}}, \citenamefont
  {{Basu}}, \citenamefont {{Erler}}, \citenamefont {{Ferraro}}, \citenamefont
  {{Hernquist}}, \citenamefont {{Madhavacheril}}, \citenamefont {{McQuinn}},
  \citenamefont {{Mroczkowski}}, \citenamefont {{Nagai}}, \citenamefont
  {{Schaan}}, \citenamefont {{Somerville}}, \citenamefont {{Sunyaev}},
  \citenamefont {{Vogelsberger}},\ and\ \citenamefont {{Werk}}}]{Battaglia19}%
  \BibitemOpen
  \bibfield  {author} {\bibinfo {author} {\bibfnamefont {N.}~\bibnamefont
  {{Battaglia}}}, \bibinfo {author} {\bibfnamefont {J.~C.}\ \bibnamefont
  {{Hill}}}, \bibinfo {author} {\bibfnamefont {S.}~\bibnamefont {{Amodeo}}},
  \bibinfo {author} {\bibfnamefont {J.~G.}\ \bibnamefont {{Bartlett}}},
  \bibinfo {author} {\bibfnamefont {K.}~\bibnamefont {{Basu}}}, \bibinfo
  {author} {\bibfnamefont {J.}~\bibnamefont {{Erler}}}, \bibinfo {author}
  {\bibfnamefont {S.}~\bibnamefont {{Ferraro}}}, \bibinfo {author}
  {\bibfnamefont {L.}~\bibnamefont {{Hernquist}}}, \bibinfo {author}
  {\bibfnamefont {M.}~\bibnamefont {{Madhavacheril}}}, \bibinfo {author}
  {\bibfnamefont {M.}~\bibnamefont {{McQuinn}}}, \bibinfo {author}
  {\bibfnamefont {T.}~\bibnamefont {{Mroczkowski}}}, \bibinfo {author}
  {\bibfnamefont {D.}~\bibnamefont {{Nagai}}}, \bibinfo {author} {\bibfnamefont
  {E.}~\bibnamefont {{Schaan}}}, \bibinfo {author} {\bibfnamefont
  {R.}~\bibnamefont {{Somerville}}}, \bibinfo {author} {\bibfnamefont
  {R.}~\bibnamefont {{Sunyaev}}}, \bibinfo {author} {\bibfnamefont
  {M.}~\bibnamefont {{Vogelsberger}}}, \ and\ \bibinfo {author} {\bibfnamefont
  {J.}~\bibnamefont {{Werk}}},\ }\href@noop {} {\bibfield  {journal} {\bibinfo
  {journal} {arXiv e-prints}\ } (\bibinfo {year} {2019}{\natexlab{a}})},\
  \Eprint {http://arxiv.org/abs/1903.04647} {arXiv:1903.04647} \BibitemShut
  {NoStop}%
\bibitem [{\citenamefont {{Crichton}}\ \emph {et~al.}(2016)\citenamefont
  {{Crichton}}, \citenamefont {{Gralla}}, \citenamefont {{Hall}}, \citenamefont
  {{Marriage}}, \citenamefont {{Zakamska}}, \citenamefont {{Battaglia}},
  \citenamefont {{Bond}}, \citenamefont {{Devlin}}, \citenamefont {{Hill}},
  \citenamefont {{Hilton}}, \citenamefont {{Hincks}}, \citenamefont
  {{Huffenberger}}, \citenamefont {{Hughes}}, \citenamefont {{Kosowsky}},
  \citenamefont {{Moodley}}, \citenamefont {{Niemack}}, \citenamefont {{Page}},
  \citenamefont {{Partridge}}, \citenamefont {{Sievers}}, \citenamefont
  {{Sif{\'o}n}}, \citenamefont {{Staggs}}, \citenamefont {{Viero}},\ and\
  \citenamefont {{Wollack}}}]{Crichton16}%
  \BibitemOpen
  \bibfield  {author} {\bibinfo {author} {\bibfnamefont {D.}~\bibnamefont
  {{Crichton}}}, \bibinfo {author} {\bibfnamefont {M.~B.}\ \bibnamefont
  {{Gralla}}}, \bibinfo {author} {\bibfnamefont {K.}~\bibnamefont {{Hall}}},
  \bibinfo {author} {\bibfnamefont {T.~A.}\ \bibnamefont {{Marriage}}},
  \bibinfo {author} {\bibfnamefont {N.~L.}\ \bibnamefont {{Zakamska}}},
  \bibinfo {author} {\bibfnamefont {N.}~\bibnamefont {{Battaglia}}}, \bibinfo
  {author} {\bibfnamefont {J.~R.}\ \bibnamefont {{Bond}}}, \bibinfo {author}
  {\bibfnamefont {M.~J.}\ \bibnamefont {{Devlin}}}, \bibinfo {author}
  {\bibfnamefont {J.~C.}\ \bibnamefont {{Hill}}}, \bibinfo {author}
  {\bibfnamefont {M.}~\bibnamefont {{Hilton}}}, \bibinfo {author}
  {\bibfnamefont {A.~D.}\ \bibnamefont {{Hincks}}}, \bibinfo {author}
  {\bibfnamefont {K.~M.}\ \bibnamefont {{Huffenberger}}}, \bibinfo {author}
  {\bibfnamefont {J.~P.}\ \bibnamefont {{Hughes}}}, \bibinfo {author}
  {\bibfnamefont {A.}~\bibnamefont {{Kosowsky}}}, \bibinfo {author}
  {\bibfnamefont {K.}~\bibnamefont {{Moodley}}}, \bibinfo {author}
  {\bibfnamefont {M.~D.}\ \bibnamefont {{Niemack}}}, \bibinfo {author}
  {\bibfnamefont {L.~A.}\ \bibnamefont {{Page}}}, \bibinfo {author}
  {\bibfnamefont {B.}~\bibnamefont {{Partridge}}}, \bibinfo {author}
  {\bibfnamefont {J.~L.}\ \bibnamefont {{Sievers}}}, \bibinfo {author}
  {\bibfnamefont {C.}~\bibnamefont {{Sif{\'o}n}}}, \bibinfo {author}
  {\bibfnamefont {S.~T.}\ \bibnamefont {{Staggs}}}, \bibinfo {author}
  {\bibfnamefont {M.~P.}\ \bibnamefont {{Viero}}}, \ and\ \bibinfo {author}
  {\bibfnamefont {E.~J.}\ \bibnamefont {{Wollack}}},\ }\href {\doibase
  10.1093/mnras/stw344} {\bibfield  {journal} {\bibinfo  {journal} {\mnras}\
  }\textbf {\bibinfo {volume} {458}},\ \bibinfo {pages} {1478} (\bibinfo {year}
  {2016})},\ \Eprint {http://arxiv.org/abs/1510.05656} {arXiv:1510.05656}
  \BibitemShut {NoStop}%
\bibitem [{\citenamefont {{Soergel}}\ \emph {et~al.}(2017)\citenamefont
  {{Soergel}}, \citenamefont {{Giannantonio}}, \citenamefont {{Efstathiou}},
  \citenamefont {{Puchwein}},\ and\ \citenamefont {{Sijacki}}}]{Soergel17}%
  \BibitemOpen
  \bibfield  {author} {\bibinfo {author} {\bibfnamefont {B.}~\bibnamefont
  {{Soergel}}}, \bibinfo {author} {\bibfnamefont {T.}~\bibnamefont
  {{Giannantonio}}}, \bibinfo {author} {\bibfnamefont {G.}~\bibnamefont
  {{Efstathiou}}}, \bibinfo {author} {\bibfnamefont {E.}~\bibnamefont
  {{Puchwein}}}, \ and\ \bibinfo {author} {\bibfnamefont {D.}~\bibnamefont
  {{Sijacki}}},\ }\href {\doibase 10.1093/mnras/stx492} {\bibfield  {journal}
  {\bibinfo  {journal} {\mnras}\ }\textbf {\bibinfo {volume} {468}},\ \bibinfo
  {pages} {577} (\bibinfo {year} {2017})},\ \Eprint
  {http://arxiv.org/abs/1612.06296} {arXiv:1612.06296} \BibitemShut {NoStop}%
\bibitem [{McQ()}]{McQuinn14}%
  \BibitemOpen
  \href@noop {} {\ }\BibitemShut {NoStop}%
\bibitem [{\citenamefont {{David}}\ \emph {et~al.}(1993)\citenamefont
  {{David}}, \citenamefont {{Slyz}}, \citenamefont {{Jones}}, \citenamefont
  {{Forman}}, \citenamefont {{Vrtilek}},\ and\ \citenamefont
  {{Arnaud}}}]{david93}%
  \BibitemOpen
  \bibfield  {author} {\bibinfo {author} {\bibfnamefont {L.~P.}\ \bibnamefont
  {{David}}}, \bibinfo {author} {\bibfnamefont {A.}~\bibnamefont {{Slyz}}},
  \bibinfo {author} {\bibfnamefont {C.}~\bibnamefont {{Jones}}}, \bibinfo
  {author} {\bibfnamefont {W.}~\bibnamefont {{Forman}}}, \bibinfo {author}
  {\bibfnamefont {S.~D.}\ \bibnamefont {{Vrtilek}}}, \ and\ \bibinfo {author}
  {\bibfnamefont {K.~A.}\ \bibnamefont {{Arnaud}}},\ }\href {\doibase
  10.1086/172936} {\bibfield  {journal} {\bibinfo  {journal} {\apj}\ }\textbf
  {\bibinfo {volume} {412}},\ \bibinfo {pages} {479} (\bibinfo {year}
  {1993})}\BibitemShut {NoStop}%
\bibitem [{\citenamefont {Mohr}\ and\ \citenamefont {Evrard}(1997)}]{mohr97}%
  \BibitemOpen
  \bibfield  {author} {\bibinfo {author} {\bibfnamefont {J.~J.}\ \bibnamefont
  {Mohr}}\ and\ \bibinfo {author} {\bibfnamefont {A.~E.}\ \bibnamefont
  {Evrard}},\ }\href@noop {} {\bibfield  {journal} {\bibinfo  {journal} {\apj}\
  }\textbf {\bibinfo {volume} {491}},\ \bibinfo {pages} {38} (\bibinfo {year}
  {1997})}\BibitemShut {NoStop}%
\bibitem [{\citenamefont {Mohr}\ \emph {et~al.}(1999)\citenamefont {Mohr},
  \citenamefont {Mathiesen},\ and\ \citenamefont {Evrard}}]{mohr99}%
  \BibitemOpen
  \bibfield  {author} {\bibinfo {author} {\bibfnamefont {J.~J.}\ \bibnamefont
  {Mohr}}, \bibinfo {author} {\bibfnamefont {B.}~\bibnamefont {Mathiesen}}, \
  and\ \bibinfo {author} {\bibfnamefont {A.~E.}\ \bibnamefont {Evrard}},\
  }\href
  {http://adsabs.harvard.edu/cgi-bin/nph-bib{\_}query?bibcode=1999ApJ...517..627M{\&}db{\_}key=AST}
  {\bibfield  {journal} {\bibinfo  {journal} {\apj}\ }\textbf {\bibinfo
  {volume} {517}},\ \bibinfo {pages} {627} (\bibinfo {year}
  {1999})}\BibitemShut {NoStop}%
\bibitem [{\citenamefont {{Bulbul}}\ \emph {et~al.}(2019)\citenamefont
  {{Bulbul}}, \citenamefont {{Chiu}}, \citenamefont {{Mohr}}, \citenamefont
  {{McDonald}}, \citenamefont {{Benson}}, \citenamefont {{Bautz}},
  \citenamefont {{Bayliss}}, \citenamefont {{Bleem}}, \citenamefont
  {{Brodwin}}, \citenamefont {{Bocquet}}, \citenamefont {{Capasso}},
  \citenamefont {{Dietrich}}, \citenamefont {{Forman}}, \citenamefont
  {{Hlavacek-Larrondo}}, \citenamefont {{Holzapfel}}, \citenamefont
  {{Khullar}}, \citenamefont {{Klein}}, \citenamefont {{Kraft}}, \citenamefont
  {{Miller}}, \citenamefont {{Reichardt}}, \citenamefont {{Saro}},
  \citenamefont {{Sharon}}, \citenamefont {{Stalder}}, \citenamefont
  {{Schrabback}},\ and\ \citenamefont {{Stanford}}}]{bulbul19}%
  \BibitemOpen
  \bibfield  {author} {\bibinfo {author} {\bibfnamefont {E.}~\bibnamefont
  {{Bulbul}}}, \bibinfo {author} {\bibfnamefont {I.-N.}\ \bibnamefont
  {{Chiu}}}, \bibinfo {author} {\bibfnamefont {J.~J.}\ \bibnamefont {{Mohr}}},
  \bibinfo {author} {\bibfnamefont {M.}~\bibnamefont {{McDonald}}}, \bibinfo
  {author} {\bibfnamefont {B.}~\bibnamefont {{Benson}}}, \bibinfo {author}
  {\bibfnamefont {M.~W.}\ \bibnamefont {{Bautz}}}, \bibinfo {author}
  {\bibfnamefont {M.}~\bibnamefont {{Bayliss}}}, \bibinfo {author}
  {\bibfnamefont {L.}~\bibnamefont {{Bleem}}}, \bibinfo {author} {\bibfnamefont
  {M.}~\bibnamefont {{Brodwin}}}, \bibinfo {author} {\bibfnamefont
  {S.}~\bibnamefont {{Bocquet}}}, \bibinfo {author} {\bibfnamefont
  {R.}~\bibnamefont {{Capasso}}}, \bibinfo {author} {\bibfnamefont {J.~P.}\
  \bibnamefont {{Dietrich}}}, \bibinfo {author} {\bibfnamefont
  {B.}~\bibnamefont {{Forman}}}, \bibinfo {author} {\bibfnamefont
  {J.}~\bibnamefont {{Hlavacek-Larrondo}}}, \bibinfo {author} {\bibfnamefont
  {W.~L.}\ \bibnamefont {{Holzapfel}}}, \bibinfo {author} {\bibfnamefont
  {G.}~\bibnamefont {{Khullar}}}, \bibinfo {author} {\bibfnamefont
  {M.}~\bibnamefont {{Klein}}}, \bibinfo {author} {\bibfnamefont
  {R.}~\bibnamefont {{Kraft}}}, \bibinfo {author} {\bibfnamefont {E.~D.}\
  \bibnamefont {{Miller}}}, \bibinfo {author} {\bibfnamefont {C.}~\bibnamefont
  {{Reichardt}}}, \bibinfo {author} {\bibfnamefont {A.}~\bibnamefont {{Saro}}},
  \bibinfo {author} {\bibfnamefont {K.}~\bibnamefont {{Sharon}}}, \bibinfo
  {author} {\bibfnamefont {B.}~\bibnamefont {{Stalder}}}, \bibinfo {author}
  {\bibfnamefont {T.}~\bibnamefont {{Schrabback}}}, \ and\ \bibinfo {author}
  {\bibfnamefont {A.}~\bibnamefont {{Stanford}}},\ }\href {\doibase
  10.3847/1538-4357/aaf230} {\bibfield  {journal} {\bibinfo  {journal} {\apj}\
  }\textbf {\bibinfo {volume} {871}},\ \bibinfo {eid} {50} (\bibinfo {year}
  {2019})},\ \Eprint {http://arxiv.org/abs/1807.02556} {arXiv:1807.02556}
  \BibitemShut {NoStop}%
\bibitem [{\citenamefont {{Cavaliere}}\ \emph {et~al.}(1998)\citenamefont
  {{Cavaliere}}, \citenamefont {{Menci}},\ and\ \citenamefont
  {{Tozzi}}}]{cavaliere98}%
  \BibitemOpen
  \bibfield  {author} {\bibinfo {author} {\bibfnamefont {A.}~\bibnamefont
  {{Cavaliere}}}, \bibinfo {author} {\bibfnamefont {N.}~\bibnamefont
  {{Menci}}}, \ and\ \bibinfo {author} {\bibfnamefont {P.}~\bibnamefont
  {{Tozzi}}},\ }\href {\doibase 10.1086/305839} {\bibfield  {journal} {\bibinfo
   {journal} {\apj}\ }\textbf {\bibinfo {volume} {501}},\ \bibinfo {pages}
  {493} (\bibinfo {year} {1998})},\ \Eprint
  {http://arxiv.org/abs/astro-ph/9802185} {astro-ph/9802185} \BibitemShut
  {NoStop}%
\bibitem [{\citenamefont {{Ponman}}\ \emph {et~al.}(1999)\citenamefont
  {{Ponman}}, \citenamefont {{Cannon}},\ and\ \citenamefont
  {{Navarro}}}]{ponman99}%
  \BibitemOpen
  \bibfield  {author} {\bibinfo {author} {\bibfnamefont {T.~J.}\ \bibnamefont
  {{Ponman}}}, \bibinfo {author} {\bibfnamefont {D.~B.}\ \bibnamefont
  {{Cannon}}}, \ and\ \bibinfo {author} {\bibfnamefont {J.~F.}\ \bibnamefont
  {{Navarro}}},\ }\href {\doibase 10.1038/16410} {\bibfield  {journal}
  {\bibinfo  {journal} {\nat}\ }\textbf {\bibinfo {volume} {397}},\ \bibinfo
  {pages} {135} (\bibinfo {year} {1999})},\ \Eprint
  {http://arxiv.org/abs/astro-ph/9810359} {astro-ph/9810359} \BibitemShut
  {NoStop}%
\bibitem [{\citenamefont {{Cavagnolo}}\ \emph {et~al.}(2010)\citenamefont
  {{Cavagnolo}}, \citenamefont {{McNamara}}, \citenamefont {{Nulsen}},
  \citenamefont {{Carilli}}, \citenamefont {{Jones}},\ and\ \citenamefont
  {{B{\^i}rzan}}}]{cavagnolo10}%
  \BibitemOpen
  \bibfield  {author} {\bibinfo {author} {\bibfnamefont {K.~W.}\ \bibnamefont
  {{Cavagnolo}}}, \bibinfo {author} {\bibfnamefont {B.~R.}\ \bibnamefont
  {{McNamara}}}, \bibinfo {author} {\bibfnamefont {P.~E.~J.}\ \bibnamefont
  {{Nulsen}}}, \bibinfo {author} {\bibfnamefont {C.~L.}\ \bibnamefont
  {{Carilli}}}, \bibinfo {author} {\bibfnamefont {C.}~\bibnamefont {{Jones}}},
  \ and\ \bibinfo {author} {\bibfnamefont {L.}~\bibnamefont {{B{\^i}rzan}}},\
  }\href {\doibase 10.1088/0004-637X/720/2/1066} {\bibfield  {journal}
  {\bibinfo  {journal} {\apj}\ }\textbf {\bibinfo {volume} {720}},\ \bibinfo
  {pages} {1066} (\bibinfo {year} {2010})},\ \Eprint
  {http://arxiv.org/abs/1006.5699} {arXiv:1006.5699} \BibitemShut {NoStop}%
\bibitem [{\citenamefont {Vikhlinin}\ \emph {et~al.}(2009)\citenamefont
  {Vikhlinin}, \citenamefont {Kravtsov}, \citenamefont {Burenin}, \citenamefont
  {Ebeling}, \citenamefont {Forman}, \citenamefont {Hornstrup}, \citenamefont
  {Jones}, \citenamefont {Murray}, \citenamefont {Nagai}, \citenamefont
  {Quintana},\ and\ \citenamefont {Voevodkin}}]{vikhlinin09b}%
  \BibitemOpen
  \bibfield  {author} {\bibinfo {author} {\bibfnamefont {A.}~\bibnamefont
  {Vikhlinin}}, \bibinfo {author} {\bibfnamefont {A.}~\bibnamefont {Kravtsov}},
  \bibinfo {author} {\bibfnamefont {R.}~\bibnamefont {Burenin}}, \bibinfo
  {author} {\bibfnamefont {H.}~\bibnamefont {Ebeling}}, \bibinfo {author}
  {\bibfnamefont {W.}~\bibnamefont {Forman}}, \bibinfo {author} {\bibfnamefont
  {A.}~\bibnamefont {Hornstrup}}, \bibinfo {author} {\bibfnamefont
  {C.}~\bibnamefont {Jones}}, \bibinfo {author} {\bibfnamefont
  {S.}~\bibnamefont {Murray}}, \bibinfo {author} {\bibfnamefont
  {D.}~\bibnamefont {Nagai}}, \bibinfo {author} {\bibfnamefont
  {H.}~\bibnamefont {Quintana}}, \ and\ \bibinfo {author} {\bibfnamefont
  {A.}~\bibnamefont {Voevodkin}},\ }\href {\doibase
  10.1088/0004-637X/692/2/1060} {\bibfield  {journal} {\bibinfo  {journal}
  {\apj}\ }\textbf {\bibinfo {volume} {692}},\ \bibinfo {pages} {1060}
  (\bibinfo {year} {2009})},\ \Eprint {http://arxiv.org/abs/0812.2720}
  {arXiv:0812.2720} \BibitemShut {NoStop}%
\bibitem [{\citenamefont {{Chiu}}\ \emph {et~al.}(2016)\citenamefont {{Chiu}},
  \citenamefont {{Mohr}}, \citenamefont {{McDonald}}, \citenamefont
  {{Bocquet}}, \citenamefont {{Ashby}}, \citenamefont {{Bayliss}},
  \citenamefont {{Benson}}, \citenamefont {{Bleem}}, \citenamefont {{Brodwin}},
  \citenamefont {{Desai}}, \citenamefont {{Dietrich}}, \citenamefont
  {{Forman}}, \citenamefont {{Gangkofner}}, \citenamefont {{Gonzalez}},
  \citenamefont {{Hennig}}, \citenamefont {{Liu}}, \citenamefont {{Reichardt}},
  \citenamefont {{Saro}}, \citenamefont {{Stalder}}, \citenamefont
  {{Stanford}}, \citenamefont {{Song}}, \citenamefont {{Schrabback}},
  \citenamefont {{{\v S}uhada}}, \citenamefont {{Strazzullo}},\ and\
  \citenamefont {{Zenteno}}}]{chiu16b}%
  \BibitemOpen
  \bibfield  {author} {\bibinfo {author} {\bibfnamefont {I.}~\bibnamefont
  {{Chiu}}}, \bibinfo {author} {\bibfnamefont {J.}~\bibnamefont {{Mohr}}},
  \bibinfo {author} {\bibfnamefont {M.}~\bibnamefont {{McDonald}}}, \bibinfo
  {author} {\bibfnamefont {S.}~\bibnamefont {{Bocquet}}}, \bibinfo {author}
  {\bibfnamefont {M.~L.~N.}\ \bibnamefont {{Ashby}}}, \bibinfo {author}
  {\bibfnamefont {M.}~\bibnamefont {{Bayliss}}}, \bibinfo {author}
  {\bibfnamefont {B.~A.}\ \bibnamefont {{Benson}}}, \bibinfo {author}
  {\bibfnamefont {L.~E.}\ \bibnamefont {{Bleem}}}, \bibinfo {author}
  {\bibfnamefont {M.}~\bibnamefont {{Brodwin}}}, \bibinfo {author}
  {\bibfnamefont {S.}~\bibnamefont {{Desai}}}, \bibinfo {author} {\bibfnamefont
  {J.~P.}\ \bibnamefont {{Dietrich}}}, \bibinfo {author} {\bibfnamefont
  {W.~R.}\ \bibnamefont {{Forman}}}, \bibinfo {author} {\bibfnamefont
  {C.}~\bibnamefont {{Gangkofner}}}, \bibinfo {author} {\bibfnamefont {A.~H.}\
  \bibnamefont {{Gonzalez}}}, \bibinfo {author} {\bibfnamefont
  {C.}~\bibnamefont {{Hennig}}}, \bibinfo {author} {\bibfnamefont
  {J.}~\bibnamefont {{Liu}}}, \bibinfo {author} {\bibfnamefont {C.~L.}\
  \bibnamefont {{Reichardt}}}, \bibinfo {author} {\bibfnamefont
  {A.}~\bibnamefont {{Saro}}}, \bibinfo {author} {\bibfnamefont
  {B.}~\bibnamefont {{Stalder}}}, \bibinfo {author} {\bibfnamefont {S.~A.}\
  \bibnamefont {{Stanford}}}, \bibinfo {author} {\bibfnamefont
  {J.}~\bibnamefont {{Song}}}, \bibinfo {author} {\bibfnamefont
  {T.}~\bibnamefont {{Schrabback}}}, \bibinfo {author} {\bibfnamefont
  {R.}~\bibnamefont {{{\v S}uhada}}}, \bibinfo {author} {\bibfnamefont
  {V.}~\bibnamefont {{Strazzullo}}}, \ and\ \bibinfo {author} {\bibfnamefont
  {A.}~\bibnamefont {{Zenteno}}},\ }\href {\doibase 10.1093/mnras/stv2303}
  {\bibfield  {journal} {\bibinfo  {journal} {\mnras}\ }\textbf {\bibinfo
  {volume} {455}},\ \bibinfo {pages} {258} (\bibinfo {year} {2016})},\ \Eprint
  {http://arxiv.org/abs/1412.7823} {arXiv:1412.7823} \BibitemShut {NoStop}%
\bibitem [{\citenamefont {Mantz}\ \emph {et~al.}(2016)\citenamefont {Mantz},
  \citenamefont {Allen}, \citenamefont {Morris},\ and\ \citenamefont
  {Schmidt}}]{mantz16}%
  \BibitemOpen
  \bibfield  {author} {\bibinfo {author} {\bibfnamefont {A.}~\bibnamefont
  {Mantz}}, \bibinfo {author} {\bibfnamefont {S.}~\bibnamefont {Allen}},
  \bibinfo {author} {\bibfnamefont {R.}~\bibnamefont {Morris}}, \ and\ \bibinfo
  {author} {\bibfnamefont {R.}~\bibnamefont {Schmidt}},\ }\href {\doibase
  10.1093/mnras/stv2899} {\bibfield  {journal} {\bibinfo  {journal} {\mnras}\
  }\textbf {\bibinfo {volume} {456}},\ \bibinfo {pages} {4020} (\bibinfo {year}
  {2016})},\ \Eprint {http://arxiv.org/abs/1509.01322} {arXiv:1509.01322}
  \BibitemShut {NoStop}%
\bibitem [{\citenamefont {{Chiu}}\ \emph {et~al.}(2018)\citenamefont {{Chiu}},
  \citenamefont {{Mohr}}, \citenamefont {{McDonald}}, \citenamefont
  {{Bocquet}}, \citenamefont {{Desai}}, \citenamefont {{Klein}}, \citenamefont
  {{Israel}}, \citenamefont {{Ashby}}, \citenamefont {{Stanford}},
  \citenamefont {{Benson}}, \citenamefont {{Brodwin}}, \citenamefont
  {{Abbott}}, \citenamefont {{Abdalla}}, \citenamefont {{Allam}}, \citenamefont
  {{Annis}}, \citenamefont {{Bayliss}}, \citenamefont {{Benoit-L{\'e}vy}},
  \citenamefont {{Bertin}}, \citenamefont {{Bleem}}, \citenamefont {{Brooks}},
  \citenamefont {{Buckley-Geer}}, \citenamefont {{Bulbul}}, \citenamefont
  {{Capasso}}, \citenamefont {{Carlstrom}}, \citenamefont {{Rosell}},
  \citenamefont {{Carretero}}, \citenamefont {{Castander}}, \citenamefont
  {{Cunha}}, \citenamefont {{D'Andrea}}, \citenamefont {{da Costa}},
  \citenamefont {{Davis}}, \citenamefont {{Diehl}}, \citenamefont {{Dietrich}},
  \citenamefont {{Doel}}, \citenamefont {{Drlica-Wagner}}, \citenamefont
  {{Eifler}}, \citenamefont {{Evrard}}, \citenamefont {{Flaugher}},
  \citenamefont {{Garc{\'{\i}}a-Bellido}}, \citenamefont {{Garmire}},
  \citenamefont {{Gaztanaga}}, \citenamefont {{Gerdes}}, \citenamefont
  {{Gonzalez}}, \citenamefont {{Gruen}}, \citenamefont {{Gruendl}},
  \citenamefont {{Gschwend}}, \citenamefont {{Gupta}}, \citenamefont
  {{Gutierrez}}, \citenamefont {{Hlavacek-L}}, \citenamefont {{Honscheid}},
  \citenamefont {{James}}, \citenamefont {{Jeltema}}, \citenamefont {{Kraft}},
  \citenamefont {{Krause}}, \citenamefont {{Kuehn}}, \citenamefont
  {{Kuhlmann}}, \citenamefont {{Kuropatkin}}, \citenamefont {{Lahav}},
  \citenamefont {{Lima}}, \citenamefont {{Maia}}, \citenamefont {{Marshall}},
  \citenamefont {{Melchior}}, \citenamefont {{Menanteau}}, \citenamefont
  {{Miquel}}, \citenamefont {{Murray}}, \citenamefont {{Nord}}, \citenamefont
  {{Ogando}}, \citenamefont {{Plazas}}, \citenamefont {{Rapetti}},
  \citenamefont {{Reichardt}}, \citenamefont {{Romer}}, \citenamefont
  {{Roodman}}, \citenamefont {{Sanchez}}, \citenamefont {{Saro}}, \citenamefont
  {{Scarpine}}, \citenamefont {{Schindler}}, \citenamefont {{Schubnell}},
  \citenamefont {{Sharon}}, \citenamefont {{Smith}}, \citenamefont {{Smith}},
  \citenamefont {{Soares-Santos}}, \citenamefont {{Sobreira}}, \citenamefont
  {{Stalder}}, \citenamefont {{Stern}}, \citenamefont {{Strazzullo}},
  \citenamefont {{Suchyta}}, \citenamefont {{Swanson}}, \citenamefont
  {{Tarle}}, \citenamefont {{Vikram}}, \citenamefont {{Walker}}, \citenamefont
  {{Weller}},\ and\ \citenamefont {{Zhang}}}]{chiu18}%
  \BibitemOpen
  \bibfield  {author} {\bibinfo {author} {\bibfnamefont {I.}~\bibnamefont
  {{Chiu}}}, \bibinfo {author} {\bibfnamefont {J.~J.}\ \bibnamefont {{Mohr}}},
  \bibinfo {author} {\bibfnamefont {M.}~\bibnamefont {{McDonald}}}, \bibinfo
  {author} {\bibfnamefont {S.}~\bibnamefont {{Bocquet}}}, \bibinfo {author}
  {\bibfnamefont {S.}~\bibnamefont {{Desai}}}, \bibinfo {author} {\bibfnamefont
  {M.}~\bibnamefont {{Klein}}}, \bibinfo {author} {\bibfnamefont
  {H.}~\bibnamefont {{Israel}}}, \bibinfo {author} {\bibfnamefont {M.~L.~N.}\
  \bibnamefont {{Ashby}}}, \bibinfo {author} {\bibfnamefont {A.}~\bibnamefont
  {{Stanford}}}, \bibinfo {author} {\bibfnamefont {B.~A.}\ \bibnamefont
  {{Benson}}}, \bibinfo {author} {\bibfnamefont {M.}~\bibnamefont {{Brodwin}}},
  \bibinfo {author} {\bibfnamefont {T.~M.~C.}\ \bibnamefont {{Abbott}}},
  \bibinfo {author} {\bibfnamefont {F.~B.}\ \bibnamefont {{Abdalla}}}, \bibinfo
  {author} {\bibfnamefont {S.}~\bibnamefont {{Allam}}}, \bibinfo {author}
  {\bibfnamefont {J.}~\bibnamefont {{Annis}}}, \bibinfo {author} {\bibfnamefont
  {M.}~\bibnamefont {{Bayliss}}}, \bibinfo {author} {\bibfnamefont
  {A.}~\bibnamefont {{Benoit-L{\'e}vy}}}, \bibinfo {author} {\bibfnamefont
  {E.}~\bibnamefont {{Bertin}}}, \bibinfo {author} {\bibfnamefont
  {L.}~\bibnamefont {{Bleem}}}, \bibinfo {author} {\bibfnamefont
  {D.}~\bibnamefont {{Brooks}}}, \bibinfo {author} {\bibfnamefont
  {E.}~\bibnamefont {{Buckley-Geer}}}, \bibinfo {author} {\bibfnamefont
  {E.}~\bibnamefont {{Bulbul}}}, \bibinfo {author} {\bibfnamefont
  {R.}~\bibnamefont {{Capasso}}}, \bibinfo {author} {\bibfnamefont {J.~E.}\
  \bibnamefont {{Carlstrom}}}, \bibinfo {author} {\bibfnamefont {A.~C.}\
  \bibnamefont {{Rosell}}}, \bibinfo {author} {\bibfnamefont {J.}~\bibnamefont
  {{Carretero}}}, \bibinfo {author} {\bibfnamefont {F.~J.}\ \bibnamefont
  {{Castander}}}, \bibinfo {author} {\bibfnamefont {C.~E.}\ \bibnamefont
  {{Cunha}}}, \bibinfo {author} {\bibfnamefont {C.~B.}\ \bibnamefont
  {{D'Andrea}}}, \bibinfo {author} {\bibfnamefont {L.~N.}\ \bibnamefont {{da
  Costa}}}, \bibinfo {author} {\bibfnamefont {C.}~\bibnamefont {{Davis}}},
  \bibinfo {author} {\bibfnamefont {H.~T.}\ \bibnamefont {{Diehl}}}, \bibinfo
  {author} {\bibfnamefont {J.~P.}\ \bibnamefont {{Dietrich}}}, \bibinfo
  {author} {\bibfnamefont {P.}~\bibnamefont {{Doel}}}, \bibinfo {author}
  {\bibfnamefont {A.}~\bibnamefont {{Drlica-Wagner}}}, \bibinfo {author}
  {\bibfnamefont {T.~F.}\ \bibnamefont {{Eifler}}}, \bibinfo {author}
  {\bibfnamefont {A.~E.}\ \bibnamefont {{Evrard}}}, \bibinfo {author}
  {\bibfnamefont {B.}~\bibnamefont {{Flaugher}}}, \bibinfo {author}
  {\bibfnamefont {J.}~\bibnamefont {{Garc{\'{\i}}a-Bellido}}}, \bibinfo
  {author} {\bibfnamefont {G.}~\bibnamefont {{Garmire}}}, \bibinfo {author}
  {\bibfnamefont {E.}~\bibnamefont {{Gaztanaga}}}, \bibinfo {author}
  {\bibfnamefont {D.~W.}\ \bibnamefont {{Gerdes}}}, \bibinfo {author}
  {\bibfnamefont {A.}~\bibnamefont {{Gonzalez}}}, \bibinfo {author}
  {\bibfnamefont {D.}~\bibnamefont {{Gruen}}}, \bibinfo {author} {\bibfnamefont
  {R.~A.}\ \bibnamefont {{Gruendl}}}, \bibinfo {author} {\bibfnamefont
  {J.}~\bibnamefont {{Gschwend}}}, \bibinfo {author} {\bibfnamefont
  {N.}~\bibnamefont {{Gupta}}}, \bibinfo {author} {\bibfnamefont
  {G.}~\bibnamefont {{Gutierrez}}}, \bibinfo {author} {\bibfnamefont
  {J.}~\bibnamefont {{Hlavacek-L}}}, \bibinfo {author} {\bibfnamefont
  {K.}~\bibnamefont {{Honscheid}}}, \bibinfo {author} {\bibfnamefont {D.~J.}\
  \bibnamefont {{James}}}, \bibinfo {author} {\bibfnamefont {T.}~\bibnamefont
  {{Jeltema}}}, \bibinfo {author} {\bibfnamefont {R.}~\bibnamefont {{Kraft}}},
  \bibinfo {author} {\bibfnamefont {E.}~\bibnamefont {{Krause}}}, \bibinfo
  {author} {\bibfnamefont {K.}~\bibnamefont {{Kuehn}}}, \bibinfo {author}
  {\bibfnamefont {S.}~\bibnamefont {{Kuhlmann}}}, \bibinfo {author}
  {\bibfnamefont {N.}~\bibnamefont {{Kuropatkin}}}, \bibinfo {author}
  {\bibfnamefont {O.}~\bibnamefont {{Lahav}}}, \bibinfo {author} {\bibfnamefont
  {M.}~\bibnamefont {{Lima}}}, \bibinfo {author} {\bibfnamefont {M.~A.~G.}\
  \bibnamefont {{Maia}}}, \bibinfo {author} {\bibfnamefont {J.~L.}\
  \bibnamefont {{Marshall}}}, \bibinfo {author} {\bibfnamefont
  {P.}~\bibnamefont {{Melchior}}}, \bibinfo {author} {\bibfnamefont
  {F.}~\bibnamefont {{Menanteau}}}, \bibinfo {author} {\bibfnamefont
  {R.}~\bibnamefont {{Miquel}}}, \bibinfo {author} {\bibfnamefont
  {S.}~\bibnamefont {{Murray}}}, \bibinfo {author} {\bibfnamefont
  {B.}~\bibnamefont {{Nord}}}, \bibinfo {author} {\bibfnamefont {R.~L.~C.}\
  \bibnamefont {{Ogando}}}, \bibinfo {author} {\bibfnamefont {A.~A.}\
  \bibnamefont {{Plazas}}}, \bibinfo {author} {\bibfnamefont {D.}~\bibnamefont
  {{Rapetti}}}, \bibinfo {author} {\bibfnamefont {C.~L.}\ \bibnamefont
  {{Reichardt}}}, \bibinfo {author} {\bibfnamefont {A.~K.}\ \bibnamefont
  {{Romer}}}, \bibinfo {author} {\bibfnamefont {A.}~\bibnamefont {{Roodman}}},
  \bibinfo {author} {\bibfnamefont {E.}~\bibnamefont {{Sanchez}}}, \bibinfo
  {author} {\bibfnamefont {A.}~\bibnamefont {{Saro}}}, \bibinfo {author}
  {\bibfnamefont {V.}~\bibnamefont {{Scarpine}}}, \bibinfo {author}
  {\bibfnamefont {R.}~\bibnamefont {{Schindler}}}, \bibinfo {author}
  {\bibfnamefont {M.}~\bibnamefont {{Schubnell}}}, \bibinfo {author}
  {\bibfnamefont {K.}~\bibnamefont {{Sharon}}}, \bibinfo {author}
  {\bibfnamefont {R.~C.}\ \bibnamefont {{Smith}}}, \bibinfo {author}
  {\bibfnamefont {M.}~\bibnamefont {{Smith}}}, \bibinfo {author} {\bibfnamefont
  {M.}~\bibnamefont {{Soares-Santos}}}, \bibinfo {author} {\bibfnamefont
  {F.}~\bibnamefont {{Sobreira}}}, \bibinfo {author} {\bibfnamefont
  {B.}~\bibnamefont {{Stalder}}}, \bibinfo {author} {\bibfnamefont
  {C.}~\bibnamefont {{Stern}}}, \bibinfo {author} {\bibfnamefont
  {V.}~\bibnamefont {{Strazzullo}}}, \bibinfo {author} {\bibfnamefont
  {E.}~\bibnamefont {{Suchyta}}}, \bibinfo {author} {\bibfnamefont {M.~E.~C.}\
  \bibnamefont {{Swanson}}}, \bibinfo {author} {\bibfnamefont {G.}~\bibnamefont
  {{Tarle}}}, \bibinfo {author} {\bibfnamefont {V.}~\bibnamefont {{Vikram}}},
  \bibinfo {author} {\bibfnamefont {A.~R.}\ \bibnamefont {{Walker}}}, \bibinfo
  {author} {\bibfnamefont {J.}~\bibnamefont {{Weller}}}, \ and\ \bibinfo
  {author} {\bibfnamefont {Y.}~\bibnamefont {{Zhang}}},\ }\href {\doibase
  10.1093/mnras/sty1284} {\bibfield  {journal} {\bibinfo  {journal} {\mnras}\
  }\textbf {\bibinfo {volume} {478}},\ \bibinfo {pages} {3072} (\bibinfo {year}
  {2018})},\ \Eprint {http://arxiv.org/abs/1711.00917} {arXiv:1711.00917}
  \BibitemShut {NoStop}%
\bibitem [{\citenamefont {{McDonald}}\ \emph {et~al.}(2014)\citenamefont
  {{McDonald}}, \citenamefont {{Benson}}, \citenamefont {{Vikhlinin}},
  \citenamefont {{Aird}}, \citenamefont {{Allen}}, \citenamefont {{Bautz}},
  \citenamefont {{Bayliss}}, \citenamefont {{Bleem}}, \citenamefont
  {{Bocquet}}, \citenamefont {{Brodwin}}, \citenamefont {{Carlstrom}},
  \citenamefont {{Chang}}, \citenamefont {{Cho}}, \citenamefont
  {{Clocchiatti}}, \citenamefont {{Crawford}}, \citenamefont {{Crites}},
  \citenamefont {{de Haan}}, \citenamefont {{Dobbs}}, \citenamefont {{Foley}},
  \citenamefont {{Forman}}, \citenamefont {{George}}, \citenamefont
  {{Gladders}}, \citenamefont {{Gonzalez}}, \citenamefont {{Halverson}},
  \citenamefont {{Hlavacek-Larrondo}}, \citenamefont {{Holder}}, \citenamefont
  {{Holzapfel}}, \citenamefont {{Hrubes}}, \citenamefont {{Jones}},
  \citenamefont {{Keisler}}, \citenamefont {{Knox}}, \citenamefont {{Lee}},
  \citenamefont {{Leitch}}, \citenamefont {{Liu}}, \citenamefont {{Lueker}},
  \citenamefont {{Luong-Van}}, \citenamefont {{Mantz}}, \citenamefont
  {{Marrone}}, \citenamefont {{McMahon}}, \citenamefont {{Meyer}},
  \citenamefont {{Miller}}, \citenamefont {{Mocanu}}, \citenamefont {{Mohr}},
  \citenamefont {{Murray}}, \citenamefont {{Padin}}, \citenamefont {{Pryke}},
  \citenamefont {{Reichardt}}, \citenamefont {{Rest}}, \citenamefont {{Ruhl}},
  \citenamefont {{Saliwanchik}}, \citenamefont {{Saro}}, \citenamefont
  {{Sayre}}, \citenamefont {{Schaffer}}, \citenamefont {{Shirokoff}},
  \citenamefont {{Spieler}}, \citenamefont {{Stalder}}, \citenamefont
  {{Stanford}}, \citenamefont {{Staniszewski}}, \citenamefont {{Stark}},
  \citenamefont {{Story}}, \citenamefont {{Stubbs}}, \citenamefont
  {{Vanderlinde}}, \citenamefont {{Vieira}}, \citenamefont {{Williamson}},
  \citenamefont {{Zahn}},\ and\ \citenamefont {{Zenteno}}}]{mcdonald14}%
  \BibitemOpen
  \bibfield  {author} {\bibinfo {author} {\bibfnamefont {M.}~\bibnamefont
  {{McDonald}}}, \bibinfo {author} {\bibfnamefont {B.~A.}\ \bibnamefont
  {{Benson}}}, \bibinfo {author} {\bibfnamefont {A.}~\bibnamefont
  {{Vikhlinin}}}, \bibinfo {author} {\bibfnamefont {K.~A.}\ \bibnamefont
  {{Aird}}}, \bibinfo {author} {\bibfnamefont {S.~W.}\ \bibnamefont {{Allen}}},
  \bibinfo {author} {\bibfnamefont {M.}~\bibnamefont {{Bautz}}}, \bibinfo
  {author} {\bibfnamefont {M.}~\bibnamefont {{Bayliss}}}, \bibinfo {author}
  {\bibfnamefont {L.~E.}\ \bibnamefont {{Bleem}}}, \bibinfo {author}
  {\bibfnamefont {S.}~\bibnamefont {{Bocquet}}}, \bibinfo {author}
  {\bibfnamefont {M.}~\bibnamefont {{Brodwin}}}, \bibinfo {author}
  {\bibfnamefont {J.~E.}\ \bibnamefont {{Carlstrom}}}, \bibinfo {author}
  {\bibfnamefont {C.~L.}\ \bibnamefont {{Chang}}}, \bibinfo {author}
  {\bibfnamefont {H.~M.}\ \bibnamefont {{Cho}}}, \bibinfo {author}
  {\bibfnamefont {A.}~\bibnamefont {{Clocchiatti}}}, \bibinfo {author}
  {\bibfnamefont {T.~M.}\ \bibnamefont {{Crawford}}}, \bibinfo {author}
  {\bibfnamefont {A.~T.}\ \bibnamefont {{Crites}}}, \bibinfo {author}
  {\bibfnamefont {T.}~\bibnamefont {{de Haan}}}, \bibinfo {author}
  {\bibfnamefont {M.~A.}\ \bibnamefont {{Dobbs}}}, \bibinfo {author}
  {\bibfnamefont {R.~J.}\ \bibnamefont {{Foley}}}, \bibinfo {author}
  {\bibfnamefont {W.~R.}\ \bibnamefont {{Forman}}}, \bibinfo {author}
  {\bibfnamefont {E.~M.}\ \bibnamefont {{George}}}, \bibinfo {author}
  {\bibfnamefont {M.~D.}\ \bibnamefont {{Gladders}}}, \bibinfo {author}
  {\bibfnamefont {A.~H.}\ \bibnamefont {{Gonzalez}}}, \bibinfo {author}
  {\bibfnamefont {N.~W.}\ \bibnamefont {{Halverson}}}, \bibinfo {author}
  {\bibfnamefont {J.}~\bibnamefont {{Hlavacek-Larrondo}}}, \bibinfo {author}
  {\bibfnamefont {G.~P.}\ \bibnamefont {{Holder}}}, \bibinfo {author}
  {\bibfnamefont {W.~L.}\ \bibnamefont {{Holzapfel}}}, \bibinfo {author}
  {\bibfnamefont {J.~D.}\ \bibnamefont {{Hrubes}}}, \bibinfo {author}
  {\bibfnamefont {C.}~\bibnamefont {{Jones}}}, \bibinfo {author} {\bibfnamefont
  {R.}~\bibnamefont {{Keisler}}}, \bibinfo {author} {\bibfnamefont
  {L.}~\bibnamefont {{Knox}}}, \bibinfo {author} {\bibfnamefont {A.~T.}\
  \bibnamefont {{Lee}}}, \bibinfo {author} {\bibfnamefont {E.~M.}\ \bibnamefont
  {{Leitch}}}, \bibinfo {author} {\bibfnamefont {J.}~\bibnamefont {{Liu}}},
  \bibinfo {author} {\bibfnamefont {M.}~\bibnamefont {{Lueker}}}, \bibinfo
  {author} {\bibfnamefont {D.}~\bibnamefont {{Luong-Van}}}, \bibinfo {author}
  {\bibfnamefont {A.}~\bibnamefont {{Mantz}}}, \bibinfo {author} {\bibfnamefont
  {D.~P.}\ \bibnamefont {{Marrone}}}, \bibinfo {author} {\bibfnamefont {J.~J.}\
  \bibnamefont {{McMahon}}}, \bibinfo {author} {\bibfnamefont {S.~S.}\
  \bibnamefont {{Meyer}}}, \bibinfo {author} {\bibfnamefont {E.~D.}\
  \bibnamefont {{Miller}}}, \bibinfo {author} {\bibfnamefont {L.}~\bibnamefont
  {{Mocanu}}}, \bibinfo {author} {\bibfnamefont {J.~J.}\ \bibnamefont
  {{Mohr}}}, \bibinfo {author} {\bibfnamefont {S.~S.}\ \bibnamefont
  {{Murray}}}, \bibinfo {author} {\bibfnamefont {S.}~\bibnamefont {{Padin}}},
  \bibinfo {author} {\bibfnamefont {C.}~\bibnamefont {{Pryke}}}, \bibinfo
  {author} {\bibfnamefont {C.~L.}\ \bibnamefont {{Reichardt}}}, \bibinfo
  {author} {\bibfnamefont {A.}~\bibnamefont {{Rest}}}, \bibinfo {author}
  {\bibfnamefont {J.~E.}\ \bibnamefont {{Ruhl}}}, \bibinfo {author}
  {\bibfnamefont {B.~R.}\ \bibnamefont {{Saliwanchik}}}, \bibinfo {author}
  {\bibfnamefont {A.}~\bibnamefont {{Saro}}}, \bibinfo {author} {\bibfnamefont
  {J.~T.}\ \bibnamefont {{Sayre}}}, \bibinfo {author} {\bibfnamefont {K.~K.}\
  \bibnamefont {{Schaffer}}}, \bibinfo {author} {\bibfnamefont
  {E.}~\bibnamefont {{Shirokoff}}}, \bibinfo {author} {\bibfnamefont {H.~G.}\
  \bibnamefont {{Spieler}}}, \bibinfo {author} {\bibfnamefont {B.}~\bibnamefont
  {{Stalder}}}, \bibinfo {author} {\bibfnamefont {S.~A.}\ \bibnamefont
  {{Stanford}}}, \bibinfo {author} {\bibfnamefont {Z.}~\bibnamefont
  {{Staniszewski}}}, \bibinfo {author} {\bibfnamefont {A.~A.}\ \bibnamefont
  {{Stark}}}, \bibinfo {author} {\bibfnamefont {K.~T.}\ \bibnamefont
  {{Story}}}, \bibinfo {author} {\bibfnamefont {C.~W.}\ \bibnamefont
  {{Stubbs}}}, \bibinfo {author} {\bibfnamefont {K.}~\bibnamefont
  {{Vanderlinde}}}, \bibinfo {author} {\bibfnamefont {J.~D.}\ \bibnamefont
  {{Vieira}}}, \bibinfo {author} {\bibfnamefont {R.}~\bibnamefont
  {{Williamson}}}, \bibinfo {author} {\bibfnamefont {O.}~\bibnamefont
  {{Zahn}}}, \ and\ \bibinfo {author} {\bibfnamefont {A.}~\bibnamefont
  {{Zenteno}}},\ }\href {\doibase 10.1088/0004-637X/794/1/67} {\bibfield
  {journal} {\bibinfo  {journal} {\apj}\ }\textbf {\bibinfo {volume} {794}},\
  \bibinfo {eid} {67} (\bibinfo {year} {2014})},\ \Eprint
  {http://arxiv.org/abs/1404.6250} {arXiv:1404.6250 [astro-ph.HE]} \BibitemShut
  {NoStop}%
\bibitem [{\citenamefont {{Barnes}}\ \emph {et~al.}(2017)\citenamefont
  {{Barnes}}, \citenamefont {{Kay}}, \citenamefont {{Bah{\'e}}}, \citenamefont
  {{Dalla Vecchia}}, \citenamefont {{McCarthy}}, \citenamefont {{Schaye}},
  \citenamefont {{Bower}}, \citenamefont {{Jenkins}}, \citenamefont {{Thomas}},
  \citenamefont {{Schaller}}, \citenamefont {{Crain}}, \citenamefont
  {{Theuns}},\ and\ \citenamefont {{White}}}]{barnes17}%
  \BibitemOpen
  \bibfield  {author} {\bibinfo {author} {\bibfnamefont {D.~J.}\ \bibnamefont
  {{Barnes}}}, \bibinfo {author} {\bibfnamefont {S.~T.}\ \bibnamefont {{Kay}}},
  \bibinfo {author} {\bibfnamefont {Y.~M.}\ \bibnamefont {{Bah{\'e}}}},
  \bibinfo {author} {\bibfnamefont {C.}~\bibnamefont {{Dalla Vecchia}}},
  \bibinfo {author} {\bibfnamefont {I.~G.}\ \bibnamefont {{McCarthy}}},
  \bibinfo {author} {\bibfnamefont {J.}~\bibnamefont {{Schaye}}}, \bibinfo
  {author} {\bibfnamefont {R.~G.}\ \bibnamefont {{Bower}}}, \bibinfo {author}
  {\bibfnamefont {A.}~\bibnamefont {{Jenkins}}}, \bibinfo {author}
  {\bibfnamefont {P.~A.}\ \bibnamefont {{Thomas}}}, \bibinfo {author}
  {\bibfnamefont {M.}~\bibnamefont {{Schaller}}}, \bibinfo {author}
  {\bibfnamefont {R.~A.}\ \bibnamefont {{Crain}}}, \bibinfo {author}
  {\bibfnamefont {T.}~\bibnamefont {{Theuns}}}, \ and\ \bibinfo {author}
  {\bibfnamefont {S.~D.~M.}\ \bibnamefont {{White}}},\ }\href {\doibase
  10.1093/mnras/stx1647} {\bibfield  {journal} {\bibinfo  {journal} {\mnras}\
  }\textbf {\bibinfo {volume} {471}},\ \bibinfo {pages} {1088} (\bibinfo {year}
  {2017})},\ \Eprint {http://arxiv.org/abs/1703.10907} {arXiv:1703.10907}
  \BibitemShut {NoStop}%
\bibitem [{\citenamefont {Pratt}\ \emph {et~al.}(2009)\citenamefont {Pratt},
  \citenamefont {Croston}, \citenamefont {Arnaud},\ and\ \citenamefont
  {B{\"{o}}hringer}}]{pratt09}%
  \BibitemOpen
  \bibfield  {author} {\bibinfo {author} {\bibfnamefont {G.}~\bibnamefont
  {Pratt}}, \bibinfo {author} {\bibfnamefont {J.}~\bibnamefont {Croston}},
  \bibinfo {author} {\bibfnamefont {M.}~\bibnamefont {Arnaud}}, \ and\ \bibinfo
  {author} {\bibfnamefont {H.}~\bibnamefont {B{\"{o}}hringer}},\ }\href
  {\doibase 10.1051/0004-6361/200810994} {\bibfield  {journal} {\bibinfo
  {journal} {\aap}\ }\textbf {\bibinfo {volume} {498}},\ \bibinfo {pages} {361}
  (\bibinfo {year} {2009})},\ \Eprint {http://arxiv.org/abs/0809.3784}
  {arXiv:0809.3784} \BibitemShut {NoStop}%
\bibitem [{\citenamefont {Bleem}\ \emph {et~al.}(2015)\citenamefont {Bleem},
  \citenamefont {Stalder}, \citenamefont {de~Haan}, \citenamefont {Aird},
  \citenamefont {Allen}, \citenamefont {Applegate}, \citenamefont {Ashby},
  \citenamefont {Bautz}, \citenamefont {Bayliss}, \citenamefont {Benson},
  \citenamefont {Bocquet}, \citenamefont {Brodwin}, \citenamefont {Carlstrom},
  \citenamefont {Chang}, \citenamefont {Chiu}, \citenamefont {Cho},
  \citenamefont {Clocchiatti}, \citenamefont {Crawford}, \citenamefont
  {Crites}, \citenamefont {Desai}, \citenamefont {Dietrich}, \citenamefont
  {Dobbs}, \citenamefont {Foley}, \citenamefont {Forman}, \citenamefont
  {George}, \citenamefont {Gladders}, \citenamefont {Gonzalez}, \citenamefont
  {Halverson}, \citenamefont {Hennig}, \citenamefont {Hoekstra}, \citenamefont
  {Holder}, \citenamefont {Holzapfel}, \citenamefont {Hrubes}, \citenamefont
  {Jones}, \citenamefont {Keisler}, \citenamefont {Knox}, \citenamefont {Lee},
  \citenamefont {Leitch}, \citenamefont {Liu}, \citenamefont {Lueker},
  \citenamefont {Luong-Van}, \citenamefont {Mantz}, \citenamefont {Marrone},
  \citenamefont {McDonald}, \citenamefont {McMahon}, \citenamefont {Meyer},
  \citenamefont {Mocanu}, \citenamefont {Mohr}, \citenamefont {Murray},
  \citenamefont {Padin}, \citenamefont {Pryke}, \citenamefont {Reichardt},
  \citenamefont {Rest}, \citenamefont {Ruel}, \citenamefont {Ruhl},
  \citenamefont {Saliwanchik}, \citenamefont {Saro}, \citenamefont {Sayre},
  \citenamefont {Schaffer}, \citenamefont {Schrabback}, \citenamefont
  {Shirokoff}, \citenamefont {Song}, \citenamefont {Spieler}, \citenamefont
  {Stanford}, \citenamefont {Staniszewski}, \citenamefont {Stark},
  \citenamefont {Story}, \citenamefont {Stubbs}, \citenamefont {Vanderlinde},
  \citenamefont {Vieira}, \citenamefont {Vikhlinin}, \citenamefont
  {Williamson}, \citenamefont {Zahn},\ and\ \citenamefont {Zenteno}}]{bleem15}%
  \BibitemOpen
  \bibfield  {author} {\bibinfo {author} {\bibfnamefont {L.}~\bibnamefont
  {Bleem}}, \bibinfo {author} {\bibfnamefont {B.}~\bibnamefont {Stalder}},
  \bibinfo {author} {\bibfnamefont {T.}~\bibnamefont {de~Haan}}, \bibinfo
  {author} {\bibfnamefont {K.}~\bibnamefont {Aird}}, \bibinfo {author}
  {\bibfnamefont {S.}~\bibnamefont {Allen}}, \bibinfo {author} {\bibfnamefont
  {D.}~\bibnamefont {Applegate}}, \bibinfo {author} {\bibfnamefont
  {M.}~\bibnamefont {Ashby}}, \bibinfo {author} {\bibfnamefont
  {M.}~\bibnamefont {Bautz}}, \bibinfo {author} {\bibfnamefont
  {M.}~\bibnamefont {Bayliss}}, \bibinfo {author} {\bibfnamefont
  {B.}~\bibnamefont {Benson}}, \bibinfo {author} {\bibfnamefont
  {S.}~\bibnamefont {Bocquet}}, \bibinfo {author} {\bibfnamefont
  {M.}~\bibnamefont {Brodwin}}, \bibinfo {author} {\bibfnamefont
  {J.}~\bibnamefont {Carlstrom}}, \bibinfo {author} {\bibfnamefont
  {C.}~\bibnamefont {Chang}}, \bibinfo {author} {\bibfnamefont
  {I.}~\bibnamefont {Chiu}}, \bibinfo {author} {\bibfnamefont {H.}~\bibnamefont
  {Cho}}, \bibinfo {author} {\bibfnamefont {A.}~\bibnamefont {Clocchiatti}},
  \bibinfo {author} {\bibfnamefont {T.}~\bibnamefont {Crawford}}, \bibinfo
  {author} {\bibfnamefont {A.}~\bibnamefont {Crites}}, \bibinfo {author}
  {\bibfnamefont {S.}~\bibnamefont {Desai}}, \bibinfo {author} {\bibfnamefont
  {J.}~\bibnamefont {Dietrich}}, \bibinfo {author} {\bibfnamefont
  {M.}~\bibnamefont {Dobbs}}, \bibinfo {author} {\bibfnamefont
  {R.}~\bibnamefont {Foley}}, \bibinfo {author} {\bibfnamefont
  {W.}~\bibnamefont {Forman}}, \bibinfo {author} {\bibfnamefont
  {E.}~\bibnamefont {George}}, \bibinfo {author} {\bibfnamefont
  {M.}~\bibnamefont {Gladders}}, \bibinfo {author} {\bibfnamefont
  {A.}~\bibnamefont {Gonzalez}}, \bibinfo {author} {\bibfnamefont
  {N.}~\bibnamefont {Halverson}}, \bibinfo {author} {\bibfnamefont
  {C.}~\bibnamefont {Hennig}}, \bibinfo {author} {\bibfnamefont
  {H.}~\bibnamefont {Hoekstra}}, \bibinfo {author} {\bibfnamefont
  {G.}~\bibnamefont {Holder}}, \bibinfo {author} {\bibfnamefont
  {W.}~\bibnamefont {Holzapfel}}, \bibinfo {author} {\bibfnamefont
  {J.}~\bibnamefont {Hrubes}}, \bibinfo {author} {\bibfnamefont
  {C.}~\bibnamefont {Jones}}, \bibinfo {author} {\bibfnamefont
  {R.}~\bibnamefont {Keisler}}, \bibinfo {author} {\bibfnamefont
  {L.}~\bibnamefont {Knox}}, \bibinfo {author} {\bibfnamefont {A.}~\bibnamefont
  {Lee}}, \bibinfo {author} {\bibfnamefont {E.}~\bibnamefont {Leitch}},
  \bibinfo {author} {\bibfnamefont {J.}~\bibnamefont {Liu}}, \bibinfo {author}
  {\bibfnamefont {M.}~\bibnamefont {Lueker}}, \bibinfo {author} {\bibfnamefont
  {D.}~\bibnamefont {Luong-Van}}, \bibinfo {author} {\bibfnamefont
  {A.}~\bibnamefont {Mantz}}, \bibinfo {author} {\bibfnamefont
  {D.}~\bibnamefont {Marrone}}, \bibinfo {author} {\bibfnamefont
  {M.}~\bibnamefont {McDonald}}, \bibinfo {author} {\bibfnamefont
  {J.}~\bibnamefont {McMahon}}, \bibinfo {author} {\bibfnamefont
  {S.}~\bibnamefont {Meyer}}, \bibinfo {author} {\bibfnamefont
  {L.}~\bibnamefont {Mocanu}}, \bibinfo {author} {\bibfnamefont
  {J.}~\bibnamefont {Mohr}}, \bibinfo {author} {\bibfnamefont {S.}~\bibnamefont
  {Murray}}, \bibinfo {author} {\bibfnamefont {S.}~\bibnamefont {Padin}},
  \bibinfo {author} {\bibfnamefont {C.}~\bibnamefont {Pryke}}, \bibinfo
  {author} {\bibfnamefont {C.}~\bibnamefont {Reichardt}}, \bibinfo {author}
  {\bibfnamefont {A.}~\bibnamefont {Rest}}, \bibinfo {author} {\bibfnamefont
  {J.}~\bibnamefont {Ruel}}, \bibinfo {author} {\bibfnamefont {J.}~\bibnamefont
  {Ruhl}}, \bibinfo {author} {\bibfnamefont {B.}~\bibnamefont {Saliwanchik}},
  \bibinfo {author} {\bibfnamefont {A.}~\bibnamefont {Saro}}, \bibinfo {author}
  {\bibfnamefont {J.}~\bibnamefont {Sayre}}, \bibinfo {author} {\bibfnamefont
  {K.}~\bibnamefont {Schaffer}}, \bibinfo {author} {\bibfnamefont
  {T.}~\bibnamefont {Schrabback}}, \bibinfo {author} {\bibfnamefont
  {E.}~\bibnamefont {Shirokoff}}, \bibinfo {author} {\bibfnamefont
  {J.}~\bibnamefont {Song}}, \bibinfo {author} {\bibfnamefont {H.}~\bibnamefont
  {Spieler}}, \bibinfo {author} {\bibfnamefont {S.}~\bibnamefont {Stanford}},
  \bibinfo {author} {\bibfnamefont {Z.}~\bibnamefont {Staniszewski}}, \bibinfo
  {author} {\bibfnamefont {A.}~\bibnamefont {Stark}}, \bibinfo {author}
  {\bibfnamefont {K.}~\bibnamefont {Story}}, \bibinfo {author} {\bibfnamefont
  {C.}~\bibnamefont {Stubbs}}, \bibinfo {author} {\bibfnamefont
  {K.}~\bibnamefont {Vanderlinde}}, \bibinfo {author} {\bibfnamefont
  {J.}~\bibnamefont {Vieira}}, \bibinfo {author} {\bibfnamefont
  {A.}~\bibnamefont {Vikhlinin}}, \bibinfo {author} {\bibfnamefont
  {R.}~\bibnamefont {Williamson}}, \bibinfo {author} {\bibfnamefont
  {O.}~\bibnamefont {Zahn}}, \ and\ \bibinfo {author} {\bibfnamefont
  {A.}~\bibnamefont {Zenteno}},\ }\href {\doibase 10.1088/0067-0049/216/2/27}
  {\bibfield  {journal} {\bibinfo  {journal} {\apjs}\ }\textbf {\bibinfo
  {volume} {216}},\ \bibinfo {pages} {27} (\bibinfo {year} {2015})},\ \Eprint
  {http://arxiv.org/abs/1409.0850} {arXiv:1409.0850} \BibitemShut {NoStop}%
\bibitem [{\citenamefont {{Planck Collaboration}}\ \emph
  {et~al.}(2016{\natexlab{d}})\citenamefont {{Planck Collaboration}},
  \citenamefont {{Ade}}, \citenamefont {{Aghanim}}, \citenamefont {{Arnaud}},
  \citenamefont {{Ashdown}}, \citenamefont {{Aumont}}, \citenamefont
  {{Baccigalupi}}, \citenamefont {{Banday}}, \citenamefont {{Barreiro}},
  \citenamefont {{Barrena}},\ and\ \citenamefont {et~al.}}]{PlanckSZ16}%
  \BibitemOpen
  \bibfield  {author} {\bibinfo {author} {\bibnamefont {{Planck
  Collaboration}}}, \bibinfo {author} {\bibfnamefont {P.~A.~R.}\ \bibnamefont
  {{Ade}}}, \bibinfo {author} {\bibfnamefont {N.}~\bibnamefont {{Aghanim}}},
  \bibinfo {author} {\bibfnamefont {M.}~\bibnamefont {{Arnaud}}}, \bibinfo
  {author} {\bibfnamefont {M.}~\bibnamefont {{Ashdown}}}, \bibinfo {author}
  {\bibfnamefont {J.}~\bibnamefont {{Aumont}}}, \bibinfo {author}
  {\bibfnamefont {C.}~\bibnamefont {{Baccigalupi}}}, \bibinfo {author}
  {\bibfnamefont {A.~J.}\ \bibnamefont {{Banday}}}, \bibinfo {author}
  {\bibfnamefont {R.~B.}\ \bibnamefont {{Barreiro}}}, \bibinfo {author}
  {\bibfnamefont {R.}~\bibnamefont {{Barrena}}}, \ and\ \bibinfo {author}
  {\bibnamefont {et~al.}},\ }\href {\doibase 10.1051/0004-6361/201525823}
  {\bibfield  {journal} {\bibinfo  {journal} {\aap}\ }\textbf {\bibinfo
  {volume} {594}},\ \bibinfo {eid} {A27} (\bibinfo {year}
  {2016}{\natexlab{d}})},\ \Eprint {http://arxiv.org/abs/1502.01598}
  {arXiv:1502.01598} \BibitemShut {NoStop}%
\bibitem [{\citenamefont {{Rykoff}}\ \emph {et~al.}(2016)\citenamefont
  {{Rykoff}}, \citenamefont {{Rozo}}, \citenamefont {{Hollowood}},
  \citenamefont {{Bermeo-Hernandez}}, \citenamefont {{Jeltema}}, \citenamefont
  {{Mayers}}, \citenamefont {{Romer}}, \citenamefont {{Rooney}}, \citenamefont
  {{Saro}}, \citenamefont {{Vergara Cervantes}}, \citenamefont {{Wechsler}},
  \citenamefont {{Wilcox}}, \citenamefont {{Abbott}}, \citenamefont
  {{Abdalla}}, \citenamefont {{Allam}}, \citenamefont {{Annis}}, \citenamefont
  {{Benoit-L{\'e}vy}}, \citenamefont {{Bernstein}}, \citenamefont {{Bertin}},
  \citenamefont {{Brooks}}, \citenamefont {{Burke}}, \citenamefont {{Capozzi}},
  \citenamefont {{Carnero Rosell}}, \citenamefont {{Carrasco Kind}},
  \citenamefont {{Castander}}, \citenamefont {{Childress}}, \citenamefont
  {{Collins}}, \citenamefont {{Cunha}}, \citenamefont {{D'Andrea}},
  \citenamefont {{da Costa}}, \citenamefont {{Davis}}, \citenamefont {{Desai}},
  \citenamefont {{Diehl}}, \citenamefont {{Dietrich}}, \citenamefont {{Doel}},
  \citenamefont {{Evrard}}, \citenamefont {{Finley}}, \citenamefont
  {{Flaugher}}, \citenamefont {{Fosalba}}, \citenamefont {{Frieman}},
  \citenamefont {{Glazebrook}}, \citenamefont {{Goldstein}}, \citenamefont
  {{Gruen}}, \citenamefont {{Gruendl}}, \citenamefont {{Gutierrez}},
  \citenamefont {{Hilton}}, \citenamefont {{Honscheid}}, \citenamefont
  {{Hoyle}}, \citenamefont {{James}}, \citenamefont {{Kay}}, \citenamefont
  {{Kuehn}}, \citenamefont {{Kuropatkin}}, \citenamefont {{Lahav}},
  \citenamefont {{Lewis}}, \citenamefont {{Lidman}}, \citenamefont {{Lima}},
  \citenamefont {{Maia}}, \citenamefont {{Mann}}, \citenamefont {{Marshall}},
  \citenamefont {{Martini}}, \citenamefont {{Melchior}}, \citenamefont
  {{Miller}}, \citenamefont {{Miquel}}, \citenamefont {{Mohr}}, \citenamefont
  {{Nichol}}, \citenamefont {{Nord}}, \citenamefont {{Ogando}}, \citenamefont
  {{Plazas}}, \citenamefont {{Reil}}, \citenamefont {{Sahl{\'e}n}},
  \citenamefont {{Sanchez}}, \citenamefont {{Santiago}}, \citenamefont
  {{Scarpine}}, \citenamefont {{Schubnell}}, \citenamefont {{Sevilla-Noarbe}},
  \citenamefont {{Smith}}, \citenamefont {{Soares-Santos}}, \citenamefont
  {{Sobreira}}, \citenamefont {{Stott}}, \citenamefont {{Suchyta}},
  \citenamefont {{Swanson}}, \citenamefont {{Tarle}}, \citenamefont {{Thomas}},
  \citenamefont {{Tucker}}, \citenamefont {{Uddin}}, \citenamefont {{Viana}},
  \citenamefont {{Vikram}}, \citenamefont {{Walker}}, \citenamefont {{Zhang}},\
  and\ \citenamefont {{DES Collaboration}}}]{rykoff16}%
  \BibitemOpen
  \bibfield  {author} {\bibinfo {author} {\bibfnamefont {E.~S.}\ \bibnamefont
  {{Rykoff}}}, \bibinfo {author} {\bibfnamefont {E.}~\bibnamefont {{Rozo}}},
  \bibinfo {author} {\bibfnamefont {D.}~\bibnamefont {{Hollowood}}}, \bibinfo
  {author} {\bibfnamefont {A.}~\bibnamefont {{Bermeo-Hernandez}}}, \bibinfo
  {author} {\bibfnamefont {T.}~\bibnamefont {{Jeltema}}}, \bibinfo {author}
  {\bibfnamefont {J.}~\bibnamefont {{Mayers}}}, \bibinfo {author}
  {\bibfnamefont {A.~K.}\ \bibnamefont {{Romer}}}, \bibinfo {author}
  {\bibfnamefont {P.}~\bibnamefont {{Rooney}}}, \bibinfo {author}
  {\bibfnamefont {A.}~\bibnamefont {{Saro}}}, \bibinfo {author} {\bibfnamefont
  {C.}~\bibnamefont {{Vergara Cervantes}}}, \bibinfo {author} {\bibfnamefont
  {R.~H.}\ \bibnamefont {{Wechsler}}}, \bibinfo {author} {\bibfnamefont
  {H.}~\bibnamefont {{Wilcox}}}, \bibinfo {author} {\bibfnamefont {T.~M.~C.}\
  \bibnamefont {{Abbott}}}, \bibinfo {author} {\bibfnamefont {F.~B.}\
  \bibnamefont {{Abdalla}}}, \bibinfo {author} {\bibfnamefont {S.}~\bibnamefont
  {{Allam}}}, \bibinfo {author} {\bibfnamefont {J.}~\bibnamefont {{Annis}}},
  \bibinfo {author} {\bibfnamefont {A.}~\bibnamefont {{Benoit-L{\'e}vy}}},
  \bibinfo {author} {\bibfnamefont {G.~M.}\ \bibnamefont {{Bernstein}}},
  \bibinfo {author} {\bibfnamefont {E.}~\bibnamefont {{Bertin}}}, \bibinfo
  {author} {\bibfnamefont {D.}~\bibnamefont {{Brooks}}}, \bibinfo {author}
  {\bibfnamefont {D.~L.}\ \bibnamefont {{Burke}}}, \bibinfo {author}
  {\bibfnamefont {D.}~\bibnamefont {{Capozzi}}}, \bibinfo {author}
  {\bibfnamefont {A.}~\bibnamefont {{Carnero Rosell}}}, \bibinfo {author}
  {\bibfnamefont {M.}~\bibnamefont {{Carrasco Kind}}}, \bibinfo {author}
  {\bibfnamefont {F.~J.}\ \bibnamefont {{Castander}}}, \bibinfo {author}
  {\bibfnamefont {M.}~\bibnamefont {{Childress}}}, \bibinfo {author}
  {\bibfnamefont {C.~A.}\ \bibnamefont {{Collins}}}, \bibinfo {author}
  {\bibfnamefont {C.~E.}\ \bibnamefont {{Cunha}}}, \bibinfo {author}
  {\bibfnamefont {C.~B.}\ \bibnamefont {{D'Andrea}}}, \bibinfo {author}
  {\bibfnamefont {L.~N.}\ \bibnamefont {{da Costa}}}, \bibinfo {author}
  {\bibfnamefont {T.~M.}\ \bibnamefont {{Davis}}}, \bibinfo {author}
  {\bibfnamefont {S.}~\bibnamefont {{Desai}}}, \bibinfo {author} {\bibfnamefont
  {H.~T.}\ \bibnamefont {{Diehl}}}, \bibinfo {author} {\bibfnamefont {J.~P.}\
  \bibnamefont {{Dietrich}}}, \bibinfo {author} {\bibfnamefont
  {P.}~\bibnamefont {{Doel}}}, \bibinfo {author} {\bibfnamefont {A.~E.}\
  \bibnamefont {{Evrard}}}, \bibinfo {author} {\bibfnamefont {D.~A.}\
  \bibnamefont {{Finley}}}, \bibinfo {author} {\bibfnamefont {B.}~\bibnamefont
  {{Flaugher}}}, \bibinfo {author} {\bibfnamefont {P.}~\bibnamefont
  {{Fosalba}}}, \bibinfo {author} {\bibfnamefont {J.}~\bibnamefont
  {{Frieman}}}, \bibinfo {author} {\bibfnamefont {K.}~\bibnamefont
  {{Glazebrook}}}, \bibinfo {author} {\bibfnamefont {D.~A.}\ \bibnamefont
  {{Goldstein}}}, \bibinfo {author} {\bibfnamefont {D.}~\bibnamefont
  {{Gruen}}}, \bibinfo {author} {\bibfnamefont {R.~A.}\ \bibnamefont
  {{Gruendl}}}, \bibinfo {author} {\bibfnamefont {G.}~\bibnamefont
  {{Gutierrez}}}, \bibinfo {author} {\bibfnamefont {M.}~\bibnamefont
  {{Hilton}}}, \bibinfo {author} {\bibfnamefont {K.}~\bibnamefont
  {{Honscheid}}}, \bibinfo {author} {\bibfnamefont {B.}~\bibnamefont
  {{Hoyle}}}, \bibinfo {author} {\bibfnamefont {D.~J.}\ \bibnamefont
  {{James}}}, \bibinfo {author} {\bibfnamefont {S.~T.}\ \bibnamefont {{Kay}}},
  \bibinfo {author} {\bibfnamefont {K.}~\bibnamefont {{Kuehn}}}, \bibinfo
  {author} {\bibfnamefont {N.}~\bibnamefont {{Kuropatkin}}}, \bibinfo {author}
  {\bibfnamefont {O.}~\bibnamefont {{Lahav}}}, \bibinfo {author} {\bibfnamefont
  {G.~F.}\ \bibnamefont {{Lewis}}}, \bibinfo {author} {\bibfnamefont
  {C.}~\bibnamefont {{Lidman}}}, \bibinfo {author} {\bibfnamefont
  {M.}~\bibnamefont {{Lima}}}, \bibinfo {author} {\bibfnamefont {M.~A.~G.}\
  \bibnamefont {{Maia}}}, \bibinfo {author} {\bibfnamefont {R.~G.}\
  \bibnamefont {{Mann}}}, \bibinfo {author} {\bibfnamefont {J.~L.}\
  \bibnamefont {{Marshall}}}, \bibinfo {author} {\bibfnamefont
  {P.}~\bibnamefont {{Martini}}}, \bibinfo {author} {\bibfnamefont
  {P.}~\bibnamefont {{Melchior}}}, \bibinfo {author} {\bibfnamefont {C.~J.}\
  \bibnamefont {{Miller}}}, \bibinfo {author} {\bibfnamefont {R.}~\bibnamefont
  {{Miquel}}}, \bibinfo {author} {\bibfnamefont {J.~J.}\ \bibnamefont
  {{Mohr}}}, \bibinfo {author} {\bibfnamefont {R.~C.}\ \bibnamefont
  {{Nichol}}}, \bibinfo {author} {\bibfnamefont {B.}~\bibnamefont {{Nord}}},
  \bibinfo {author} {\bibfnamefont {R.}~\bibnamefont {{Ogando}}}, \bibinfo
  {author} {\bibfnamefont {A.~A.}\ \bibnamefont {{Plazas}}}, \bibinfo {author}
  {\bibfnamefont {K.}~\bibnamefont {{Reil}}}, \bibinfo {author} {\bibfnamefont
  {M.}~\bibnamefont {{Sahl{\'e}n}}}, \bibinfo {author} {\bibfnamefont
  {E.}~\bibnamefont {{Sanchez}}}, \bibinfo {author} {\bibfnamefont
  {B.}~\bibnamefont {{Santiago}}}, \bibinfo {author} {\bibfnamefont
  {V.}~\bibnamefont {{Scarpine}}}, \bibinfo {author} {\bibfnamefont
  {M.}~\bibnamefont {{Schubnell}}}, \bibinfo {author} {\bibfnamefont
  {I.}~\bibnamefont {{Sevilla-Noarbe}}}, \bibinfo {author} {\bibfnamefont
  {R.~C.}\ \bibnamefont {{Smith}}}, \bibinfo {author} {\bibfnamefont
  {M.}~\bibnamefont {{Soares-Santos}}}, \bibinfo {author} {\bibfnamefont
  {F.}~\bibnamefont {{Sobreira}}}, \bibinfo {author} {\bibfnamefont {J.~P.}\
  \bibnamefont {{Stott}}}, \bibinfo {author} {\bibfnamefont {E.}~\bibnamefont
  {{Suchyta}}}, \bibinfo {author} {\bibfnamefont {M.~E.~C.}\ \bibnamefont
  {{Swanson}}}, \bibinfo {author} {\bibfnamefont {G.}~\bibnamefont {{Tarle}}},
  \bibinfo {author} {\bibfnamefont {D.}~\bibnamefont {{Thomas}}}, \bibinfo
  {author} {\bibfnamefont {D.}~\bibnamefont {{Tucker}}}, \bibinfo {author}
  {\bibfnamefont {S.}~\bibnamefont {{Uddin}}}, \bibinfo {author} {\bibfnamefont
  {P.~T.~P.}\ \bibnamefont {{Viana}}}, \bibinfo {author} {\bibfnamefont
  {V.}~\bibnamefont {{Vikram}}}, \bibinfo {author} {\bibfnamefont {A.~R.}\
  \bibnamefont {{Walker}}}, \bibinfo {author} {\bibfnamefont {Y.}~\bibnamefont
  {{Zhang}}}, \ and\ \bibinfo {author} {\bibnamefont {{DES Collaboration}}},\
  }\href {\doibase 10.3847/0067-0049/224/1/1} {\bibfield  {journal} {\bibinfo
  {journal} {\apjs}\ }\textbf {\bibinfo {volume} {224}},\ \bibinfo {eid} {1}
  (\bibinfo {year} {2016})},\ \Eprint {http://arxiv.org/abs/1601.00621}
  {arXiv:1601.00621} \BibitemShut {NoStop}%
\bibitem [{\citenamefont {{Hilton}}\ \emph {et~al.}(2018)\citenamefont
  {{Hilton}}, \citenamefont {{Hasselfield}}, \citenamefont {{Sif{\'o}n}},
  \citenamefont {{Battaglia}}, \citenamefont {{Aiola}}, \citenamefont
  {{Bharadwaj}}, \citenamefont {{Bond}}, \citenamefont {{Choi}}, \citenamefont
  {{Crichton}}, \citenamefont {{Datta}}, \citenamefont {{Devlin}},
  \citenamefont {{Dunkley}}, \citenamefont {{D{\"u}nner}}, \citenamefont
  {{Gallardo}}, \citenamefont {{Gralla}}, \citenamefont {{Hincks}},
  \citenamefont {{Ho}}, \citenamefont {{Hubmayr}}, \citenamefont
  {{Huffenberger}}, \citenamefont {{Hughes}}, \citenamefont {{Koopman}},
  \citenamefont {{Kosowsky}}, \citenamefont {{Louis}}, \citenamefont
  {{Madhavacheril}}, \citenamefont {{Marriage}}, \citenamefont {{Maurin}},
  \citenamefont {{McMahon}}, \citenamefont {{Miyatake}}, \citenamefont
  {{Moodley}}, \citenamefont {{N{\ae}ss}}, \citenamefont {{Nati}},
  \citenamefont {{Newburgh}}, \citenamefont {{Niemack}}, \citenamefont
  {{Oguri}}, \citenamefont {{Page}}, \citenamefont {{Partridge}}, \citenamefont
  {{Schmitt}}, \citenamefont {{Sievers}}, \citenamefont {{Spergel}},
  \citenamefont {{Staggs}}, \citenamefont {{Trac}}, \citenamefont {{van
  Engelen}}, \citenamefont {{Vavagiakis}},\ and\ \citenamefont
  {{Wollack}}}]{hilton18}%
  \BibitemOpen
  \bibfield  {author} {\bibinfo {author} {\bibfnamefont {M.}~\bibnamefont
  {{Hilton}}}, \bibinfo {author} {\bibfnamefont {M.}~\bibnamefont
  {{Hasselfield}}}, \bibinfo {author} {\bibfnamefont {C.}~\bibnamefont
  {{Sif{\'o}n}}}, \bibinfo {author} {\bibfnamefont {N.}~\bibnamefont
  {{Battaglia}}}, \bibinfo {author} {\bibfnamefont {S.}~\bibnamefont
  {{Aiola}}}, \bibinfo {author} {\bibfnamefont {V.}~\bibnamefont
  {{Bharadwaj}}}, \bibinfo {author} {\bibfnamefont {J.~R.}\ \bibnamefont
  {{Bond}}}, \bibinfo {author} {\bibfnamefont {S.~K.}\ \bibnamefont {{Choi}}},
  \bibinfo {author} {\bibfnamefont {D.}~\bibnamefont {{Crichton}}}, \bibinfo
  {author} {\bibfnamefont {R.}~\bibnamefont {{Datta}}}, \bibinfo {author}
  {\bibfnamefont {M.~J.}\ \bibnamefont {{Devlin}}}, \bibinfo {author}
  {\bibfnamefont {J.}~\bibnamefont {{Dunkley}}}, \bibinfo {author}
  {\bibfnamefont {R.}~\bibnamefont {{D{\"u}nner}}}, \bibinfo {author}
  {\bibfnamefont {P.~A.}\ \bibnamefont {{Gallardo}}}, \bibinfo {author}
  {\bibfnamefont {M.}~\bibnamefont {{Gralla}}}, \bibinfo {author}
  {\bibfnamefont {A.~D.}\ \bibnamefont {{Hincks}}}, \bibinfo {author}
  {\bibfnamefont {S.-P.~P.}\ \bibnamefont {{Ho}}}, \bibinfo {author}
  {\bibfnamefont {J.}~\bibnamefont {{Hubmayr}}}, \bibinfo {author}
  {\bibfnamefont {K.~M.}\ \bibnamefont {{Huffenberger}}}, \bibinfo {author}
  {\bibfnamefont {J.~P.}\ \bibnamefont {{Hughes}}}, \bibinfo {author}
  {\bibfnamefont {B.~J.}\ \bibnamefont {{Koopman}}}, \bibinfo {author}
  {\bibfnamefont {A.}~\bibnamefont {{Kosowsky}}}, \bibinfo {author}
  {\bibfnamefont {T.}~\bibnamefont {{Louis}}}, \bibinfo {author} {\bibfnamefont
  {M.~S.}\ \bibnamefont {{Madhavacheril}}}, \bibinfo {author} {\bibfnamefont
  {T.~A.}\ \bibnamefont {{Marriage}}}, \bibinfo {author} {\bibfnamefont
  {L.}~\bibnamefont {{Maurin}}}, \bibinfo {author} {\bibfnamefont
  {J.}~\bibnamefont {{McMahon}}}, \bibinfo {author} {\bibfnamefont
  {H.}~\bibnamefont {{Miyatake}}}, \bibinfo {author} {\bibfnamefont
  {K.}~\bibnamefont {{Moodley}}}, \bibinfo {author} {\bibfnamefont
  {S.}~\bibnamefont {{N{\ae}ss}}}, \bibinfo {author} {\bibfnamefont
  {F.}~\bibnamefont {{Nati}}}, \bibinfo {author} {\bibfnamefont
  {L.}~\bibnamefont {{Newburgh}}}, \bibinfo {author} {\bibfnamefont {M.~D.}\
  \bibnamefont {{Niemack}}}, \bibinfo {author} {\bibfnamefont {M.}~\bibnamefont
  {{Oguri}}}, \bibinfo {author} {\bibfnamefont {L.~A.}\ \bibnamefont {{Page}}},
  \bibinfo {author} {\bibfnamefont {B.}~\bibnamefont {{Partridge}}}, \bibinfo
  {author} {\bibfnamefont {B.~L.}\ \bibnamefont {{Schmitt}}}, \bibinfo {author}
  {\bibfnamefont {J.}~\bibnamefont {{Sievers}}}, \bibinfo {author}
  {\bibfnamefont {D.~N.}\ \bibnamefont {{Spergel}}}, \bibinfo {author}
  {\bibfnamefont {S.~T.}\ \bibnamefont {{Staggs}}}, \bibinfo {author}
  {\bibfnamefont {H.}~\bibnamefont {{Trac}}}, \bibinfo {author} {\bibfnamefont
  {A.}~\bibnamefont {{van Engelen}}}, \bibinfo {author} {\bibfnamefont {E.~M.}\
  \bibnamefont {{Vavagiakis}}}, \ and\ \bibinfo {author} {\bibfnamefont
  {E.~J.}\ \bibnamefont {{Wollack}}},\ }\href {\doibase
  10.3847/1538-4365/aaa6cb} {\bibfield  {journal} {\bibinfo  {journal} {\apjs}\
  }\textbf {\bibinfo {volume} {235}},\ \bibinfo {eid} {20} (\bibinfo {year}
  {2018})},\ \Eprint {http://arxiv.org/abs/1709.05600} {arXiv:1709.05600}
  \BibitemShut {NoStop}%
\bibitem [{\citenamefont {{Klein}}\ \emph {et~al.}(2018)\citenamefont
  {{Klein}}, \citenamefont {{Grandis}}, \citenamefont {{Mohr}}, \citenamefont
  {{Paulus}},\ and\ \citenamefont {{the DES Collaboration}}}]{klein18}%
  \BibitemOpen
  \bibfield  {author} {\bibinfo {author} {\bibfnamefont {M.}~\bibnamefont
  {{Klein}}}, \bibinfo {author} {\bibfnamefont {S.}~\bibnamefont {{Grandis}}},
  \bibinfo {author} {\bibfnamefont {J.}~\bibnamefont {{Mohr}}}, \bibinfo
  {author} {\bibfnamefont {M.}~\bibnamefont {{Paulus}}}, \ and\ \bibinfo
  {author} {\bibnamefont {{the DES Collaboration}}},\ }\href@noop {} {\bibfield
   {journal} {\bibinfo  {journal} {arXiv e-prints}\ } (\bibinfo {year}
  {2018})},\ \Eprint {http://arxiv.org/abs/1812.09956} {arXiv:1812.09956}
  \BibitemShut {NoStop}%
\bibitem [{\citenamefont {Mantz}\ \emph {et~al.}(2010)\citenamefont {Mantz},
  \citenamefont {Allen}, \citenamefont {Ebeling}, \citenamefont {Rapetti},\
  and\ \citenamefont {Drlica-Wagner}}]{mantz10b}%
  \BibitemOpen
  \bibfield  {author} {\bibinfo {author} {\bibfnamefont {A.}~\bibnamefont
  {Mantz}}, \bibinfo {author} {\bibfnamefont {S.}~\bibnamefont {Allen}},
  \bibinfo {author} {\bibfnamefont {H.}~\bibnamefont {Ebeling}}, \bibinfo
  {author} {\bibfnamefont {D.}~\bibnamefont {Rapetti}}, \ and\ \bibinfo
  {author} {\bibfnamefont {A.}~\bibnamefont {Drlica-Wagner}},\ }\href {\doibase
  10.1111/j.1365-2966.2010.16993.x} {\bibfield  {journal} {\bibinfo  {journal}
  {\mnras}\ }\textbf {\bibinfo {volume} {406}},\ \bibinfo {pages} {1773}
  (\bibinfo {year} {2010})},\ \Eprint {http://arxiv.org/abs/0909.3099}
  {arXiv:0909.3099 [astro-ph.CO]} \BibitemShut {NoStop}%
\bibitem [{\citenamefont {{de Haan}}\ \emph {et~al.}(2016)\citenamefont {{de
  Haan}}, \citenamefont {{Benson}}, \citenamefont {{Bleem}}, \citenamefont
  {{Allen}}, \citenamefont {{Applegate}}, \citenamefont {{Ashby}},
  \citenamefont {{Bautz}}, \citenamefont {{Bayliss}}, \citenamefont
  {{Bocquet}}, \citenamefont {{Brodwin}}, \citenamefont {{Carlstrom}},
  \citenamefont {{Chang}}, \citenamefont {{Chiu}}, \citenamefont {{Cho}},
  \citenamefont {{Clocchiatti}}, \citenamefont {{Crawford}}, \citenamefont
  {{Crites}}, \citenamefont {{Desai}}, \citenamefont {{Dietrich}},
  \citenamefont {{Dobbs}}, \citenamefont {{Doucouliagos}}, \citenamefont
  {{Foley}}, \citenamefont {{Forman}}, \citenamefont {{Garmire}}, \citenamefont
  {{George}}, \citenamefont {{Gladders}}, \citenamefont {{Gonzalez}},
  \citenamefont {{Gupta}}, \citenamefont {{Halverson}}, \citenamefont
  {{Hlavacek-Larrondo}}, \citenamefont {{Hoekstra}}, \citenamefont {{Holder}},
  \citenamefont {{Holzapfel}}, \citenamefont {{Hou}}, \citenamefont {{Hrubes}},
  \citenamefont {{Huang}}, \citenamefont {{Jones}}, \citenamefont {{Keisler}},
  \citenamefont {{Knox}}, \citenamefont {{Lee}}, \citenamefont {{Leitch}},
  \citenamefont {{von der Linden}}, \citenamefont {{Luong-Van}}, \citenamefont
  {{Mantz}}, \citenamefont {{Marrone}}, \citenamefont {{McDonald}},
  \citenamefont {{McMahon}}, \citenamefont {{Meyer}}, \citenamefont {{Mocanu}},
  \citenamefont {{Mohr}}, \citenamefont {{Murray}}, \citenamefont {{Padin}},
  \citenamefont {{Pryke}}, \citenamefont {{Rapetti}}, \citenamefont
  {{Reichardt}}, \citenamefont {{Rest}}, \citenamefont {{Ruel}}, \citenamefont
  {{Ruhl}}, \citenamefont {{Saliwanchik}}, \citenamefont {{Saro}},
  \citenamefont {{Sayre}}, \citenamefont {{Schaffer}}, \citenamefont
  {{Schrabback}}, \citenamefont {{Shirokoff}}, \citenamefont {{Song}},
  \citenamefont {{Spieler}}, \citenamefont {{Stalder}}, \citenamefont
  {{Stanford}}, \citenamefont {{Staniszewski}}, \citenamefont {{Stark}},
  \citenamefont {{Story}}, \citenamefont {{Stubbs}}, \citenamefont
  {{Vanderlinde}}, \citenamefont {{Vieira}}, \citenamefont {{Vikhlinin}},
  \citenamefont {{Williamson}},\ and\ \citenamefont {{Zenteno}}}]{dehaan16}%
  \BibitemOpen
  \bibfield  {author} {\bibinfo {author} {\bibfnamefont {T.}~\bibnamefont {{de
  Haan}}}, \bibinfo {author} {\bibfnamefont {B.~A.}\ \bibnamefont {{Benson}}},
  \bibinfo {author} {\bibfnamefont {L.~E.}\ \bibnamefont {{Bleem}}}, \bibinfo
  {author} {\bibfnamefont {S.~W.}\ \bibnamefont {{Allen}}}, \bibinfo {author}
  {\bibfnamefont {D.~E.}\ \bibnamefont {{Applegate}}}, \bibinfo {author}
  {\bibfnamefont {M.~L.~N.}\ \bibnamefont {{Ashby}}}, \bibinfo {author}
  {\bibfnamefont {M.}~\bibnamefont {{Bautz}}}, \bibinfo {author} {\bibfnamefont
  {M.}~\bibnamefont {{Bayliss}}}, \bibinfo {author} {\bibfnamefont
  {S.}~\bibnamefont {{Bocquet}}}, \bibinfo {author} {\bibfnamefont
  {M.}~\bibnamefont {{Brodwin}}}, \bibinfo {author} {\bibfnamefont {J.~E.}\
  \bibnamefont {{Carlstrom}}}, \bibinfo {author} {\bibfnamefont {C.~L.}\
  \bibnamefont {{Chang}}}, \bibinfo {author} {\bibfnamefont {I.}~\bibnamefont
  {{Chiu}}}, \bibinfo {author} {\bibfnamefont {H.-M.}\ \bibnamefont {{Cho}}},
  \bibinfo {author} {\bibfnamefont {A.}~\bibnamefont {{Clocchiatti}}}, \bibinfo
  {author} {\bibfnamefont {T.~M.}\ \bibnamefont {{Crawford}}}, \bibinfo
  {author} {\bibfnamefont {A.~T.}\ \bibnamefont {{Crites}}}, \bibinfo {author}
  {\bibfnamefont {S.}~\bibnamefont {{Desai}}}, \bibinfo {author} {\bibfnamefont
  {J.~P.}\ \bibnamefont {{Dietrich}}}, \bibinfo {author} {\bibfnamefont
  {M.~A.}\ \bibnamefont {{Dobbs}}}, \bibinfo {author} {\bibfnamefont {A.~N.}\
  \bibnamefont {{Doucouliagos}}}, \bibinfo {author} {\bibfnamefont {R.~J.}\
  \bibnamefont {{Foley}}}, \bibinfo {author} {\bibfnamefont {W.~R.}\
  \bibnamefont {{Forman}}}, \bibinfo {author} {\bibfnamefont {G.~P.}\
  \bibnamefont {{Garmire}}}, \bibinfo {author} {\bibfnamefont {E.~M.}\
  \bibnamefont {{George}}}, \bibinfo {author} {\bibfnamefont {M.~D.}\
  \bibnamefont {{Gladders}}}, \bibinfo {author} {\bibfnamefont {A.~H.}\
  \bibnamefont {{Gonzalez}}}, \bibinfo {author} {\bibfnamefont
  {N.}~\bibnamefont {{Gupta}}}, \bibinfo {author} {\bibfnamefont {N.~W.}\
  \bibnamefont {{Halverson}}}, \bibinfo {author} {\bibfnamefont
  {J.}~\bibnamefont {{Hlavacek-Larrondo}}}, \bibinfo {author} {\bibfnamefont
  {H.}~\bibnamefont {{Hoekstra}}}, \bibinfo {author} {\bibfnamefont {G.~P.}\
  \bibnamefont {{Holder}}}, \bibinfo {author} {\bibfnamefont {W.~L.}\
  \bibnamefont {{Holzapfel}}}, \bibinfo {author} {\bibfnamefont
  {Z.}~\bibnamefont {{Hou}}}, \bibinfo {author} {\bibfnamefont {J.~D.}\
  \bibnamefont {{Hrubes}}}, \bibinfo {author} {\bibfnamefont {N.}~\bibnamefont
  {{Huang}}}, \bibinfo {author} {\bibfnamefont {C.}~\bibnamefont {{Jones}}},
  \bibinfo {author} {\bibfnamefont {R.}~\bibnamefont {{Keisler}}}, \bibinfo
  {author} {\bibfnamefont {L.}~\bibnamefont {{Knox}}}, \bibinfo {author}
  {\bibfnamefont {A.~T.}\ \bibnamefont {{Lee}}}, \bibinfo {author}
  {\bibfnamefont {E.~M.}\ \bibnamefont {{Leitch}}}, \bibinfo {author}
  {\bibfnamefont {A.}~\bibnamefont {{von der Linden}}}, \bibinfo {author}
  {\bibfnamefont {D.}~\bibnamefont {{Luong-Van}}}, \bibinfo {author}
  {\bibfnamefont {A.}~\bibnamefont {{Mantz}}}, \bibinfo {author} {\bibfnamefont
  {D.~P.}\ \bibnamefont {{Marrone}}}, \bibinfo {author} {\bibfnamefont
  {M.}~\bibnamefont {{McDonald}}}, \bibinfo {author} {\bibfnamefont {J.~J.}\
  \bibnamefont {{McMahon}}}, \bibinfo {author} {\bibfnamefont {S.~S.}\
  \bibnamefont {{Meyer}}}, \bibinfo {author} {\bibfnamefont {L.~M.}\
  \bibnamefont {{Mocanu}}}, \bibinfo {author} {\bibfnamefont {J.~J.}\
  \bibnamefont {{Mohr}}}, \bibinfo {author} {\bibfnamefont {S.~S.}\
  \bibnamefont {{Murray}}}, \bibinfo {author} {\bibfnamefont {S.}~\bibnamefont
  {{Padin}}}, \bibinfo {author} {\bibfnamefont {C.}~\bibnamefont {{Pryke}}},
  \bibinfo {author} {\bibfnamefont {D.}~\bibnamefont {{Rapetti}}}, \bibinfo
  {author} {\bibfnamefont {C.~L.}\ \bibnamefont {{Reichardt}}}, \bibinfo
  {author} {\bibfnamefont {A.}~\bibnamefont {{Rest}}}, \bibinfo {author}
  {\bibfnamefont {J.}~\bibnamefont {{Ruel}}}, \bibinfo {author} {\bibfnamefont
  {J.~E.}\ \bibnamefont {{Ruhl}}}, \bibinfo {author} {\bibfnamefont {B.~R.}\
  \bibnamefont {{Saliwanchik}}}, \bibinfo {author} {\bibfnamefont
  {A.}~\bibnamefont {{Saro}}}, \bibinfo {author} {\bibfnamefont {J.~T.}\
  \bibnamefont {{Sayre}}}, \bibinfo {author} {\bibfnamefont {K.~K.}\
  \bibnamefont {{Schaffer}}}, \bibinfo {author} {\bibfnamefont
  {T.}~\bibnamefont {{Schrabback}}}, \bibinfo {author} {\bibfnamefont
  {E.}~\bibnamefont {{Shirokoff}}}, \bibinfo {author} {\bibfnamefont
  {J.}~\bibnamefont {{Song}}}, \bibinfo {author} {\bibfnamefont {H.~G.}\
  \bibnamefont {{Spieler}}}, \bibinfo {author} {\bibfnamefont {B.}~\bibnamefont
  {{Stalder}}}, \bibinfo {author} {\bibfnamefont {S.~A.}\ \bibnamefont
  {{Stanford}}}, \bibinfo {author} {\bibfnamefont {Z.}~\bibnamefont
  {{Staniszewski}}}, \bibinfo {author} {\bibfnamefont {A.~A.}\ \bibnamefont
  {{Stark}}}, \bibinfo {author} {\bibfnamefont {K.~T.}\ \bibnamefont
  {{Story}}}, \bibinfo {author} {\bibfnamefont {C.~W.}\ \bibnamefont
  {{Stubbs}}}, \bibinfo {author} {\bibfnamefont {K.}~\bibnamefont
  {{Vanderlinde}}}, \bibinfo {author} {\bibfnamefont {J.~D.}\ \bibnamefont
  {{Vieira}}}, \bibinfo {author} {\bibfnamefont {A.}~\bibnamefont
  {{Vikhlinin}}}, \bibinfo {author} {\bibfnamefont {R.}~\bibnamefont
  {{Williamson}}}, \ and\ \bibinfo {author} {\bibfnamefont {A.}~\bibnamefont
  {{Zenteno}}},\ }\href {\doibase 10.3847/0004-637X/832/1/95} {\bibfield
  {journal} {\bibinfo  {journal} {\apj}\ }\textbf {\bibinfo {volume} {832}},\
  \bibinfo {eid} {95} (\bibinfo {year} {2016})},\ \Eprint
  {http://arxiv.org/abs/1603.06522} {arXiv:1603.06522} \BibitemShut {NoStop}%
\bibitem [{\citenamefont {{Bocquet}}\ \emph {et~al.}(2018)\citenamefont
  {{Bocquet}}, \citenamefont {{Dietrich}}, \citenamefont {{Schrabback}},
  \citenamefont {{Bleem}}, \citenamefont {{Klein}}, \citenamefont {{Allen}},
  \citenamefont {{Applegate}}, \citenamefont {{Ashby}}, \citenamefont
  {{Bautz}}, \citenamefont {{Bayliss}}, \citenamefont {{Benson}}, \citenamefont
  {{Brodwin}}, \citenamefont {{Bulbul}}, \citenamefont {{Canning}},
  \citenamefont {{Capasso}}, \citenamefont {{Carlstrom}}, \citenamefont
  {{Chang}}, \citenamefont {{Chiu}}, \citenamefont {{Cho}}, \citenamefont
  {{Clocchiatti}}, \citenamefont {{Crawford}}, \citenamefont {{Crites}},
  \citenamefont {{de Haan}}, \citenamefont {{Desai}}, \citenamefont {{Dobbs}},
  \citenamefont {{Foley}}, \citenamefont {{Forman}}, \citenamefont {{Garmire}},
  \citenamefont {{George}}, \citenamefont {{Gladders}}, \citenamefont
  {{Gonzalez}}, \citenamefont {{Grandis}}, \citenamefont {{Gupta}},
  \citenamefont {{Halverson}}, \citenamefont {{Hlavacek-Larrondo}},
  \citenamefont {{Hoekstra}}, \citenamefont {{Holder}}, \citenamefont
  {{Holzapfel}}, \citenamefont {{Hou}}, \citenamefont {{Hrubes}}, \citenamefont
  {{Huang}}, \citenamefont {{Jones}}, \citenamefont {{Khullar}}, \citenamefont
  {{Knox}}, \citenamefont {{Kraft}}, \citenamefont {{Lee}}, \citenamefont {{von
  der Linden}}, \citenamefont {{Luong-Van}}, \citenamefont {{Mantz}},
  \citenamefont {{Marrone}}, \citenamefont {{McDonald}}, \citenamefont
  {{McMahon}}, \citenamefont {{Meyer}}, \citenamefont {{Mocanu}}, \citenamefont
  {{Mohr}}, \citenamefont {{Morris}}, \citenamefont {{Padin}}, \citenamefont
  {{Patil}}, \citenamefont {{Pryke}}, \citenamefont {{Rapetti}}, \citenamefont
  {{Reichardt}}, \citenamefont {{Rest}}, \citenamefont {{Ruhl}}, \citenamefont
  {{Saliwanchik}}, \citenamefont {{Saro}}, \citenamefont {{Sayre}},
  \citenamefont {{Schaffer}}, \citenamefont {{Shirokoff}}, \citenamefont
  {{Stalder}}, \citenamefont {{Stanford}}, \citenamefont {{Staniszewski}},
  \citenamefont {{Stark}}, \citenamefont {{Story}}, \citenamefont
  {{Strazzullo}}, \citenamefont {{Stubbs}}, \citenamefont {{Vanderlinde}},
  \citenamefont {{Vieira}}, \citenamefont {{Vikhlinin}}, \citenamefont
  {{Williamson}},\ and\ \citenamefont {{Zenteno}}}]{bocquet18}%
  \BibitemOpen
  \bibfield  {author} {\bibinfo {author} {\bibfnamefont {S.}~\bibnamefont
  {{Bocquet}}}, \bibinfo {author} {\bibfnamefont {J.~P.}\ \bibnamefont
  {{Dietrich}}}, \bibinfo {author} {\bibfnamefont {T.}~\bibnamefont
  {{Schrabback}}}, \bibinfo {author} {\bibfnamefont {L.~E.}\ \bibnamefont
  {{Bleem}}}, \bibinfo {author} {\bibfnamefont {M.}~\bibnamefont {{Klein}}},
  \bibinfo {author} {\bibfnamefont {S.~W.}\ \bibnamefont {{Allen}}}, \bibinfo
  {author} {\bibfnamefont {D.~E.}\ \bibnamefont {{Applegate}}}, \bibinfo
  {author} {\bibfnamefont {M.~L.~N.}\ \bibnamefont {{Ashby}}}, \bibinfo
  {author} {\bibfnamefont {M.}~\bibnamefont {{Bautz}}}, \bibinfo {author}
  {\bibfnamefont {M.}~\bibnamefont {{Bayliss}}}, \bibinfo {author}
  {\bibfnamefont {B.~A.}\ \bibnamefont {{Benson}}}, \bibinfo {author}
  {\bibfnamefont {M.}~\bibnamefont {{Brodwin}}}, \bibinfo {author}
  {\bibfnamefont {E.}~\bibnamefont {{Bulbul}}}, \bibinfo {author}
  {\bibfnamefont {R.~E.~A.}\ \bibnamefont {{Canning}}}, \bibinfo {author}
  {\bibfnamefont {R.}~\bibnamefont {{Capasso}}}, \bibinfo {author}
  {\bibfnamefont {J.~E.}\ \bibnamefont {{Carlstrom}}}, \bibinfo {author}
  {\bibfnamefont {C.~L.}\ \bibnamefont {{Chang}}}, \bibinfo {author}
  {\bibfnamefont {I.}~\bibnamefont {{Chiu}}}, \bibinfo {author} {\bibfnamefont
  {H.}~\bibnamefont {{Cho}}}, \bibinfo {author} {\bibfnamefont
  {A.}~\bibnamefont {{Clocchiatti}}}, \bibinfo {author} {\bibfnamefont {T.~M.}\
  \bibnamefont {{Crawford}}}, \bibinfo {author} {\bibfnamefont {A.~T.}\
  \bibnamefont {{Crites}}}, \bibinfo {author} {\bibfnamefont {T.}~\bibnamefont
  {{de Haan}}}, \bibinfo {author} {\bibfnamefont {S.}~\bibnamefont {{Desai}}},
  \bibinfo {author} {\bibfnamefont {M.~A.}\ \bibnamefont {{Dobbs}}}, \bibinfo
  {author} {\bibfnamefont {R.~J.}\ \bibnamefont {{Foley}}}, \bibinfo {author}
  {\bibfnamefont {W.~R.}\ \bibnamefont {{Forman}}}, \bibinfo {author}
  {\bibfnamefont {G.~P.}\ \bibnamefont {{Garmire}}}, \bibinfo {author}
  {\bibfnamefont {E.~M.}\ \bibnamefont {{George}}}, \bibinfo {author}
  {\bibfnamefont {M.~D.}\ \bibnamefont {{Gladders}}}, \bibinfo {author}
  {\bibfnamefont {A.~H.}\ \bibnamefont {{Gonzalez}}}, \bibinfo {author}
  {\bibfnamefont {S.}~\bibnamefont {{Grandis}}}, \bibinfo {author}
  {\bibfnamefont {N.}~\bibnamefont {{Gupta}}}, \bibinfo {author} {\bibfnamefont
  {N.~W.}\ \bibnamefont {{Halverson}}}, \bibinfo {author} {\bibfnamefont
  {J.}~\bibnamefont {{Hlavacek-Larrondo}}}, \bibinfo {author} {\bibfnamefont
  {H.}~\bibnamefont {{Hoekstra}}}, \bibinfo {author} {\bibfnamefont {G.~P.}\
  \bibnamefont {{Holder}}}, \bibinfo {author} {\bibfnamefont {W.~L.}\
  \bibnamefont {{Holzapfel}}}, \bibinfo {author} {\bibfnamefont
  {Z.}~\bibnamefont {{Hou}}}, \bibinfo {author} {\bibfnamefont {J.~D.}\
  \bibnamefont {{Hrubes}}}, \bibinfo {author} {\bibfnamefont {N.}~\bibnamefont
  {{Huang}}}, \bibinfo {author} {\bibfnamefont {C.}~\bibnamefont {{Jones}}},
  \bibinfo {author} {\bibfnamefont {G.}~\bibnamefont {{Khullar}}}, \bibinfo
  {author} {\bibfnamefont {L.}~\bibnamefont {{Knox}}}, \bibinfo {author}
  {\bibfnamefont {R.}~\bibnamefont {{Kraft}}}, \bibinfo {author} {\bibfnamefont
  {A.~T.}\ \bibnamefont {{Lee}}}, \bibinfo {author} {\bibfnamefont
  {A.}~\bibnamefont {{von der Linden}}}, \bibinfo {author} {\bibfnamefont
  {D.}~\bibnamefont {{Luong-Van}}}, \bibinfo {author} {\bibfnamefont
  {A.}~\bibnamefont {{Mantz}}}, \bibinfo {author} {\bibfnamefont {D.~P.}\
  \bibnamefont {{Marrone}}}, \bibinfo {author} {\bibfnamefont {M.}~\bibnamefont
  {{McDonald}}}, \bibinfo {author} {\bibfnamefont {J.~J.}\ \bibnamefont
  {{McMahon}}}, \bibinfo {author} {\bibfnamefont {S.~S.}\ \bibnamefont
  {{Meyer}}}, \bibinfo {author} {\bibfnamefont {L.~M.}\ \bibnamefont
  {{Mocanu}}}, \bibinfo {author} {\bibfnamefont {J.~J.}\ \bibnamefont
  {{Mohr}}}, \bibinfo {author} {\bibfnamefont {R.~G.}\ \bibnamefont
  {{Morris}}}, \bibinfo {author} {\bibfnamefont {S.}~\bibnamefont {{Padin}}},
  \bibinfo {author} {\bibfnamefont {S.}~\bibnamefont {{Patil}}}, \bibinfo
  {author} {\bibfnamefont {C.}~\bibnamefont {{Pryke}}}, \bibinfo {author}
  {\bibfnamefont {D.}~\bibnamefont {{Rapetti}}}, \bibinfo {author}
  {\bibfnamefont {C.~L.}\ \bibnamefont {{Reichardt}}}, \bibinfo {author}
  {\bibfnamefont {A.}~\bibnamefont {{Rest}}}, \bibinfo {author} {\bibfnamefont
  {J.~E.}\ \bibnamefont {{Ruhl}}}, \bibinfo {author} {\bibfnamefont {B.~R.}\
  \bibnamefont {{Saliwanchik}}}, \bibinfo {author} {\bibfnamefont
  {A.}~\bibnamefont {{Saro}}}, \bibinfo {author} {\bibfnamefont {J.~T.}\
  \bibnamefont {{Sayre}}}, \bibinfo {author} {\bibfnamefont {K.~K.}\
  \bibnamefont {{Schaffer}}}, \bibinfo {author} {\bibfnamefont
  {E.}~\bibnamefont {{Shirokoff}}}, \bibinfo {author} {\bibfnamefont
  {B.}~\bibnamefont {{Stalder}}}, \bibinfo {author} {\bibfnamefont {S.~A.}\
  \bibnamefont {{Stanford}}}, \bibinfo {author} {\bibfnamefont
  {Z.}~\bibnamefont {{Staniszewski}}}, \bibinfo {author} {\bibfnamefont
  {A.~A.}\ \bibnamefont {{Stark}}}, \bibinfo {author} {\bibfnamefont {K.~T.}\
  \bibnamefont {{Story}}}, \bibinfo {author} {\bibfnamefont {V.}~\bibnamefont
  {{Strazzullo}}}, \bibinfo {author} {\bibfnamefont {C.~W.}\ \bibnamefont
  {{Stubbs}}}, \bibinfo {author} {\bibfnamefont {K.}~\bibnamefont
  {{Vanderlinde}}}, \bibinfo {author} {\bibfnamefont {J.~D.}\ \bibnamefont
  {{Vieira}}}, \bibinfo {author} {\bibfnamefont {A.}~\bibnamefont
  {{Vikhlinin}}}, \bibinfo {author} {\bibfnamefont {R.}~\bibnamefont
  {{Williamson}}}, \ and\ \bibinfo {author} {\bibfnamefont {A.}~\bibnamefont
  {{Zenteno}}},\ }\href@noop {} {\bibfield  {journal} {\bibinfo  {journal}
  {arXiv e-prints}\ } (\bibinfo {year} {2018})},\ \Eprint
  {http://arxiv.org/abs/1812.01679} {arXiv:1812.01679} \BibitemShut {NoStop}%
\bibitem [{\citenamefont {Predehl}\ \emph {et~al.}(2010)\citenamefont
  {Predehl}, \citenamefont {Andritschke}, \citenamefont {B{\"{o}}hringer},
  \citenamefont {Bornemann}, \citenamefont {Br{\"{a}}uninger}, \citenamefont
  {Brunner}, \citenamefont {Brusa}, \citenamefont {Burkert}, \citenamefont
  {Burwitz}, \citenamefont {Cappelluti}, \citenamefont {Churazov},
  \citenamefont {Dennerl}, \citenamefont {Eder}, \citenamefont {Elbs},
  \citenamefont {Freyberg}, \citenamefont {Friedrich}, \citenamefont
  {F{\"{u}}rmetz}, \citenamefont {Gaida}, \citenamefont {H{\"{a}}lker},
  \citenamefont {Hartner}, \citenamefont {Hasinger}, \citenamefont {Hermann},
  \citenamefont {Huber}, \citenamefont {Kendziorra}, \citenamefont {von
  Kienlin}, \citenamefont {Kink}, \citenamefont {Kreykenbohm}, \citenamefont
  {Lamer}, \citenamefont {Lapchov}, \citenamefont {Lehmann}, \citenamefont
  {Meidinger}, \citenamefont {Mican}, \citenamefont {Mohr}, \citenamefont
  {M{\"{u}}hlegger}, \citenamefont {M{\"{u}}ller}, \citenamefont {Nandra},
  \citenamefont {Pavlinsky}, \citenamefont {Pfeffermann}, \citenamefont
  {Reiprich}, \citenamefont {Robrade}, \citenamefont {Roh{\'{e}}},
  \citenamefont {Santangelo}, \citenamefont {Sch{\"{a}}chner}, \citenamefont
  {Schanz}, \citenamefont {Schmid}, \citenamefont {Schmitt}, \citenamefont
  {Schreib}, \citenamefont {Schrey}, \citenamefont {Schwope}, \citenamefont
  {Steinmetz}, \citenamefont {Str{\"{u}}der}, \citenamefont {Sunyaev},
  \citenamefont {Tenzer}, \citenamefont {Tiedemann}, \citenamefont {Vongehr},\
  and\ \citenamefont {Wilms}}]{predehl10}%
  \BibitemOpen
  \bibfield  {author} {\bibinfo {author} {\bibfnamefont {P.}~\bibnamefont
  {Predehl}}, \bibinfo {author} {\bibfnamefont {R.}~\bibnamefont
  {Andritschke}}, \bibinfo {author} {\bibfnamefont {H.}~\bibnamefont
  {B{\"{o}}hringer}}, \bibinfo {author} {\bibfnamefont {W.}~\bibnamefont
  {Bornemann}}, \bibinfo {author} {\bibfnamefont {H.}~\bibnamefont
  {Br{\"{a}}uninger}}, \bibinfo {author} {\bibfnamefont {H.}~\bibnamefont
  {Brunner}}, \bibinfo {author} {\bibfnamefont {M.}~\bibnamefont {Brusa}},
  \bibinfo {author} {\bibfnamefont {W.}~\bibnamefont {Burkert}}, \bibinfo
  {author} {\bibfnamefont {V.}~\bibnamefont {Burwitz}}, \bibinfo {author}
  {\bibfnamefont {N.}~\bibnamefont {Cappelluti}}, \bibinfo {author}
  {\bibfnamefont {E.}~\bibnamefont {Churazov}}, \bibinfo {author}
  {\bibfnamefont {K.}~\bibnamefont {Dennerl}}, \bibinfo {author} {\bibfnamefont
  {J.}~\bibnamefont {Eder}}, \bibinfo {author} {\bibfnamefont {J.}~\bibnamefont
  {Elbs}}, \bibinfo {author} {\bibfnamefont {M.}~\bibnamefont {Freyberg}},
  \bibinfo {author} {\bibfnamefont {P.}~\bibnamefont {Friedrich}}, \bibinfo
  {author} {\bibfnamefont {M.}~\bibnamefont {F{\"{u}}rmetz}}, \bibinfo {author}
  {\bibfnamefont {R.}~\bibnamefont {Gaida}}, \bibinfo {author} {\bibfnamefont
  {O.}~\bibnamefont {H{\"{a}}lker}}, \bibinfo {author} {\bibfnamefont
  {G.}~\bibnamefont {Hartner}}, \bibinfo {author} {\bibfnamefont
  {G.}~\bibnamefont {Hasinger}}, \bibinfo {author} {\bibfnamefont
  {S.}~\bibnamefont {Hermann}}, \bibinfo {author} {\bibfnamefont
  {H.}~\bibnamefont {Huber}}, \bibinfo {author} {\bibfnamefont
  {E.}~\bibnamefont {Kendziorra}}, \bibinfo {author} {\bibfnamefont
  {A.}~\bibnamefont {von Kienlin}}, \bibinfo {author} {\bibfnamefont
  {W.}~\bibnamefont {Kink}}, \bibinfo {author} {\bibfnamefont {I.}~\bibnamefont
  {Kreykenbohm}}, \bibinfo {author} {\bibfnamefont {G.}~\bibnamefont {Lamer}},
  \bibinfo {author} {\bibfnamefont {I.}~\bibnamefont {Lapchov}}, \bibinfo
  {author} {\bibfnamefont {K.}~\bibnamefont {Lehmann}}, \bibinfo {author}
  {\bibfnamefont {N.}~\bibnamefont {Meidinger}}, \bibinfo {author}
  {\bibfnamefont {B.}~\bibnamefont {Mican}}, \bibinfo {author} {\bibfnamefont
  {J.}~\bibnamefont {Mohr}}, \bibinfo {author} {\bibfnamefont {M.}~\bibnamefont
  {M{\"{u}}hlegger}}, \bibinfo {author} {\bibfnamefont {S.}~\bibnamefont
  {M{\"{u}}ller}}, \bibinfo {author} {\bibfnamefont {K.}~\bibnamefont
  {Nandra}}, \bibinfo {author} {\bibfnamefont {M.}~\bibnamefont {Pavlinsky}},
  \bibinfo {author} {\bibfnamefont {E.}~\bibnamefont {Pfeffermann}}, \bibinfo
  {author} {\bibfnamefont {T.}~\bibnamefont {Reiprich}}, \bibinfo {author}
  {\bibfnamefont {J.}~\bibnamefont {Robrade}}, \bibinfo {author} {\bibfnamefont
  {C.}~\bibnamefont {Roh{\'{e}}}}, \bibinfo {author} {\bibfnamefont
  {A.}~\bibnamefont {Santangelo}}, \bibinfo {author} {\bibfnamefont
  {G.}~\bibnamefont {Sch{\"{a}}chner}}, \bibinfo {author} {\bibfnamefont
  {T.}~\bibnamefont {Schanz}}, \bibinfo {author} {\bibfnamefont
  {C.}~\bibnamefont {Schmid}}, \bibinfo {author} {\bibfnamefont
  {J.}~\bibnamefont {Schmitt}}, \bibinfo {author} {\bibfnamefont
  {R.}~\bibnamefont {Schreib}}, \bibinfo {author} {\bibfnamefont
  {F.}~\bibnamefont {Schrey}}, \bibinfo {author} {\bibfnamefont
  {A.}~\bibnamefont {Schwope}}, \bibinfo {author} {\bibfnamefont
  {M.}~\bibnamefont {Steinmetz}}, \bibinfo {author} {\bibfnamefont
  {L.}~\bibnamefont {Str{\"{u}}der}}, \bibinfo {author} {\bibfnamefont
  {R.}~\bibnamefont {Sunyaev}}, \bibinfo {author} {\bibfnamefont
  {C.}~\bibnamefont {Tenzer}}, \bibinfo {author} {\bibfnamefont
  {L.}~\bibnamefont {Tiedemann}}, \bibinfo {author} {\bibfnamefont
  {M.}~\bibnamefont {Vongehr}}, \ and\ \bibinfo {author} {\bibfnamefont
  {J.}~\bibnamefont {Wilms}},\ }in\ \href {\doibase 10.1117/12.856577} {\emph
  {\bibinfo {booktitle} {Society of Photo-Optical Instrumentation Engineers
  (SPIE) Conference Series}}},\ \bibinfo {series} {Society of Photo-Optical
  Instrumentation Engineers (SPIE) Conference Series}, Vol.\ \bibinfo {volume}
  {7732}\ (\bibinfo {year} {2010})\ \Eprint {http://arxiv.org/abs/1001.2502}
  {arXiv:1001.2502 [astro-ph.CO]} \BibitemShut {NoStop}%
\bibitem [{\citenamefont {{Merloni}}\ \emph {et~al.}(2012)\citenamefont
  {{Merloni}}, \citenamefont {{Predehl}}, \citenamefont {{Becker}},
  \citenamefont {{B{\"o}hringer}}, \citenamefont {{Boller}}, \citenamefont
  {{Brunner}}, \citenamefont {{Brusa}}, \citenamefont {{Dennerl}},
  \citenamefont {{Freyberg}}, \citenamefont {{Friedrich}}, \citenamefont
  {{Georgakakis}}, \citenamefont {{Haberl}}, \citenamefont {{Hasinger}},
  \citenamefont {{Meidinger}}, \citenamefont {{Mohr}}, \citenamefont
  {{Nandra}}, \citenamefont {{Rau}}, \citenamefont {{Reiprich}}, \citenamefont
  {{Robrade}}, \citenamefont {{Salvato}}, \citenamefont {{Santangelo}},
  \citenamefont {{Sasaki}}, \citenamefont {{Schwope}}, \citenamefont
  {{Wilms}},\ and\ \citenamefont {{German eROSITA Consortium}}}]{merloni12}%
  \BibitemOpen
  \bibfield  {author} {\bibinfo {author} {\bibfnamefont {A.}~\bibnamefont
  {{Merloni}}}, \bibinfo {author} {\bibfnamefont {P.}~\bibnamefont
  {{Predehl}}}, \bibinfo {author} {\bibfnamefont {W.}~\bibnamefont {{Becker}}},
  \bibinfo {author} {\bibfnamefont {H.}~\bibnamefont {{B{\"o}hringer}}},
  \bibinfo {author} {\bibfnamefont {T.}~\bibnamefont {{Boller}}}, \bibinfo
  {author} {\bibfnamefont {H.}~\bibnamefont {{Brunner}}}, \bibinfo {author}
  {\bibfnamefont {M.}~\bibnamefont {{Brusa}}}, \bibinfo {author} {\bibfnamefont
  {K.}~\bibnamefont {{Dennerl}}}, \bibinfo {author} {\bibfnamefont
  {M.}~\bibnamefont {{Freyberg}}}, \bibinfo {author} {\bibfnamefont
  {P.}~\bibnamefont {{Friedrich}}}, \bibinfo {author} {\bibfnamefont
  {A.}~\bibnamefont {{Georgakakis}}}, \bibinfo {author} {\bibfnamefont
  {F.}~\bibnamefont {{Haberl}}}, \bibinfo {author} {\bibfnamefont
  {G.}~\bibnamefont {{Hasinger}}}, \bibinfo {author} {\bibfnamefont
  {N.}~\bibnamefont {{Meidinger}}}, \bibinfo {author} {\bibfnamefont
  {J.}~\bibnamefont {{Mohr}}}, \bibinfo {author} {\bibfnamefont
  {K.}~\bibnamefont {{Nandra}}}, \bibinfo {author} {\bibfnamefont
  {A.}~\bibnamefont {{Rau}}}, \bibinfo {author} {\bibfnamefont {T.~H.}\
  \bibnamefont {{Reiprich}}}, \bibinfo {author} {\bibfnamefont
  {J.}~\bibnamefont {{Robrade}}}, \bibinfo {author} {\bibfnamefont
  {M.}~\bibnamefont {{Salvato}}}, \bibinfo {author} {\bibfnamefont
  {A.}~\bibnamefont {{Santangelo}}}, \bibinfo {author} {\bibfnamefont
  {M.}~\bibnamefont {{Sasaki}}}, \bibinfo {author} {\bibfnamefont
  {A.}~\bibnamefont {{Schwope}}}, \bibinfo {author} {\bibfnamefont
  {J.}~\bibnamefont {{Wilms}}}, \ and\ \bibinfo {author} {\bibfnamefont
  {t.}~\bibnamefont {{German eROSITA Consortium}}},\ }\href@noop {} {\bibfield
  {journal} {\bibinfo  {journal} {ArXiv e-prints}\ } (\bibinfo {year}
  {2012})},\ \Eprint {http://arxiv.org/abs/1209.3114} {arXiv:1209.3114
  [astro-ph.HE]} \BibitemShut {NoStop}%
\bibitem [{\citenamefont {{Grandis}}\ \emph {et~al.}(2018)\citenamefont
  {{Grandis}}, \citenamefont {{Mohr}}, \citenamefont {{Dietrich}},
  \citenamefont {{Bocquet}}, \citenamefont {{Saro}}, \citenamefont {{Klein}},
  \citenamefont {{Paulus}},\ and\ \citenamefont {{Capasso}}}]{grandis18}%
  \BibitemOpen
  \bibfield  {author} {\bibinfo {author} {\bibfnamefont {S.}~\bibnamefont
  {{Grandis}}}, \bibinfo {author} {\bibfnamefont {J.~J.}\ \bibnamefont
  {{Mohr}}}, \bibinfo {author} {\bibfnamefont {J.~P.}\ \bibnamefont
  {{Dietrich}}}, \bibinfo {author} {\bibfnamefont {S.}~\bibnamefont
  {{Bocquet}}}, \bibinfo {author} {\bibfnamefont {A.}~\bibnamefont {{Saro}}},
  \bibinfo {author} {\bibfnamefont {M.}~\bibnamefont {{Klein}}}, \bibinfo
  {author} {\bibfnamefont {M.}~\bibnamefont {{Paulus}}}, \ and\ \bibinfo
  {author} {\bibfnamefont {R.}~\bibnamefont {{Capasso}}},\ }\href@noop {}
  {\bibfield  {journal} {\bibinfo  {journal} {arXiv e-prints}\ } (\bibinfo
  {year} {2018})},\ \Eprint {http://arxiv.org/abs/1810.10553}
  {arXiv:1810.10553} \BibitemShut {NoStop}%
\bibitem [{\citenamefont {Benson}\ \emph {et~al.}(2014)\citenamefont {Benson},
  \citenamefont {Ade}, \citenamefont {Ahmed}, \citenamefont {Allen},
  \citenamefont {Arnold}, \citenamefont {Austermann}, \citenamefont {Bender},
  \citenamefont {Bleem}, \citenamefont {Carlstrom}, \citenamefont {Chang},
  \citenamefont {Cho}, \citenamefont {Cliche}, \citenamefont {Crawford},
  \citenamefont {Cukierman}, \citenamefont {de~Haan}, \citenamefont {Dobbs},
  \citenamefont {Dutcher}, \citenamefont {Everett}, \citenamefont {Gilbert},
  \citenamefont {Halverson}, \citenamefont {Hanson}, \citenamefont
  {Harrington}, \citenamefont {Hattori}, \citenamefont {Henning}, \citenamefont
  {Hilton}, \citenamefont {Holder}, \citenamefont {Holzapfel}, \citenamefont
  {Irwin}, \citenamefont {Keisler}, \citenamefont {Knox}, \citenamefont
  {Kubik}, \citenamefont {Kuo}, \citenamefont {Lee}, \citenamefont {Leitch},
  \citenamefont {Li}, \citenamefont {McDonald}, \citenamefont {Meyer},
  \citenamefont {Montgomery}, \citenamefont {Myers}, \citenamefont {Natoli},
  \citenamefont {Nguyen}, \citenamefont {Novosad}, \citenamefont {Padin},
  \citenamefont {Pan}, \citenamefont {Pearson}, \citenamefont {Reichardt},
  \citenamefont {Ruhl}, \citenamefont {Saliwanchik}, \citenamefont {Simard},
  \citenamefont {Smecher}, \citenamefont {Sayre}, \citenamefont {Shirokoff},
  \citenamefont {Stark}, \citenamefont {Story}, \citenamefont {Suzuki},
  \citenamefont {Thompson}, \citenamefont {Tucker}, \citenamefont
  {Vanderlinde}, \citenamefont {Vieira}, \citenamefont {Vikhlinin},
  \citenamefont {Wang}, \citenamefont {Yefremenko},\ and\ \citenamefont
  {Yoon}}]{benson14}%
  \BibitemOpen
  \bibfield  {author} {\bibinfo {author} {\bibfnamefont {B.}~\bibnamefont
  {Benson}}, \bibinfo {author} {\bibfnamefont {P.}~\bibnamefont {Ade}},
  \bibinfo {author} {\bibfnamefont {Z.}~\bibnamefont {Ahmed}}, \bibinfo
  {author} {\bibfnamefont {S.}~\bibnamefont {Allen}}, \bibinfo {author}
  {\bibfnamefont {K.}~\bibnamefont {Arnold}}, \bibinfo {author} {\bibfnamefont
  {J.}~\bibnamefont {Austermann}}, \bibinfo {author} {\bibfnamefont
  {A.}~\bibnamefont {Bender}}, \bibinfo {author} {\bibfnamefont
  {L.}~\bibnamefont {Bleem}}, \bibinfo {author} {\bibfnamefont
  {J.}~\bibnamefont {Carlstrom}}, \bibinfo {author} {\bibfnamefont
  {C.}~\bibnamefont {Chang}}, \bibinfo {author} {\bibfnamefont
  {H.}~\bibnamefont {Cho}}, \bibinfo {author} {\bibfnamefont {J.}~\bibnamefont
  {Cliche}}, \bibinfo {author} {\bibfnamefont {T.}~\bibnamefont {Crawford}},
  \bibinfo {author} {\bibfnamefont {A.}~\bibnamefont {Cukierman}}, \bibinfo
  {author} {\bibfnamefont {T.}~\bibnamefont {de~Haan}}, \bibinfo {author}
  {\bibfnamefont {M.}~\bibnamefont {Dobbs}}, \bibinfo {author} {\bibfnamefont
  {D.}~\bibnamefont {Dutcher}}, \bibinfo {author} {\bibfnamefont
  {W.}~\bibnamefont {Everett}}, \bibinfo {author} {\bibfnamefont
  {A.}~\bibnamefont {Gilbert}}, \bibinfo {author} {\bibfnamefont
  {N.}~\bibnamefont {Halverson}}, \bibinfo {author} {\bibfnamefont
  {D.}~\bibnamefont {Hanson}}, \bibinfo {author} {\bibfnamefont
  {N.}~\bibnamefont {Harrington}}, \bibinfo {author} {\bibfnamefont
  {K.}~\bibnamefont {Hattori}}, \bibinfo {author} {\bibfnamefont
  {J.}~\bibnamefont {Henning}}, \bibinfo {author} {\bibfnamefont
  {G.}~\bibnamefont {Hilton}}, \bibinfo {author} {\bibfnamefont
  {G.}~\bibnamefont {Holder}}, \bibinfo {author} {\bibfnamefont
  {W.}~\bibnamefont {Holzapfel}}, \bibinfo {author} {\bibfnamefont
  {K.}~\bibnamefont {Irwin}}, \bibinfo {author} {\bibfnamefont
  {R.}~\bibnamefont {Keisler}}, \bibinfo {author} {\bibfnamefont
  {L.}~\bibnamefont {Knox}}, \bibinfo {author} {\bibfnamefont {D.}~\bibnamefont
  {Kubik}}, \bibinfo {author} {\bibfnamefont {C.}~\bibnamefont {Kuo}}, \bibinfo
  {author} {\bibfnamefont {A.}~\bibnamefont {Lee}}, \bibinfo {author}
  {\bibfnamefont {E.}~\bibnamefont {Leitch}}, \bibinfo {author} {\bibfnamefont
  {D.}~\bibnamefont {Li}}, \bibinfo {author} {\bibfnamefont {M.}~\bibnamefont
  {McDonald}}, \bibinfo {author} {\bibfnamefont {S.}~\bibnamefont {Meyer}},
  \bibinfo {author} {\bibfnamefont {J.}~\bibnamefont {Montgomery}}, \bibinfo
  {author} {\bibfnamefont {M.}~\bibnamefont {Myers}}, \bibinfo {author}
  {\bibfnamefont {T.}~\bibnamefont {Natoli}}, \bibinfo {author} {\bibfnamefont
  {H.}~\bibnamefont {Nguyen}}, \bibinfo {author} {\bibfnamefont
  {V.}~\bibnamefont {Novosad}}, \bibinfo {author} {\bibfnamefont
  {S.}~\bibnamefont {Padin}}, \bibinfo {author} {\bibfnamefont
  {Z.}~\bibnamefont {Pan}}, \bibinfo {author} {\bibfnamefont {J.}~\bibnamefont
  {Pearson}}, \bibinfo {author} {\bibfnamefont {C.}~\bibnamefont {Reichardt}},
  \bibinfo {author} {\bibfnamefont {J.}~\bibnamefont {Ruhl}}, \bibinfo {author}
  {\bibfnamefont {B.}~\bibnamefont {Saliwanchik}}, \bibinfo {author}
  {\bibfnamefont {G.}~\bibnamefont {Simard}}, \bibinfo {author} {\bibfnamefont
  {G.}~\bibnamefont {Smecher}}, \bibinfo {author} {\bibfnamefont
  {J.}~\bibnamefont {Sayre}}, \bibinfo {author} {\bibfnamefont
  {E.}~\bibnamefont {Shirokoff}}, \bibinfo {author} {\bibfnamefont
  {A.}~\bibnamefont {Stark}}, \bibinfo {author} {\bibfnamefont
  {K.}~\bibnamefont {Story}}, \bibinfo {author} {\bibfnamefont
  {A.}~\bibnamefont {Suzuki}}, \bibinfo {author} {\bibfnamefont
  {K.}~\bibnamefont {Thompson}}, \bibinfo {author} {\bibfnamefont
  {C.}~\bibnamefont {Tucker}}, \bibinfo {author} {\bibfnamefont
  {K.}~\bibnamefont {Vanderlinde}}, \bibinfo {author} {\bibfnamefont
  {J.}~\bibnamefont {Vieira}}, \bibinfo {author} {\bibfnamefont
  {A.}~\bibnamefont {Vikhlinin}}, \bibinfo {author} {\bibfnamefont
  {G.}~\bibnamefont {Wang}}, \bibinfo {author} {\bibfnamefont {V.}~\bibnamefont
  {Yefremenko}}, \ and\ \bibinfo {author} {\bibfnamefont {K.}~\bibnamefont
  {Yoon}},\ }in\ \href {\doibase 10.1117/12.2057305} {\emph {\bibinfo
  {booktitle} {Millimeter, Submillimeter, and Far-Infrared Detectors and
  Instrumentation for Astronomy VII}}},\ \bibinfo {series} {\procspie}, Vol.\
  \bibinfo {volume} {9153}\ (\bibinfo {year} {2014})\ p.\ \bibinfo {pages}
  {91531P},\ \Eprint {http://arxiv.org/abs/1407.2973} {arXiv:1407.2973
  [astro-ph.IM]} \BibitemShut {NoStop}%
\bibitem [{\citenamefont {{Choi}}\ \emph {et~al.}(2018)\citenamefont {{Choi}},
  \citenamefont {{Austermann}}, \citenamefont {{Beall}}, \citenamefont
  {{Crowley}}, \citenamefont {{Datta}}, \citenamefont {{Duff}}, \citenamefont
  {{Gallardo}}, \citenamefont {{Ho}}, \citenamefont {{Hubmayr}}, \citenamefont
  {{Koopman}}, \citenamefont {{Li}}, \citenamefont {{Nati}}, \citenamefont
  {{Niemack}}, \citenamefont {{Page}}, \citenamefont {{Salatino}},
  \citenamefont {{Simon}}, \citenamefont {{Staggs}}, \citenamefont {{Stevens}},
  \citenamefont {{Ullom}},\ and\ \citenamefont {{Wollack}}}]{AdvACTpol18}%
  \BibitemOpen
  \bibfield  {author} {\bibinfo {author} {\bibfnamefont {S.~K.}\ \bibnamefont
  {{Choi}}}, \bibinfo {author} {\bibfnamefont {J.}~\bibnamefont
  {{Austermann}}}, \bibinfo {author} {\bibfnamefont {J.~A.}\ \bibnamefont
  {{Beall}}}, \bibinfo {author} {\bibfnamefont {K.~T.}\ \bibnamefont
  {{Crowley}}}, \bibinfo {author} {\bibfnamefont {R.}~\bibnamefont {{Datta}}},
  \bibinfo {author} {\bibfnamefont {S.~M.}\ \bibnamefont {{Duff}}}, \bibinfo
  {author} {\bibfnamefont {P.~A.}\ \bibnamefont {{Gallardo}}}, \bibinfo
  {author} {\bibfnamefont {S.~P.}\ \bibnamefont {{Ho}}}, \bibinfo {author}
  {\bibfnamefont {J.}~\bibnamefont {{Hubmayr}}}, \bibinfo {author}
  {\bibfnamefont {B.~J.}\ \bibnamefont {{Koopman}}}, \bibinfo {author}
  {\bibfnamefont {Y.}~\bibnamefont {{Li}}}, \bibinfo {author} {\bibfnamefont
  {F.}~\bibnamefont {{Nati}}}, \bibinfo {author} {\bibfnamefont {M.~D.}\
  \bibnamefont {{Niemack}}}, \bibinfo {author} {\bibfnamefont {L.~A.}\
  \bibnamefont {{Page}}}, \bibinfo {author} {\bibfnamefont {M.}~\bibnamefont
  {{Salatino}}}, \bibinfo {author} {\bibfnamefont {S.~M.}\ \bibnamefont
  {{Simon}}}, \bibinfo {author} {\bibfnamefont {S.~T.}\ \bibnamefont
  {{Staggs}}}, \bibinfo {author} {\bibfnamefont {J.}~\bibnamefont {{Stevens}}},
  \bibinfo {author} {\bibfnamefont {J.}~\bibnamefont {{Ullom}}}, \ and\
  \bibinfo {author} {\bibfnamefont {E.~J.}\ \bibnamefont {{Wollack}}},\ }\href
  {\doibase 10.1007/s10909-018-1982-4} {\bibfield  {journal} {\bibinfo
  {journal} {Journal of Low Temperature Physics}\ }\textbf {\bibinfo {volume}
  {193}},\ \bibinfo {pages} {267} (\bibinfo {year} {2018})},\ \Eprint
  {http://arxiv.org/abs/1711.04841} {arXiv:1711.04841 [astro-ph.IM]}
  \BibitemShut {NoStop}%
\bibitem [{\citenamefont {{Ade}}\ \emph
  {et~al.}(2019{\natexlab{a}})\citenamefont {{Ade}}, \citenamefont {{Aguirre}},
  \citenamefont {{Ahmed}}, \citenamefont {{Aiola}}, \citenamefont {{Ali}},
  \citenamefont {{Alonso}}, \citenamefont {{Alvarez}}, \citenamefont
  {{Arnold}}, \citenamefont {{Ashton}}, \citenamefont {{Austermann}},\ and\
  \citenamefont {et~al.}}]{SO19}%
  \BibitemOpen
  \bibfield  {author} {\bibinfo {author} {\bibfnamefont {P.}~\bibnamefont
  {{Ade}}}, \bibinfo {author} {\bibfnamefont {J.}~\bibnamefont {{Aguirre}}},
  \bibinfo {author} {\bibfnamefont {Z.}~\bibnamefont {{Ahmed}}}, \bibinfo
  {author} {\bibfnamefont {S.}~\bibnamefont {{Aiola}}}, \bibinfo {author}
  {\bibfnamefont {A.}~\bibnamefont {{Ali}}}, \bibinfo {author} {\bibfnamefont
  {D.}~\bibnamefont {{Alonso}}}, \bibinfo {author} {\bibfnamefont {M.~A.}\
  \bibnamefont {{Alvarez}}}, \bibinfo {author} {\bibfnamefont {K.}~\bibnamefont
  {{Arnold}}}, \bibinfo {author} {\bibfnamefont {P.}~\bibnamefont {{Ashton}}},
  \bibinfo {author} {\bibfnamefont {J.}~\bibnamefont {{Austermann}}}, \ and\
  \bibinfo {author} {\bibnamefont {et~al.}},\ }\href {\doibase
  10.1088/1475-7516/2019/02/056} {\bibfield  {journal} {\bibinfo  {journal}
  {\jcap}\ }\textbf {\bibinfo {volume} {2}},\ \bibinfo {eid} {056} (\bibinfo
  {year} {2019}{\natexlab{a}})},\ \Eprint {http://arxiv.org/abs/1808.07445}
  {arXiv:1808.07445} \BibitemShut {NoStop}%
\bibitem [{\citenamefont {{Mroczkowski}}\ \emph {et~al.}(2018)\citenamefont
  {{Mroczkowski}}, \citenamefont {{Nagai}}, \citenamefont {{Basu}},
  \citenamefont {{Chluba}}, \citenamefont {{Sayers}}, \citenamefont {{Adam}},
  \citenamefont {{Churazov}}, \citenamefont {{Crites}}, \citenamefont {{Di
  Mascolo}}, \citenamefont {{Eckert}}, \citenamefont {{Macias-Perez}},
  \citenamefont {{Mayet}}, \citenamefont {{Perotto}}, \citenamefont
  {{Pointecouteau}}, \citenamefont {{Romero}}, \citenamefont {{Ruppin}},
  \citenamefont {{Scannapieco}},\ and\ \citenamefont
  {{ZuHone}}}]{Mroczkowski2018}%
  \BibitemOpen
  \bibfield  {author} {\bibinfo {author} {\bibfnamefont {T.}~\bibnamefont
  {{Mroczkowski}}}, \bibinfo {author} {\bibfnamefont {D.}~\bibnamefont
  {{Nagai}}}, \bibinfo {author} {\bibfnamefont {K.}~\bibnamefont {{Basu}}},
  \bibinfo {author} {\bibfnamefont {J.}~\bibnamefont {{Chluba}}}, \bibinfo
  {author} {\bibfnamefont {J.}~\bibnamefont {{Sayers}}}, \bibinfo {author}
  {\bibfnamefont {R.}~\bibnamefont {{Adam}}}, \bibinfo {author} {\bibfnamefont
  {E.}~\bibnamefont {{Churazov}}}, \bibinfo {author} {\bibfnamefont
  {A.}~\bibnamefont {{Crites}}}, \bibinfo {author} {\bibfnamefont
  {L.}~\bibnamefont {{Di Mascolo}}}, \bibinfo {author} {\bibfnamefont
  {D.}~\bibnamefont {{Eckert}}}, \bibinfo {author} {\bibfnamefont
  {J.}~\bibnamefont {{Macias-Perez}}}, \bibinfo {author} {\bibfnamefont
  {F.}~\bibnamefont {{Mayet}}}, \bibinfo {author} {\bibfnamefont
  {L.}~\bibnamefont {{Perotto}}}, \bibinfo {author} {\bibfnamefont
  {E.}~\bibnamefont {{Pointecouteau}}}, \bibinfo {author} {\bibfnamefont
  {C.}~\bibnamefont {{Romero}}}, \bibinfo {author} {\bibfnamefont
  {F.}~\bibnamefont {{Ruppin}}}, \bibinfo {author} {\bibfnamefont
  {E.}~\bibnamefont {{Scannapieco}}}, \ and\ \bibinfo {author} {\bibfnamefont
  {J.}~\bibnamefont {{ZuHone}}},\ }\href@noop {} {\bibfield  {journal}
  {\bibinfo  {journal} {arXiv e-prints}\ ,\ \bibinfo {eid} {arXiv:1811.02310}}
  (\bibinfo {year} {2018})},\ \Eprint {http://arxiv.org/abs/1811.02310}
  {arXiv:1811.02310 [astro-ph.CO]} \BibitemShut {NoStop}%
\bibitem [{\citenamefont {{Battaglia}}\ \emph
  {et~al.}(2019{\natexlab{b}})\citenamefont {{Battaglia}}, \citenamefont
  {{Hill}}, \citenamefont {{Amodeo}}, \citenamefont {{Bartlett}}, \citenamefont
  {{Basu}}, \citenamefont {{Erler}}, \citenamefont {{Ferraro}}, \citenamefont
  {{Hernquist}}, \citenamefont {{Madhavacheril}}, \citenamefont {{McQuinn}},
  \citenamefont {{Mroczkowski}}, \citenamefont {{Nagai}}, \citenamefont
  {{Schaan}}, \citenamefont {{Somerville}}, \citenamefont {{Sunyaev}},
  \citenamefont {{Vogelsberger}},\ and\ \citenamefont
  {{Werk}}}]{Battaglia2019}%
  \BibitemOpen
  \bibfield  {author} {\bibinfo {author} {\bibfnamefont {N.}~\bibnamefont
  {{Battaglia}}}, \bibinfo {author} {\bibfnamefont {J.~C.}\ \bibnamefont
  {{Hill}}}, \bibinfo {author} {\bibfnamefont {S.}~\bibnamefont {{Amodeo}}},
  \bibinfo {author} {\bibfnamefont {J.~G.}\ \bibnamefont {{Bartlett}}},
  \bibinfo {author} {\bibfnamefont {K.}~\bibnamefont {{Basu}}}, \bibinfo
  {author} {\bibfnamefont {J.}~\bibnamefont {{Erler}}}, \bibinfo {author}
  {\bibfnamefont {S.}~\bibnamefont {{Ferraro}}}, \bibinfo {author}
  {\bibfnamefont {L.}~\bibnamefont {{Hernquist}}}, \bibinfo {author}
  {\bibfnamefont {M.}~\bibnamefont {{Madhavacheril}}}, \bibinfo {author}
  {\bibfnamefont {M.}~\bibnamefont {{McQuinn}}}, \bibinfo {author}
  {\bibfnamefont {T.}~\bibnamefont {{Mroczkowski}}}, \bibinfo {author}
  {\bibfnamefont {D.}~\bibnamefont {{Nagai}}}, \bibinfo {author} {\bibfnamefont
  {E.}~\bibnamefont {{Schaan}}}, \bibinfo {author} {\bibfnamefont
  {R.}~\bibnamefont {{Somerville}}}, \bibinfo {author} {\bibfnamefont
  {R.}~\bibnamefont {{Sunyaev}}}, \bibinfo {author} {\bibfnamefont
  {M.}~\bibnamefont {{Vogelsberger}}}, \ and\ \bibinfo {author} {\bibfnamefont
  {J.}~\bibnamefont {{Werk}}},\ }\href@noop {} {\bibfield  {journal} {\bibinfo
  {journal} {arXiv e-prints}\ ,\ \bibinfo {eid} {arXiv:1903.04647}} (\bibinfo
  {year} {2019}{\natexlab{b}})},\ \Eprint {http://arxiv.org/abs/1903.04647}
  {arXiv:1903.04647 [astro-ph.CO]} \BibitemShut {NoStop}%
\bibitem [{\citenamefont {{Vikram}}\ \emph {et~al.}(2017)\citenamefont
  {{Vikram}}, \citenamefont {{Lidz}},\ and\ \citenamefont
  {{Jain}}}]{Vikram2017}%
  \BibitemOpen
  \bibfield  {author} {\bibinfo {author} {\bibfnamefont {V.}~\bibnamefont
  {{Vikram}}}, \bibinfo {author} {\bibfnamefont {A.}~\bibnamefont {{Lidz}}}, \
  and\ \bibinfo {author} {\bibfnamefont {B.}~\bibnamefont {{Jain}}},\
  }\href@noop {} {\bibfield  {journal} {\bibinfo  {journal} {\mnras}\ }\textbf
  {\bibinfo {volume} {467}},\ \bibinfo {pages} {2315} (\bibinfo {year}
  {2017})}\BibitemShut {NoStop}%
\bibitem [{\citenamefont {{Hill}}\ \emph {et~al.}(2018)\citenamefont {{Hill}},
  \citenamefont {{Baxter}}, \citenamefont {{Lidz}}, \citenamefont {{Greco}},\
  and\ \citenamefont {{Jain}}}]{Hill2018}%
  \BibitemOpen
  \bibfield  {author} {\bibinfo {author} {\bibfnamefont {J.~C.}\ \bibnamefont
  {{Hill}}}, \bibinfo {author} {\bibfnamefont {E.~J.}\ \bibnamefont
  {{Baxter}}}, \bibinfo {author} {\bibfnamefont {A.}~\bibnamefont {{Lidz}}},
  \bibinfo {author} {\bibfnamefont {J.~P.}\ \bibnamefont {{Greco}}}, \ and\
  \bibinfo {author} {\bibfnamefont {B.}~\bibnamefont {{Jain}}},\ }\href@noop {}
  {\bibfield  {journal} {\bibinfo  {journal} {\prd}\ }\textbf {\bibinfo
  {volume} {97}},\ \bibinfo {pages} {083501} (\bibinfo {year}
  {2018})}\BibitemShut {NoStop}%
\bibitem [{\citenamefont {{Tanimura}}\ \emph {et~al.}(2019)\citenamefont
  {{Tanimura}}, \citenamefont {{Hinshaw}}, \citenamefont {{McCarthy}},
  \citenamefont {{Van Waerbeke}}, \citenamefont {{Aghanim}}, \citenamefont
  {{Ma}}, \citenamefont {{Mead}}, \citenamefont {{Hojjati}},\ and\
  \citenamefont {{Tr{\"o}ster}}}]{Tanimura2019}%
  \BibitemOpen
  \bibfield  {author} {\bibinfo {author} {\bibfnamefont {H.}~\bibnamefont
  {{Tanimura}}}, \bibinfo {author} {\bibfnamefont {G.}~\bibnamefont
  {{Hinshaw}}}, \bibinfo {author} {\bibfnamefont {I.~G.}\ \bibnamefont
  {{McCarthy}}}, \bibinfo {author} {\bibfnamefont {L.}~\bibnamefont {{Van
  Waerbeke}}}, \bibinfo {author} {\bibfnamefont {N.}~\bibnamefont {{Aghanim}}},
  \bibinfo {author} {\bibfnamefont {Y.-Z.}\ \bibnamefont {{Ma}}}, \bibinfo
  {author} {\bibfnamefont {A.}~\bibnamefont {{Mead}}}, \bibinfo {author}
  {\bibfnamefont {A.}~\bibnamefont {{Hojjati}}}, \ and\ \bibinfo {author}
  {\bibfnamefont {T.}~\bibnamefont {{Tr{\"o}ster}}},\ }\href {\doibase
  10.1093/mnras/sty3118} {\bibfield  {journal} {\bibinfo  {journal} {\mnras}\
  }\textbf {\bibinfo {volume} {483}},\ \bibinfo {pages} {223} (\bibinfo {year}
  {2019})},\ \Eprint {http://arxiv.org/abs/1709.05024} {arXiv:1709.05024
  [astro-ph.CO]} \BibitemShut {NoStop}%
\bibitem [{\citenamefont {{Hojjati}}\ \emph {et~al.}(2017)\citenamefont
  {{Hojjati}}, \citenamefont {{Tr{\"o}ster}}, \citenamefont
  {{Harnois-D{\'e}raps}}, \citenamefont {{McCarthy}}, \citenamefont {{van
  Waerbeke}}, \citenamefont {{Choi}}, \citenamefont {{Erben}}, \citenamefont
  {{Heymans}}, \citenamefont {{Hildebrandt}}, \citenamefont {{Hinshaw}},
  \citenamefont {{Ma}}, \citenamefont {{Miller}}, \citenamefont {{Viola}},\
  and\ \citenamefont {{Tanimura}}}]{Hojjati2017}%
  \BibitemOpen
  \bibfield  {author} {\bibinfo {author} {\bibfnamefont {A.}~\bibnamefont
  {{Hojjati}}}, \bibinfo {author} {\bibfnamefont {T.}~\bibnamefont
  {{Tr{\"o}ster}}}, \bibinfo {author} {\bibfnamefont {J.}~\bibnamefont
  {{Harnois-D{\'e}raps}}}, \bibinfo {author} {\bibfnamefont {I.~G.}\
  \bibnamefont {{McCarthy}}}, \bibinfo {author} {\bibfnamefont
  {L.}~\bibnamefont {{van Waerbeke}}}, \bibinfo {author} {\bibfnamefont
  {A.}~\bibnamefont {{Choi}}}, \bibinfo {author} {\bibfnamefont
  {T.}~\bibnamefont {{Erben}}}, \bibinfo {author} {\bibfnamefont
  {C.}~\bibnamefont {{Heymans}}}, \bibinfo {author} {\bibfnamefont
  {H.}~\bibnamefont {{Hildebrandt}}}, \bibinfo {author} {\bibfnamefont
  {G.}~\bibnamefont {{Hinshaw}}}, \bibinfo {author} {\bibfnamefont {Y.-Z.}\
  \bibnamefont {{Ma}}}, \bibinfo {author} {\bibfnamefont {L.}~\bibnamefont
  {{Miller}}}, \bibinfo {author} {\bibfnamefont {M.}~\bibnamefont {{Viola}}}, \
  and\ \bibinfo {author} {\bibfnamefont {H.}~\bibnamefont {{Tanimura}}},\
  }\href@noop {} {\bibfield  {journal} {\bibinfo  {journal} {\mnras}\ }\textbf
  {\bibinfo {volume} {471}},\ \bibinfo {pages} {1565} (\bibinfo {year}
  {2017})}\BibitemShut {NoStop}%
\bibitem [{\citenamefont {{Pandey}}\ \emph {et~al.}(2019)\citenamefont
  {{Pandey}}, \citenamefont {{Baxter}}, \citenamefont {{Xu}}, \citenamefont
  {{Orlowski-Scherer}}, \citenamefont {{Zhu}}, \citenamefont {{Lidz}},
  \citenamefont {{Aguirre}}, \citenamefont {{DeRose}}, \citenamefont
  {{Devlin}}, \citenamefont {{Hill}}, \citenamefont {{Jain}}, \citenamefont
  {{Sheth}}, \citenamefont {{Avila}}, \citenamefont {{Bertin}}, \citenamefont
  {{Brooks}}, \citenamefont {{Buckley-Geer}}, \citenamefont {{Carnero Rosell}},
  \citenamefont {{Carrasco Kind}}, \citenamefont {{Carretero}}, \citenamefont
  {{Castander}}, \citenamefont {{Cawthon}}, \citenamefont {{da Costa}},
  \citenamefont {{De Vicente}}, \citenamefont {{Desai}}, \citenamefont
  {{Diehl}}, \citenamefont {{Dietrich}}, \citenamefont {{Doel}}, \citenamefont
  {{Evrard}}, \citenamefont {{Flaugher}}, \citenamefont {{Fosalba}},
  \citenamefont {{Frieman}}, \citenamefont {{Garc{\'\i}a-Bellido}},
  \citenamefont {{Gerdes}}, \citenamefont {{Giannantonio}}, \citenamefont
  {{Gruendl}}, \citenamefont {{Gschwend}}, \citenamefont {{Hartley}},
  \citenamefont {{Hollowood}}, \citenamefont {{James}}, \citenamefont
  {{Krause}}, \citenamefont {{Kuehn}}, \citenamefont {{Kuropatkin}},
  \citenamefont {{Maia}}, \citenamefont {{Marshall}}, \citenamefont
  {{Melchior}}, \citenamefont {{Menanteau}}, \citenamefont {{Miquel}},
  \citenamefont {{Plazas}}, \citenamefont {{Roodman}}, \citenamefont
  {{Sanchez}}, \citenamefont {{Serrano}}, \citenamefont {{Sevilla-Noarbe}},
  \citenamefont {{Smith}}, \citenamefont {{Soares-Santos}}, \citenamefont
  {{Sobreira}}, \citenamefont {{Suchyta}}, \citenamefont {{Swanson}},
  \citenamefont {{Tarle}},\ and\ \citenamefont {{Wechsler}}}]{Pandey19}%
  \BibitemOpen
  \bibfield  {author} {\bibinfo {author} {\bibfnamefont {S.}~\bibnamefont
  {{Pandey}}}, \bibinfo {author} {\bibfnamefont {E.~J.}\ \bibnamefont
  {{Baxter}}}, \bibinfo {author} {\bibfnamefont {Z.}~\bibnamefont {{Xu}}},
  \bibinfo {author} {\bibfnamefont {J.}~\bibnamefont {{Orlowski-Scherer}}},
  \bibinfo {author} {\bibfnamefont {N.}~\bibnamefont {{Zhu}}}, \bibinfo
  {author} {\bibfnamefont {A.}~\bibnamefont {{Lidz}}}, \bibinfo {author}
  {\bibfnamefont {J.}~\bibnamefont {{Aguirre}}}, \bibinfo {author}
  {\bibfnamefont {J.}~\bibnamefont {{DeRose}}}, \bibinfo {author}
  {\bibfnamefont {M.}~\bibnamefont {{Devlin}}}, \bibinfo {author}
  {\bibfnamefont {J.~C.}\ \bibnamefont {{Hill}}}, \bibinfo {author}
  {\bibfnamefont {B.}~\bibnamefont {{Jain}}}, \bibinfo {author} {\bibfnamefont
  {R.~K.}\ \bibnamefont {{Sheth}}}, \bibinfo {author} {\bibfnamefont
  {S.}~\bibnamefont {{Avila}}}, \bibinfo {author} {\bibfnamefont
  {E.}~\bibnamefont {{Bertin}}}, \bibinfo {author} {\bibfnamefont
  {D.}~\bibnamefont {{Brooks}}}, \bibinfo {author} {\bibfnamefont
  {E.}~\bibnamefont {{Buckley-Geer}}}, \bibinfo {author} {\bibfnamefont
  {A.}~\bibnamefont {{Carnero Rosell}}}, \bibinfo {author} {\bibfnamefont
  {M.}~\bibnamefont {{Carrasco Kind}}}, \bibinfo {author} {\bibfnamefont
  {J.}~\bibnamefont {{Carretero}}}, \bibinfo {author} {\bibfnamefont {F.~J.}\
  \bibnamefont {{Castander}}}, \bibinfo {author} {\bibfnamefont
  {R.}~\bibnamefont {{Cawthon}}}, \bibinfo {author} {\bibfnamefont {L.~N.}\
  \bibnamefont {{da Costa}}}, \bibinfo {author} {\bibfnamefont
  {J.}~\bibnamefont {{De Vicente}}}, \bibinfo {author} {\bibfnamefont
  {S.}~\bibnamefont {{Desai}}}, \bibinfo {author} {\bibfnamefont {H.~T.}\
  \bibnamefont {{Diehl}}}, \bibinfo {author} {\bibfnamefont {J.~P.}\
  \bibnamefont {{Dietrich}}}, \bibinfo {author} {\bibfnamefont
  {P.}~\bibnamefont {{Doel}}}, \bibinfo {author} {\bibfnamefont {A.~E.}\
  \bibnamefont {{Evrard}}}, \bibinfo {author} {\bibfnamefont {B.}~\bibnamefont
  {{Flaugher}}}, \bibinfo {author} {\bibfnamefont {P.}~\bibnamefont
  {{Fosalba}}}, \bibinfo {author} {\bibfnamefont {J.}~\bibnamefont
  {{Frieman}}}, \bibinfo {author} {\bibfnamefont {J.}~\bibnamefont
  {{Garc{\'\i}a-Bellido}}}, \bibinfo {author} {\bibfnamefont {D.~W.}\
  \bibnamefont {{Gerdes}}}, \bibinfo {author} {\bibfnamefont {T.}~\bibnamefont
  {{Giannantonio}}}, \bibinfo {author} {\bibfnamefont {R.~A.}\ \bibnamefont
  {{Gruendl}}}, \bibinfo {author} {\bibfnamefont {J.}~\bibnamefont
  {{Gschwend}}}, \bibinfo {author} {\bibfnamefont {W.~G.}\ \bibnamefont
  {{Hartley}}}, \bibinfo {author} {\bibfnamefont {D.~L.}\ \bibnamefont
  {{Hollowood}}}, \bibinfo {author} {\bibfnamefont {D.~J.}\ \bibnamefont
  {{James}}}, \bibinfo {author} {\bibfnamefont {E.}~\bibnamefont {{Krause}}},
  \bibinfo {author} {\bibfnamefont {K.}~\bibnamefont {{Kuehn}}}, \bibinfo
  {author} {\bibfnamefont {N.}~\bibnamefont {{Kuropatkin}}}, \bibinfo {author}
  {\bibfnamefont {M.~A.~G.}\ \bibnamefont {{Maia}}}, \bibinfo {author}
  {\bibfnamefont {J.~L.}\ \bibnamefont {{Marshall}}}, \bibinfo {author}
  {\bibfnamefont {P.}~\bibnamefont {{Melchior}}}, \bibinfo {author}
  {\bibfnamefont {F.}~\bibnamefont {{Menanteau}}}, \bibinfo {author}
  {\bibfnamefont {R.}~\bibnamefont {{Miquel}}}, \bibinfo {author}
  {\bibfnamefont {A.~A.}\ \bibnamefont {{Plazas}}}, \bibinfo {author}
  {\bibfnamefont {A.}~\bibnamefont {{Roodman}}}, \bibinfo {author}
  {\bibfnamefont {E.}~\bibnamefont {{Sanchez}}}, \bibinfo {author}
  {\bibfnamefont {S.}~\bibnamefont {{Serrano}}}, \bibinfo {author}
  {\bibfnamefont {I.}~\bibnamefont {{Sevilla-Noarbe}}}, \bibinfo {author}
  {\bibfnamefont {M.}~\bibnamefont {{Smith}}}, \bibinfo {author} {\bibfnamefont
  {M.}~\bibnamefont {{Soares-Santos}}}, \bibinfo {author} {\bibfnamefont
  {F.}~\bibnamefont {{Sobreira}}}, \bibinfo {author} {\bibfnamefont
  {E.}~\bibnamefont {{Suchyta}}}, \bibinfo {author} {\bibfnamefont {M.~E.~C.}\
  \bibnamefont {{Swanson}}}, \bibinfo {author} {\bibfnamefont {G.}~\bibnamefont
  {{Tarle}}}, \ and\ \bibinfo {author} {\bibfnamefont {R.~H.}\ \bibnamefont
  {{Wechsler}}},\ }\href@noop {} {\bibfield  {journal} {\bibinfo  {journal}
  {arXiv e-prints}\ ,\ \bibinfo {eid} {arXiv:1904.13347}} (\bibinfo {year}
  {2019})},\ \Eprint {http://arxiv.org/abs/1904.13347} {arXiv:1904.13347
  [astro-ph.CO]} \BibitemShut {NoStop}%
\bibitem [{\citenamefont {{Ade}}\ \emph
  {et~al.}(2019{\natexlab{b}})\citenamefont {{Ade}}, \citenamefont {{Aguirre}},
  \citenamefont {{Ahmed}}, \citenamefont {{Aiola}}, \citenamefont {{Ali}},
  \citenamefont {{Alonso}}, \citenamefont {{Alvarez}}, \citenamefont
  {{Arnold}}, \citenamefont {{Ashton}}, \citenamefont {{Austermann}},
  \citenamefont {{Awan}}, \citenamefont {{Baccigalupi}}, \citenamefont
  {{Baildon}}, \citenamefont {{Barron}}, \citenamefont {{Battaglia}},
  \citenamefont {{Battye}}, \citenamefont {{Baxter}}, \citenamefont
  {{Bazarko}}, \citenamefont {{Beall}}, \citenamefont {{Bean}}, \citenamefont
  {{Beck}}, \citenamefont {{Beckman}}, \citenamefont {{Beringue}},
  \citenamefont {{Bianchini}}, \citenamefont {{Boada}}, \citenamefont
  {{Boettger}}, \citenamefont {{Bond}}, \citenamefont {{Borrill}},
  \citenamefont {{Brown}}, \citenamefont {{Bruno}}, \citenamefont {{Bryan}},
  \citenamefont {{Calabrese}}, \citenamefont {{Calafut}}, \citenamefont
  {{Calisse}}, \citenamefont {{Carron}}, \citenamefont {{Challinor}},
  \citenamefont {{Chesmore}}, \citenamefont {{Chinone}}, \citenamefont
  {{Chluba}}, \citenamefont {{Cho}}, \citenamefont {{Choi}}, \citenamefont
  {{Coppi}}, \citenamefont {{Cothard}}, \citenamefont {{Coughlin}},
  \citenamefont {{Crichton}}, \citenamefont {{Crowley}}, \citenamefont
  {{Crowley}}, \citenamefont {{Cukierman}}, \citenamefont {{D'Ewart}},
  \citenamefont {{D{\"u}nner}}, \citenamefont {{de Haan}}, \citenamefont
  {{Devlin}}, \citenamefont {{Dicker}}, \citenamefont {{Didier}}, \citenamefont
  {{Dobbs}}, \citenamefont {{Dober}}, \citenamefont {{Duell}}, \citenamefont
  {{Duff}}, \citenamefont {{Duivenvoorden}}, \citenamefont {{Dunkley}},
  \citenamefont {{Dusatko}}, \citenamefont {{Errard}}, \citenamefont
  {{Fabbian}}, \citenamefont {{Feeney}}, \citenamefont {{Ferraro}},
  \citenamefont {{Flux{\`a}}}, \citenamefont {{Freese}}, \citenamefont
  {{Frisch}}, \citenamefont {{Frolov}}, \citenamefont {{Fuller}}, \citenamefont
  {{Fuzia}}, \citenamefont {{Galitzki}}, \citenamefont {{Gallardo}},
  \citenamefont {{Tomas Galvez Ghersi}}, \citenamefont {{Gao}}, \citenamefont
  {{Gawiser}}, \citenamefont {{Gerbino}}, \citenamefont {{Gluscevic}},
  \citenamefont {{Goeckner-Wald}}, \citenamefont {{Golec}}, \citenamefont
  {{Gordon}}, \citenamefont {{Gralla}}, \citenamefont {{Green}}, \citenamefont
  {{Grigorian}}, \citenamefont {{Groh}}, \citenamefont {{Groppi}},
  \citenamefont {{Guan}}, \citenamefont {{Gudmundsson}}, \citenamefont {{Han}},
  \citenamefont {{Hargrave}}, \citenamefont {{Hasegawa}}, \citenamefont
  {{Hasselfield}}, \citenamefont {{Hattori}}, \citenamefont {{Haynes}},
  \citenamefont {{Hazumi}}, \citenamefont {{He}}, \citenamefont {{Healy}},
  \citenamefont {{Henderson}}, \citenamefont {{Hervias-Caimapo}}, \citenamefont
  {{Hill}}, \citenamefont {{Hill}}, \citenamefont {{Hilton}}, \citenamefont
  {{Hilton}}, \citenamefont {{Hincks}}, \citenamefont {{Hinshaw}},
  \citenamefont {{Hlo{\v{z}}ek}}, \citenamefont {{Ho}}, \citenamefont {{Ho}},
  \citenamefont {{Howe}}, \citenamefont {{Huang}}, \citenamefont {{Hubmayr}},
  \citenamefont {{Huffenberger}}, \citenamefont {{Hughes}}, \citenamefont
  {{Ijjas}}, \citenamefont {{Ikape}}, \citenamefont {{Irwin}}, \citenamefont
  {{Jaffe}}, \citenamefont {{Jain}}, \citenamefont {{Jeong}}, \citenamefont
  {{Kaneko}}, \citenamefont {{Karpel}}, \citenamefont {{Katayama}},
  \citenamefont {{Keating}}, \citenamefont {{Kernasovskiy}}, \citenamefont
  {{Keskitalo}}, \citenamefont {{Kisner}}, \citenamefont {{Kiuchi}},
  \citenamefont {{Klein}}, \citenamefont {{Knowles}}, \citenamefont
  {{Koopman}}, \citenamefont {{Kosowsky}}, \citenamefont {{Krachmalnicoff}},
  \citenamefont {{Kuenstner}}, \citenamefont {{Kuo}}, \citenamefont {{Kusaka}},
  \citenamefont {{Lashner}}, \citenamefont {{Lee}}, \citenamefont {{Lee}},
  \citenamefont {{Leon}}, \citenamefont {{Leung}}, \citenamefont {{Lewis}},
  \citenamefont {{Li}}, \citenamefont {{Li}}, \citenamefont {{Limon}},
  \citenamefont {{Linder}}, \citenamefont {{Lopez-Caraballo}}, \citenamefont
  {{Louis}}, \citenamefont {{Lowry}}, \citenamefont {{Lungu}}, \citenamefont
  {{Madhavacheril}}, \citenamefont {{Mak}}, \citenamefont {{Maldonado}},
  \citenamefont {{Mani}}, \citenamefont {{Mates}}, \citenamefont {{Matsuda}},
  \citenamefont {{Maurin}}, \citenamefont {{Mauskopf}}, \citenamefont {{May}},
  \citenamefont {{McCallum}}, \citenamefont {{McKenney}}, \citenamefont
  {{McMahon}}, \citenamefont {{Meerburg}}, \citenamefont {{Meyers}},
  \citenamefont {{Miller}}, \citenamefont {{Mirmelstein}}, \citenamefont
  {{Moodley}}, \citenamefont {{Munchmeyer}}, \citenamefont {{Munson}},
  \citenamefont {{Naess}}, \citenamefont {{Nati}}, \citenamefont {{Navaroli}},
  \citenamefont {{Newburgh}}, \citenamefont {{Nguyen}}, \citenamefont
  {{Niemack}}, \citenamefont {{Nishino}}, \citenamefont {{Orlowski-Scherer}},
  \citenamefont {{Page}}, \citenamefont {{Partridge}}, \citenamefont
  {{Peloton}}, \citenamefont {{Perrotta}}, \citenamefont {{Piccirillo}},
  \citenamefont {{Pisano}}, \citenamefont {{Poletti}}, \citenamefont {{Puddu}},
  \citenamefont {{Puglisi}}, \citenamefont {{Raum}}, \citenamefont
  {{Reichardt}}, \citenamefont {{Remazeilles}}, \citenamefont {{Rephaeli}},
  \citenamefont {{Riechers}}, \citenamefont {{Rojas}}, \citenamefont {{Roy}},
  \citenamefont {{Sadeh}}, \citenamefont {{Sakurai}}, \citenamefont
  {{Salatino}}, \citenamefont {{Sathyanarayana Rao}}, \citenamefont {{Schaan}},
  \citenamefont {{Schmittfull}}, \citenamefont {{Sehgal}}, \citenamefont
  {{Seibert}}, \citenamefont {{Seljak}}, \citenamefont {{Sherwin}},
  \citenamefont {{Shimon}}, \citenamefont {{Sierra}}, \citenamefont
  {{Sievers}}, \citenamefont {{Sikhosana}}, \citenamefont {{Silva-Feaver}},
  \citenamefont {{Simon}}, \citenamefont {{Sinclair}}, \citenamefont
  {{Siritanasak}}, \citenamefont {{Smith}}, \citenamefont {{Smith}},
  \citenamefont {{Spergel}}, \citenamefont {{Staggs}}, \citenamefont {{Stein}},
  \citenamefont {{Stevens}}, \citenamefont {{Stompor}}, \citenamefont
  {{Suzuki}}, \citenamefont {{Tajima}}, \citenamefont {{Takakura}},
  \citenamefont {{Teply}}, \citenamefont {{Thomas}}, \citenamefont {{Thorne}},
  \citenamefont {{Thornton}}, \citenamefont {{Trac}}, \citenamefont {{Tsai}},
  \citenamefont {{Tucker}}, \citenamefont {{Ullom}}, \citenamefont
  {{Vagnozzi}}, \citenamefont {{van Engelen}}, \citenamefont {{Van Lanen}},
  \citenamefont {{Van Winkle}}, \citenamefont {{Vavagiakis}}, \citenamefont
  {{Verg{\`e}s}}, \citenamefont {{Vissers}}, \citenamefont {{Wagoner}},
  \citenamefont {{Walker}}, \citenamefont {{Ward}}, \citenamefont
  {{Westbrook}}, \citenamefont {{Whitehorn}}, \citenamefont {{Williams}},
  \citenamefont {{Williams}}, \citenamefont {{Wollack}}, \citenamefont {{Xu}},
  \citenamefont {{Yu}}, \citenamefont {{Yu}}, \citenamefont {{Zago}},
  \citenamefont {{Zhang}}, \citenamefont {{Zhu}},\ and\ \citenamefont {{The
  Simons Observatory collaboration}}}]{2019JCAP...02..056A}%
  \BibitemOpen
  \bibfield  {author} {\bibinfo {author} {\bibfnamefont {P.}~\bibnamefont
  {{Ade}}}, \bibinfo {author} {\bibfnamefont {J.}~\bibnamefont {{Aguirre}}},
  \bibinfo {author} {\bibfnamefont {Z.}~\bibnamefont {{Ahmed}}}, \bibinfo
  {author} {\bibfnamefont {S.}~\bibnamefont {{Aiola}}}, \bibinfo {author}
  {\bibfnamefont {A.}~\bibnamefont {{Ali}}}, \bibinfo {author} {\bibfnamefont
  {D.}~\bibnamefont {{Alonso}}}, \bibinfo {author} {\bibfnamefont {M.~A.}\
  \bibnamefont {{Alvarez}}}, \bibinfo {author} {\bibfnamefont {K.}~\bibnamefont
  {{Arnold}}}, \bibinfo {author} {\bibfnamefont {P.}~\bibnamefont {{Ashton}}},
  \bibinfo {author} {\bibfnamefont {J.}~\bibnamefont {{Austermann}}}, \bibinfo
  {author} {\bibfnamefont {H.}~\bibnamefont {{Awan}}}, \bibinfo {author}
  {\bibfnamefont {C.}~\bibnamefont {{Baccigalupi}}}, \bibinfo {author}
  {\bibfnamefont {T.}~\bibnamefont {{Baildon}}}, \bibinfo {author}
  {\bibfnamefont {D.}~\bibnamefont {{Barron}}}, \bibinfo {author}
  {\bibfnamefont {N.}~\bibnamefont {{Battaglia}}}, \bibinfo {author}
  {\bibfnamefont {R.}~\bibnamefont {{Battye}}}, \bibinfo {author}
  {\bibfnamefont {E.}~\bibnamefont {{Baxter}}}, \bibinfo {author}
  {\bibfnamefont {A.}~\bibnamefont {{Bazarko}}}, \bibinfo {author}
  {\bibfnamefont {J.~A.}\ \bibnamefont {{Beall}}}, \bibinfo {author}
  {\bibfnamefont {R.}~\bibnamefont {{Bean}}}, \bibinfo {author} {\bibfnamefont
  {D.}~\bibnamefont {{Beck}}}, \bibinfo {author} {\bibfnamefont
  {S.}~\bibnamefont {{Beckman}}}, \bibinfo {author} {\bibfnamefont
  {B.}~\bibnamefont {{Beringue}}}, \bibinfo {author} {\bibfnamefont
  {F.}~\bibnamefont {{Bianchini}}}, \bibinfo {author} {\bibfnamefont
  {S.}~\bibnamefont {{Boada}}}, \bibinfo {author} {\bibfnamefont
  {D.}~\bibnamefont {{Boettger}}}, \bibinfo {author} {\bibfnamefont {J.~R.}\
  \bibnamefont {{Bond}}}, \bibinfo {author} {\bibfnamefont {J.}~\bibnamefont
  {{Borrill}}}, \bibinfo {author} {\bibfnamefont {M.~L.}\ \bibnamefont
  {{Brown}}}, \bibinfo {author} {\bibfnamefont {S.~M.}\ \bibnamefont
  {{Bruno}}}, \bibinfo {author} {\bibfnamefont {S.}~\bibnamefont {{Bryan}}},
  \bibinfo {author} {\bibfnamefont {E.}~\bibnamefont {{Calabrese}}}, \bibinfo
  {author} {\bibfnamefont {V.}~\bibnamefont {{Calafut}}}, \bibinfo {author}
  {\bibfnamefont {P.}~\bibnamefont {{Calisse}}}, \bibinfo {author}
  {\bibfnamefont {J.}~\bibnamefont {{Carron}}}, \bibinfo {author}
  {\bibfnamefont {A.}~\bibnamefont {{Challinor}}}, \bibinfo {author}
  {\bibfnamefont {G.}~\bibnamefont {{Chesmore}}}, \bibinfo {author}
  {\bibfnamefont {Y.}~\bibnamefont {{Chinone}}}, \bibinfo {author}
  {\bibfnamefont {J.}~\bibnamefont {{Chluba}}}, \bibinfo {author}
  {\bibfnamefont {H.-M.~S.}\ \bibnamefont {{Cho}}}, \bibinfo {author}
  {\bibfnamefont {S.}~\bibnamefont {{Choi}}}, \bibinfo {author} {\bibfnamefont
  {G.}~\bibnamefont {{Coppi}}}, \bibinfo {author} {\bibfnamefont {N.~F.}\
  \bibnamefont {{Cothard}}}, \bibinfo {author} {\bibfnamefont {K.}~\bibnamefont
  {{Coughlin}}}, \bibinfo {author} {\bibfnamefont {D.}~\bibnamefont
  {{Crichton}}}, \bibinfo {author} {\bibfnamefont {K.~D.}\ \bibnamefont
  {{Crowley}}}, \bibinfo {author} {\bibfnamefont {K.~T.}\ \bibnamefont
  {{Crowley}}}, \bibinfo {author} {\bibfnamefont {A.}~\bibnamefont
  {{Cukierman}}}, \bibinfo {author} {\bibfnamefont {J.~M.}\ \bibnamefont
  {{D'Ewart}}}, \bibinfo {author} {\bibfnamefont {R.}~\bibnamefont
  {{D{\"u}nner}}}, \bibinfo {author} {\bibfnamefont {T.}~\bibnamefont {{de
  Haan}}}, \bibinfo {author} {\bibfnamefont {M.}~\bibnamefont {{Devlin}}},
  \bibinfo {author} {\bibfnamefont {S.}~\bibnamefont {{Dicker}}}, \bibinfo
  {author} {\bibfnamefont {J.}~\bibnamefont {{Didier}}}, \bibinfo {author}
  {\bibfnamefont {M.}~\bibnamefont {{Dobbs}}}, \bibinfo {author} {\bibfnamefont
  {B.}~\bibnamefont {{Dober}}}, \bibinfo {author} {\bibfnamefont {C.~J.}\
  \bibnamefont {{Duell}}}, \bibinfo {author} {\bibfnamefont {S.}~\bibnamefont
  {{Duff}}}, \bibinfo {author} {\bibfnamefont {A.}~\bibnamefont
  {{Duivenvoorden}}}, \bibinfo {author} {\bibfnamefont {J.}~\bibnamefont
  {{Dunkley}}}, \bibinfo {author} {\bibfnamefont {J.}~\bibnamefont
  {{Dusatko}}}, \bibinfo {author} {\bibfnamefont {J.}~\bibnamefont {{Errard}}},
  \bibinfo {author} {\bibfnamefont {G.}~\bibnamefont {{Fabbian}}}, \bibinfo
  {author} {\bibfnamefont {S.}~\bibnamefont {{Feeney}}}, \bibinfo {author}
  {\bibfnamefont {S.}~\bibnamefont {{Ferraro}}}, \bibinfo {author}
  {\bibfnamefont {P.}~\bibnamefont {{Flux{\`a}}}}, \bibinfo {author}
  {\bibfnamefont {K.}~\bibnamefont {{Freese}}}, \bibinfo {author}
  {\bibfnamefont {J.~C.}\ \bibnamefont {{Frisch}}}, \bibinfo {author}
  {\bibfnamefont {A.}~\bibnamefont {{Frolov}}}, \bibinfo {author}
  {\bibfnamefont {G.}~\bibnamefont {{Fuller}}}, \bibinfo {author}
  {\bibfnamefont {B.}~\bibnamefont {{Fuzia}}}, \bibinfo {author} {\bibfnamefont
  {N.}~\bibnamefont {{Galitzki}}}, \bibinfo {author} {\bibfnamefont {P.~A.}\
  \bibnamefont {{Gallardo}}}, \bibinfo {author} {\bibfnamefont
  {J.}~\bibnamefont {{Tomas Galvez Ghersi}}}, \bibinfo {author} {\bibfnamefont
  {J.}~\bibnamefont {{Gao}}}, \bibinfo {author} {\bibfnamefont
  {E.}~\bibnamefont {{Gawiser}}}, \bibinfo {author} {\bibfnamefont
  {M.}~\bibnamefont {{Gerbino}}}, \bibinfo {author} {\bibfnamefont
  {V.}~\bibnamefont {{Gluscevic}}}, \bibinfo {author} {\bibfnamefont
  {N.}~\bibnamefont {{Goeckner-Wald}}}, \bibinfo {author} {\bibfnamefont
  {J.}~\bibnamefont {{Golec}}}, \bibinfo {author} {\bibfnamefont
  {S.}~\bibnamefont {{Gordon}}}, \bibinfo {author} {\bibfnamefont
  {M.}~\bibnamefont {{Gralla}}}, \bibinfo {author} {\bibfnamefont
  {D.}~\bibnamefont {{Green}}}, \bibinfo {author} {\bibfnamefont
  {A.}~\bibnamefont {{Grigorian}}}, \bibinfo {author} {\bibfnamefont
  {J.}~\bibnamefont {{Groh}}}, \bibinfo {author} {\bibfnamefont
  {C.}~\bibnamefont {{Groppi}}}, \bibinfo {author} {\bibfnamefont
  {Y.}~\bibnamefont {{Guan}}}, \bibinfo {author} {\bibfnamefont {J.~E.}\
  \bibnamefont {{Gudmundsson}}}, \bibinfo {author} {\bibfnamefont
  {D.}~\bibnamefont {{Han}}}, \bibinfo {author} {\bibfnamefont
  {P.}~\bibnamefont {{Hargrave}}}, \bibinfo {author} {\bibfnamefont
  {M.}~\bibnamefont {{Hasegawa}}}, \bibinfo {author} {\bibfnamefont
  {M.}~\bibnamefont {{Hasselfield}}}, \bibinfo {author} {\bibfnamefont
  {M.}~\bibnamefont {{Hattori}}}, \bibinfo {author} {\bibfnamefont
  {V.}~\bibnamefont {{Haynes}}}, \bibinfo {author} {\bibfnamefont
  {M.}~\bibnamefont {{Hazumi}}}, \bibinfo {author} {\bibfnamefont
  {Y.}~\bibnamefont {{He}}}, \bibinfo {author} {\bibfnamefont {E.}~\bibnamefont
  {{Healy}}}, \bibinfo {author} {\bibfnamefont {S.~W.}\ \bibnamefont
  {{Henderson}}}, \bibinfo {author} {\bibfnamefont {C.}~\bibnamefont
  {{Hervias-Caimapo}}}, \bibinfo {author} {\bibfnamefont {C.~A.}\ \bibnamefont
  {{Hill}}}, \bibinfo {author} {\bibfnamefont {J.~C.}\ \bibnamefont {{Hill}}},
  \bibinfo {author} {\bibfnamefont {G.}~\bibnamefont {{Hilton}}}, \bibinfo
  {author} {\bibfnamefont {M.}~\bibnamefont {{Hilton}}}, \bibinfo {author}
  {\bibfnamefont {A.~D.}\ \bibnamefont {{Hincks}}}, \bibinfo {author}
  {\bibfnamefont {G.}~\bibnamefont {{Hinshaw}}}, \bibinfo {author}
  {\bibfnamefont {R.}~\bibnamefont {{Hlo{\v{z}}ek}}}, \bibinfo {author}
  {\bibfnamefont {S.}~\bibnamefont {{Ho}}}, \bibinfo {author} {\bibfnamefont
  {S.-P.~P.}\ \bibnamefont {{Ho}}}, \bibinfo {author} {\bibfnamefont
  {L.}~\bibnamefont {{Howe}}}, \bibinfo {author} {\bibfnamefont
  {Z.}~\bibnamefont {{Huang}}}, \bibinfo {author} {\bibfnamefont
  {J.}~\bibnamefont {{Hubmayr}}}, \bibinfo {author} {\bibfnamefont
  {K.}~\bibnamefont {{Huffenberger}}}, \bibinfo {author} {\bibfnamefont
  {J.~P.}\ \bibnamefont {{Hughes}}}, \bibinfo {author} {\bibfnamefont
  {A.}~\bibnamefont {{Ijjas}}}, \bibinfo {author} {\bibfnamefont
  {M.}~\bibnamefont {{Ikape}}}, \bibinfo {author} {\bibfnamefont
  {K.}~\bibnamefont {{Irwin}}}, \bibinfo {author} {\bibfnamefont {A.~H.}\
  \bibnamefont {{Jaffe}}}, \bibinfo {author} {\bibfnamefont {B.}~\bibnamefont
  {{Jain}}}, \bibinfo {author} {\bibfnamefont {O.}~\bibnamefont {{Jeong}}},
  \bibinfo {author} {\bibfnamefont {D.}~\bibnamefont {{Kaneko}}}, \bibinfo
  {author} {\bibfnamefont {E.~D.}\ \bibnamefont {{Karpel}}}, \bibinfo {author}
  {\bibfnamefont {N.}~\bibnamefont {{Katayama}}}, \bibinfo {author}
  {\bibfnamefont {B.}~\bibnamefont {{Keating}}}, \bibinfo {author}
  {\bibfnamefont {S.~S.}\ \bibnamefont {{Kernasovskiy}}}, \bibinfo {author}
  {\bibfnamefont {R.}~\bibnamefont {{Keskitalo}}}, \bibinfo {author}
  {\bibfnamefont {T.}~\bibnamefont {{Kisner}}}, \bibinfo {author}
  {\bibfnamefont {K.}~\bibnamefont {{Kiuchi}}}, \bibinfo {author}
  {\bibfnamefont {J.}~\bibnamefont {{Klein}}}, \bibinfo {author} {\bibfnamefont
  {K.}~\bibnamefont {{Knowles}}}, \bibinfo {author} {\bibfnamefont
  {B.}~\bibnamefont {{Koopman}}}, \bibinfo {author} {\bibfnamefont
  {A.}~\bibnamefont {{Kosowsky}}}, \bibinfo {author} {\bibfnamefont
  {N.}~\bibnamefont {{Krachmalnicoff}}}, \bibinfo {author} {\bibfnamefont
  {S.~E.}\ \bibnamefont {{Kuenstner}}}, \bibinfo {author} {\bibfnamefont
  {C.-L.}\ \bibnamefont {{Kuo}}}, \bibinfo {author} {\bibfnamefont
  {A.}~\bibnamefont {{Kusaka}}}, \bibinfo {author} {\bibfnamefont
  {J.}~\bibnamefont {{Lashner}}}, \bibinfo {author} {\bibfnamefont
  {A.}~\bibnamefont {{Lee}}}, \bibinfo {author} {\bibfnamefont
  {E.}~\bibnamefont {{Lee}}}, \bibinfo {author} {\bibfnamefont
  {D.}~\bibnamefont {{Leon}}}, \bibinfo {author} {\bibfnamefont {J.~S.~Y.}\
  \bibnamefont {{Leung}}}, \bibinfo {author} {\bibfnamefont {A.}~\bibnamefont
  {{Lewis}}}, \bibinfo {author} {\bibfnamefont {Y.}~\bibnamefont {{Li}}},
  \bibinfo {author} {\bibfnamefont {Z.}~\bibnamefont {{Li}}}, \bibinfo {author}
  {\bibfnamefont {M.}~\bibnamefont {{Limon}}}, \bibinfo {author} {\bibfnamefont
  {E.}~\bibnamefont {{Linder}}}, \bibinfo {author} {\bibfnamefont
  {C.}~\bibnamefont {{Lopez-Caraballo}}}, \bibinfo {author} {\bibfnamefont
  {T.}~\bibnamefont {{Louis}}}, \bibinfo {author} {\bibfnamefont
  {L.}~\bibnamefont {{Lowry}}}, \bibinfo {author} {\bibfnamefont
  {M.}~\bibnamefont {{Lungu}}}, \bibinfo {author} {\bibfnamefont
  {M.}~\bibnamefont {{Madhavacheril}}}, \bibinfo {author} {\bibfnamefont
  {D.}~\bibnamefont {{Mak}}}, \bibinfo {author} {\bibfnamefont
  {F.}~\bibnamefont {{Maldonado}}}, \bibinfo {author} {\bibfnamefont
  {H.}~\bibnamefont {{Mani}}}, \bibinfo {author} {\bibfnamefont
  {B.}~\bibnamefont {{Mates}}}, \bibinfo {author} {\bibfnamefont
  {F.}~\bibnamefont {{Matsuda}}}, \bibinfo {author} {\bibfnamefont
  {L.}~\bibnamefont {{Maurin}}}, \bibinfo {author} {\bibfnamefont
  {P.}~\bibnamefont {{Mauskopf}}}, \bibinfo {author} {\bibfnamefont
  {A.}~\bibnamefont {{May}}}, \bibinfo {author} {\bibfnamefont
  {N.}~\bibnamefont {{McCallum}}}, \bibinfo {author} {\bibfnamefont
  {C.}~\bibnamefont {{McKenney}}}, \bibinfo {author} {\bibfnamefont
  {J.}~\bibnamefont {{McMahon}}}, \bibinfo {author} {\bibfnamefont {P.~D.}\
  \bibnamefont {{Meerburg}}}, \bibinfo {author} {\bibfnamefont
  {J.}~\bibnamefont {{Meyers}}}, \bibinfo {author} {\bibfnamefont
  {A.}~\bibnamefont {{Miller}}}, \bibinfo {author} {\bibfnamefont
  {M.}~\bibnamefont {{Mirmelstein}}}, \bibinfo {author} {\bibfnamefont
  {K.}~\bibnamefont {{Moodley}}}, \bibinfo {author} {\bibfnamefont
  {M.}~\bibnamefont {{Munchmeyer}}}, \bibinfo {author} {\bibfnamefont
  {C.}~\bibnamefont {{Munson}}}, \bibinfo {author} {\bibfnamefont
  {S.}~\bibnamefont {{Naess}}}, \bibinfo {author} {\bibfnamefont
  {F.}~\bibnamefont {{Nati}}}, \bibinfo {author} {\bibfnamefont
  {M.}~\bibnamefont {{Navaroli}}}, \bibinfo {author} {\bibfnamefont
  {L.}~\bibnamefont {{Newburgh}}}, \bibinfo {author} {\bibfnamefont {H.~N.}\
  \bibnamefont {{Nguyen}}}, \bibinfo {author} {\bibfnamefont {M.}~\bibnamefont
  {{Niemack}}}, \bibinfo {author} {\bibfnamefont {H.}~\bibnamefont
  {{Nishino}}}, \bibinfo {author} {\bibfnamefont {J.}~\bibnamefont
  {{Orlowski-Scherer}}}, \bibinfo {author} {\bibfnamefont {L.}~\bibnamefont
  {{Page}}}, \bibinfo {author} {\bibfnamefont {B.}~\bibnamefont {{Partridge}}},
  \bibinfo {author} {\bibfnamefont {J.}~\bibnamefont {{Peloton}}}, \bibinfo
  {author} {\bibfnamefont {F.}~\bibnamefont {{Perrotta}}}, \bibinfo {author}
  {\bibfnamefont {L.}~\bibnamefont {{Piccirillo}}}, \bibinfo {author}
  {\bibfnamefont {G.}~\bibnamefont {{Pisano}}}, \bibinfo {author}
  {\bibfnamefont {D.}~\bibnamefont {{Poletti}}}, \bibinfo {author}
  {\bibfnamefont {R.}~\bibnamefont {{Puddu}}}, \bibinfo {author} {\bibfnamefont
  {G.}~\bibnamefont {{Puglisi}}}, \bibinfo {author} {\bibfnamefont
  {C.}~\bibnamefont {{Raum}}}, \bibinfo {author} {\bibfnamefont {C.~L.}\
  \bibnamefont {{Reichardt}}}, \bibinfo {author} {\bibfnamefont
  {M.}~\bibnamefont {{Remazeilles}}}, \bibinfo {author} {\bibfnamefont
  {Y.}~\bibnamefont {{Rephaeli}}}, \bibinfo {author} {\bibfnamefont
  {D.}~\bibnamefont {{Riechers}}}, \bibinfo {author} {\bibfnamefont
  {F.}~\bibnamefont {{Rojas}}}, \bibinfo {author} {\bibfnamefont
  {A.}~\bibnamefont {{Roy}}}, \bibinfo {author} {\bibfnamefont
  {S.}~\bibnamefont {{Sadeh}}}, \bibinfo {author} {\bibfnamefont
  {Y.}~\bibnamefont {{Sakurai}}}, \bibinfo {author} {\bibfnamefont
  {M.}~\bibnamefont {{Salatino}}}, \bibinfo {author} {\bibfnamefont
  {M.}~\bibnamefont {{Sathyanarayana Rao}}}, \bibinfo {author} {\bibfnamefont
  {E.}~\bibnamefont {{Schaan}}}, \bibinfo {author} {\bibfnamefont
  {M.}~\bibnamefont {{Schmittfull}}}, \bibinfo {author} {\bibfnamefont
  {N.}~\bibnamefont {{Sehgal}}}, \bibinfo {author} {\bibfnamefont
  {J.}~\bibnamefont {{Seibert}}}, \bibinfo {author} {\bibfnamefont
  {U.}~\bibnamefont {{Seljak}}}, \bibinfo {author} {\bibfnamefont
  {B.}~\bibnamefont {{Sherwin}}}, \bibinfo {author} {\bibfnamefont
  {M.}~\bibnamefont {{Shimon}}}, \bibinfo {author} {\bibfnamefont
  {C.}~\bibnamefont {{Sierra}}}, \bibinfo {author} {\bibfnamefont
  {J.}~\bibnamefont {{Sievers}}}, \bibinfo {author} {\bibfnamefont
  {P.}~\bibnamefont {{Sikhosana}}}, \bibinfo {author} {\bibfnamefont
  {M.}~\bibnamefont {{Silva-Feaver}}}, \bibinfo {author} {\bibfnamefont
  {S.~M.}\ \bibnamefont {{Simon}}}, \bibinfo {author} {\bibfnamefont
  {A.}~\bibnamefont {{Sinclair}}}, \bibinfo {author} {\bibfnamefont
  {P.}~\bibnamefont {{Siritanasak}}}, \bibinfo {author} {\bibfnamefont
  {K.}~\bibnamefont {{Smith}}}, \bibinfo {author} {\bibfnamefont {S.~R.}\
  \bibnamefont {{Smith}}}, \bibinfo {author} {\bibfnamefont {D.}~\bibnamefont
  {{Spergel}}}, \bibinfo {author} {\bibfnamefont {S.~T.}\ \bibnamefont
  {{Staggs}}}, \bibinfo {author} {\bibfnamefont {G.}~\bibnamefont {{Stein}}},
  \bibinfo {author} {\bibfnamefont {J.~R.}\ \bibnamefont {{Stevens}}}, \bibinfo
  {author} {\bibfnamefont {R.}~\bibnamefont {{Stompor}}}, \bibinfo {author}
  {\bibfnamefont {A.}~\bibnamefont {{Suzuki}}}, \bibinfo {author}
  {\bibfnamefont {O.}~\bibnamefont {{Tajima}}}, \bibinfo {author}
  {\bibfnamefont {S.}~\bibnamefont {{Takakura}}}, \bibinfo {author}
  {\bibfnamefont {G.}~\bibnamefont {{Teply}}}, \bibinfo {author} {\bibfnamefont
  {D.~B.}\ \bibnamefont {{Thomas}}}, \bibinfo {author} {\bibfnamefont
  {B.}~\bibnamefont {{Thorne}}}, \bibinfo {author} {\bibfnamefont
  {R.}~\bibnamefont {{Thornton}}}, \bibinfo {author} {\bibfnamefont
  {H.}~\bibnamefont {{Trac}}}, \bibinfo {author} {\bibfnamefont
  {C.}~\bibnamefont {{Tsai}}}, \bibinfo {author} {\bibfnamefont
  {C.}~\bibnamefont {{Tucker}}}, \bibinfo {author} {\bibfnamefont
  {J.}~\bibnamefont {{Ullom}}}, \bibinfo {author} {\bibfnamefont
  {S.}~\bibnamefont {{Vagnozzi}}}, \bibinfo {author} {\bibfnamefont
  {A.}~\bibnamefont {{van Engelen}}}, \bibinfo {author} {\bibfnamefont
  {J.}~\bibnamefont {{Van Lanen}}}, \bibinfo {author} {\bibfnamefont {D.~D.}\
  \bibnamefont {{Van Winkle}}}, \bibinfo {author} {\bibfnamefont {E.~M.}\
  \bibnamefont {{Vavagiakis}}}, \bibinfo {author} {\bibfnamefont
  {C.}~\bibnamefont {{Verg{\`e}s}}}, \bibinfo {author} {\bibfnamefont
  {M.}~\bibnamefont {{Vissers}}}, \bibinfo {author} {\bibfnamefont
  {K.}~\bibnamefont {{Wagoner}}}, \bibinfo {author} {\bibfnamefont
  {S.}~\bibnamefont {{Walker}}}, \bibinfo {author} {\bibfnamefont
  {J.}~\bibnamefont {{Ward}}}, \bibinfo {author} {\bibfnamefont
  {B.}~\bibnamefont {{Westbrook}}}, \bibinfo {author} {\bibfnamefont
  {N.}~\bibnamefont {{Whitehorn}}}, \bibinfo {author} {\bibfnamefont
  {J.}~\bibnamefont {{Williams}}}, \bibinfo {author} {\bibfnamefont
  {J.}~\bibnamefont {{Williams}}}, \bibinfo {author} {\bibfnamefont {E.~J.}\
  \bibnamefont {{Wollack}}}, \bibinfo {author} {\bibfnamefont {Z.}~\bibnamefont
  {{Xu}}}, \bibinfo {author} {\bibfnamefont {B.}~\bibnamefont {{Yu}}}, \bibinfo
  {author} {\bibfnamefont {C.}~\bibnamefont {{Yu}}}, \bibinfo {author}
  {\bibfnamefont {F.}~\bibnamefont {{Zago}}}, \bibinfo {author} {\bibfnamefont
  {H.}~\bibnamefont {{Zhang}}}, \bibinfo {author} {\bibfnamefont
  {N.}~\bibnamefont {{Zhu}}}, \ and\ \bibinfo {author} {\bibnamefont {{The
  Simons Observatory collaboration}}},\ }\href {\doibase
  10.1088/1475-7516/2019/02/056} {\bibfield  {journal} {\bibinfo  {journal}
  {Journal of Cosmology and Astro-Particle Physics}\ }\textbf {\bibinfo
  {volume} {2019}},\ \bibinfo {eid} {056} (\bibinfo {year}
  {2019}{\natexlab{b}})},\ \Eprint {http://arxiv.org/abs/1808.07445}
  {arXiv:1808.07445 [astro-ph.CO]} \BibitemShut {NoStop}%
\bibitem [{\citenamefont {{Abazajian}}\ \emph {et~al.}(2016)\citenamefont
  {{Abazajian}}, \citenamefont {{Adshead}}, \citenamefont {{Ahmed}},
  \citenamefont {{Allen}}, \citenamefont {{Alonso}}, \citenamefont {{Arnold}},
  \citenamefont {{Baccigalupi}}, \citenamefont {{Bartlett}}, \citenamefont
  {{Battaglia}}, \citenamefont {{Benson}}, \citenamefont {{Bischoff}},
  \citenamefont {{Borrill}}, \citenamefont {{Buza}}, \citenamefont
  {{Calabrese}}, \citenamefont {{Caldwell}}, \citenamefont {{Carlstrom}},
  \citenamefont {{Chang}}, \citenamefont {{Crawford}}, \citenamefont
  {{Cyr-Racine}}, \citenamefont {{De Bernardis}}, \citenamefont {{de Haan}},
  \citenamefont {{di Serego Alighieri}}, \citenamefont {{Dunkley}},
  \citenamefont {{Dvorkin}}, \citenamefont {{Errard}}, \citenamefont
  {{Fabbian}}, \citenamefont {{Feeney}}, \citenamefont {{Ferraro}},
  \citenamefont {{Filippini}}, \citenamefont {{Flauger}}, \citenamefont
  {{Fuller}}, \citenamefont {{Gluscevic}}, \citenamefont {{Green}},
  \citenamefont {{Grin}}, \citenamefont {{Grohs}}, \citenamefont {{Henning}},
  \citenamefont {{Hill}}, \citenamefont {{Hlozek}}, \citenamefont {{Holder}},
  \citenamefont {{Holzapfel}}, \citenamefont {{Hu}}, \citenamefont
  {{Huffenberger}}, \citenamefont {{Keskitalo}}, \citenamefont {{Knox}},
  \citenamefont {{Kosowsky}}, \citenamefont {{Kovac}}, \citenamefont
  {{Kovetz}}, \citenamefont {{Kuo}}, \citenamefont {{Kusaka}}, \citenamefont
  {{Le Jeune}}, \citenamefont {{Lee}}, \citenamefont {{Lilley}}, \citenamefont
  {{Loverde}}, \citenamefont {{Madhavacheril}}, \citenamefont {{Mantz}},
  \citenamefont {{Marsh}}, \citenamefont {{McMahon}}, \citenamefont
  {{Meerburg}}, \citenamefont {{Meyers}}, \citenamefont {{Miller}},
  \citenamefont {{Munoz}}, \citenamefont {{Nguyen}}, \citenamefont {{Niemack}},
  \citenamefont {{Peloso}}, \citenamefont {{Peloton}}, \citenamefont
  {{Pogosian}}, \citenamefont {{Pryke}}, \citenamefont {{Raveri}},
  \citenamefont {{Reichardt}}, \citenamefont {{Rocha}}, \citenamefont
  {{Rotti}}, \citenamefont {{Schaan}}, \citenamefont {{Schmittfull}},
  \citenamefont {{Scott}}, \citenamefont {{Sehgal}}, \citenamefont
  {{Shandera}}, \citenamefont {{Sherwin}}, \citenamefont {{Smith}},
  \citenamefont {{Sorbo}}, \citenamefont {{Starkman}}, \citenamefont {{Story}},
  \citenamefont {{van Engelen}}, \citenamefont {{Vieira}}, \citenamefont
  {{Watson}}, \citenamefont {{Whitehorn}},\ and\ \citenamefont {{Kimmy
  Wu}}}]{2016arXiv161002743A}%
  \BibitemOpen
  \bibfield  {author} {\bibinfo {author} {\bibfnamefont {K.~N.}\ \bibnamefont
  {{Abazajian}}}, \bibinfo {author} {\bibfnamefont {P.}~\bibnamefont
  {{Adshead}}}, \bibinfo {author} {\bibfnamefont {Z.}~\bibnamefont {{Ahmed}}},
  \bibinfo {author} {\bibfnamefont {S.~W.}\ \bibnamefont {{Allen}}}, \bibinfo
  {author} {\bibfnamefont {D.}~\bibnamefont {{Alonso}}}, \bibinfo {author}
  {\bibfnamefont {K.~S.}\ \bibnamefont {{Arnold}}}, \bibinfo {author}
  {\bibfnamefont {C.}~\bibnamefont {{Baccigalupi}}}, \bibinfo {author}
  {\bibfnamefont {J.~G.}\ \bibnamefont {{Bartlett}}}, \bibinfo {author}
  {\bibfnamefont {N.}~\bibnamefont {{Battaglia}}}, \bibinfo {author}
  {\bibfnamefont {B.~A.}\ \bibnamefont {{Benson}}}, \bibinfo {author}
  {\bibfnamefont {C.~A.}\ \bibnamefont {{Bischoff}}}, \bibinfo {author}
  {\bibfnamefont {J.}~\bibnamefont {{Borrill}}}, \bibinfo {author}
  {\bibfnamefont {V.}~\bibnamefont {{Buza}}}, \bibinfo {author} {\bibfnamefont
  {E.}~\bibnamefont {{Calabrese}}}, \bibinfo {author} {\bibfnamefont
  {R.}~\bibnamefont {{Caldwell}}}, \bibinfo {author} {\bibfnamefont {J.~E.}\
  \bibnamefont {{Carlstrom}}}, \bibinfo {author} {\bibfnamefont {C.~L.}\
  \bibnamefont {{Chang}}}, \bibinfo {author} {\bibfnamefont {T.~M.}\
  \bibnamefont {{Crawford}}}, \bibinfo {author} {\bibfnamefont {F.-Y.}\
  \bibnamefont {{Cyr-Racine}}}, \bibinfo {author} {\bibfnamefont
  {F.}~\bibnamefont {{De Bernardis}}}, \bibinfo {author} {\bibfnamefont
  {T.}~\bibnamefont {{de Haan}}}, \bibinfo {author} {\bibfnamefont
  {S.}~\bibnamefont {{di Serego Alighieri}}}, \bibinfo {author} {\bibfnamefont
  {J.}~\bibnamefont {{Dunkley}}}, \bibinfo {author} {\bibfnamefont
  {C.}~\bibnamefont {{Dvorkin}}}, \bibinfo {author} {\bibfnamefont
  {J.}~\bibnamefont {{Errard}}}, \bibinfo {author} {\bibfnamefont
  {G.}~\bibnamefont {{Fabbian}}}, \bibinfo {author} {\bibfnamefont
  {S.}~\bibnamefont {{Feeney}}}, \bibinfo {author} {\bibfnamefont
  {S.}~\bibnamefont {{Ferraro}}}, \bibinfo {author} {\bibfnamefont {J.~P.}\
  \bibnamefont {{Filippini}}}, \bibinfo {author} {\bibfnamefont
  {R.}~\bibnamefont {{Flauger}}}, \bibinfo {author} {\bibfnamefont {G.~M.}\
  \bibnamefont {{Fuller}}}, \bibinfo {author} {\bibfnamefont {V.}~\bibnamefont
  {{Gluscevic}}}, \bibinfo {author} {\bibfnamefont {D.}~\bibnamefont
  {{Green}}}, \bibinfo {author} {\bibfnamefont {D.}~\bibnamefont {{Grin}}},
  \bibinfo {author} {\bibfnamefont {E.}~\bibnamefont {{Grohs}}}, \bibinfo
  {author} {\bibfnamefont {J.~W.}\ \bibnamefont {{Henning}}}, \bibinfo {author}
  {\bibfnamefont {J.~C.}\ \bibnamefont {{Hill}}}, \bibinfo {author}
  {\bibfnamefont {R.}~\bibnamefont {{Hlozek}}}, \bibinfo {author}
  {\bibfnamefont {G.}~\bibnamefont {{Holder}}}, \bibinfo {author}
  {\bibfnamefont {W.}~\bibnamefont {{Holzapfel}}}, \bibinfo {author}
  {\bibfnamefont {W.}~\bibnamefont {{Hu}}}, \bibinfo {author} {\bibfnamefont
  {K.~M.}\ \bibnamefont {{Huffenberger}}}, \bibinfo {author} {\bibfnamefont
  {R.}~\bibnamefont {{Keskitalo}}}, \bibinfo {author} {\bibfnamefont
  {L.}~\bibnamefont {{Knox}}}, \bibinfo {author} {\bibfnamefont
  {A.}~\bibnamefont {{Kosowsky}}}, \bibinfo {author} {\bibfnamefont
  {J.}~\bibnamefont {{Kovac}}}, \bibinfo {author} {\bibfnamefont {E.~D.}\
  \bibnamefont {{Kovetz}}}, \bibinfo {author} {\bibfnamefont {C.-L.}\
  \bibnamefont {{Kuo}}}, \bibinfo {author} {\bibfnamefont {A.}~\bibnamefont
  {{Kusaka}}}, \bibinfo {author} {\bibfnamefont {M.}~\bibnamefont {{Le
  Jeune}}}, \bibinfo {author} {\bibfnamefont {A.~T.}\ \bibnamefont {{Lee}}},
  \bibinfo {author} {\bibfnamefont {M.}~\bibnamefont {{Lilley}}}, \bibinfo
  {author} {\bibfnamefont {M.}~\bibnamefont {{Loverde}}}, \bibinfo {author}
  {\bibfnamefont {M.~S.}\ \bibnamefont {{Madhavacheril}}}, \bibinfo {author}
  {\bibfnamefont {A.}~\bibnamefont {{Mantz}}}, \bibinfo {author} {\bibfnamefont
  {D.~J.~E.}\ \bibnamefont {{Marsh}}}, \bibinfo {author} {\bibfnamefont
  {J.}~\bibnamefont {{McMahon}}}, \bibinfo {author} {\bibfnamefont {P.~D.}\
  \bibnamefont {{Meerburg}}}, \bibinfo {author} {\bibfnamefont
  {J.}~\bibnamefont {{Meyers}}}, \bibinfo {author} {\bibfnamefont {A.~D.}\
  \bibnamefont {{Miller}}}, \bibinfo {author} {\bibfnamefont {J.~B.}\
  \bibnamefont {{Munoz}}}, \bibinfo {author} {\bibfnamefont {H.~N.}\
  \bibnamefont {{Nguyen}}}, \bibinfo {author} {\bibfnamefont {M.~D.}\
  \bibnamefont {{Niemack}}}, \bibinfo {author} {\bibfnamefont {M.}~\bibnamefont
  {{Peloso}}}, \bibinfo {author} {\bibfnamefont {J.}~\bibnamefont {{Peloton}}},
  \bibinfo {author} {\bibfnamefont {L.}~\bibnamefont {{Pogosian}}}, \bibinfo
  {author} {\bibfnamefont {C.}~\bibnamefont {{Pryke}}}, \bibinfo {author}
  {\bibfnamefont {M.}~\bibnamefont {{Raveri}}}, \bibinfo {author}
  {\bibfnamefont {C.~L.}\ \bibnamefont {{Reichardt}}}, \bibinfo {author}
  {\bibfnamefont {G.}~\bibnamefont {{Rocha}}}, \bibinfo {author} {\bibfnamefont
  {A.}~\bibnamefont {{Rotti}}}, \bibinfo {author} {\bibfnamefont
  {E.}~\bibnamefont {{Schaan}}}, \bibinfo {author} {\bibfnamefont {M.~M.}\
  \bibnamefont {{Schmittfull}}}, \bibinfo {author} {\bibfnamefont
  {D.}~\bibnamefont {{Scott}}}, \bibinfo {author} {\bibfnamefont
  {N.}~\bibnamefont {{Sehgal}}}, \bibinfo {author} {\bibfnamefont
  {S.}~\bibnamefont {{Shandera}}}, \bibinfo {author} {\bibfnamefont {B.~D.}\
  \bibnamefont {{Sherwin}}}, \bibinfo {author} {\bibfnamefont {T.~L.}\
  \bibnamefont {{Smith}}}, \bibinfo {author} {\bibfnamefont {L.}~\bibnamefont
  {{Sorbo}}}, \bibinfo {author} {\bibfnamefont {G.~D.}\ \bibnamefont
  {{Starkman}}}, \bibinfo {author} {\bibfnamefont {K.~T.}\ \bibnamefont
  {{Story}}}, \bibinfo {author} {\bibfnamefont {A.}~\bibnamefont {{van
  Engelen}}}, \bibinfo {author} {\bibfnamefont {J.~D.}\ \bibnamefont
  {{Vieira}}}, \bibinfo {author} {\bibfnamefont {S.}~\bibnamefont {{Watson}}},
  \bibinfo {author} {\bibfnamefont {N.}~\bibnamefont {{Whitehorn}}}, \ and\
  \bibinfo {author} {\bibfnamefont {W.~L.}\ \bibnamefont {{Kimmy Wu}}},\
  }\href@noop {} {\bibfield  {journal} {\bibinfo  {journal} {arXiv e-prints}\
  ,\ \bibinfo {eid} {arXiv:1610.02743}} (\bibinfo {year} {2016})},\ \Eprint
  {http://arxiv.org/abs/1610.02743} {arXiv:1610.02743 [astro-ph.CO]}
  \BibitemShut {NoStop}%
\bibitem [{\citenamefont {{Battaglia}}\ \emph {et~al.}(2017)\citenamefont
  {{Battaglia}}, \citenamefont {{Ferraro}}, \citenamefont {{Schaan}},\ and\
  \citenamefont {{Spergel}}}]{2017JCAP...11..040B}%
  \BibitemOpen
  \bibfield  {author} {\bibinfo {author} {\bibfnamefont {N.}~\bibnamefont
  {{Battaglia}}}, \bibinfo {author} {\bibfnamefont {S.}~\bibnamefont
  {{Ferraro}}}, \bibinfo {author} {\bibfnamefont {E.}~\bibnamefont {{Schaan}}},
  \ and\ \bibinfo {author} {\bibfnamefont {D.~N.}\ \bibnamefont {{Spergel}}},\
  }\href {\doibase 10.1088/1475-7516/2017/11/040} {\bibfield  {journal}
  {\bibinfo  {journal} {\jcap}\ }\textbf {\bibinfo {volume} {11}},\ \bibinfo
  {eid} {040} (\bibinfo {year} {2017})},\ \Eprint
  {http://arxiv.org/abs/1705.05881} {arXiv:1705.05881} \BibitemShut {NoStop}%
\bibitem [{\citenamefont {{Hill}}\ and\ \citenamefont
  {{Sherwin}}(2013)}]{Hill13}%
  \BibitemOpen
  \bibfield  {author} {\bibinfo {author} {\bibfnamefont {J.~C.}\ \bibnamefont
  {{Hill}}}\ and\ \bibinfo {author} {\bibfnamefont {B.~D.}\ \bibnamefont
  {{Sherwin}}},\ }\href {\doibase 10.1103/PhysRevD.87.023527} {\bibfield
  {journal} {\bibinfo  {journal} {\prd}\ }\textbf {\bibinfo {volume} {87}},\
  \bibinfo {eid} {023527} (\bibinfo {year} {2013})},\ \Eprint
  {http://arxiv.org/abs/1205.5794} {arXiv:1205.5794 [astro-ph.CO]} \BibitemShut
  {NoStop}%
\bibitem [{\citenamefont {{Hill}}\ and\ \citenamefont
  {{Pajer}}(2013)}]{Hill13b}%
  \BibitemOpen
  \bibfield  {author} {\bibinfo {author} {\bibfnamefont {J.~C.}\ \bibnamefont
  {{Hill}}}\ and\ \bibinfo {author} {\bibfnamefont {E.}~\bibnamefont
  {{Pajer}}},\ }\href {\doibase 10.1103/PhysRevD.88.063526} {\bibfield
  {journal} {\bibinfo  {journal} {\prd}\ }\textbf {\bibinfo {volume} {88}},\
  \bibinfo {eid} {063526} (\bibinfo {year} {2013})},\ \Eprint
  {http://arxiv.org/abs/1303.4726} {arXiv:1303.4726 [astro-ph.CO]} \BibitemShut
  {NoStop}%
\bibitem [{\citenamefont {{van Waerbeke}}\ \emph {et~al.}(2014)\citenamefont
  {{van Waerbeke}}, \citenamefont {{Hinshaw}},\ and\ \citenamefont
  {{Murray}}}]{vanWaerbeke2014}%
  \BibitemOpen
  \bibfield  {author} {\bibinfo {author} {\bibfnamefont {L.}~\bibnamefont {{van
  Waerbeke}}}, \bibinfo {author} {\bibfnamefont {G.}~\bibnamefont {{Hinshaw}}},
  \ and\ \bibinfo {author} {\bibfnamefont {N.}~\bibnamefont {{Murray}}},\
  }\href {\doibase 10.1103/PhysRevD.89.023508} {\bibfield  {journal} {\bibinfo
  {journal} {\prd}\ }\textbf {\bibinfo {volume} {89}},\ \bibinfo {eid} {023508}
  (\bibinfo {year} {2014})},\ \Eprint {http://arxiv.org/abs/1310.5721}
  {arXiv:1310.5721 [astro-ph.CO]} \BibitemShut {NoStop}%
\bibitem [{\citenamefont {{Hill}}\ and\ \citenamefont
  {{Spergel}}(2014)}]{Hill2014}%
  \BibitemOpen
  \bibfield  {author} {\bibinfo {author} {\bibfnamefont {J.~C.}\ \bibnamefont
  {{Hill}}}\ and\ \bibinfo {author} {\bibfnamefont {D.~N.}\ \bibnamefont
  {{Spergel}}},\ }\href {\doibase 10.1088/1475-7516/2014/02/030} {\bibfield
  {journal} {\bibinfo  {journal} {\jcap}\ }\textbf {\bibinfo {volume} {2}},\
  \bibinfo {eid} {030} (\bibinfo {year} {2014})},\ \Eprint
  {http://arxiv.org/abs/1312.4525} {arXiv:1312.4525} \BibitemShut {NoStop}%
\bibitem [{\citenamefont {{Planck Collaboration}}\ \emph
  {et~al.}(2016{\natexlab{e}})\citenamefont {{Planck Collaboration}},
  \citenamefont {{Ade}}, \citenamefont {{Aghanim}}, \citenamefont {{Arnaud}},
  \citenamefont {{Ashdown}}, \citenamefont {{Aumont}}, \citenamefont
  {{Baccigalupi}}, \citenamefont {{Banday}}, \citenamefont {{Barreiro}},
  \citenamefont {{Bartlett}}, \citenamefont {{Bartolo}}, \citenamefont
  {{Battaner}}, \citenamefont {{Battye}}, \citenamefont {{Benabed}},
  \citenamefont {{Beno{\^\i}t}}, \citenamefont {{Benoit-L{\'e}vy}},
  \citenamefont {{Bernard}}, \citenamefont {{Bersanelli}}, \citenamefont
  {{Bielewicz}}, \citenamefont {{Bock}}, \citenamefont {{Bonaldi}},
  \citenamefont {{Bonavera}}, \citenamefont {{Bond}}, \citenamefont
  {{Borrill}}, \citenamefont {{Bouchet}}, \citenamefont {{Bucher}},
  \citenamefont {{Burigana}}, \citenamefont {{Butler}}, \citenamefont
  {{Calabrese}}, \citenamefont {{Cardoso}}, \citenamefont {{Catalano}},
  \citenamefont {{Challinor}}, \citenamefont {{Chamballu}}, \citenamefont
  {{Chary}}, \citenamefont {{Chiang}}, \citenamefont {{Christensen}},
  \citenamefont {{Church}}, \citenamefont {{Clements}}, \citenamefont
  {{Colombi}}, \citenamefont {{Colombo}}, \citenamefont {{Combet}},
  \citenamefont {{Comis}}, \citenamefont {{Couchot}}, \citenamefont
  {{Coulais}}, \citenamefont {{Crill}}, \citenamefont {{Curto}}, \citenamefont
  {{Cuttaia}}, \citenamefont {{Danese}}, \citenamefont {{Davies}},
  \citenamefont {{Davis}}, \citenamefont {{de Bernardis}}, \citenamefont {{de
  Rosa}}, \citenamefont {{de Zotti}}, \citenamefont {{Delabrouille}},
  \citenamefont {{D{\'e}sert}}, \citenamefont {{Diego}}, \citenamefont
  {{Dolag}}, \citenamefont {{Dole}}, \citenamefont {{Donzelli}}, \citenamefont
  {{Dor{\'e}}}, \citenamefont {{Douspis}}, \citenamefont {{Ducout}},
  \citenamefont {{Dupac}}, \citenamefont {{Efstathiou}}, \citenamefont
  {{Elsner}}, \citenamefont {{En{\ss}lin}}, \citenamefont {{Eriksen}},
  \citenamefont {{Falgarone}}, \citenamefont {{Fergusson}}, \citenamefont
  {{Finelli}}, \citenamefont {{Forni}}, \citenamefont {{Frailis}},
  \citenamefont {{Fraisse}}, \citenamefont {{Franceschi}}, \citenamefont
  {{Frejsel}}, \citenamefont {{Galeotta}}, \citenamefont {{Galli}},
  \citenamefont {{Ganga}}, \citenamefont {{Giard}}, \citenamefont
  {{Giraud-H{\'e}raud}}, \citenamefont {{Gjerl{\o}w}}, \citenamefont
  {{Gonz{\'a}lez-Nuevo}}, \citenamefont {{G{\'o}rski}}, \citenamefont
  {{Gratton}}, \citenamefont {{Gregorio}}, \citenamefont {{Gruppuso}},
  \citenamefont {{Gudmundsson}}, \citenamefont {{Hansen}}, \citenamefont
  {{Hanson}}, \citenamefont {{Harrison}}, \citenamefont
  {{Henrot-Versill{\'e}}}, \citenamefont {{Hern{\'a}ndez-Monteagudo}},
  \citenamefont {{Herranz}}, \citenamefont {{Hildebrandt}}, \citenamefont
  {{Hivon}}, \citenamefont {{Hobson}}, \citenamefont {{Holmes}}, \citenamefont
  {{Hornstrup}}, \citenamefont {{Hovest}}, \citenamefont {{Huffenberger}},
  \citenamefont {{Hurier}}, \citenamefont {{Jaffe}}, \citenamefont {{Jaffe}},
  \citenamefont {{Jones}}, \citenamefont {{Juvela}}, \citenamefont
  {{Keih{\"a}nen}}, \citenamefont {{Keskitalo}}, \citenamefont {{Kisner}},
  \citenamefont {{Kneissl}}, \citenamefont {{Knoche}}, \citenamefont {{Kunz}},
  \citenamefont {{Kurki-Suonio}}, \citenamefont {{Lagache}}, \citenamefont
  {{L{\"a}hteenm{\"a}ki}}, \citenamefont {{Lamarre}}, \citenamefont
  {{Lasenby}}, \citenamefont {{Lattanzi}}, \citenamefont {{Lawrence}},
  \citenamefont {{Leonardi}}, \citenamefont {{Lesgourgues}}, \citenamefont
  {{Levrier}}, \citenamefont {{Liguori}}, \citenamefont {{Lilje}},
  \citenamefont {{Linden-V{\o}rnle}}, \citenamefont {{L{\'o}pez-Caniego}},
  \citenamefont {{Lubin}}, \citenamefont {{Mac{\'\i}as-P{\'e}rez}},
  \citenamefont {{Maggio}}, \citenamefont {{Maino}}, \citenamefont {{Mand
  olesi}}, \citenamefont {{Mangilli}}, \citenamefont {{Maris}}, \citenamefont
  {{Martin}}, \citenamefont {{Mart{\'\i}nez-Gonz{\'a}lez}}, \citenamefont
  {{Masi}}, \citenamefont {{Matarrese}}, \citenamefont {{McGehee}},
  \citenamefont {{Meinhold}}, \citenamefont {{Melchiorri}}, \citenamefont
  {{Melin}}, \citenamefont {{Mendes}}, \citenamefont {{Mennella}},
  \citenamefont {{Migliaccio}}, \citenamefont {{Mitra}}, \citenamefont
  {{Miville-Desch{\^e}nes}}, \citenamefont {{Moneti}}, \citenamefont
  {{Montier}}, \citenamefont {{Morgante}}, \citenamefont {{Mortlock}},
  \citenamefont {{Moss}}, \citenamefont {{Munshi}}, \citenamefont {{Murphy}},
  \citenamefont {{Naselsky}}, \citenamefont {{Nati}}, \citenamefont {{Natoli}},
  \citenamefont {{Netterfield}}, \citenamefont {{N{\o}rgaard-Nielsen}},
  \citenamefont {{Noviello}}, \citenamefont {{Novikov}}, \citenamefont
  {{Novikov}}, \citenamefont {{Oxborrow}}, \citenamefont {{Paci}},
  \citenamefont {{Pagano}}, \citenamefont {{Pajot}}, \citenamefont
  {{Paoletti}}, \citenamefont {{Partridge}}, \citenamefont {{Pasian}},
  \citenamefont {{Patanchon}}, \citenamefont {{Pearson}}, \citenamefont
  {{Perdereau}}, \citenamefont {{Perotto}}, \citenamefont {{Perrotta}},
  \citenamefont {{Pettorino}}, \citenamefont {{Piacentini}}, \citenamefont
  {{Piat}}, \citenamefont {{Pierpaoli}}, \citenamefont {{Pietrobon}},
  \citenamefont {{Plaszczynski}}, \citenamefont {{Pointecouteau}},
  \citenamefont {{Polenta}}, \citenamefont {{Popa}}, \citenamefont {{Pratt}},
  \citenamefont {{Pr{\'e}zeau}}, \citenamefont {{Prunet}}, \citenamefont
  {{Puget}}, \citenamefont {{Rachen}}, \citenamefont {{Rebolo}}, \citenamefont
  {{Reinecke}}, \citenamefont {{Remazeilles}}, \citenamefont {{Renault}},
  \citenamefont {{Renzi}}, \citenamefont {{Ristorcelli}}, \citenamefont
  {{Rocha}}, \citenamefont {{Roman}}, \citenamefont {{Rosset}}, \citenamefont
  {{Rossetti}}, \citenamefont {{Roudier}}, \citenamefont
  {{Rubi{\~n}o-Mart{\'\i}n}}, \citenamefont {{Rusholme}}, \citenamefont
  {{Sandri}}, \citenamefont {{Santos}}, \citenamefont {{Savelainen}},
  \citenamefont {{Savini}}, \citenamefont {{Scott}}, \citenamefont
  {{Seiffert}}, \citenamefont {{Shellard}}, \citenamefont {{Spencer}},
  \citenamefont {{Stolyarov}}, \citenamefont {{Stompor}}, \citenamefont
  {{Sudiwala}}, \citenamefont {{Sunyaev}}, \citenamefont {{Sutton}},
  \citenamefont {{Suur-Uski}}, \citenamefont {{Sygnet}}, \citenamefont
  {{Tauber}}, \citenamefont {{Terenzi}}, \citenamefont {{Toffolatti}},
  \citenamefont {{Tomasi}}, \citenamefont {{Tristram}}, \citenamefont
  {{Tucci}}, \citenamefont {{Tuovinen}}, \citenamefont {{T{\"u}rler}},
  \citenamefont {{Umana}}, \citenamefont {{Valenziano}}, \citenamefont
  {{Valiviita}}, \citenamefont {{Van Tent}}, \citenamefont {{Vielva}},
  \citenamefont {{Villa}}, \citenamefont {{Wade}}, \citenamefont {{Wandelt}},
  \citenamefont {{Wehus}}, \citenamefont {{Weller}}, \citenamefont {{White}},
  \citenamefont {{Yvon}}, \citenamefont {{Zacchei}},\ and\ \citenamefont
  {{Zonca}}}]{Planck16b}%
  \BibitemOpen
  \bibfield  {author} {\bibinfo {author} {\bibnamefont {{Planck
  Collaboration}}}, \bibinfo {author} {\bibfnamefont {P.~A.~R.}\ \bibnamefont
  {{Ade}}}, \bibinfo {author} {\bibfnamefont {N.}~\bibnamefont {{Aghanim}}},
  \bibinfo {author} {\bibfnamefont {M.}~\bibnamefont {{Arnaud}}}, \bibinfo
  {author} {\bibfnamefont {M.}~\bibnamefont {{Ashdown}}}, \bibinfo {author}
  {\bibfnamefont {J.}~\bibnamefont {{Aumont}}}, \bibinfo {author}
  {\bibfnamefont {C.}~\bibnamefont {{Baccigalupi}}}, \bibinfo {author}
  {\bibfnamefont {A.~J.}\ \bibnamefont {{Banday}}}, \bibinfo {author}
  {\bibfnamefont {R.~B.}\ \bibnamefont {{Barreiro}}}, \bibinfo {author}
  {\bibfnamefont {J.~G.}\ \bibnamefont {{Bartlett}}}, \bibinfo {author}
  {\bibfnamefont {N.}~\bibnamefont {{Bartolo}}}, \bibinfo {author}
  {\bibfnamefont {E.}~\bibnamefont {{Battaner}}}, \bibinfo {author}
  {\bibfnamefont {R.}~\bibnamefont {{Battye}}}, \bibinfo {author}
  {\bibfnamefont {K.}~\bibnamefont {{Benabed}}}, \bibinfo {author}
  {\bibfnamefont {A.}~\bibnamefont {{Beno{\^\i}t}}}, \bibinfo {author}
  {\bibfnamefont {A.}~\bibnamefont {{Benoit-L{\'e}vy}}}, \bibinfo {author}
  {\bibfnamefont {J.~P.}\ \bibnamefont {{Bernard}}}, \bibinfo {author}
  {\bibfnamefont {M.}~\bibnamefont {{Bersanelli}}}, \bibinfo {author}
  {\bibfnamefont {P.}~\bibnamefont {{Bielewicz}}}, \bibinfo {author}
  {\bibfnamefont {J.~J.}\ \bibnamefont {{Bock}}}, \bibinfo {author}
  {\bibfnamefont {A.}~\bibnamefont {{Bonaldi}}}, \bibinfo {author}
  {\bibfnamefont {L.}~\bibnamefont {{Bonavera}}}, \bibinfo {author}
  {\bibfnamefont {J.~R.}\ \bibnamefont {{Bond}}}, \bibinfo {author}
  {\bibfnamefont {J.}~\bibnamefont {{Borrill}}}, \bibinfo {author}
  {\bibfnamefont {F.~R.}\ \bibnamefont {{Bouchet}}}, \bibinfo {author}
  {\bibfnamefont {M.}~\bibnamefont {{Bucher}}}, \bibinfo {author}
  {\bibfnamefont {C.}~\bibnamefont {{Burigana}}}, \bibinfo {author}
  {\bibfnamefont {R.~C.}\ \bibnamefont {{Butler}}}, \bibinfo {author}
  {\bibfnamefont {E.}~\bibnamefont {{Calabrese}}}, \bibinfo {author}
  {\bibfnamefont {J.~F.}\ \bibnamefont {{Cardoso}}}, \bibinfo {author}
  {\bibfnamefont {A.}~\bibnamefont {{Catalano}}}, \bibinfo {author}
  {\bibfnamefont {A.}~\bibnamefont {{Challinor}}}, \bibinfo {author}
  {\bibfnamefont {A.}~\bibnamefont {{Chamballu}}}, \bibinfo {author}
  {\bibfnamefont {R.~R.}\ \bibnamefont {{Chary}}}, \bibinfo {author}
  {\bibfnamefont {H.~C.}\ \bibnamefont {{Chiang}}}, \bibinfo {author}
  {\bibfnamefont {P.~R.}\ \bibnamefont {{Christensen}}}, \bibinfo {author}
  {\bibfnamefont {S.}~\bibnamefont {{Church}}}, \bibinfo {author}
  {\bibfnamefont {D.~L.}\ \bibnamefont {{Clements}}}, \bibinfo {author}
  {\bibfnamefont {S.}~\bibnamefont {{Colombi}}}, \bibinfo {author}
  {\bibfnamefont {L.~P.~L.}\ \bibnamefont {{Colombo}}}, \bibinfo {author}
  {\bibfnamefont {C.}~\bibnamefont {{Combet}}}, \bibinfo {author}
  {\bibfnamefont {B.}~\bibnamefont {{Comis}}}, \bibinfo {author} {\bibfnamefont
  {F.}~\bibnamefont {{Couchot}}}, \bibinfo {author} {\bibfnamefont
  {A.}~\bibnamefont {{Coulais}}}, \bibinfo {author} {\bibfnamefont {B.~P.}\
  \bibnamefont {{Crill}}}, \bibinfo {author} {\bibfnamefont {A.}~\bibnamefont
  {{Curto}}}, \bibinfo {author} {\bibfnamefont {F.}~\bibnamefont {{Cuttaia}}},
  \bibinfo {author} {\bibfnamefont {L.}~\bibnamefont {{Danese}}}, \bibinfo
  {author} {\bibfnamefont {R.~D.}\ \bibnamefont {{Davies}}}, \bibinfo {author}
  {\bibfnamefont {R.~J.}\ \bibnamefont {{Davis}}}, \bibinfo {author}
  {\bibfnamefont {P.}~\bibnamefont {{de Bernardis}}}, \bibinfo {author}
  {\bibfnamefont {A.}~\bibnamefont {{de Rosa}}}, \bibinfo {author}
  {\bibfnamefont {G.}~\bibnamefont {{de Zotti}}}, \bibinfo {author}
  {\bibfnamefont {J.}~\bibnamefont {{Delabrouille}}}, \bibinfo {author}
  {\bibfnamefont {F.~X.}\ \bibnamefont {{D{\'e}sert}}}, \bibinfo {author}
  {\bibfnamefont {J.~M.}\ \bibnamefont {{Diego}}}, \bibinfo {author}
  {\bibfnamefont {K.}~\bibnamefont {{Dolag}}}, \bibinfo {author} {\bibfnamefont
  {H.}~\bibnamefont {{Dole}}}, \bibinfo {author} {\bibfnamefont
  {S.}~\bibnamefont {{Donzelli}}}, \bibinfo {author} {\bibfnamefont
  {O.}~\bibnamefont {{Dor{\'e}}}}, \bibinfo {author} {\bibfnamefont
  {M.}~\bibnamefont {{Douspis}}}, \bibinfo {author} {\bibfnamefont
  {A.}~\bibnamefont {{Ducout}}}, \bibinfo {author} {\bibfnamefont
  {X.}~\bibnamefont {{Dupac}}}, \bibinfo {author} {\bibfnamefont
  {G.}~\bibnamefont {{Efstathiou}}}, \bibinfo {author} {\bibfnamefont
  {F.}~\bibnamefont {{Elsner}}}, \bibinfo {author} {\bibfnamefont {T.~A.}\
  \bibnamefont {{En{\ss}lin}}}, \bibinfo {author} {\bibfnamefont {H.~K.}\
  \bibnamefont {{Eriksen}}}, \bibinfo {author} {\bibfnamefont {E.}~\bibnamefont
  {{Falgarone}}}, \bibinfo {author} {\bibfnamefont {J.}~\bibnamefont
  {{Fergusson}}}, \bibinfo {author} {\bibfnamefont {F.}~\bibnamefont
  {{Finelli}}}, \bibinfo {author} {\bibfnamefont {O.}~\bibnamefont {{Forni}}},
  \bibinfo {author} {\bibfnamefont {M.}~\bibnamefont {{Frailis}}}, \bibinfo
  {author} {\bibfnamefont {A.~A.}\ \bibnamefont {{Fraisse}}}, \bibinfo {author}
  {\bibfnamefont {E.}~\bibnamefont {{Franceschi}}}, \bibinfo {author}
  {\bibfnamefont {A.}~\bibnamefont {{Frejsel}}}, \bibinfo {author}
  {\bibfnamefont {S.}~\bibnamefont {{Galeotta}}}, \bibinfo {author}
  {\bibfnamefont {S.}~\bibnamefont {{Galli}}}, \bibinfo {author} {\bibfnamefont
  {K.}~\bibnamefont {{Ganga}}}, \bibinfo {author} {\bibfnamefont
  {M.}~\bibnamefont {{Giard}}}, \bibinfo {author} {\bibfnamefont
  {Y.}~\bibnamefont {{Giraud-H{\'e}raud}}}, \bibinfo {author} {\bibfnamefont
  {E.}~\bibnamefont {{Gjerl{\o}w}}}, \bibinfo {author} {\bibfnamefont
  {J.}~\bibnamefont {{Gonz{\'a}lez-Nuevo}}}, \bibinfo {author} {\bibfnamefont
  {K.~M.}\ \bibnamefont {{G{\'o}rski}}}, \bibinfo {author} {\bibfnamefont
  {S.}~\bibnamefont {{Gratton}}}, \bibinfo {author} {\bibfnamefont
  {A.}~\bibnamefont {{Gregorio}}}, \bibinfo {author} {\bibfnamefont
  {A.}~\bibnamefont {{Gruppuso}}}, \bibinfo {author} {\bibfnamefont {J.~E.}\
  \bibnamefont {{Gudmundsson}}}, \bibinfo {author} {\bibfnamefont {F.~K.}\
  \bibnamefont {{Hansen}}}, \bibinfo {author} {\bibfnamefont {D.}~\bibnamefont
  {{Hanson}}}, \bibinfo {author} {\bibfnamefont {D.~L.}\ \bibnamefont
  {{Harrison}}}, \bibinfo {author} {\bibfnamefont {S.}~\bibnamefont
  {{Henrot-Versill{\'e}}}}, \bibinfo {author} {\bibfnamefont {C.}~\bibnamefont
  {{Hern{\'a}ndez-Monteagudo}}}, \bibinfo {author} {\bibfnamefont
  {D.}~\bibnamefont {{Herranz}}}, \bibinfo {author} {\bibfnamefont {S.~R.}\
  \bibnamefont {{Hildebrandt}}}, \bibinfo {author} {\bibfnamefont
  {E.}~\bibnamefont {{Hivon}}}, \bibinfo {author} {\bibfnamefont
  {M.}~\bibnamefont {{Hobson}}}, \bibinfo {author} {\bibfnamefont {W.~A.}\
  \bibnamefont {{Holmes}}}, \bibinfo {author} {\bibfnamefont {A.}~\bibnamefont
  {{Hornstrup}}}, \bibinfo {author} {\bibfnamefont {W.}~\bibnamefont
  {{Hovest}}}, \bibinfo {author} {\bibfnamefont {K.~M.}\ \bibnamefont
  {{Huffenberger}}}, \bibinfo {author} {\bibfnamefont {G.}~\bibnamefont
  {{Hurier}}}, \bibinfo {author} {\bibfnamefont {A.~H.}\ \bibnamefont
  {{Jaffe}}}, \bibinfo {author} {\bibfnamefont {T.~R.}\ \bibnamefont
  {{Jaffe}}}, \bibinfo {author} {\bibfnamefont {W.~C.}\ \bibnamefont
  {{Jones}}}, \bibinfo {author} {\bibfnamefont {M.}~\bibnamefont {{Juvela}}},
  \bibinfo {author} {\bibfnamefont {E.}~\bibnamefont {{Keih{\"a}nen}}},
  \bibinfo {author} {\bibfnamefont {R.}~\bibnamefont {{Keskitalo}}}, \bibinfo
  {author} {\bibfnamefont {T.~S.}\ \bibnamefont {{Kisner}}}, \bibinfo {author}
  {\bibfnamefont {R.}~\bibnamefont {{Kneissl}}}, \bibinfo {author}
  {\bibfnamefont {J.}~\bibnamefont {{Knoche}}}, \bibinfo {author}
  {\bibfnamefont {M.}~\bibnamefont {{Kunz}}}, \bibinfo {author} {\bibfnamefont
  {H.}~\bibnamefont {{Kurki-Suonio}}}, \bibinfo {author} {\bibfnamefont
  {G.}~\bibnamefont {{Lagache}}}, \bibinfo {author} {\bibfnamefont
  {A.}~\bibnamefont {{L{\"a}hteenm{\"a}ki}}}, \bibinfo {author} {\bibfnamefont
  {J.~M.}\ \bibnamefont {{Lamarre}}}, \bibinfo {author} {\bibfnamefont
  {A.}~\bibnamefont {{Lasenby}}}, \bibinfo {author} {\bibfnamefont
  {M.}~\bibnamefont {{Lattanzi}}}, \bibinfo {author} {\bibfnamefont {C.~R.}\
  \bibnamefont {{Lawrence}}}, \bibinfo {author} {\bibfnamefont
  {R.}~\bibnamefont {{Leonardi}}}, \bibinfo {author} {\bibfnamefont
  {J.}~\bibnamefont {{Lesgourgues}}}, \bibinfo {author} {\bibfnamefont
  {F.}~\bibnamefont {{Levrier}}}, \bibinfo {author} {\bibfnamefont
  {M.}~\bibnamefont {{Liguori}}}, \bibinfo {author} {\bibfnamefont {P.~B.}\
  \bibnamefont {{Lilje}}}, \bibinfo {author} {\bibfnamefont {M.}~\bibnamefont
  {{Linden-V{\o}rnle}}}, \bibinfo {author} {\bibfnamefont {M.}~\bibnamefont
  {{L{\'o}pez-Caniego}}}, \bibinfo {author} {\bibfnamefont {P.~M.}\
  \bibnamefont {{Lubin}}}, \bibinfo {author} {\bibfnamefont {J.~F.}\
  \bibnamefont {{Mac{\'\i}as-P{\'e}rez}}}, \bibinfo {author} {\bibfnamefont
  {G.}~\bibnamefont {{Maggio}}}, \bibinfo {author} {\bibfnamefont
  {D.}~\bibnamefont {{Maino}}}, \bibinfo {author} {\bibfnamefont
  {N.}~\bibnamefont {{Mand olesi}}}, \bibinfo {author} {\bibfnamefont
  {A.}~\bibnamefont {{Mangilli}}}, \bibinfo {author} {\bibfnamefont
  {M.}~\bibnamefont {{Maris}}}, \bibinfo {author} {\bibfnamefont {P.~G.}\
  \bibnamefont {{Martin}}}, \bibinfo {author} {\bibfnamefont {E.}~\bibnamefont
  {{Mart{\'\i}nez-Gonz{\'a}lez}}}, \bibinfo {author} {\bibfnamefont
  {S.}~\bibnamefont {{Masi}}}, \bibinfo {author} {\bibfnamefont
  {S.}~\bibnamefont {{Matarrese}}}, \bibinfo {author} {\bibfnamefont
  {P.}~\bibnamefont {{McGehee}}}, \bibinfo {author} {\bibfnamefont {P.~R.}\
  \bibnamefont {{Meinhold}}}, \bibinfo {author} {\bibfnamefont
  {A.}~\bibnamefont {{Melchiorri}}}, \bibinfo {author} {\bibfnamefont {J.~B.}\
  \bibnamefont {{Melin}}}, \bibinfo {author} {\bibfnamefont {L.}~\bibnamefont
  {{Mendes}}}, \bibinfo {author} {\bibfnamefont {A.}~\bibnamefont
  {{Mennella}}}, \bibinfo {author} {\bibfnamefont {M.}~\bibnamefont
  {{Migliaccio}}}, \bibinfo {author} {\bibfnamefont {S.}~\bibnamefont
  {{Mitra}}}, \bibinfo {author} {\bibfnamefont {M.~A.}\ \bibnamefont
  {{Miville-Desch{\^e}nes}}}, \bibinfo {author} {\bibfnamefont
  {A.}~\bibnamefont {{Moneti}}}, \bibinfo {author} {\bibfnamefont
  {L.}~\bibnamefont {{Montier}}}, \bibinfo {author} {\bibfnamefont
  {G.}~\bibnamefont {{Morgante}}}, \bibinfo {author} {\bibfnamefont
  {D.}~\bibnamefont {{Mortlock}}}, \bibinfo {author} {\bibfnamefont
  {A.}~\bibnamefont {{Moss}}}, \bibinfo {author} {\bibfnamefont
  {D.}~\bibnamefont {{Munshi}}}, \bibinfo {author} {\bibfnamefont {J.~A.}\
  \bibnamefont {{Murphy}}}, \bibinfo {author} {\bibfnamefont {P.}~\bibnamefont
  {{Naselsky}}}, \bibinfo {author} {\bibfnamefont {F.}~\bibnamefont {{Nati}}},
  \bibinfo {author} {\bibfnamefont {P.}~\bibnamefont {{Natoli}}}, \bibinfo
  {author} {\bibfnamefont {C.~B.}\ \bibnamefont {{Netterfield}}}, \bibinfo
  {author} {\bibfnamefont {H.~U.}\ \bibnamefont {{N{\o}rgaard-Nielsen}}},
  \bibinfo {author} {\bibfnamefont {F.}~\bibnamefont {{Noviello}}}, \bibinfo
  {author} {\bibfnamefont {D.}~\bibnamefont {{Novikov}}}, \bibinfo {author}
  {\bibfnamefont {I.}~\bibnamefont {{Novikov}}}, \bibinfo {author}
  {\bibfnamefont {C.~A.}\ \bibnamefont {{Oxborrow}}}, \bibinfo {author}
  {\bibfnamefont {F.}~\bibnamefont {{Paci}}}, \bibinfo {author} {\bibfnamefont
  {L.}~\bibnamefont {{Pagano}}}, \bibinfo {author} {\bibfnamefont
  {F.}~\bibnamefont {{Pajot}}}, \bibinfo {author} {\bibfnamefont
  {D.}~\bibnamefont {{Paoletti}}}, \bibinfo {author} {\bibfnamefont
  {B.}~\bibnamefont {{Partridge}}}, \bibinfo {author} {\bibfnamefont
  {F.}~\bibnamefont {{Pasian}}}, \bibinfo {author} {\bibfnamefont
  {G.}~\bibnamefont {{Patanchon}}}, \bibinfo {author} {\bibfnamefont {T.~J.}\
  \bibnamefont {{Pearson}}}, \bibinfo {author} {\bibfnamefont {O.}~\bibnamefont
  {{Perdereau}}}, \bibinfo {author} {\bibfnamefont {L.}~\bibnamefont
  {{Perotto}}}, \bibinfo {author} {\bibfnamefont {F.}~\bibnamefont
  {{Perrotta}}}, \bibinfo {author} {\bibfnamefont {V.}~\bibnamefont
  {{Pettorino}}}, \bibinfo {author} {\bibfnamefont {F.}~\bibnamefont
  {{Piacentini}}}, \bibinfo {author} {\bibfnamefont {M.}~\bibnamefont
  {{Piat}}}, \bibinfo {author} {\bibfnamefont {E.}~\bibnamefont {{Pierpaoli}}},
  \bibinfo {author} {\bibfnamefont {D.}~\bibnamefont {{Pietrobon}}}, \bibinfo
  {author} {\bibfnamefont {S.}~\bibnamefont {{Plaszczynski}}}, \bibinfo
  {author} {\bibfnamefont {E.}~\bibnamefont {{Pointecouteau}}}, \bibinfo
  {author} {\bibfnamefont {G.}~\bibnamefont {{Polenta}}}, \bibinfo {author}
  {\bibfnamefont {L.}~\bibnamefont {{Popa}}}, \bibinfo {author} {\bibfnamefont
  {G.~W.}\ \bibnamefont {{Pratt}}}, \bibinfo {author} {\bibfnamefont
  {G.}~\bibnamefont {{Pr{\'e}zeau}}}, \bibinfo {author} {\bibfnamefont
  {S.}~\bibnamefont {{Prunet}}}, \bibinfo {author} {\bibfnamefont {J.~L.}\
  \bibnamefont {{Puget}}}, \bibinfo {author} {\bibfnamefont {J.~P.}\
  \bibnamefont {{Rachen}}}, \bibinfo {author} {\bibfnamefont {R.}~\bibnamefont
  {{Rebolo}}}, \bibinfo {author} {\bibfnamefont {M.}~\bibnamefont
  {{Reinecke}}}, \bibinfo {author} {\bibfnamefont {M.}~\bibnamefont
  {{Remazeilles}}}, \bibinfo {author} {\bibfnamefont {C.}~\bibnamefont
  {{Renault}}}, \bibinfo {author} {\bibfnamefont {A.}~\bibnamefont {{Renzi}}},
  \bibinfo {author} {\bibfnamefont {I.}~\bibnamefont {{Ristorcelli}}}, \bibinfo
  {author} {\bibfnamefont {G.}~\bibnamefont {{Rocha}}}, \bibinfo {author}
  {\bibfnamefont {M.}~\bibnamefont {{Roman}}}, \bibinfo {author} {\bibfnamefont
  {C.}~\bibnamefont {{Rosset}}}, \bibinfo {author} {\bibfnamefont
  {M.}~\bibnamefont {{Rossetti}}}, \bibinfo {author} {\bibfnamefont
  {G.}~\bibnamefont {{Roudier}}}, \bibinfo {author} {\bibfnamefont {J.~A.}\
  \bibnamefont {{Rubi{\~n}o-Mart{\'\i}n}}}, \bibinfo {author} {\bibfnamefont
  {B.}~\bibnamefont {{Rusholme}}}, \bibinfo {author} {\bibfnamefont
  {M.}~\bibnamefont {{Sandri}}}, \bibinfo {author} {\bibfnamefont
  {D.}~\bibnamefont {{Santos}}}, \bibinfo {author} {\bibfnamefont
  {M.}~\bibnamefont {{Savelainen}}}, \bibinfo {author} {\bibfnamefont
  {G.}~\bibnamefont {{Savini}}}, \bibinfo {author} {\bibfnamefont
  {D.}~\bibnamefont {{Scott}}}, \bibinfo {author} {\bibfnamefont {M.~D.}\
  \bibnamefont {{Seiffert}}}, \bibinfo {author} {\bibfnamefont {E.~P.~S.}\
  \bibnamefont {{Shellard}}}, \bibinfo {author} {\bibfnamefont {L.~D.}\
  \bibnamefont {{Spencer}}}, \bibinfo {author} {\bibfnamefont {V.}~\bibnamefont
  {{Stolyarov}}}, \bibinfo {author} {\bibfnamefont {R.}~\bibnamefont
  {{Stompor}}}, \bibinfo {author} {\bibfnamefont {R.}~\bibnamefont
  {{Sudiwala}}}, \bibinfo {author} {\bibfnamefont {R.}~\bibnamefont
  {{Sunyaev}}}, \bibinfo {author} {\bibfnamefont {D.}~\bibnamefont {{Sutton}}},
  \bibinfo {author} {\bibfnamefont {A.~S.}\ \bibnamefont {{Suur-Uski}}},
  \bibinfo {author} {\bibfnamefont {J.~F.}\ \bibnamefont {{Sygnet}}}, \bibinfo
  {author} {\bibfnamefont {J.~A.}\ \bibnamefont {{Tauber}}}, \bibinfo {author}
  {\bibfnamefont {L.}~\bibnamefont {{Terenzi}}}, \bibinfo {author}
  {\bibfnamefont {L.}~\bibnamefont {{Toffolatti}}}, \bibinfo {author}
  {\bibfnamefont {M.}~\bibnamefont {{Tomasi}}}, \bibinfo {author}
  {\bibfnamefont {M.}~\bibnamefont {{Tristram}}}, \bibinfo {author}
  {\bibfnamefont {M.}~\bibnamefont {{Tucci}}}, \bibinfo {author} {\bibfnamefont
  {J.}~\bibnamefont {{Tuovinen}}}, \bibinfo {author} {\bibfnamefont
  {M.}~\bibnamefont {{T{\"u}rler}}}, \bibinfo {author} {\bibfnamefont
  {G.}~\bibnamefont {{Umana}}}, \bibinfo {author} {\bibfnamefont
  {L.}~\bibnamefont {{Valenziano}}}, \bibinfo {author} {\bibfnamefont
  {J.}~\bibnamefont {{Valiviita}}}, \bibinfo {author} {\bibfnamefont
  {B.}~\bibnamefont {{Van Tent}}}, \bibinfo {author} {\bibfnamefont
  {P.}~\bibnamefont {{Vielva}}}, \bibinfo {author} {\bibfnamefont
  {F.}~\bibnamefont {{Villa}}}, \bibinfo {author} {\bibfnamefont {L.~A.}\
  \bibnamefont {{Wade}}}, \bibinfo {author} {\bibfnamefont {B.~D.}\
  \bibnamefont {{Wandelt}}}, \bibinfo {author} {\bibfnamefont {I.~K.}\
  \bibnamefont {{Wehus}}}, \bibinfo {author} {\bibfnamefont {J.}~\bibnamefont
  {{Weller}}}, \bibinfo {author} {\bibfnamefont {S.~D.~M.}\ \bibnamefont
  {{White}}}, \bibinfo {author} {\bibfnamefont {D.}~\bibnamefont {{Yvon}}},
  \bibinfo {author} {\bibfnamefont {A.}~\bibnamefont {{Zacchei}}}, \ and\
  \bibinfo {author} {\bibfnamefont {A.}~\bibnamefont {{Zonca}}},\ }\href
  {\doibase 10.1051/0004-6361/201525833} {\bibfield  {journal} {\bibinfo
  {journal} {\aap}\ }\textbf {\bibinfo {volume} {594}},\ \bibinfo {eid} {A24}
  (\bibinfo {year} {2016}{\natexlab{e}})},\ \Eprint
  {http://arxiv.org/abs/1502.01597} {arXiv:1502.01597 [astro-ph.CO]}
  \BibitemShut {NoStop}%
\bibitem [{\citenamefont {{Horowitz}}\ and\ \citenamefont
  {{Seljak}}(2017)}]{Horowitz2017}%
  \BibitemOpen
  \bibfield  {author} {\bibinfo {author} {\bibfnamefont {B.}~\bibnamefont
  {{Horowitz}}}\ and\ \bibinfo {author} {\bibfnamefont {U.}~\bibnamefont
  {{Seljak}}},\ }\href {\doibase 10.1093/mnras/stx766} {\bibfield  {journal}
  {\bibinfo  {journal} {\mnras}\ }\textbf {\bibinfo {volume} {469}},\ \bibinfo
  {pages} {394} (\bibinfo {year} {2017})},\ \Eprint
  {http://arxiv.org/abs/1609.01850} {arXiv:1609.01850 [astro-ph.CO]}
  \BibitemShut {NoStop}%
\bibitem [{\citenamefont {{Salvati}}\ \emph {et~al.}(2018)\citenamefont
  {{Salvati}}, \citenamefont {{Douspis}},\ and\ \citenamefont
  {{Aghanim}}}]{Salvati18}%
  \BibitemOpen
  \bibfield  {author} {\bibinfo {author} {\bibfnamefont {L.}~\bibnamefont
  {{Salvati}}}, \bibinfo {author} {\bibfnamefont {M.}~\bibnamefont
  {{Douspis}}}, \ and\ \bibinfo {author} {\bibfnamefont {N.}~\bibnamefont
  {{Aghanim}}},\ }\href {\doibase 10.1051/0004-6361/201731990} {\bibfield
  {journal} {\bibinfo  {journal} {\aap}\ }\textbf {\bibinfo {volume} {614}},\
  \bibinfo {eid} {A13} (\bibinfo {year} {2018})},\ \Eprint
  {http://arxiv.org/abs/1708.00697} {arXiv:1708.00697} \BibitemShut {NoStop}%
\bibitem [{\citenamefont {{Heitmann}}\ \emph {et~al.}(2009)\citenamefont
  {{Heitmann}}, \citenamefont {{Higdon}}, \citenamefont {{White}},
  \citenamefont {{Habib}}, \citenamefont {{Williams}}, \citenamefont
  {{Lawrence}},\ and\ \citenamefont {{Wagner}}}]{Heitmann09}%
  \BibitemOpen
  \bibfield  {author} {\bibinfo {author} {\bibfnamefont {K.}~\bibnamefont
  {{Heitmann}}}, \bibinfo {author} {\bibfnamefont {D.}~\bibnamefont
  {{Higdon}}}, \bibinfo {author} {\bibfnamefont {M.}~\bibnamefont {{White}}},
  \bibinfo {author} {\bibfnamefont {S.}~\bibnamefont {{Habib}}}, \bibinfo
  {author} {\bibfnamefont {B.~J.}\ \bibnamefont {{Williams}}}, \bibinfo
  {author} {\bibfnamefont {E.}~\bibnamefont {{Lawrence}}}, \ and\ \bibinfo
  {author} {\bibfnamefont {C.}~\bibnamefont {{Wagner}}},\ }\href {\doibase
  10.1088/0004-637X/705/1/156} {\bibfield  {journal} {\bibinfo  {journal}
  {\apj}\ }\textbf {\bibinfo {volume} {705}},\ \bibinfo {pages} {156} (\bibinfo
  {year} {2009})},\ \Eprint {http://arxiv.org/abs/0902.0429} {arXiv:0902.0429
  [astro-ph.CO]} \BibitemShut {NoStop}%
\bibitem [{\citenamefont {{van Daalen}}\ and\ \citenamefont
  {{Schaye}}(2015)}]{vanDaalen2015}%
  \BibitemOpen
  \bibfield  {author} {\bibinfo {author} {\bibfnamefont {M.~P.}\ \bibnamefont
  {{van Daalen}}}\ and\ \bibinfo {author} {\bibfnamefont {J.}~\bibnamefont
  {{Schaye}}},\ }\href {\doibase 10.1093/mnras/stv1456} {\bibfield  {journal}
  {\bibinfo  {journal} {\mnras}\ }\textbf {\bibinfo {volume} {452}},\ \bibinfo
  {pages} {2247} (\bibinfo {year} {2015})},\ \Eprint
  {http://arxiv.org/abs/1501.05950} {arXiv:1501.05950} \BibitemShut {NoStop}%
\bibitem [{\citenamefont {{Smith}}\ and\ \citenamefont
  {{Watts}}(2005)}]{Smith2005}%
  \BibitemOpen
  \bibfield  {author} {\bibinfo {author} {\bibfnamefont {R.~E.}\ \bibnamefont
  {{Smith}}}\ and\ \bibinfo {author} {\bibfnamefont {P.~I.~R.}\ \bibnamefont
  {{Watts}}},\ }\href {\doibase 10.1111/j.1365-2966.2005.09053.x} {\bibfield
  {journal} {\bibinfo  {journal} {\mnras}\ }\textbf {\bibinfo {volume} {360}},\
  \bibinfo {pages} {203} (\bibinfo {year} {2005})},\ \Eprint
  {http://arxiv.org/abs/astro-ph/0412441} {astro-ph/0412441} \BibitemShut
  {NoStop}%
\bibitem [{\citenamefont {{Smith}}\ \emph {et~al.}(2007)\citenamefont
  {{Smith}}, \citenamefont {{Scoccimarro}},\ and\ \citenamefont
  {{Sheth}}}]{Smith2007}%
  \BibitemOpen
  \bibfield  {author} {\bibinfo {author} {\bibfnamefont {R.~E.}\ \bibnamefont
  {{Smith}}}, \bibinfo {author} {\bibfnamefont {R.}~\bibnamefont
  {{Scoccimarro}}}, \ and\ \bibinfo {author} {\bibfnamefont {R.~K.}\
  \bibnamefont {{Sheth}}},\ }\href {\doibase 10.1103/PhysRevD.75.063512}
  {\bibfield  {journal} {\bibinfo  {journal} {\prd}\ }\textbf {\bibinfo
  {volume} {75}},\ \bibinfo {eid} {063512} (\bibinfo {year} {2007})},\ \Eprint
  {http://arxiv.org/abs/astro-ph/0609547} {astro-ph/0609547} \BibitemShut
  {NoStop}%
\bibitem [{\citenamefont {{Giocoli}}\ \emph {et~al.}(2010)\citenamefont
  {{Giocoli}}, \citenamefont {{Bartelmann}}, \citenamefont {{Sheth}},\ and\
  \citenamefont {{Cacciato}}}]{Giocoli2010}%
  \BibitemOpen
  \bibfield  {author} {\bibinfo {author} {\bibfnamefont {C.}~\bibnamefont
  {{Giocoli}}}, \bibinfo {author} {\bibfnamefont {M.}~\bibnamefont
  {{Bartelmann}}}, \bibinfo {author} {\bibfnamefont {R.~K.}\ \bibnamefont
  {{Sheth}}}, \ and\ \bibinfo {author} {\bibfnamefont {M.}~\bibnamefont
  {{Cacciato}}},\ }\href {\doibase 10.1111/j.1365-2966.2010.17108.x} {\bibfield
   {journal} {\bibinfo  {journal} {\mnras}\ }\textbf {\bibinfo {volume}
  {408}},\ \bibinfo {pages} {300} (\bibinfo {year} {2010})},\ \Eprint
  {http://arxiv.org/abs/1003.4740} {arXiv:1003.4740} \BibitemShut {NoStop}%
\bibitem [{\citenamefont {{Smith}}\ \emph {et~al.}(2011)\citenamefont
  {{Smith}}, \citenamefont {{Desjacques}},\ and\ \citenamefont
  {{Marian}}}]{Smith2011}%
  \BibitemOpen
  \bibfield  {author} {\bibinfo {author} {\bibfnamefont {R.~E.}\ \bibnamefont
  {{Smith}}}, \bibinfo {author} {\bibfnamefont {V.}~\bibnamefont
  {{Desjacques}}}, \ and\ \bibinfo {author} {\bibfnamefont {L.}~\bibnamefont
  {{Marian}}},\ }\href {\doibase 10.1103/PhysRevD.83.043526} {\bibfield
  {journal} {\bibinfo  {journal} {\prd}\ }\textbf {\bibinfo {volume} {83}},\
  \bibinfo {eid} {043526} (\bibinfo {year} {2011})},\ \Eprint
  {http://arxiv.org/abs/1009.5085} {arXiv:1009.5085 [astro-ph.CO]} \BibitemShut
  {NoStop}%
\bibitem [{\citenamefont {{Valageas}}\ and\ \citenamefont
  {{Nishimichi}}(2011)}]{Valageas2011}%
  \BibitemOpen
  \bibfield  {author} {\bibinfo {author} {\bibfnamefont {P.}~\bibnamefont
  {{Valageas}}}\ and\ \bibinfo {author} {\bibfnamefont {T.}~\bibnamefont
  {{Nishimichi}}},\ }\href {\doibase 10.1051/0004-6361/201015685} {\bibfield
  {journal} {\bibinfo  {journal} {\aap}\ }\textbf {\bibinfo {volume} {527}},\
  \bibinfo {eid} {A87} (\bibinfo {year} {2011})},\ \Eprint
  {http://arxiv.org/abs/1009.0597} {arXiv:1009.0597} \BibitemShut {NoStop}%
\bibitem [{\citenamefont {{Mohammed}}\ and\ \citenamefont
  {{Seljak}}(2014)}]{Mohammed2014a}%
  \BibitemOpen
  \bibfield  {author} {\bibinfo {author} {\bibfnamefont {I.}~\bibnamefont
  {{Mohammed}}}\ and\ \bibinfo {author} {\bibfnamefont {U.}~\bibnamefont
  {{Seljak}}},\ }\href {\doibase 10.1093/mnras/stu1972} {\bibfield  {journal}
  {\bibinfo  {journal} {\mnras}\ }\textbf {\bibinfo {volume} {445}},\ \bibinfo
  {pages} {3382} (\bibinfo {year} {2014})},\ \Eprint
  {http://arxiv.org/abs/1407.0060} {arXiv:1407.0060} \BibitemShut {NoStop}%
\bibitem [{\citenamefont {{Mandelbaum}}\ \emph {et~al.}(2005)\citenamefont
  {{Mandelbaum}}, \citenamefont {{Tasitsiomi}}, \citenamefont {{Seljak}},
  \citenamefont {{Kravtsov}},\ and\ \citenamefont
  {{Wechsler}}}]{Mandelbaum2005}%
  \BibitemOpen
  \bibfield  {author} {\bibinfo {author} {\bibfnamefont {R.}~\bibnamefont
  {{Mandelbaum}}}, \bibinfo {author} {\bibfnamefont {A.}~\bibnamefont
  {{Tasitsiomi}}}, \bibinfo {author} {\bibfnamefont {U.}~\bibnamefont
  {{Seljak}}}, \bibinfo {author} {\bibfnamefont {A.~V.}\ \bibnamefont
  {{Kravtsov}}}, \ and\ \bibinfo {author} {\bibfnamefont {R.~H.}\ \bibnamefont
  {{Wechsler}}},\ }\href {\doibase 10.1111/j.1365-2966.2005.09417.x} {\bibfield
   {journal} {\bibinfo  {journal} {\mnras}\ }\textbf {\bibinfo {volume}
  {362}},\ \bibinfo {pages} {1451} (\bibinfo {year} {2005})},\ \Eprint
  {http://arxiv.org/abs/astro-ph/0410711} {astro-ph/0410711} \BibitemShut
  {NoStop}%
\bibitem [{\citenamefont {{van den Bosch}}\ \emph {et~al.}(2013)\citenamefont
  {{van den Bosch}}, \citenamefont {{More}}, \citenamefont {{Cacciato}},
  \citenamefont {{Mo}},\ and\ \citenamefont {{Yang}}}]{vandenBosch2013}%
  \BibitemOpen
  \bibfield  {author} {\bibinfo {author} {\bibfnamefont {F.~C.}\ \bibnamefont
  {{van den Bosch}}}, \bibinfo {author} {\bibfnamefont {S.}~\bibnamefont
  {{More}}}, \bibinfo {author} {\bibfnamefont {M.}~\bibnamefont {{Cacciato}}},
  \bibinfo {author} {\bibfnamefont {H.}~\bibnamefont {{Mo}}}, \ and\ \bibinfo
  {author} {\bibfnamefont {X.}~\bibnamefont {{Yang}}},\ }\href {\doibase
  10.1093/mnras/sts006} {\bibfield  {journal} {\bibinfo  {journal} {\mnras}\
  }\textbf {\bibinfo {volume} {430}},\ \bibinfo {pages} {725} (\bibinfo {year}
  {2013})},\ \Eprint {http://arxiv.org/abs/1206.6890} {arXiv:1206.6890}
  \BibitemShut {NoStop}%
\bibitem [{\citenamefont {{Kwan}}\ \emph {et~al.}(2015)\citenamefont {{Kwan}},
  \citenamefont {{Heitmann}}, \citenamefont {{Habib}}, \citenamefont
  {{Padmanabhan}}, \citenamefont {{Lawrence}}, \citenamefont {{Finkel}},
  \citenamefont {{Frontiere}},\ and\ \citenamefont {{Pope}}}]{Kwan2015}%
  \BibitemOpen
  \bibfield  {author} {\bibinfo {author} {\bibfnamefont {J.}~\bibnamefont
  {{Kwan}}}, \bibinfo {author} {\bibfnamefont {K.}~\bibnamefont {{Heitmann}}},
  \bibinfo {author} {\bibfnamefont {S.}~\bibnamefont {{Habib}}}, \bibinfo
  {author} {\bibfnamefont {N.}~\bibnamefont {{Padmanabhan}}}, \bibinfo {author}
  {\bibfnamefont {E.}~\bibnamefont {{Lawrence}}}, \bibinfo {author}
  {\bibfnamefont {H.}~\bibnamefont {{Finkel}}}, \bibinfo {author}
  {\bibfnamefont {N.}~\bibnamefont {{Frontiere}}}, \ and\ \bibinfo {author}
  {\bibfnamefont {A.}~\bibnamefont {{Pope}}},\ }\href {\doibase
  10.1088/0004-637X/810/1/35} {\bibfield  {journal} {\bibinfo  {journal}
  {\apj}\ }\textbf {\bibinfo {volume} {810}},\ \bibinfo {eid} {35} (\bibinfo
  {year} {2015})},\ \Eprint {http://arxiv.org/abs/1311.6444} {arXiv:1311.6444}
  \BibitemShut {NoStop}%
\bibitem [{\citenamefont {{Villaescusa-Navarro}}\ \emph
  {et~al.}(2018)\citenamefont {{Villaescusa-Navarro}}, \citenamefont {{Genel}},
  \citenamefont {{Castorina}}, \citenamefont {{Obuljen}}, \citenamefont
  {{Spergel}}, \citenamefont {{Hernquist}}, \citenamefont {{Nelson}},
  \citenamefont {{Carucci}}, \citenamefont {{Pillepich}}, \citenamefont
  {{Marinacci}}, \citenamefont {{Diemer}}, \citenamefont {{Vogelsberger}},
  \citenamefont {{Weinberger}},\ and\ \citenamefont
  {{Pakmor}}}]{Villaescusa-Navarro2018}%
  \BibitemOpen
  \bibfield  {author} {\bibinfo {author} {\bibfnamefont {F.}~\bibnamefont
  {{Villaescusa-Navarro}}}, \bibinfo {author} {\bibfnamefont {S.}~\bibnamefont
  {{Genel}}}, \bibinfo {author} {\bibfnamefont {E.}~\bibnamefont
  {{Castorina}}}, \bibinfo {author} {\bibfnamefont {A.}~\bibnamefont
  {{Obuljen}}}, \bibinfo {author} {\bibfnamefont {D.~N.}\ \bibnamefont
  {{Spergel}}}, \bibinfo {author} {\bibfnamefont {L.}~\bibnamefont
  {{Hernquist}}}, \bibinfo {author} {\bibfnamefont {D.}~\bibnamefont
  {{Nelson}}}, \bibinfo {author} {\bibfnamefont {I.~P.}\ \bibnamefont
  {{Carucci}}}, \bibinfo {author} {\bibfnamefont {A.}~\bibnamefont
  {{Pillepich}}}, \bibinfo {author} {\bibfnamefont {F.}~\bibnamefont
  {{Marinacci}}}, \bibinfo {author} {\bibfnamefont {B.}~\bibnamefont
  {{Diemer}}}, \bibinfo {author} {\bibfnamefont {M.}~\bibnamefont
  {{Vogelsberger}}}, \bibinfo {author} {\bibfnamefont {R.}~\bibnamefont
  {{Weinberger}}}, \ and\ \bibinfo {author} {\bibfnamefont {R.}~\bibnamefont
  {{Pakmor}}},\ }\href {\doibase 10.3847/1538-4357/aadba0} {\bibfield
  {journal} {\bibinfo  {journal} {\apj}\ }\textbf {\bibinfo {volume} {866}},\
  \bibinfo {eid} {135} (\bibinfo {year} {2018})},\ \Eprint
  {http://arxiv.org/abs/1804.09180} {arXiv:1804.09180} \BibitemShut {NoStop}%
\bibitem [{\citenamefont {{Barreira}}\ \emph {et~al.}(2019)\citenamefont
  {{Barreira}}, \citenamefont {{Nelson}}, \citenamefont {{Pillepich}},
  \citenamefont {{Springel}}, \citenamefont {{Schmidt}}, \citenamefont
  {{Pakmor}}, \citenamefont {{Hernquist}},\ and\ \citenamefont
  {{Vogelsberger}}}]{Barreira19}%
  \BibitemOpen
  \bibfield  {author} {\bibinfo {author} {\bibfnamefont {A.}~\bibnamefont
  {{Barreira}}}, \bibinfo {author} {\bibfnamefont {D.}~\bibnamefont
  {{Nelson}}}, \bibinfo {author} {\bibfnamefont {A.}~\bibnamefont
  {{Pillepich}}}, \bibinfo {author} {\bibfnamefont {V.}~\bibnamefont
  {{Springel}}}, \bibinfo {author} {\bibfnamefont {F.}~\bibnamefont
  {{Schmidt}}}, \bibinfo {author} {\bibfnamefont {R.}~\bibnamefont {{Pakmor}}},
  \bibinfo {author} {\bibfnamefont {L.}~\bibnamefont {{Hernquist}}}, \ and\
  \bibinfo {author} {\bibfnamefont {M.}~\bibnamefont {{Vogelsberger}}},\
  }\href@noop {} {\bibfield  {journal} {\bibinfo  {journal} {arXiv e-prints}\
  ,\ \bibinfo {eid} {arXiv:1904.02070}} (\bibinfo {year} {2019})},\ \Eprint
  {http://arxiv.org/abs/1904.02070} {arXiv:1904.02070 [astro-ph.CO]}
  \BibitemShut {NoStop}%
\bibitem [{\citenamefont {{Yang}}\ \emph {et~al.}(2013)\citenamefont {{Yang}},
  \citenamefont {{Kratochvil}}, \citenamefont {{Huffenberger}}, \citenamefont
  {{Haiman}},\ and\ \citenamefont {{May}}}]{Yang13}%
  \BibitemOpen
  \bibfield  {author} {\bibinfo {author} {\bibfnamefont {X.}~\bibnamefont
  {{Yang}}}, \bibinfo {author} {\bibfnamefont {J.~M.}\ \bibnamefont
  {{Kratochvil}}}, \bibinfo {author} {\bibfnamefont {K.}~\bibnamefont
  {{Huffenberger}}}, \bibinfo {author} {\bibfnamefont {Z.}~\bibnamefont
  {{Haiman}}}, \ and\ \bibinfo {author} {\bibfnamefont {M.}~\bibnamefont
  {{May}}},\ }\href {\doibase 10.1103/PhysRevD.87.023511} {\bibfield  {journal}
  {\bibinfo  {journal} {\prd}\ }\textbf {\bibinfo {volume} {87}},\ \bibinfo
  {eid} {023511} (\bibinfo {year} {2013})},\ \Eprint
  {http://arxiv.org/abs/1210.0608} {arXiv:1210.0608 [astro-ph.CO]} \BibitemShut
  {NoStop}%
\bibitem [{\citenamefont {{Cui}}\ \emph {et~al.}(2012)\citenamefont {{Cui}},
  \citenamefont {{Borgani}}, \citenamefont {{Dolag}}, \citenamefont
  {{Murante}},\ and\ \citenamefont {{Tornatore}}}]{Cui12}%
  \BibitemOpen
  \bibfield  {author} {\bibinfo {author} {\bibfnamefont {W.}~\bibnamefont
  {{Cui}}}, \bibinfo {author} {\bibfnamefont {S.}~\bibnamefont {{Borgani}}},
  \bibinfo {author} {\bibfnamefont {K.}~\bibnamefont {{Dolag}}}, \bibinfo
  {author} {\bibfnamefont {G.}~\bibnamefont {{Murante}}}, \ and\ \bibinfo
  {author} {\bibfnamefont {L.}~\bibnamefont {{Tornatore}}},\ }\href {\doibase
  10.1111/j.1365-2966.2012.21037.x} {\bibfield  {journal} {\bibinfo  {journal}
  {\mnras}\ }\textbf {\bibinfo {volume} {423}},\ \bibinfo {pages} {2279}
  (\bibinfo {year} {2012})},\ \Eprint {http://arxiv.org/abs/1111.3066}
  {arXiv:1111.3066 [astro-ph.CO]} \BibitemShut {NoStop}%
\bibitem [{\citenamefont {{Paillas}}\ \emph {et~al.}(2017)\citenamefont
  {{Paillas}}, \citenamefont {{Lagos}}, \citenamefont {{Padilla}},
  \citenamefont {{Tissera}}, \citenamefont {{Helly}},\ and\ \citenamefont
  {{Schaller}}}]{Paillas17}%
  \BibitemOpen
  \bibfield  {author} {\bibinfo {author} {\bibfnamefont {E.}~\bibnamefont
  {{Paillas}}}, \bibinfo {author} {\bibfnamefont {C.~D.~P.}\ \bibnamefont
  {{Lagos}}}, \bibinfo {author} {\bibfnamefont {N.}~\bibnamefont {{Padilla}}},
  \bibinfo {author} {\bibfnamefont {P.}~\bibnamefont {{Tissera}}}, \bibinfo
  {author} {\bibfnamefont {J.}~\bibnamefont {{Helly}}}, \ and\ \bibinfo
  {author} {\bibfnamefont {M.}~\bibnamefont {{Schaller}}},\ }\href {\doibase
  10.1093/mnras/stx1514} {\bibfield  {journal} {\bibinfo  {journal} {\mnras}\
  }\textbf {\bibinfo {volume} {470}},\ \bibinfo {pages} {4434} (\bibinfo {year}
  {2017})},\ \Eprint {http://arxiv.org/abs/1609.00101} {arXiv:1609.00101
  [astro-ph.CO]} \BibitemShut {NoStop}%
\bibitem [{\citenamefont {{Weiss}}\ \emph {et~al.}(2019)\citenamefont
  {{Weiss}}, \citenamefont {{Schneider}}, \citenamefont {{Sgier}},
  \citenamefont {{Kacprzak}}, \citenamefont {{Amara}},\ and\ \citenamefont
  {{Refregier}}}]{Weiss19}%
  \BibitemOpen
  \bibfield  {author} {\bibinfo {author} {\bibfnamefont {A.~J.}\ \bibnamefont
  {{Weiss}}}, \bibinfo {author} {\bibfnamefont {A.}~\bibnamefont
  {{Schneider}}}, \bibinfo {author} {\bibfnamefont {R.}~\bibnamefont
  {{Sgier}}}, \bibinfo {author} {\bibfnamefont {T.}~\bibnamefont {{Kacprzak}}},
  \bibinfo {author} {\bibfnamefont {A.}~\bibnamefont {{Amara}}}, \ and\
  \bibinfo {author} {\bibfnamefont {A.}~\bibnamefont {{Refregier}}},\
  }\href@noop {} {\bibfield  {journal} {\bibinfo  {journal} {arXiv e-prints}\
  ,\ \bibinfo {eid} {arXiv:1905.11636}} (\bibinfo {year} {2019})},\ \Eprint
  {http://arxiv.org/abs/1905.11636} {arXiv:1905.11636 [astro-ph.CO]}
  \BibitemShut {NoStop}%
\bibitem [{\citenamefont {{Fluri}}\ \emph {et~al.}(2019)\citenamefont
  {{Fluri}}, \citenamefont {{Kacprzak}}, \citenamefont {{Lucchi}},
  \citenamefont {{Refregier}}, \citenamefont {{Amara}}, \citenamefont
  {{Hofmann}},\ and\ \citenamefont {{Schneider}}}]{Fluri2019}%
  \BibitemOpen
  \bibfield  {author} {\bibinfo {author} {\bibfnamefont {J.}~\bibnamefont
  {{Fluri}}}, \bibinfo {author} {\bibfnamefont {T.}~\bibnamefont {{Kacprzak}}},
  \bibinfo {author} {\bibfnamefont {A.}~\bibnamefont {{Lucchi}}}, \bibinfo
  {author} {\bibfnamefont {A.}~\bibnamefont {{Refregier}}}, \bibinfo {author}
  {\bibfnamefont {A.}~\bibnamefont {{Amara}}}, \bibinfo {author} {\bibfnamefont
  {T.}~\bibnamefont {{Hofmann}}}, \ and\ \bibinfo {author} {\bibfnamefont
  {A.}~\bibnamefont {{Schneider}}},\ }\href@noop {} {\bibfield  {journal}
  {\bibinfo  {journal} {arXiv e-prints}\ ,\ \bibinfo {eid} {arXiv:1906.03156}}
  (\bibinfo {year} {2019})},\ \Eprint {http://arxiv.org/abs/1906.03156}
  {arXiv:1906.03156 [astro-ph.CO]} \BibitemShut {NoStop}%
\bibitem [{\citenamefont {{Rodr{\'\i}guez}}\ \emph {et~al.}(2018)\citenamefont
  {{Rodr{\'\i}guez}}, \citenamefont {{Kacprzak}}, \citenamefont {{Lucchi}},
  \citenamefont {{Amara}}, \citenamefont {{Sgier}}, \citenamefont {{Fluri}},
  \citenamefont {{Hofmann}},\ and\ \citenamefont
  {{R{\'e}fr{\'e}gier}}}]{Rodriguez2018}%
  \BibitemOpen
  \bibfield  {author} {\bibinfo {author} {\bibfnamefont {A.~C.}\ \bibnamefont
  {{Rodr{\'\i}guez}}}, \bibinfo {author} {\bibfnamefont {T.}~\bibnamefont
  {{Kacprzak}}}, \bibinfo {author} {\bibfnamefont {A.}~\bibnamefont
  {{Lucchi}}}, \bibinfo {author} {\bibfnamefont {A.}~\bibnamefont {{Amara}}},
  \bibinfo {author} {\bibfnamefont {R.}~\bibnamefont {{Sgier}}}, \bibinfo
  {author} {\bibfnamefont {J.}~\bibnamefont {{Fluri}}}, \bibinfo {author}
  {\bibfnamefont {T.}~\bibnamefont {{Hofmann}}}, \ and\ \bibinfo {author}
  {\bibfnamefont {A.}~\bibnamefont {{R{\'e}fr{\'e}gier}}},\ }\href@noop {}
  {\bibfield  {journal} {\bibinfo  {journal} {Computational Astrophysics and
  Cosmology}\ }\textbf {\bibinfo {volume} {5}},\ \bibinfo {pages} {4} (\bibinfo
  {year} {2018})}\BibitemShut {NoStop}%
\bibitem [{\citenamefont {{He}}\ \emph {et~al.}(2018)\citenamefont {{He}},
  \citenamefont {{Li}}, \citenamefont {{Feng}}, \citenamefont {{Ho}},
  \citenamefont {{Ravanbakhsh}}, \citenamefont {{Chen}},\ and\ \citenamefont
  {{P{\'o}czos}}}]{He2018}%
  \BibitemOpen
  \bibfield  {author} {\bibinfo {author} {\bibfnamefont {S.}~\bibnamefont
  {{He}}}, \bibinfo {author} {\bibfnamefont {Y.}~\bibnamefont {{Li}}}, \bibinfo
  {author} {\bibfnamefont {Y.}~\bibnamefont {{Feng}}}, \bibinfo {author}
  {\bibfnamefont {S.}~\bibnamefont {{Ho}}}, \bibinfo {author} {\bibfnamefont
  {S.}~\bibnamefont {{Ravanbakhsh}}}, \bibinfo {author} {\bibfnamefont
  {W.}~\bibnamefont {{Chen}}}, \ and\ \bibinfo {author} {\bibfnamefont
  {B.}~\bibnamefont {{P{\'o}czos}}},\ }\href@noop {} {\bibfield  {journal}
  {\bibinfo  {journal} {arXiv e-prints}\ ,\ \bibinfo {eid} {arXiv:1811.06533}}
  (\bibinfo {year} {2018})},\ \Eprint {http://arxiv.org/abs/1811.06533}
  {arXiv:1811.06533 [astro-ph.CO]} \BibitemShut {NoStop}%
\bibitem [{\citenamefont {{Zhang}}\ \emph {et~al.}(2019)\citenamefont
  {{Zhang}}, \citenamefont {{Wang}}, \citenamefont {{Zhang}}, \citenamefont
  {{Sun}}, \citenamefont {{He}}, \citenamefont {{Contardo}}, \citenamefont
  {{Villaescusa-Navarro}},\ and\ \citenamefont {{Ho}}}]{Zhang19}%
  \BibitemOpen
  \bibfield  {author} {\bibinfo {author} {\bibfnamefont {X.}~\bibnamefont
  {{Zhang}}}, \bibinfo {author} {\bibfnamefont {Y.}~\bibnamefont {{Wang}}},
  \bibinfo {author} {\bibfnamefont {W.}~\bibnamefont {{Zhang}}}, \bibinfo
  {author} {\bibfnamefont {Y.}~\bibnamefont {{Sun}}}, \bibinfo {author}
  {\bibfnamefont {S.}~\bibnamefont {{He}}}, \bibinfo {author} {\bibfnamefont
  {G.}~\bibnamefont {{Contardo}}}, \bibinfo {author} {\bibfnamefont
  {F.}~\bibnamefont {{Villaescusa-Navarro}}}, \ and\ \bibinfo {author}
  {\bibfnamefont {S.}~\bibnamefont {{Ho}}},\ }\href@noop {} {\bibfield
  {journal} {\bibinfo  {journal} {arXiv e-prints}\ ,\ \bibinfo {eid}
  {arXiv:1902.05965}} (\bibinfo {year} {2019})},\ \Eprint
  {http://arxiv.org/abs/1902.05965} {arXiv:1902.05965 [astro-ph.CO]}
  \BibitemShut {NoStop}%
\bibitem [{\citenamefont {{Tr{\"o}ster}}\ \emph {et~al.}(2019)\citenamefont
  {{Tr{\"o}ster}}, \citenamefont {{Ferguson}}, \citenamefont
  {{Harnois-D{\'e}raps}},\ and\ \citenamefont {{McCarthy}}}]{Troester2019}%
  \BibitemOpen
  \bibfield  {author} {\bibinfo {author} {\bibfnamefont {T.}~\bibnamefont
  {{Tr{\"o}ster}}}, \bibinfo {author} {\bibfnamefont {C.}~\bibnamefont
  {{Ferguson}}}, \bibinfo {author} {\bibfnamefont {J.}~\bibnamefont
  {{Harnois-D{\'e}raps}}}, \ and\ \bibinfo {author} {\bibfnamefont {I.~G.}\
  \bibnamefont {{McCarthy}}},\ }\href@noop {} {\bibfield  {journal} {\bibinfo
  {journal} {ArXiv e-prints}\ ,\ \bibinfo {eid} {arXiv:1903.12173}} (\bibinfo
  {year} {2019})},\ \Eprint {http://arxiv.org/abs/1903.12173} {arXiv:1903.12173
  [astro-ph.CO]} \BibitemShut {NoStop}%
\bibitem [{\citenamefont {{Knabenhans}}\ \emph {et~al.}(2019)\citenamefont
  {{Knabenhans}}, \citenamefont {{Stadel}}, \citenamefont {{Marelli}},
  \citenamefont {{Potter}}, \citenamefont {{Teyssier}}, \citenamefont
  {{Legrand}}, \citenamefont {{Schneider}}, \citenamefont {{Sudret}},
  \citenamefont {{Blot}}, \citenamefont {{Awan}}, \citenamefont {{Burigana}},
  \citenamefont {{Carvalho}}, \citenamefont {{Kurki-Suonio}},\ and\
  \citenamefont {{Sirri}}}]{Knabenhans19}%
  \BibitemOpen
  \bibfield  {author} {\bibinfo {author} {\bibfnamefont {M.}~\bibnamefont
  {{Knabenhans}}}, \bibinfo {author} {\bibfnamefont {J.}~\bibnamefont
  {{Stadel}}}, \bibinfo {author} {\bibfnamefont {S.}~\bibnamefont {{Marelli}}},
  \bibinfo {author} {\bibfnamefont {D.}~\bibnamefont {{Potter}}}, \bibinfo
  {author} {\bibfnamefont {R.}~\bibnamefont {{Teyssier}}}, \bibinfo {author}
  {\bibfnamefont {L.}~\bibnamefont {{Legrand}}}, \bibinfo {author}
  {\bibfnamefont {A.}~\bibnamefont {{Schneider}}}, \bibinfo {author}
  {\bibfnamefont {B.}~\bibnamefont {{Sudret}}}, \bibinfo {author}
  {\bibfnamefont {L.}~\bibnamefont {{Blot}}}, \bibinfo {author} {\bibfnamefont
  {S.}~\bibnamefont {{Awan}}}, \bibinfo {author} {\bibfnamefont
  {C.}~\bibnamefont {{Burigana}}}, \bibinfo {author} {\bibfnamefont {C.~S.}\
  \bibnamefont {{Carvalho}}}, \bibinfo {author} {\bibfnamefont
  {H.}~\bibnamefont {{Kurki-Suonio}}}, \ and\ \bibinfo {author} {\bibfnamefont
  {G.}~\bibnamefont {{Sirri}}},\ }\href {\doibase 10.1093/mnras/stz197}
  {\bibfield  {journal} {\bibinfo  {journal} {\mnras}\ }\textbf {\bibinfo
  {volume} {484}},\ \bibinfo {pages} {5509} (\bibinfo {year} {2019})},\ \Eprint
  {http://arxiv.org/abs/1809.04695} {arXiv:1809.04695} \BibitemShut {NoStop}%
\end{thebibliography}%


\end{document}